%% file: ms.tex
\documentclass[12pt,preprint]{aastex}

\newcommand{\um}{$\mu$m}

\newcommand{\Msun}{\,$M_{\odot}$}
\newcommand{\Lsun}{\,$L_{\odot}$}
\newcommand{\hii}{\mbox{$\mathrm{H\,{\scriptstyle {II}}}$}}

\newcommand{\tcol}{\mbox{$^{13}${\rmfamily CO}{(1--0)}}}
\newcommand{\MSX}{{\it MSX}}
\newcommand{\Spitzer}{{\it Spitzer}}
\newcommand{\ISO}{{\it ISO}}
\newcommand{\IRAS}{{\it IRAS}}

\newcommand{\cms}{${\rm cm}^{-2}$}
\newcommand{\cmc}{${\rm cm}^{-3}$}
\newcommand{\hh}{H$_{2}$}

\newcommand{\irdcone}{G028.37+00.07}
\newcommand{\irdctwo}{G028.23$-$00.19}
\newcommand{\irdcthree}{G035.39$-$00.33}
\newcommand{\irdcfour}{G031.97+00.07}
\newcommand{\irdcfive}{G038.95$-$0.47}
\newcommand{\irdcsix}{G022.35+00.41}
\newcommand{\irdcseven}{G019.27+00.07}
\newcommand{\irdceight}{G035.59-00.24}
\newcommand{\irdcnine}{G024.60+00.08}
\newcommand{\irdcten}{G027.97$-$00.42}
\newcommand{\irdctwelve}{G028.08+00.07}
\newcommand{\irdcforteen}{G025.04$-$00.20}
\newcommand{\irdcfifteen}{G015.31$-$00.16}
\newcommand{\irdcsixteen}{G018.82$-$00.28}
\newcommand{\irdceighteen}{G053.11+00.05}
\newcommand{\irdcnineteen}{G028.67+00.13}
\newcommand{\irdctwentyone}{G034.77$-$00.55}
\newcommand{\irdctwentysix}{G028.28$-$00.34}
\newcommand{\irdctwentyseven}{G015.05+00.07}
\newcommand{\irdctwentynine}{G030.97$-$00.14}
\newcommand{\irdctwentynineb}{G031.02$-$00.10}
\newcommand{\irdcthirty}{G028.53$-$00.25}
\newcommand{\irdcthirtyone}{G036.67$-$00.11}
\newcommand{\irdcthirtythree}{G023.60+00.00}
\newcommand{\irdcthirtyfour}{G053.25+00.04}
\newcommand{\irdcthirtynine}{G033.69$-$00.01}
\newcommand{\irdcfortytwoone}{G028.04$-$00.46}
\newcommand{\irdcfortytwotwo}{G028.10$-$00.45}
\newcommand{\irdcfortythree}{G034.43+00.24}
\newcommand{\irdcfortyfour}{G048.65$-$00.29}
\newcommand{\irdcfortyfive}{G024.08+00.04}
\newcommand{\irdcfortysix}{G027.94$-$00.47}
\newcommand{\irdcfortynine}{G022.73+00.11}
\newcommand{\irdcfiftyone}{G024.33+00.11}
\newcommand{\irdcfiftythree}{G027.75+00.16}
\newcommand{\irdcfiftyfour}{G030.57$-$00.23}
\newcommand{\irdcfiftyfive}{G053.31+00.00}
\newcommand{\irdcfiftyseven}{G030.14$-$00.06}
\newcommand{\av}{A$_{V}$}
\newcommand{\MJysr}{MJy\,sr$^{-1}$}

\slugcomment{}

\shorttitle{Cores within IRDCs}
\shortauthors{Rathborne et al.}

\begin{document}

\title{The early stages of star formation in Infrared Dark Clouds: characterizing the core dust properties}
\author{J. M. Rathborne\footnote{Current address: Departamento de Astronom\'ia, Universidad de Chile, Santiago, Chile}, J. M. Jackson, E. T. Chambers\footnote{Current address: Department of Physics and Astronomy, Northwestern University, Evanston, IL 60208, U.S.A}, I. Stojimirovic}
\affil{Institute for Astrophysical Research, Boston University, Boston, MA 02215, U.S.A; rathborn@bu.edu, jackson@bu.edu, etc1@bu.edu, irena@bu.edu}
\and
\author{R. Simon}
\affil{I.Physikalisches Institut, Universit\"at zu K\"oln, 50937 K\"oln, Germany; simonr@ph1.uni-koeln.de} 
\and
\author{R. Shipman and W. Frieswijk}
\affil{Kapteyn Astronomical Institute, University of Groningen, and Netherlands Institute for Space Research, P.O. Box 800, 9700 AV Groningen, Netherlands; russ@sron.rug.nl, frieswyk@astro.rug.nl}
\begin{abstract}
Identified as extinction features against the bright Galactic mid-infrared background, infrared 
dark clouds (IRDCs) are thought to harbor the very earliest stages of star and cluster formation. 
In order to better characterize the properties of their embedded cores, we have obtained 
new 24\,\um, 60--100\,\um, and sub-millimeter continuum data toward a sample of 38 
IRDCs. The 24\,\um\, \Spitzer\, images reveal that while the IRDCs remain dark, many of 
the cores are associated with bright 24\,\um\, emission sources, which suggests that they 
contain one or more embedded protostars. Combining the 24\,\um, 60--100\,\um, and sub-millimeter 
continuum data, we have constructed broadband spectral energy distributions (SEDs) for 157 of 
the cores within these IRDCs and, using simple gray-body fits to the SEDs, have estimated their 
dust temperatures, emissivities, opacities, bolometric luminosities, masses and densities. Based on 
their \Spitzer/IRAC 3--8\,\um\, colors and the presence of 24\,\um\, point source emission, we have 
separated cores that harbor active, high-mass star formation from cores 
that are quiescent. The active `protostellar' cores  typically have warmer dust 
temperatures and higher bolometric luminosities than the more quiescent, perhaps `pre-protostellar',
cores. Because the mass distributions of the populations are similar, however, we speculate that 
the active and quiescent cores may represent different evolutionary stages of the same underlying 
population of cores. Although we cannot rule out low-mass star-formation in the quiescent cores, 
the most massive of them are excellent candidates for the 
`high-mass starless core' phase,  the very earliest in the formation of a high-mass star.
\end{abstract}
\section{Introduction}

The earliest phase of isolated low-mass star-formation occurs within Bok Globules. Viewed 
against background stars, Bok globules are identified as isolated, well-defined patches of 
optical obscuration and have typical visual extinctions, \av, of 1--25 mag \citep{Bok47}. 
The individual precursor to a low-mass star, referred to as the `pre-protostellar core', are found 
within Bok Globules. These pre-protostellar cores are typically small ($\sim$0.05\,pc) and 
dense (10$^{5}$--10$^{6}$\,\cmc), with low temperatures ($\sim$10\,K) and low masses 
(0.5--5\,\Msun; e.g., \citealp{Myers83,Ward-Thompson94}).

Because many low-mass star-forming regions are nearby, their pre-protostellar cores and 
protostars have been studied extensively. Moreover, these studies have the added benefit 
of achieving sufficiently high spatial resolutions to distinguish the individual protostars, allowing
one to study and characterize their various evolutionary stages, i.e., Class-0, I, II, and III 
\citep{Lada84,Adams87,Andre94}. These stages are characterized by IR spectral energy 
distributions (SEDs) corresponding to increasing black-body temperatures. Moreover, differences
 in the shapes of their SEDs trace the gradual removal of circumstellar material surrounding the central
protostar as it evolves and emerges from its natal core.

While these early evolutionary stages are well-characterized for low-mass star-formation, it
has been difficult to classify high-mass protostars in a similar manner. The combination of 
large distances to high-mass star-forming regions, their rarity and rapid evolution, and
the fact that most high-mass stars form deeply embedded in dense molecular clumps and within
clusters with many lower-mass stars nearby, makes their identification and separation
difficult. However,  in a recent study combining \MSX, \IRAS, and sub-millimeter data 
toward 42 regions of high-mass star formation, \citet{Molinari08} have investigated the 
evolution of SEDs for young, high-mass protostars and has attempted to classify them via 
their SEDs. They find that objects in apparently different evolutionary stages occupy different 
areas in a bolometric  luminosity versus envelope mass diagram, in a similar manner to the 
low-mass regime and, thus, conclude that high-mass star-formation may be a scaled up 
analog to low-mass star-formation.  If this is the case, then perhaps the early stages of high-mass 
star-formation can also be characterized via differences in the SEDs of their dense molecular cores.

In order to better characterize the properties of high-mass pre-protostellar and protostellar 
cores, a large sample of cores in the very earliest stages of high-mass star-formation is required.
High-mass stars and  clusters form from cold, dense molecular clumps within giant molecular 
clouds \citep{Blitz91,Blitz99,Lada03}. Recent studies suggest that the cold precursors of warm
cluster-forming molecular clumps can be identified as `infrared dark clouds' 
(IRDCs; \citealp{Simon-msxgrs,Rathborne06}). Surveys of the Galactic Plane at mid-IR wavelengths 
made with the \ISO\, and \MSX\, satellites first identified IRDCs as dark extinction features seen in
absorption against the bright mid-IR emission arising from the Galactic background 
\citep{Perault96,Carey98,Hennebelle01}.  IRDCs are ubiquitous across the Galaxy 
\citep{Simon-catalog} and are typically long and very filamentary. They correspond to the densest
parts of much larger giant molecular clouds \citep{Simon-msxgrs} and are characterized 
by high densities ($>10^{5}$\,cm$^{-3}$), high column densities 
($\sim$10$^{23}$--10$^{25}$\,cm$^{-2}$), and low temperatures 
($<25$\,K; \citealp{Egan98,Carey98,Carey00}).

Recently, \cite{Rathborne06} conducted a survey of the 1.2\,mm continuum emission
toward 38 IRDCs using MAMBO-II on the IRAM 30\,m telescope. These IRDCs were selected 
from the catalog of \MSX\, IRDC candidates \citep{Simon-catalog} and all have known 
kinematic distances, determined via the morphological match of \tcol\, emission to the 
mid-IR extinction \citep{Simon-msxgrs,Jackson06}. In these IRDCs \cite{Rathborne06} 
found 190 cores, 140 of which are cold, compact cores. These cold, compact cores have 
typical sizes of $<$~0.5~pc and masses of $\sim$~120\,\Msun\, \citep{Rathborne06}.  
Indeed, millimeter and sub-millimeter observations toward other IRDCs suggest such compact
cores are ubiquitous within IRDCs (e.g., \citealp{Lis94,Carey00,Garay04,Ormel05,Beuther05}).  
Because IRDCs are cold, their thermal dust emission will peak in the far-IR/sub-millimeter 
regime. At these wavelengths the dust emission is optically thin, making this regime the 
best for probing their internal structure and revealing their star-forming cores. If IRDCs are 
the high-mass analogue to Bok globules and the cold precursors to cluster-forming 
molecular clumps, then these dense cores may be the precursors to the stars
\citep{Rathborne06}.

IRAM and JCMT molecular line spectra, \Spitzer\, 3--8\,\um\, and 
24\,\um\, continuum images, and GBT water and methanol maser spectra toward a sample of 
cores within IRDCs reveal that most contain little evidence for active star-formation, such cores
are called `quiescent' \citep{Chambers09}. However, some of the IRDC
cores do appear to be actively forming stars as they show broad molecular line emission, 
shocked gas, bright 24\,\um\, emission, and strong water and methanol maser emission 
(e.g., \citealp{Rathborne05,Wang06,Chambers09}). Each of these tracers provides 
independent evidence for star-formation, either indirectly (from the interaction between 
the protostar and the surrounding core traced through the broad line emission and the 
shocked gas) or more directly (from the heating of the dust surrounding the central protostar 
traced via the bright 24\,\um\, emission). It is likely, therefore, that these particular cores 
contain protostars. Indeed, a number of protostars have already been identified within 
IRDCs and span a range in mass, from low- and intermediate-mass \citep{Carey00,Redman03} 
to high-mass \citep{Beuther05,Rathborne05,Pillai06,Wang06}.

In order to characterize the cores within IRDCs, we 
have conducted a large, multi-wavelength observational survey, combining IR, sub-millimeter, 
and millimeter continuum data. These data, when combined to make broadband SEDs,
provide estimates of the core dust temperatures, dust emissivities, opacities, bolometric 
luminosities, and masses. Because the cores have SEDs that peak in the far-IR, to date, it 
has been very difficult to estimate bolometric luminosities of the individual protostars due 
to the uncertain extrapolation from much longer and/or shorter wavelengths. Indeed, many 
previous studies using \MSX\, and \IRAS\, fluxes have been forced to make assumptions 
about the relative contributions to the far-IR flux from individual objects within a star-forming
region (e.g. \citealp{Molinari08}). Thus, the inclusion of the sub-millimeter and far-IR data is 
critical to determine reliable SEDs for the cores and to better constrain these parameters. In this 
paper we provide the first mid-IR, far-IR, sub-millimeter, and millimeter  SEDs for a large sample 
of cores within IRDCs.


\section{Observations}

The source list for these observations comprise 38 IRDCs and the 190 cores identified within 
them \citep{Rathborne06}. To characterize the emission from these IRDCs and their cores, 
we have obtained continuum data at many wavelengths: 24\,\um, 60--100\,\um, 350\,\um, 
450\,\um, 850\,\um, and 1.2\,mm\footnote{For details of the 1.2\,mm continuum emission 
data see \cite{Rathborne06}.}. Table~\ref{obs-summary} gives a summary of the telescopes,
instruments, wavelength range, observing dates, angular resolution and 1 $\sigma$ noise for 
the data.

\subsection{24\,\um\, continuum images}

The 24\,\um\, continuum images were obtained using the Multi-band Imaging Photometer 
\citep[MIPS;][]{Rieke04} array on-board the {\it Spitzer Space Telescope}.  Images toward 
30 of the IRDCs were obtained as part of a \Spitzer\, cycle 1 General Observer (GO) 
proposal. The images were obtained in the raster-scanning mode during the periods 
2004 October 16--19 and 2005 April 7--13.  Because the IRDCs have complex and 
extended morphologies, the map sizes varied to cover the extent of the extinction at 
8\,\um. For 25 of the IRDCs, a 3 column $\times$ 3 row (13\arcmin $\times$13\arcmin) 
map was sufficient to cover the 8\,\um\, extinction. For four of the remaining IRDCs we 
used a 3 column $\times$ 5 row (13\arcmin $\times$19\arcmin) map, while toward the 
remaining IRDC, we used a 5 column $\times$ 5 row (19\arcmin $\times$ 19\arcmin) 
map to image the full extent of the 8\,\um\, extinction. In all cases, the raster map was stepped by 
half the array between consecutive scans. Three repeats of the map ($\sim$3 sec exposure 
time per point) were combined to produce the final image which achieved a 1 $\sigma$ 
point source sensitivity of 124\,$\mu$Jy and an extended source sensitivity of 0.13\,\MJysr\, 
(calculated using the on-line sensitivity calculator).

For the 8 IRDCs that were not part of our GO \Spitzer\, proposal, we use 24\um\, images 
from MIPSGAL (see \citealp{Carey05} for a description of the survey details). These images 
were obtained by mapping large portions of the Galactic Plane using the `fast' scanning
mode. The resulting mosaics have a 1 $\sigma$ point source sensitivity of  207\,$\mu$Jy 
and an extended source sensitivity of 0.21\,\MJysr.

All data were processed using the S13.2.0 version of the MIPS reduction pipeline. 

\subsection{60--100\,\um\, continuum spectra}

We used the SED mode of \Spitzer/MIPS to obtain long-slit, low-resolution (R$\sim$15--25) 
spectra toward a sample of these IRDC cores. Due to the saturation limits of MIPS toward the 
Galactic plane, our observations were limited to include only those cores with 1.2\,mm fluxes 
$<$\,2\,Jy.  In total, we obtained a far-IR continuum spectrum  toward 72 cores. For 65 of the 
cores, 4 repeats of a 10\,sec integration were combined to produce the final spectrum which 
achieved a 1$\sigma$ sensitivity of $\sim$ 30, 70, and 150 mJy at 60, 75, and 90\,\um\, 
respectively.  For the remaining 7 bright cores (those with 1.2\,mm  fluxes between 1--2\,Jy), 
4 repeats of a 3\,sec exposure were combined. All spectra were obtained in the pointed 
observation mode. At these wavelengths, \Spitzer\, has an angular resolution of 
$\sim$13--24\,\arcsec\, (9.8\,\arcsec\,pixels).

Because of the contamination by the second order diffracted light and an inoperative 
detector module, the wavelength coverage of the spectra was restricted to 65--97\,\um. 
The spectra were acquired in two epochs: 2006 October and 2007 May. Data from 2006 
were processed using the S14.4.0 pipeline version, while 2007 data were reduced using 
version S16.1.0. The pointed SED-mode observation provides a set of six pairs of data frames 
between the target position (`on') and nearby sky position (`off'). For all analysis, we use the 
pipeline produced post-basic calibrated data (post-BCDs), which deliver mosaic images 
for the `on' and `off' spectra.

We set the scan mirror to chop between the `on' and `off' positions with a chop throw of 
1\arcmin\, to 3\arcmin. The `off' position was selected for each individual core to be 
nearby and free from 1.2 mm continuum emission. However, because the chop distance 
and visibility of the cores were limited, many of the `off' positions fall within the 8\,\um\, 
extinction associated with the IRDC.  Although the cores dominate, we found that many 
of the `off' positions were not completely free of far-IR emission. Thus, we could not use
the pipeline-produced `on-off' spectra due to the `off' contamination. Instead, we generated
our own background and sky-subtracted spectrum using the local background immediately
surrounding the core.

The spectra were flux calibrated using data obtained from multiple observations of three
bright stars, with 70\,\um\, flux densities of 13--19\,Jy. For point sources, the fluxes 
are accurate to within $\sim$ 20\%. The typical 1 $\sigma$ noise in these data is $\sim$
100 mJy beam$^{-1}$.

\subsection{350\,\um, 450\,\um, and 850\,\um\, continuum images}

The sub-millimeter continuum images were obtained using both the Caltech 
Submillimeter Observatory (CSO) and the James Clerk Maxwell Telescope (JCMT) 
via either a `scan-mapping' or `on-the-fly' mode. Due to weather and time constraints,
not all 38 IRDCs were observed at all sub-millimeter wavelengths: in total, 37 of the 
IRDCs were observed at 350\,\um, 29 at 450\,\um\, and 8 at 850\,\um.  All 350\,\um\,
continuum images were obtained with the CSO and all 850\,\um\, continuum images 
were obtained with the JCMT. The 450\,\um\, continuum images were obtained at 
either the CSO or the JCMT.

The CSO data were obtained over three observing runs, 2004 September 5 and 9, 
2005 April 12--15, and 2006 April 25--28 while the JCMT data were obtained on 
2004 September 12--13. During the CSO observations, the $\tau_{\mathrm{225GHz}}$ 
measured from the CSO sky dipper was $\sim$ 0.07 during the 2005 and 2006 April 
observing periods and slightly higher, 0.09, for 2004 September. The measured 1 $\sigma$
noise in the images is $\sim$ 0.2\,Jy beam$^{-1}$. The JCMT data were obtained in 
better conditions which resulted in a 1 $\sigma$ noise of $\sim$0.06\,Jy beam$^{-1}$.

Where possible, the size of the individual maps were selected to cover the majority of 
the extinction feature and the 1.2\,mm\, continuum emission associated with the IRDC 
and ranged from 2\arcmin$\times$2\arcmin\, to 8\arcmin$\times$8\arcmin. Standard 
reduction methods within the software packages CRUSH (for the CSO data) and SURF 
(for the JCMT data) were used to correct for atmospheric opacity and to remove atmospheric
fluctuations. The data were flux-calibrated using either G34.3 or Uranus. Due to the 
uncertainties in accurate flux-calibration of sub-millimeter data, the calibration of the 
fluxes quoted here have errors of $\sim$ 40\%.

\section{Results}

\subsection{Continuum images}

\begin{figure}
\begin{center}
\includegraphics[width=0.46\textwidth,clip=true]{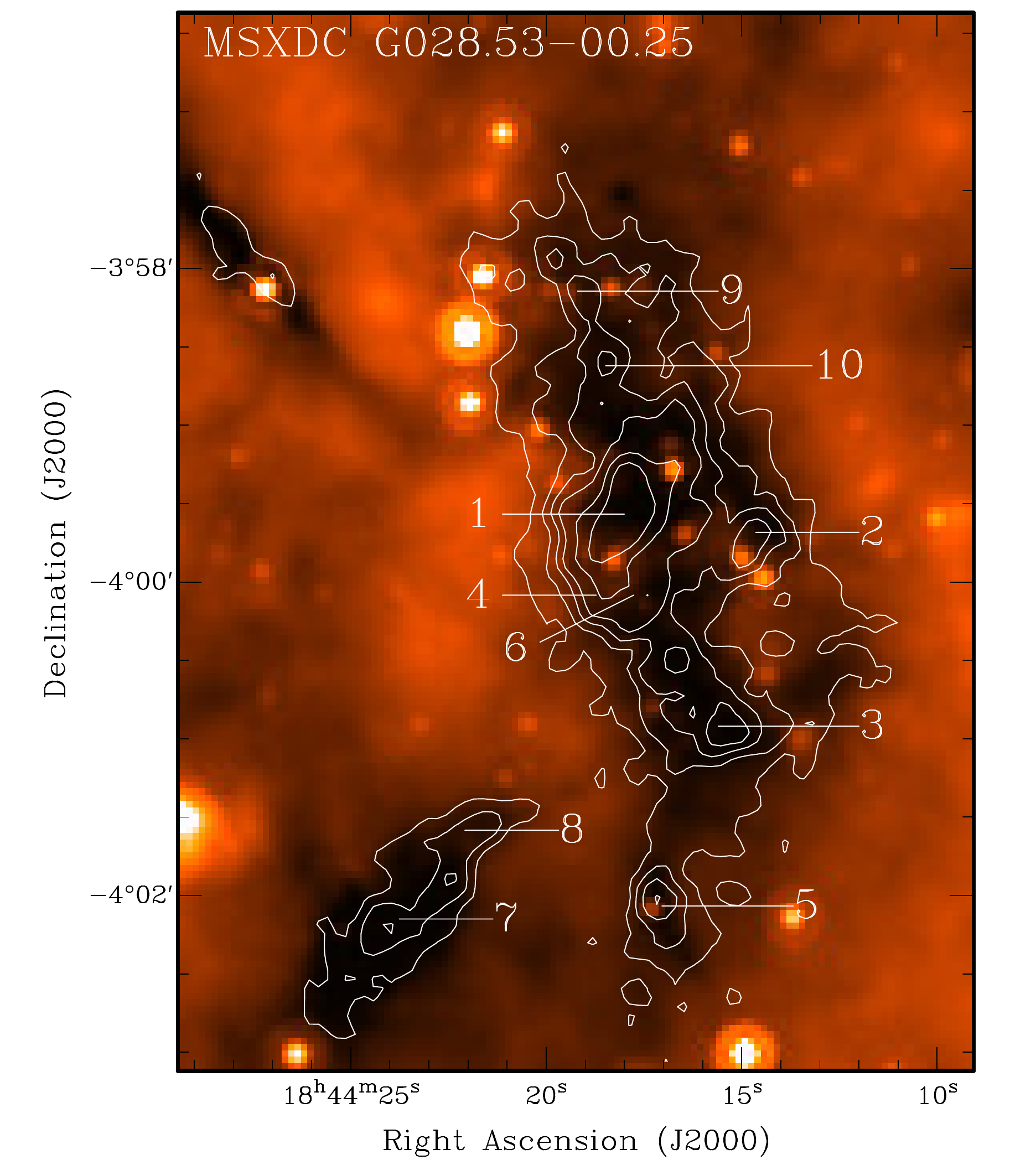}
\hspace{-0.5cm}
\includegraphics[width=0.5\textwidth,clip=true]{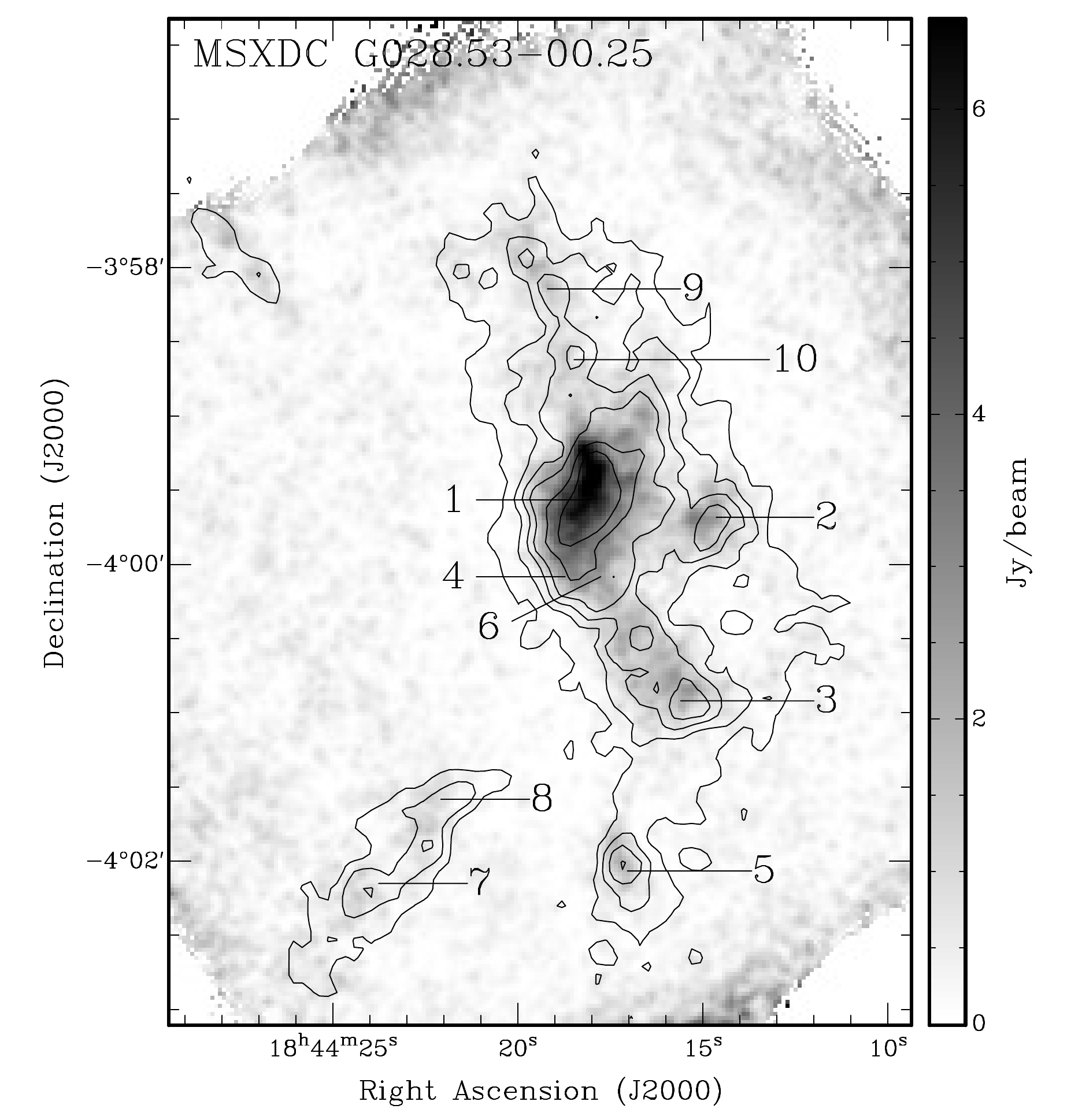}\\
\hspace{-0.5cm}
\includegraphics[width=0.5\textwidth,clip=true]{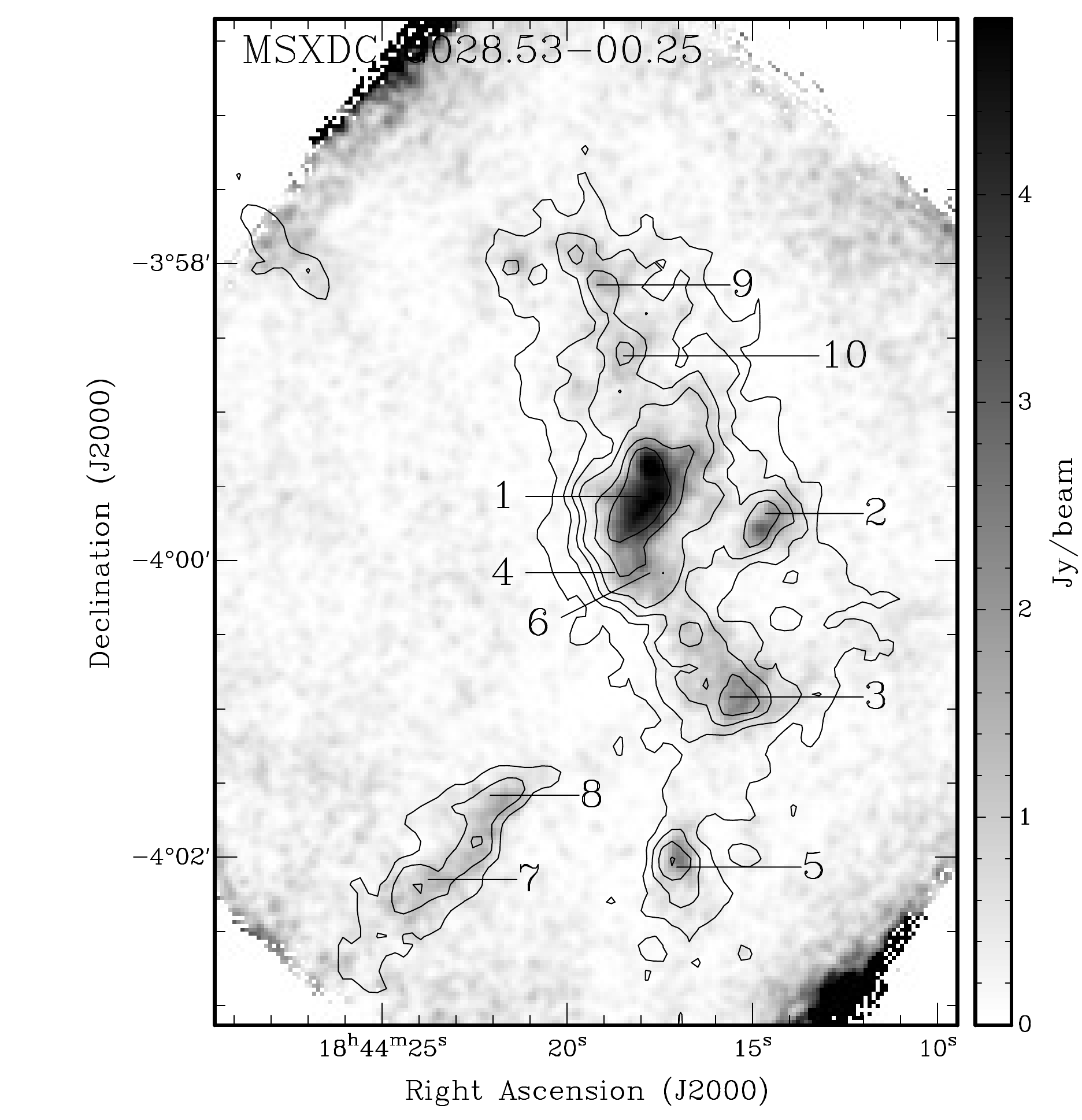}
\hspace{-0.3cm}
\includegraphics[width=0.5\textwidth,clip=true]{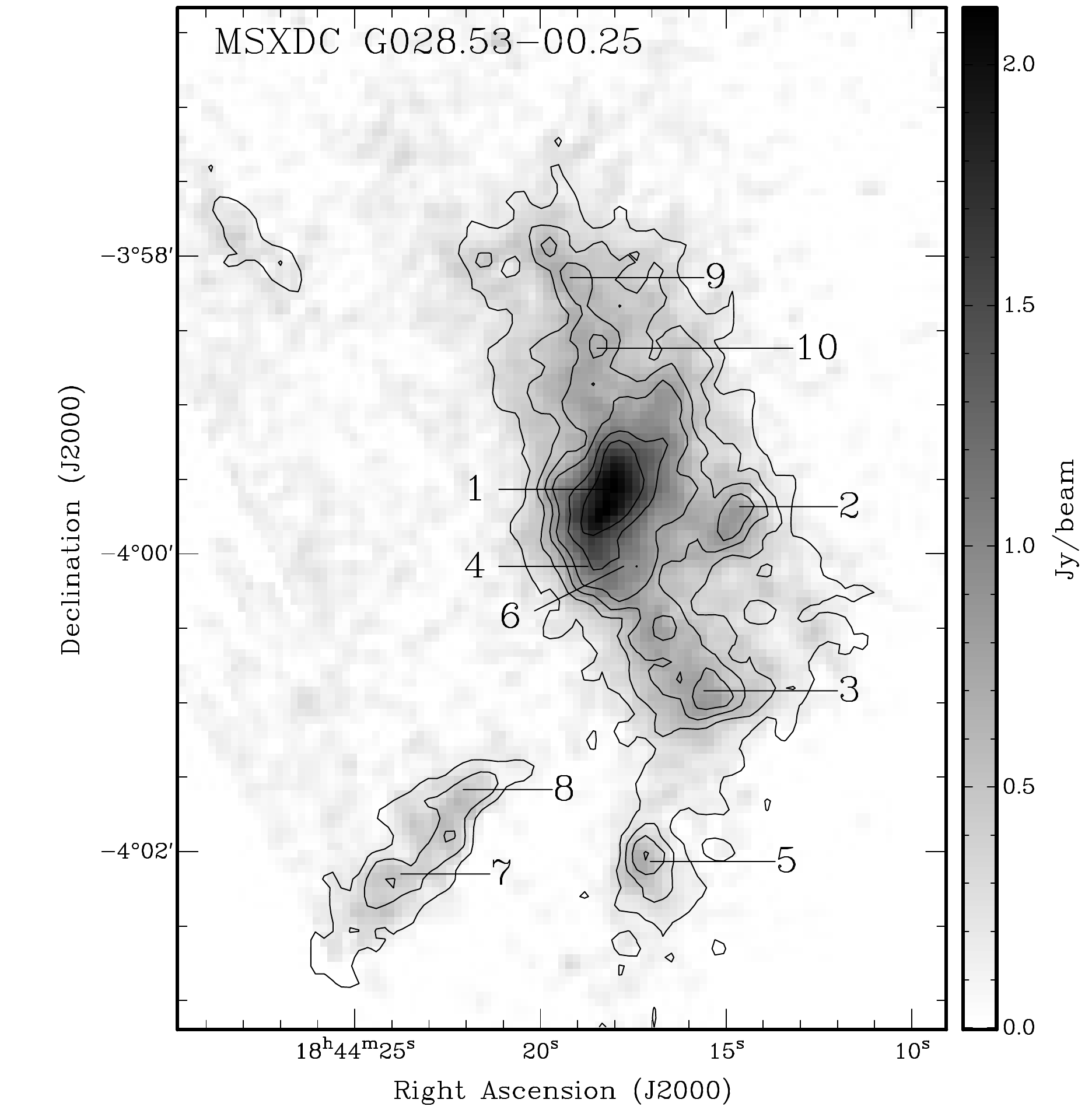}
\end{center}
\caption{\label{eg-data}An example of the continuum data obtained toward the IRDC \irdcthirty. From
    top left to bottom right the continuum images are: 24\,\um, 350\,\um, 450\,\um, and 850\,\um. In all cases the
    contours overlaid are the 1.2\,mm continuum emission from \cite{Rathborne06}. The contour levels 
    are 30 (3$\sigma$), 60, 90, 120, 180, 240 mJy beam$^{-1}$. These images reveal that while the IRDC
    remains dark at 24\,\um\, it is associated with bright emission at all sub-millimeter wavelengths. Moreover,
    the cores are clearly identified above the emission from the larger cloud. Broadband SEDs were generated for
    all the cores and their properties determined from gray-body fits. See the appendix for the SED plots and 
    the derived parameters.}
\end{figure}

Figure~\ref{eg-data} shows an example of the continuum data obtained toward one 
of the IRDCs: \irdcthirty. Here we show the 24\,\um, 850\,\um, 450\,\um, and 
350\,\um\, continuum images overlaid with contours of the 1.2\,mm continuum 
emission. The continuum data toward this IRDC is typical in that it shows that IRDCs 
remain dark at 24\,\um, which confirms their extremely high column densities and 
low temperatures (see the appendix for all the 24\,\um\, images).  Moreover, the 
24\,\um\, extinction morphologies closely match the morphologies of the 1.2\,mm 
continuum emission. This is expected as the 1.2\,mm continuum emission traces the 
cold, dense dust which is blocking the Galactic mid-IR emission.  While the IRDCs 
themselves remain dark, some IRDCs appear to contain bright, compact 24\,\um\, 
{\it {emission}} sources. In many cases these 24\,\um\, emission sources  are 
coincident with the millimeter cores (e.g., Figs.~\ref{seds-7}, \ref{seds-33}, 
\ref{seds-51}, \ref{seds-43}).

Sub-millimeter continuum emission was detected toward all the observed IRDCs and
matches well the morphology of the 1.2\,mm continuum emission (e.g.\, Fig.~\ref{eg-data}).  
While the extended sub-millimeter continuum emission from the larger IRDC is sometimes
faint, the cores are associated with bright, compact sub-millimeter continuum emission. 
Because strong emission at millimeter/sub-millimeter wavelengths can trace either temperature 
or column density enhancements, millimeter/sub-millimeter data alone are unable to
accurately indicate the physical properties of potential star-forming cores. Thus, one needs an 
additional method to distinguish high temperature cores from high column density cores.

\subsection{Spectral Energy Distributions}
\label{sed-section}

To characterize the emission arising from the cores, we have constructed broadband SEDs by 
combining the 24\,\um, 60--100\,\um, 350\,\um, 450\,\um, 850\,\um, and 1.2mm 
continuum data.  To model their SEDs we assume that the emission is 
optically thin and use a single temperature, modified gray-body function of the form,

\[F_{\nu} = \Omega_{s}\,B_{\nu}(T_{D})\,(1 - e^{-\tau_{\nu}})\] 

\noindent where  $\Omega_{s}$\, is the source solid angle, $B_{\nu}(T_{D})$ is the Planck function at 
the dust temperature ($T_{D}$), and $\tau_{\nu}$ is the optical depth \citep{Gordon95}.  
We calculate $\tau_{250}$ 
using the relation $\tau_{\nu} = \tau_{250} (250/\lambda)^\beta$, where $\beta$ is the dust 
emissivity index and $\lambda$ is the wavelength in \um. The free parameters in the fits 
are the dust temperatures ($T_{D}$), optical depths at 250\,\um\, ($\tau_{250}$), and dust 
emissivity indices, $\beta$. 

While this model may not be appropriate to characterize the emission in all cases,  it is the simplest 
and most systematic approach we can take to obtaining estimates of the global dust parameters. 
Because we are interested in the entire core which is typically larger than protostellar envelopes and
disks, this model will be sufficient and will produce a  good approximation to the dust parameters 
on $\sim$ 1 pc spatial scales. Give the limited measurements at different wavelengths
in the majority of the SEDs, complex radiation transfer modeling that includes disks, envelopes, 
protostars as well as several geometrical effects (e.g. \citealp{Robitaille07}) would not produce 
robust results. Moreover, because the \cite{Robitaille07} models 
currently do not incorporate the data at wavelengths $>$ 100\,\um\, accurately, the use of these detailed models is not
appropriate for our purposes (priv. com. T. Robitaille). 

The millimeter/sub-millimeter fluxes and sizes for each core were determined by fitting a two-dimensional 
gaussian profile with a constant background level to the emission at each core position (determined from the 
1.2\,mm continuum emission; see Table~\ref{cores} and \citealp{Rathborne06}). Each core was fit 
individually and the gaussian 
was inspected to determine that it fit well the flux from the core while excluding the underlying large-scale 
emission from the cloud. 

The 24\,\um\, fluxes were calculated by interactively fitting a gaussian profile with a constant background level 
to the radial profile of the point source emission at each core position. This fitting takes into account any 
local variations in the noise of the data or variations in the local background emission.  We assume an error 
for the 24\,\um\, flux calibration of 20\%. In the majority of cases where a  gaussian profile could not be 
fit to the emission, we have determined an upper limit to the flux by calculating the mean value in a 
small region centered at the core position. Because the emission may not be optically thin at 24\,\um, 
the  24\,\um\, fluxes are only included in the SED fitting for cores that have no 60--100\,\um\,  
continuum spectrum.  Because we wish to constrain the total luminosity of the core including any 
embedded sources, and the emission from the cores typically peaks in the far-IR, the 60--100\,\um\, data 
were crucial in determining a good fit,  particularly in the cases where no  24\,\um\, emission was 
detected. For inclusion within the SED the 60--100\,\um\, continuum spectra were separated 
into 4 bins. The bin centers are at 55, 65, 75, 85, and 95\,\um\, and are 10\,\um\, wide. The mean 
fluxes within these ranges were used to produce the points that are included within the SEDs. 
Table~\ref{cores} lists the coordinates, distances, and fluxes measured for all 190 cores.

A core SED was generated if fluxes could be extracted for at least two of the millimeter/sub-millimeter 
wavelengths. Thus, not all cores have an SED and derived parameters.  We use the least squares fitting routine 
in IDL, MPFITFUN, to determine the best fit to the data. This procedure has the advantage of allowing us to 
constrain the values for the input parameters, include the errors in the fitting, and to correctly handle the upper 
limits at 24\,\um. To model the gray-body emission we assume that the cores are isothermal and use a range 
in $T_{D}$  of 5--55\,K, $\tau_{250}$ of 0.0001--1.0, and $\beta$ of 0.5--2.5. To account for the errors we 
assumed 'standard' weighting for the data such that the weight, W = 1.0/error$^{2}$, where the error
includes both the calibration uncertainty and the image noise, added in quadrature. When the core had a 
24\,\um\, upper limit or a 60--100\,\um\,  continuum spectrum,  the weighting for the 24\,\um\, flux 
was set to 0.0 which forced  the fitting routine to ignore this data point.  In all of these cases the resulting 
fit was consistent with the upper limit (i.e. the function fell below the limit).

Figures~\ref{seds-27}--\ref{seds-34} show the 24\,\um\, image (overlaid with the 1.2\,mm continuum
emission) for each IRDC as well as the broadband SEDs for the cores. In total, broadband SEDs were 
generated for 157 of the sample of 190 cores. The millimeter/sub-millimeter/far-IR fluxes are shown on 
these plots as filled circles with the corresponding error bars. The  24\,\um\, fluxes are plotted as either 
filled circles (when included in the fitting), open circles (when excluded from the fitting), or as upper 
limit arrows. 

For the 100 cores that have millimeter/sub-millimeter emission and either  24\,\um\, emission or a 
60--100\,\um\,  continuum spectrum, we plot the grey-body function (solid line) determined from the 
best fit. Labeled on each plot are the IRDC and core name, the output $T_{D}$, $\tau$, $\beta$, 
luminosity, mass and the $\chi^{2}$ returned from the fit.  Using the output errors for the
$T_{D}$, $\tau$, $\beta$, we also include the functions that correspond to the upper and lower values 
for the each of the three parameters (dotted lines). Table~\ref{core-properties-good} lists the parameters 
determined for these cores. 

The remaining 57 cores have an upper limit at  24\,\um\,  and no  60--100\,\um\, continuum spectrum, 
i.e., only fluxes in the millimeter/sub-millimeter regime.  Because the emission from the cores peaks in the far-IR, 
the absence of either a  24\,\um\, flux measurement or far-IR continuum spectrum made the determination 
of the individual core parameters difficult. For these cores the plots show two functions (examples are 
G015.31-00.16 MM4, MM5, G025.04-00.20 MM3, MM4, MM5; Figs.~\ref{seds-15} and ~\ref{seds-14} respectively).
The first fit (solid line) is determined using only the millimeter/sub-millimeter fluxes. Because these SEDs only contain
two data points and the emission likely does not cover the peak, the parameters determined from them 
are probably lower limits. The second fit (dashed line) was determined by including the 24\,\um\, upper
limit as a real data point. In this case, the values determined
from the fitting represent upper limits. For these cores 
we include two values for the $T_{D}$, $\tau$, $\beta$, luminosity, mass and  $\chi^{2}$ on the plots; the estimated lower and upper limits.
Table~\ref{core-properties-limits} lists the lower and upper limits to the parameters for these cores.
Because these fits are unreliable the parameters derived from them are not included in any of the following analysis.

\section{Discussion}

\subsection{Evidence for active star formation: separation of cores with high-mass protostars}

To identify cores that may contain active star formation, we use a combination of their 
\Spitzer\,/IRAC 3--8\,\um\, colors and the presence or absence of compact 24\,\um\, emission. 
We use the IRAC data and classification scheme presented in \cite{Chambers09} but 
use our own 24\,\um\, flux measurements which were determined via a gaussian fit to the radial 
profile of the emission (rather than via aperture photometry as in \citealp{Chambers09}). 

The IRAC classification scheme is based on the colors of objects in the IRAC 3--8\,\um\, images 
which are grouped in to three categories: `red', `green', and `blue'. Most objects have
either red or blue colors in these bands which indicate that that the flux is either increasing or
decreasing, respectively, with wavelength. The `red' objects are associated with bright 8\,\um\, 
emission and likely correspond to \hii\, regions while the `blue' objects are associated with a region 
of bright 3.6\,\um\, emission and are predominantly unextincted stars. Those rare objects that are `green' show 
enhanced emission at 4.5\,\um\, and could correspond to either an extincted star or, 
if the emission is extended, shocked gas. These latter objects are referred to as either `green fuzzies' 
or EGOs \citep{Chambers09,Cyganowski08}, and are thought to trace the shocked gas in an 
outflow \citep{Marston04,Noriega-Crespo04}. Of the 190 cores within our sample, 
\cite{Chambers09} find that 35 are associated with `red', 47 with `green', 6 with  `blue' objects, 
while the remaining 102 cores are not associated with any significant IRAC emission. 

Emission at 24\,\um\, is an indicator of active star formation. Because this
emission traces warm dust that is the heated as material accretes from a core 
onto a central protostar, the detection of a bright 24\,\um\, point source associated with
a dense core suggests that star formation is occurring within it. We find that 93 of the 190 cores
have a detectable 24\,\um\, point source emission. 

Thus, the combination of a green fuzzy and 24\,\um\, emission toward a dense core suggests 
that there is both shocked gas and an accreting protostar within the core. Conversely, 
cores that do not contain an accreting protostar will remain cold, with dust temperatures 
too low to emit detectable 24\,\um\, emission. Because  they also lack outflows, they will also show 
no evidence of shocked gas. As a 
result, the detection of a green fuzzy and a 24\,\um\, point source coincident with a core 
indicates active star formation within that core. Conversely, the non-detection of shocked gas and 
24\,\um\, emission indicates that the core may be cold, with no (detectable)
internal exciting source. While the absence of these star forming tracers is not definitive
proof that there is no current star formation occurring within the core, it is
suggestive.  For example, low-mass protostars may lack sufficient luminosity to be detectable at 24\,\um\,
given the typical distances to these IRDCs. 
However, a lack of heated dust and no shocked gas is certainly what is expected 
for a core in the cold, pre-protostellar phase.

Using their IRAC 3--8\,\um\, colors and their 24\,\um\, emission we use the \cite{Chambers09}
technique to separate the cores
into five specific groups: quiescent cores, intermediate cores, active cores, red cores and blue cores. We define 
a `quiescent core' to be a core that contains no significant 3--8\,\um\,  nor 24\,\um\,  emission. 
`Intermediate cores' contain either a green fuzzy or a 24\,\um\,  point source, but not both. 
`Active cores' contain a green fuzzy and a 24\,\um\,  point source. `Red cores' are those 
cores associated with bright 8\,\um\, emission. Because bright emission at 8\,\um\, will 
be dominated by emission from either heated dust or UV-excited polycyclic aromatic 
hydrocarbons (PAHs), these cores likely correspond to embedded \hii\, regions where 
high-mass stars have already formed. 
Because the intermediate 
cores are associated with only one of the two proposed criteria for active star formation, 
we have separated these into a distinct group.  These cores perhaps represent a transition 
phase between the active and quiescent cores. 

Of the 190 cores, we group 35 as `red', 6 as `blue', 38 as `active', 32 as `intermediate', 79 as `quiescent'.
We generated SEDs for 25 of the `red', 3 of the `blue', 34 of `active', 29 of the  `intermediate', 
and 66 of the  `quiescent' cores.  The IRAC classifications, the presence of 24\,\um\, emission 
and the resulting core designations are included on the SED plots. 

\subsection{Dust temperatures, opacities, and dust emissivities}

\begin{figure}
\begin{center}
\includegraphics[width=0.7\textwidth,clip=true]{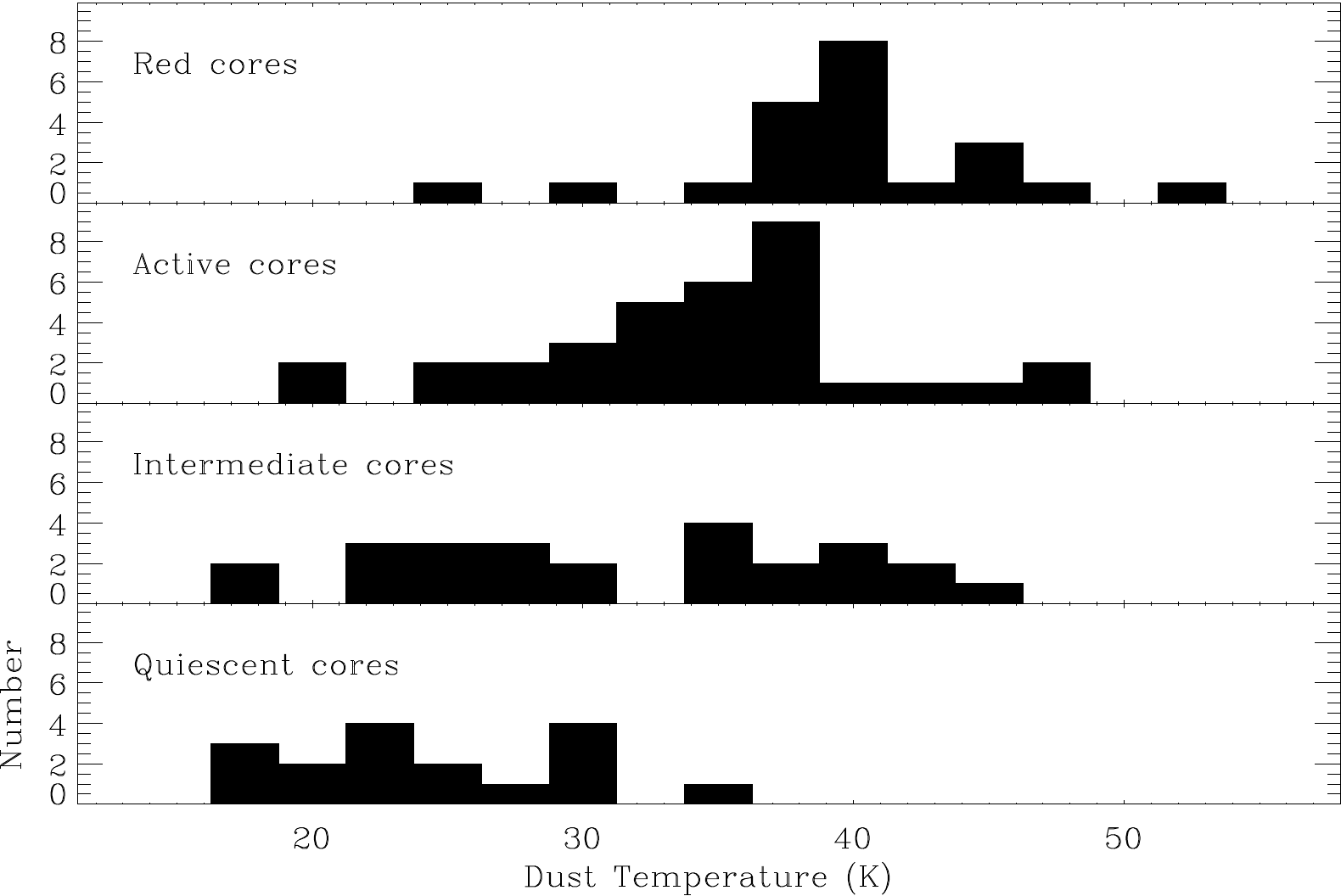}
\end{center}
\caption{\label{histograms-temp}Histograms of the dust temperature (T$_{D}$)  derived 
  from the gray-body fits to the core SEDs. The panels shows the histograms
  for the samples of the red, active, intermediate, and quiescent cores. We find that, as
  expected, the derived dust temperatures decrease from the
  red, to the active, to the intermediate, to the quiescent cores.
  Table~\ref{summary} lists the median and standard deviations
  of these distributions.}
\end{figure}

\begin{figure}
\begin{center}
\includegraphics[width=0.7\textwidth,clip=true]{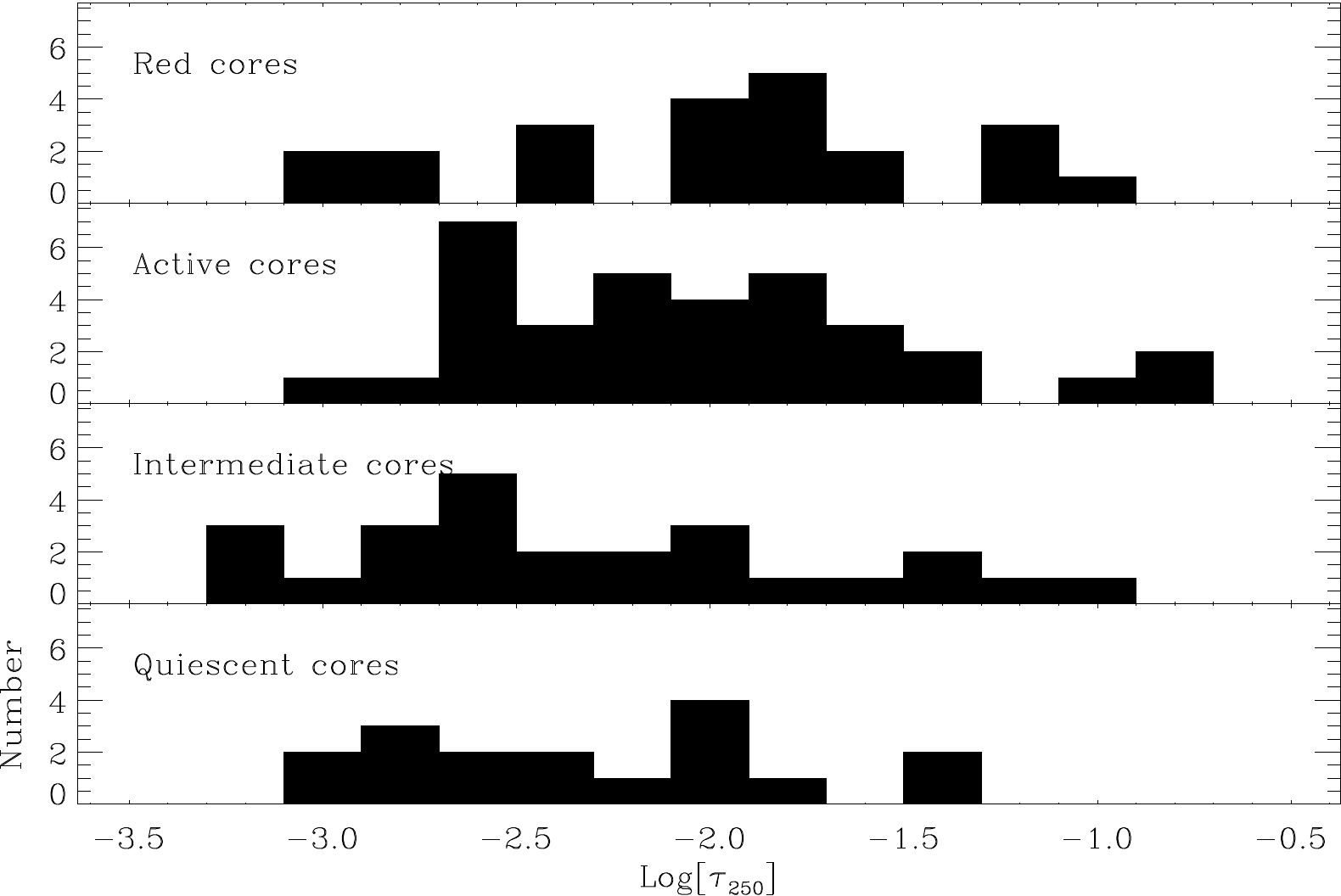}
\end{center}
\caption{\label{histograms-tau}Histograms of the 250\,\um\, opacity ($\tau_{250}$)  derived 
  from the gray-body fits to the core SEDs. The panels show the histograms for the samples of
 the red, active, intermediate, and quiescent cores. We find that the majority of the
  cores are optically thin, with low values of $\tau_{250}$ (most have $\tau_{250} <$ 0.01).
  Table~\ref{summary} lists the median and standard deviations of
  these distributions.}
\end{figure}

\begin{figure}
\begin{center}
\includegraphics[width=0.7\textwidth,clip=true]{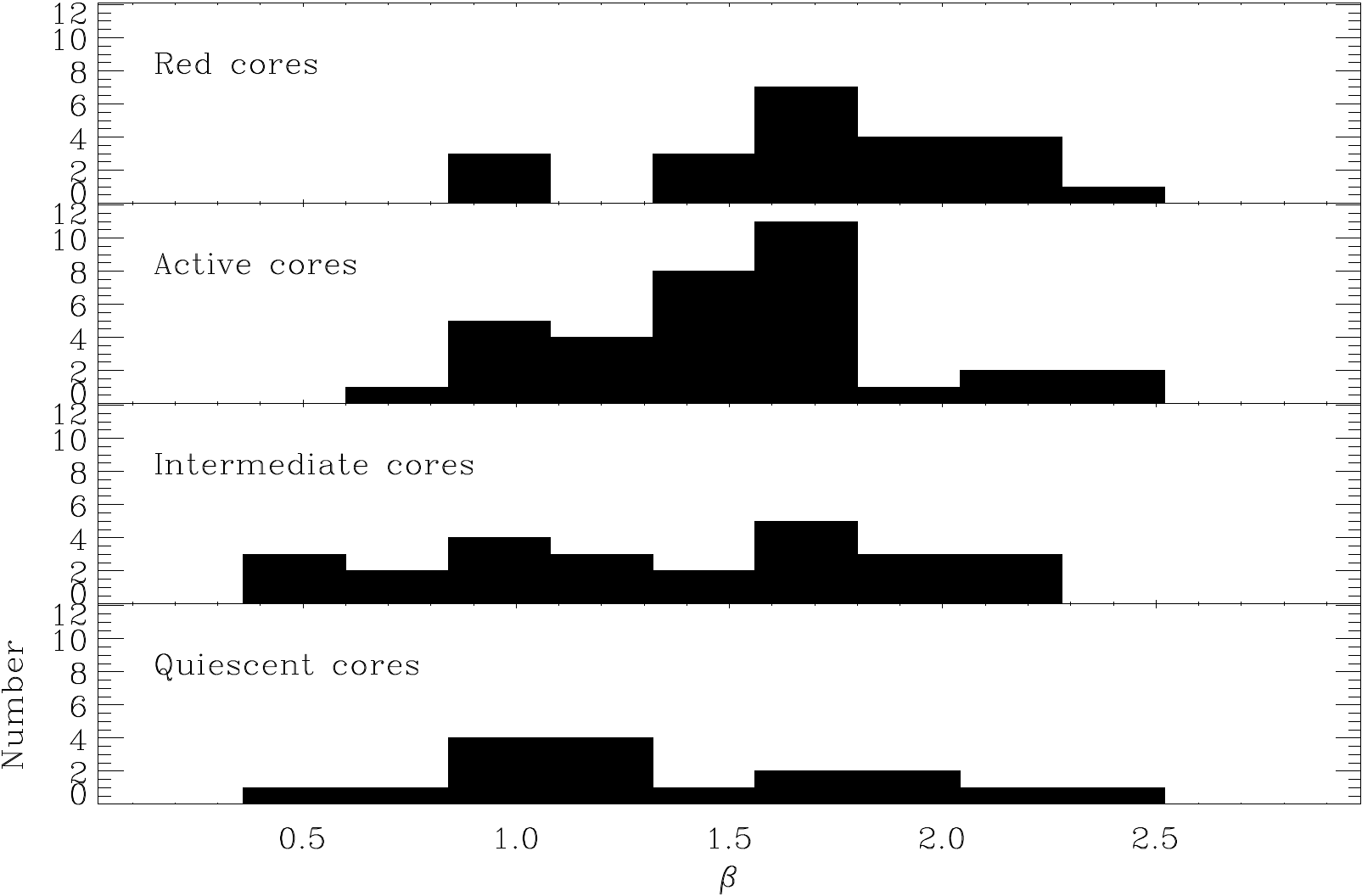}
\end{center}
\caption{\label{histograms-beta}Histograms of the dust emissivity index 
     ($\beta$) derived from the gray-body fits to the core SEDs. The panels show the histograms for the samples of
 the red, active, intermediate, and quiescent cores. Although they span
a similar range, the median of the distribution of $\beta$ for star-forming cores may be different from the median of the distribution of $\beta$ for cores with no apparent 
star-formation . Table~\ref{summary} lists the median
     and standard deviations of these distributions.}
\end{figure}

Figures~\ref{histograms-temp}--\ref{histograms-beta} show histograms for the number 
distributions of the derived T$_{D}$, $\tau_{250}$, and $\beta$ for our sample of cores. 
These figures are separated into four panels and show the number distributions for the red, 
active, intermediate, and quiescent cores for which reliable SEDs could be generated\footnote{Because
there are only 2 blue cores and they are presumably associated with foreground stars we do not include them here.}.  For these 100 cores, 22 can be classified as `red',  34 as `active', 25 as 
`intermediate', and 17 as `quiescent'.  Table~\ref{summary} lists the median and standard 
deviation measured from these distributions.  

The derivation of core dust temperatures allows us to establish whether significant heating, 
either by accretion or embedded sources, is occurring within the cores (mean radius of 
$\sim$ 0.5\,pc). As expected, Figure~\ref{histograms-temp} shows that the highest 
dust temperatures are measured toward the red cores (median T$_{D}$ of 41\,K), 
which presumably already contain a high-mass star. The measured dust temperatures 
are cooler for the samples of active, intermediate, and quiescent cores (median of  
34\,K, 31\,K and 23\,K respectively) as expected if the star formation activity is in an earlier state. 

While these dust temperatures are derived for each core individually,  one must exercise 
caution when applying these temperatures to the cores where there is clearly active star 
formation. For instance, if the red cores really do harbor a high-mass star and \hii\, region, 
then close to the protostar their internal temperatures should be heated to well above 41\,K. 
Because our angular resolution ($\sim$ 11\arcsec; 0.2\,pc) samples dust from both the 
central protostar and its surrounding larger core, the derived dust temperatures reflect a 
beam-weighted average of the actual dust temperature distribution.
If a cold $\sim$ 1 pc core contains a small volume of heated dust, the derived dust temperature will 
be larger than that of the cold envelope but smaller than that of the heated region.
While the 
general trend of higher dust temperatures for cores with active star formation compared 
to lower temperatures for cores without star formation is expected, it is likely that, within 
the star-forming cores, there is a small volume of gas which is heated to temperatures 
significantly greater than those calculated here.  Because the dust and gas are unlikely to be 
completely thermally coupled, it is possible that in these cases the gas temperatures are 
significantly higher than the dust temperatures derived from the SED fit.

The derived values of $\tau_{250}$ reveal that all cores are optically thin at 250\,\um\, 
($\tau_{250}$ $\sim$ 0.01) regardless of their star-formation activity (Fig.~\ref{histograms-tau}). 
The derived values for $\beta$ (Fig.~\ref{histograms-beta}) show a range in values 
across the complete range input into the gray-body fitting routine. Since active star formation 
may change the properties of the dust grains, one might expect the emissivity index to vary 
between the different groups. It appears that, although they have fairly similar median values, 
the distributions may be slightly different for the red and active cores compared to the distribution 
for cores with no apparent star-formation.

\subsection{Core masses, column and volume densities}

Mass estimates obtained from molecular line emission toward very cold, very high 
column density cores are often unreliable. The combination of high molecular line optical 
depths toward these cores and the potential for molecular depletion makes it difficult to 
accurately trace the internal structure of such cores using molecular line emission. 
Because the emission from cold, dense dust peaks at millimeter/sub-millimeter wavelengths  
and is optically thin, it is a superior tracer of the internal structure and the masses of these 
cold, dense regions. However, accurate mass estimates from dust continuum emission require 
a good measurement of the dust temperatures and emissivities.  Most mass estimates from 
dust emission assume a single value for $T_{D}$ and $\beta$. Because a
large sample of cores potentially spans a large range in evolutionary stages and environments, 
the true values of $T_{D}$ and $\beta$ may vary considerably within the sample. To better determine the 
masses, therefore, one requires estimates of the dust temperatures and emissivities for each core. 
We have achieved this for our sample of cores and, thus, can now obtain a more accurate  
census of their masses, column and volume densities.

To calculate masses we use 
\[M = \frac{F^{i}_{\nu} D^{2}}{\kappa_{\nu} B_{\nu} (T_{D})}\] 
\noindent \citep{Hildebrand83}, where $F^{i}_{\nu}$ is the observed integrated source flux density, 
$D$ is the distance, $\kappa_{\nu}$ is the dust opacity per gram of dust, and $B_{\nu}(T_{D})$ 
is the Planck function at the dust temperature.  In all cases, we assume a gas-to-dust mass ratio 
of 100.  We use the individual values for $T_{D}$ and $\beta$ derived from the SED for each 
core. To calculate the masses and densities we use the flux measured at 1.2\,mm and 
$\kappa_{1.2mm}$ of 1.0~cm$^2$ g$^{-1}$ \citep{Ossenkopf94}.

The beam-averaged \hh\, column density, N(\hh), was calculated using the expression
\[N(H_{2}) =  \frac{F^{p}_{\nu}}{B_{\nu} (T_{D})  \mu m_{p} \kappa_{\nu} \Omega_{b}}\]
\noindent where $F^{p}_{\nu}$ is the observed peak source flux density, $\mu$ is the 
mean molecular weight (2.8),  $m_{p}$ is the mass
of a proton, and $\Omega_{b}$ is the beam solid angle. The 
volume-averaged \hh\, density, n(\hh), was estimated using the dust masses and by 
assuming a volume of a sphere. The derived masses, column and volume 
densities for the cores are listed in Table~\ref{core-properties-good}. 

\begin{figure}
\begin{center}
\includegraphics[width=0.7\textwidth,clip=true]{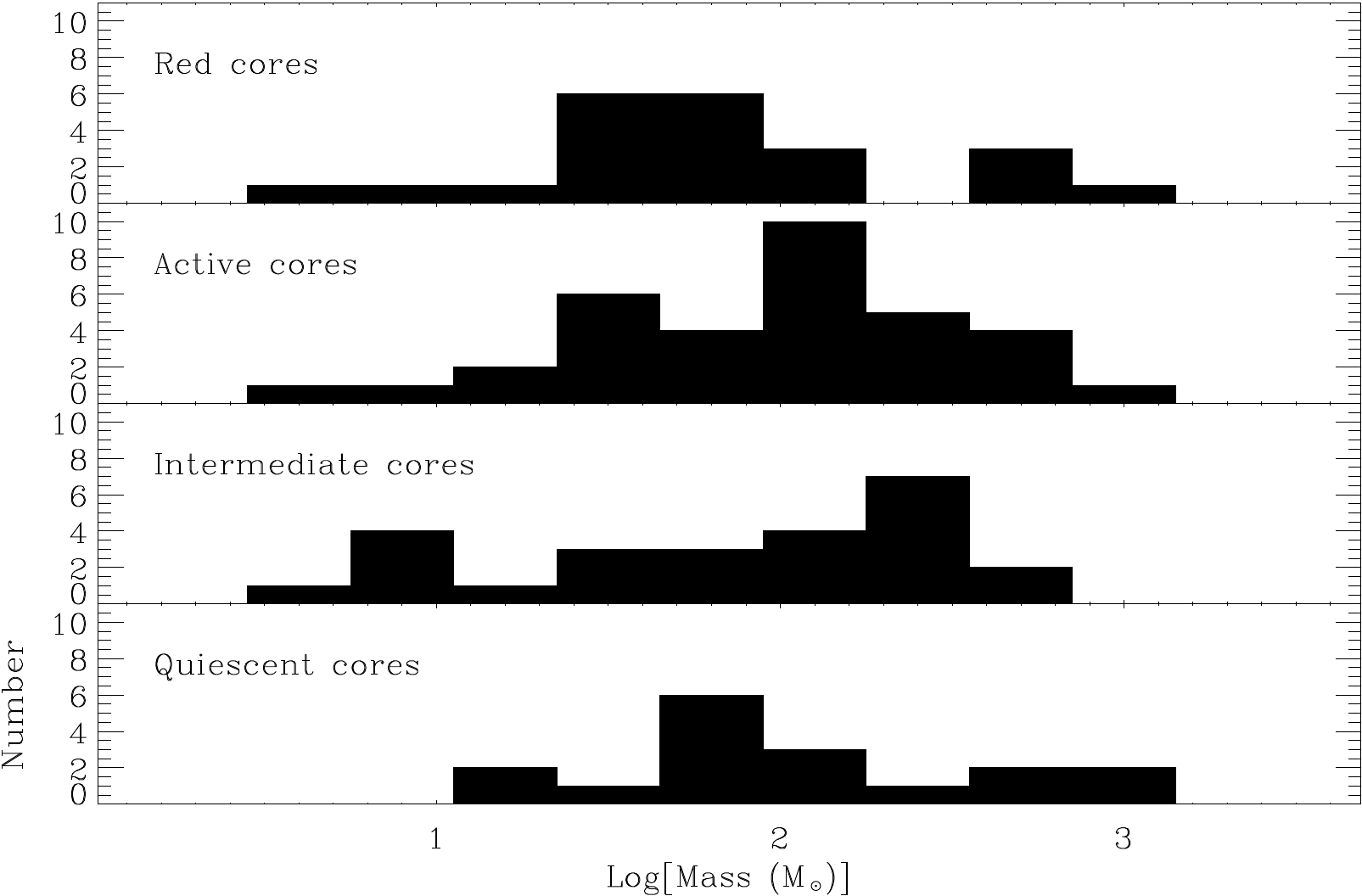}
\end{center}
\caption{\label{histograms-mass}Histograms of the derived dust mass. The panels show the histograms for the samples of
 the red, active, intermediate, and quiescent  cores. We find that the mass distributions for the cores 
  span a similar range, regardless of their apparent star-formation activity.   Table~\ref{summary}
  lists the median and standard deviations of these distributions.}
\end{figure}

\begin{figure}
\begin{center}
\includegraphics[width=0.7\textwidth,clip=true]{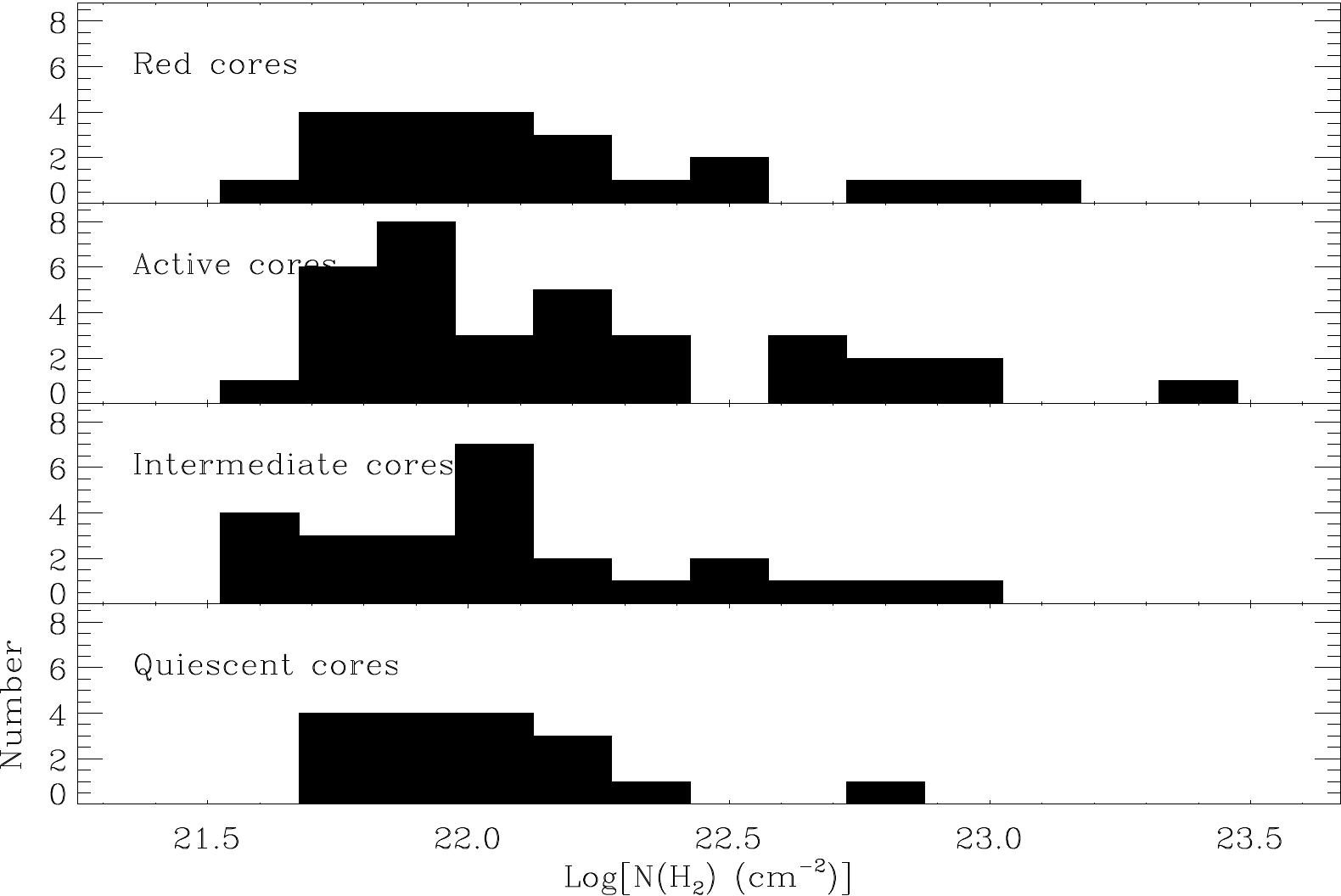}
\end{center}
\caption{\label{histograms-column}Histograms of the derived \hh\, column density, N(\hh). The panels show the histograms for the samples of
  the red, active, intermediate, and quiescent  cores. We find that the cores
  have similar column densities, regardless of their apparent
  star-formation activity.  Table~\ref{summary} lists the median and
  standard deviations of these distributions.}
\end{figure}

\begin{figure}
\begin{center}
\includegraphics[width=0.7\textwidth,clip=true]{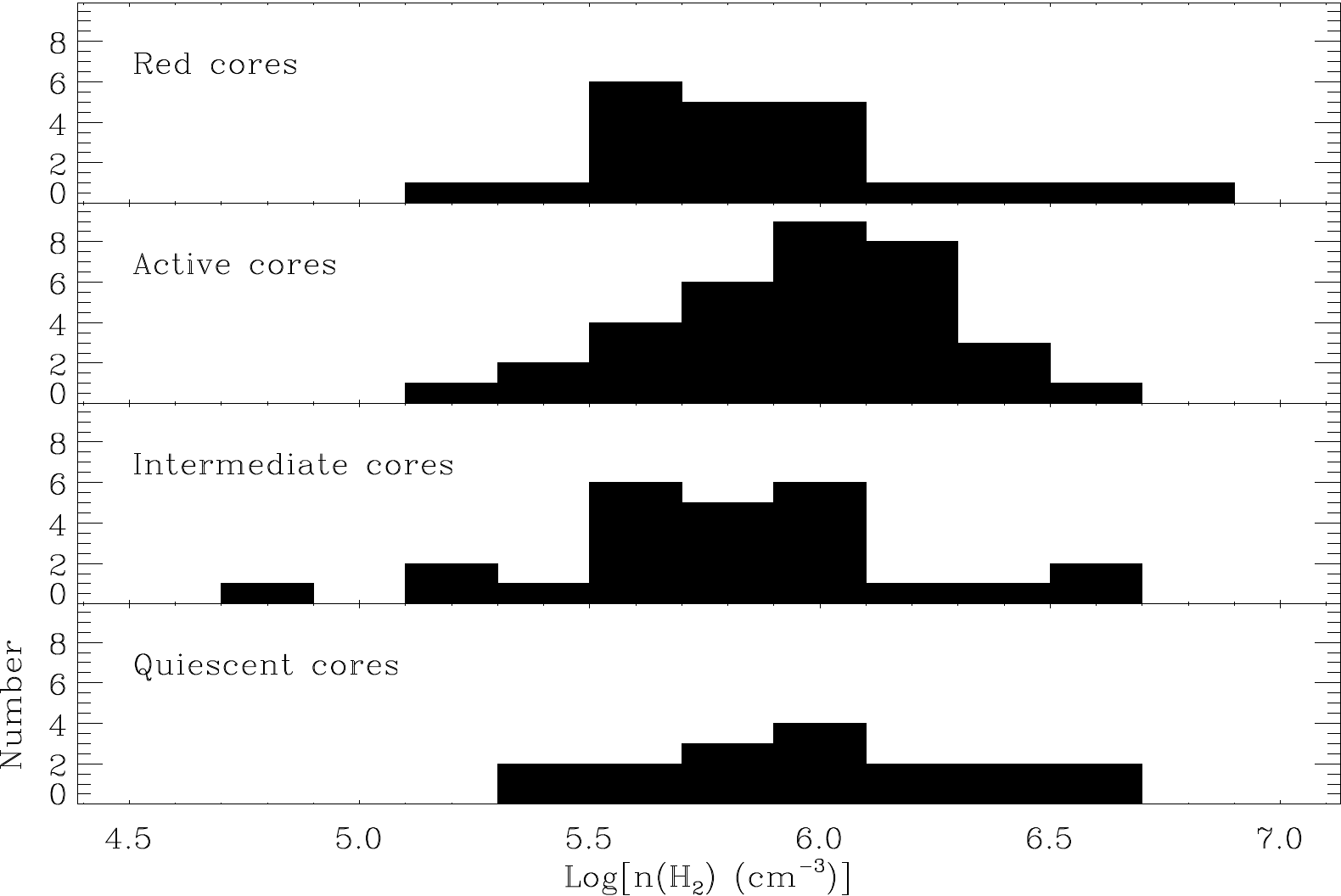}
\end{center}
\caption{\label{histograms-density}Histograms of the derived \hh\, volume density, n(\hh). The panels show the histograms for the samples of
  the red, active, intermediate, and quiescent  cores.  We find that the cores
  have similar volume densities, regardless of their apparent
  star-formation activity.  Table~\ref{summary} lists the median and
  standard deviations of these distributions.}
\end{figure}

Figures~\ref{histograms-mass}--\ref{histograms-density} show histograms of the number 
distributions for the masses, column and volume densities. A summary of the median and 
standard deviation values calculated from  these distributions is given in Table~\ref{summary}. 
We find that the median masses are comparable between the red, active, intermediate, and 
quiescent cores (Fig.~\ref{histograms-mass}). To quantify the similarities between these 
distributions, we have also 
calculated, via a Kolmogorov-Smirnov (KS) test,  the probability that these distributions arise 
from the same parent population. We find that the probability that the red and active cores
are derived from the same parent is 6\%, the probability that the red and quiescent cores are 
derived from the same parent is 84\%, while the probability that the active and quiescent
cores are derived from the same parent is 39\%.

At all wavelengths the `active' cores that show evidence for star formation activity have significantly 
more luminous emission compared to those that are more quiescent.  However, such a close correspondence 
between the mass distributions of these populations implies that the bright emission observed 
toward the more active cores may simply arise from differences in their internal temperatures 
and not because they have higher densities. Indeed, we find no significant difference between 
the distribution for the derived column and volume densities for the cores that show star 
formation activity compared to those that are more quiescent (Figs.~\ref{histograms-column} 
and \ref{histograms-density}). 


\subsection{Core luminosities: identifying high-mass protostars and high-mass starless cores}

Bolometric luminosities provide information on a core's embedded young 
stellar objects and evolutionary state. The core luminosities were estimated by 
integrating the emission under the best fit gray-body curve and using the derived kinematic 
distance. The luminosities are included on the SED plot for each core and are listed in 
Table~\ref{core-properties-good}.

\begin{figure}
\begin{center}
\includegraphics[width=0.7\textwidth,clip=true]{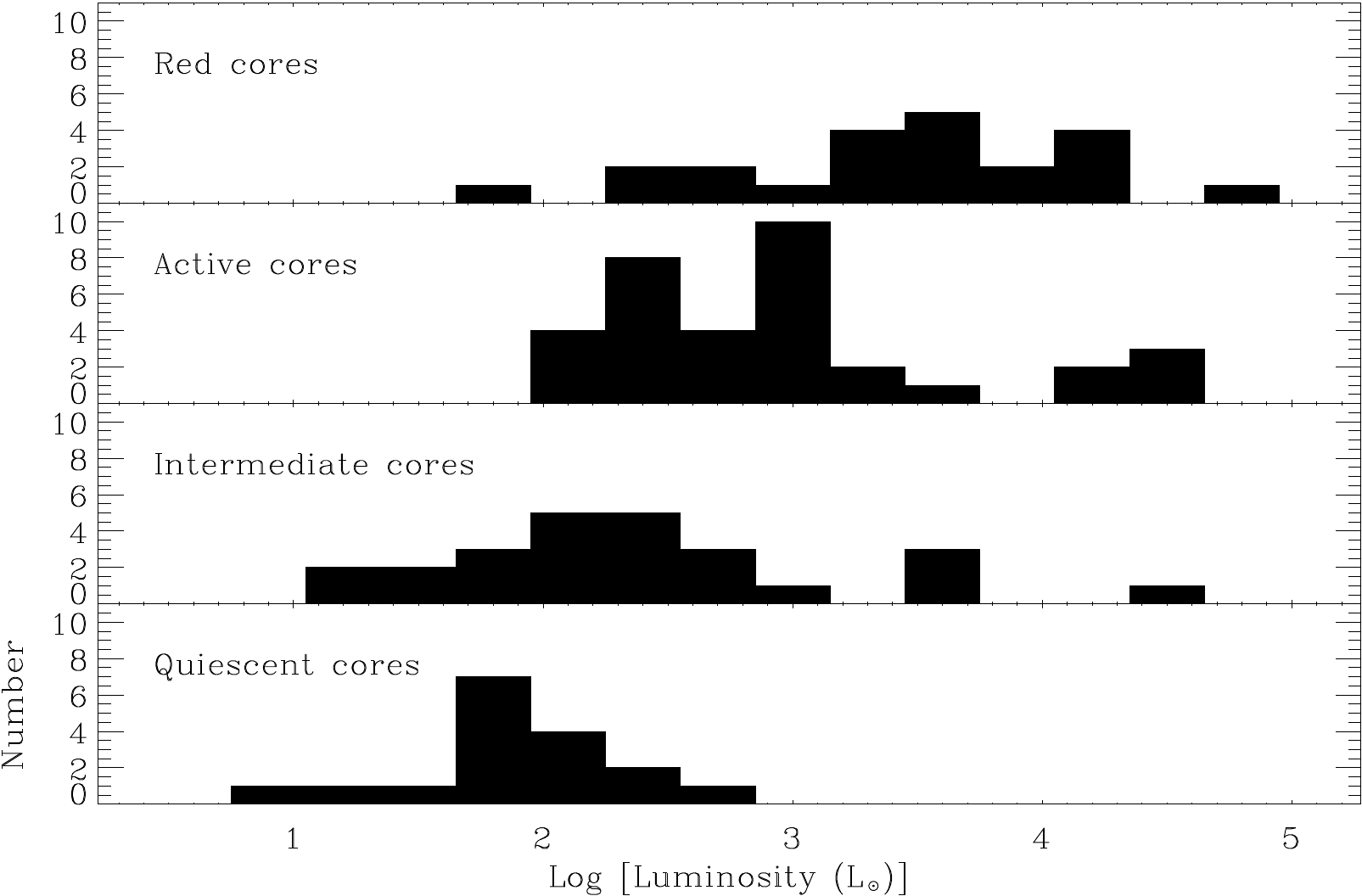}
\end{center}
\caption{\label{histograms-lum}Histograms of the bolometric luminosity derived 
  from the gray-body fits to the core SEDs. The panels show the histograms for the samples of
  the red, active, intermediate, and quiescent  cores. We find that, as
  expected, the derived luminosities decrease from the
  red, to the active, to the intermediate, to the quiescent cores.
   Table~\ref{summary} lists the median and standard deviations
  of these distributions.}
\end{figure}

Figure~\ref{histograms-lum} shows the number distributions of the derived bolometric 
luminosities for the cores (the median and standard deviation values calculated from 
these distributions are given in Table~\ref{summary}). While the complete sample of 
cores span a range in bolometric luminosities of $\sim$10--10$^{5}$\,\Lsun, there is a 
clear trend for the core luminosity to decrease from  the red cores to the quiescent cores. 
We find that the red cores have typical luminosities of $\sim$ 10$^{3.7}$\,\Lsun, while 
the active and intermediate cores have lower luminosities of $\sim$ 10$^{2.8}$\,\Lsun\, 
and $\sim$ 10$^{2.3}$\,\Lsun\, respectively. Lower still, the quiescent cores 
have typical luminosities of $\sim$ 10$^{1.9}$\,\Lsun.

A rough approximation of the final stellar mass can be made from the estimates of 
the core luminosity. Low-mass (M$<$2\,\Msun) stars never achieve luminosities 
$>$~100\,\Lsun\, in their pre--main-sequence evolution (e.g.\, \citealp{Palla90}). 
Because L $>$ 100\,\Lsun\, for the majority of the cores, it is unlikely that, in these 
cases, a single low-mass star dominates the core luminosity. Given our 
sensitivity and angular resolution, we cannot rule out the possibility, however, that 
many of the cores may comprise a cluster of unresolved low-mass stars that, together,
produce a larger core luminosity. 

On the other hand, high luminosities may arise because the core contains a 
high-mass protostar (M$>$8\,\Msun). Because the luminosity for a high-mass 
protostar ($\sim$10$^{3}$--10$^{5}$\,\Lsun)  is much larger than the luminosity 
from a low-mass protostar  ($\sim$ 100\,\Lsun), cores with luminosities 
$>$~10$^{4}$\,\Lsun\, probably harbor a high-mass protostar.  Since high-mass 
stars typically form in clusters surrounded by many lower mass stars \citep{Lada03}, 
lower mass protostars may also be forming within these cores and contributing to the 
overall observed luminosities. Regardless of the exact number of stars within the 
core, it is likely that the observed bolometric luminosity is dominated by the 
most massive protostar.

Using a bolometric luminosity of 10$^{4}$\,\Lsun\, as a rough threshold, 
we can identify cores that have sufficient luminosities to harbor a high-mass protostar. 
When applying this threshold to the sample of `active' and `intermediate' cores, we find 
that  6 cores have luminosities $>$ 10$^{4}$\,\Lsun\, and, thus, may contain a 
high-mass protostar (five from the active core sample and one from the intermediate core sample).
Indeed, high-angular resolution millimeter/sub-millimeter interferometry images toward 
two of these cores  (\irdcfiftyone~MM1 and \irdcfortythree~MM; Figs.~\ref{seds-51} and \ref{seds-43} 
respectively) show that both cores contain bright, compact structures that remain unresolved 
at $\sim$ 0.03\,pc angular resolution \citep{Rathborne07,Rathborne08}. Moreover, their 
high-angular resolution spectra show many complex molecular emission features indicative of 
hot molecular cores (HMCs); an early stage in the formation of individual high-mass stars  
(e.g.\,\citealp{Garay99}).

The majority of the remaining active and intermediate cores (45 cores; 76\%) have luminosities 
between 10$^{2}$ and 10$^{4}$\,\Lsun. If these cores contain a single protostar or 
main sequence star, then they likely correspond to an intermediate mass star 
(2 $<$ M $<$ 8\,\Msun).  The active/intermediate cores that have lower bolometric luminosities 
(i.e. L $<$ 10$^{2}$\,\Lsun) may either be in an earlier evolutionary phase or be
only forming low-mass protostars. They contain heated dust, as evidenced by the 
24\,\um\, emission, but the current luminosity appears too low for a high-mass protostar. In these cases, the 
observed luminosity may arise from the accretion of material from the core onto 
the low-mass protostars. Indeed, recent work suggests low/intermediate mass protostars may 
continue to accrete material from their surroundings and grow to become high-mass
protostars at the end of their evolution (e.g.\, \citealp{Beuther07}). Thus, we cannot rule
out the possibility that these cores may eventually give rise to high-mass stars.

As expected, the quiescent cores all have lower (L$<$10$^{3}$\,\Lsun) bolometric 
luminosities, presumably because there is no high-mass central source to heat the dust. 
While the number of quiescent cores included within these histograms is small (17), there
are 62 additional cores within these IRDCs that also meet our criteria for a lack of
active, high-mass star formation. These cores have no detectable 3--8\,\um\, nor 24\,\um\, emission
but do not have the far-IR continuum data which is invaluable for determining their properties.
As a result, they are not included in the analysis here, but are also likely pre-protostellar 
in nature. 

The quiescent cores have no apparent evidence for star-formation in the \Spitzer\, 
data. One possibility for this absence of star-formation activity is that they are young but 
will eventually form stars. On the other hand, it may be that these cores are currently forming only low-mass stars 
that are not bright enough to detect.  Thus, we cannot rule out the possibility that the 
quiescent cores are currently forming low-mass 
protostars. Indeed, using the expression from the recent work of \cite{Krumholz08} that 
predicts the luminosity to mass ratio, L/M, of a core powered by the accretion on to low mass stars,
 \[L/M = 3.6 M^{-0.33} \Sigma^{0.66} T_{b}^{0.16}\,   L_{\odot}/M_{\odot}\]
\noindent where M is the core mass in units of 100\,\Msun, $\Sigma$\, is the mass 
column density in g\,\cms, and $T_{b}$ is the background temperature in units of 10 K, 
we find that for the quiescent cores, the median value predicted for the luminosity
due to the accretion on to low-mass stars is $\sim$300\,\Lsun. There are 6 quiescent cores
within our sample that have luminosities comparable to, or greater than, 300\,\Lsun. Thus,
their observed luminosity could be powered by accretion onto low-mass stars.

At the distances to these IRDCs (typical distance of $\sim$ 4 kpc) 
low-mass protostars are difficult to detect. Thus, many of the `quiescent' cores may contain 
low-mass protostars that will not be detectable with the available sensitivity 
and angular resolution. If, on the other hand, these candidate quiescent cores are in 
fact in a pre-protostellar phase, the most massive cores are excellent examples for the 
elusive `high-mass starless cores'; the very earliest stage in the formation of a high-mass star. 

For the analysis presented in this paper, we assume that each core contains a 
single protostar. Without sensitive, high-angular resolution data, it is
difficult to determine any potential multiplicity within the cores. High-angular 
resolution ($\sim$ 2\arcsec; $\sim$ 0.03 pc) millimeter and sub-millimeter data 
toward 6 of the highest mass cores within the current sample reveal that 4 of the 
cores are resolved into multiple protostellar condensations \citep{Rathborne07,Rathborne08}.  
It may be that these cores will in fact give rise to star clusters. In all cases, however,  
it appears that one central, massive condensation most likely dominates the bolometric 
luminosity. 

Using the expression that relates the maximum stellar mass in a cluster ($m_{max}$) 
to the mass of the cluster ($M_{cluster}$), $m_{max}$ = 1.2 $M_{cluster}^{0.45}$ 
\citep{Larson03}, we can estimate the mass of a cluster than will give rise to at least 
one high-mass star. Setting $m_{max}$ to 8\,\Msun, we find that the total cluster mass 
that will harbor a high-mass star should have a mass of $\sim$ 67\,\Msun. Assuming 
a cluster star formation efficiency of $\sim$ 30\% \citep{Lada03}, we calculate that 
such a stellar cluster could form from a core with a mass of $\sim$ 225\,\Msun. Thus, 
in the pre-protostellar phase, a core that will form a star cluster that will eventually form a high-mass star 
ought to have a mass of at least $\sim$ 225\,\Msun. Five quiescent cores 
within these IRDCs meet this criteria and, thus, will probably form a cluster with a high-mass
star. In particular, the most massive of these, the cores \irdcthirty~MM1 and MM3 
(Fig.~\ref{seds-33}) have low dust temperatures ($\sim$ 20\,K) and bolometric 
luminosities ($\sim$ 10$^{2}$\,\Lsun), but high masses ($\sim$ 10$^{3}$\,\Msun), 
column densities ($\sim$ 6$\times10^{23}$ \,\cms)  and volume densities 
($\sim$ 10$^{6}$\,\cmc). These values are consistent with the expected properties 
of high-mass starless cores.

\begin{figure}
\begin{center}
\includegraphics[width=0.7\textwidth,clip=true]{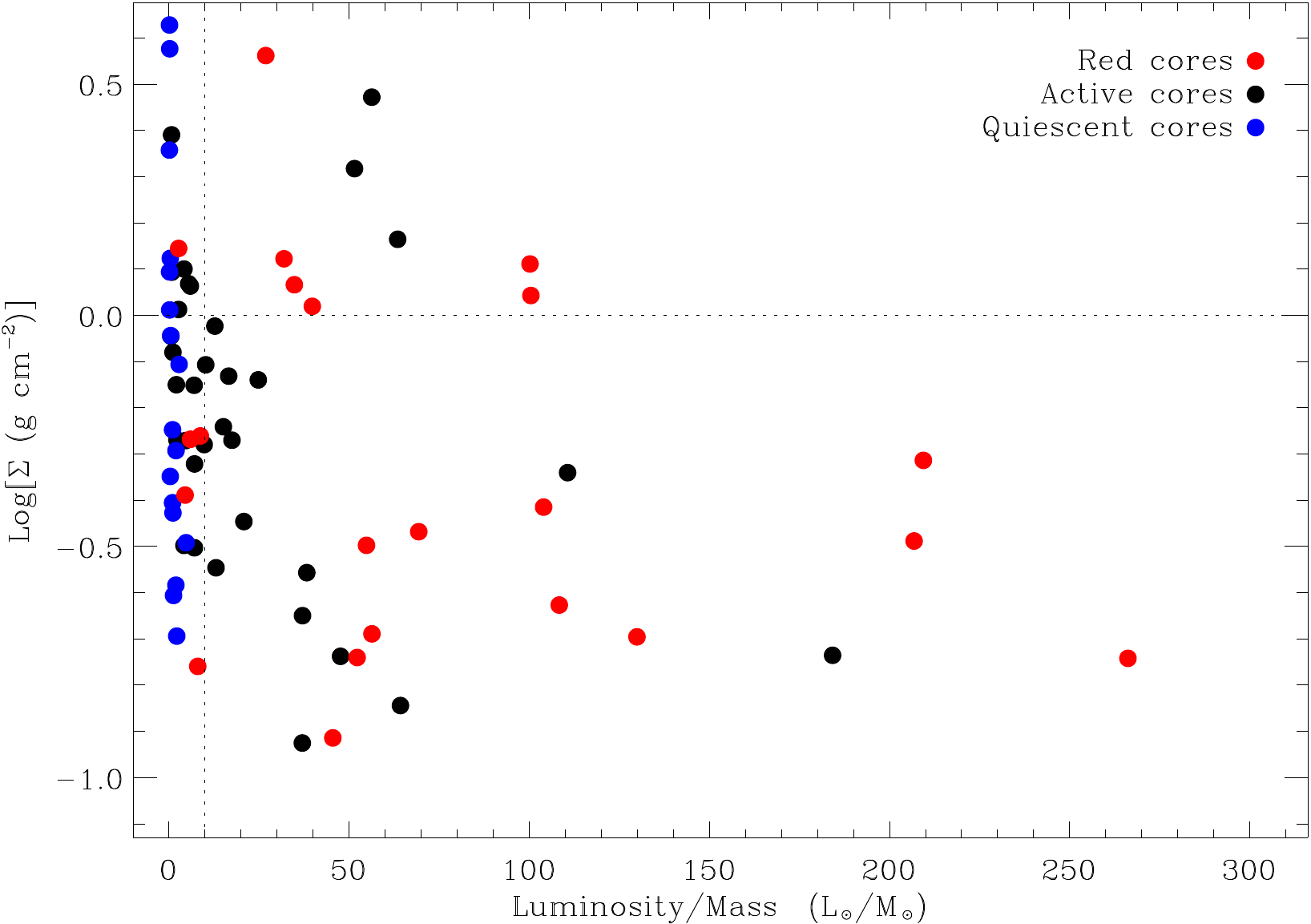}
\end{center}
\caption{\label{LM-sigma}Plot of bolometric luminosity to mass ratio (L/M) versus mass column density ($\Sigma$) for the cores. The dotted lines mark the
critical values of these parameters for the formation of a high-mass star (determined from the recent theoretical work of \citealp{Krumholz08}). These criteria 
mark the thresholds above which the fragmentation of a core is suppressed,  by the increase in temperature and Jeans mass. According to these criteria, 
we find  that nine of the cores may give rise to a high-mass protostar. For the remaining cores, it appears that the fragmentation and the formation of lower-mass
protostars is still possible. }
\end{figure}

Recent theoretical work suggests that to form a high-mass star from a core the 
cloud's critical mass column density, $\Sigma$, should be $\gtrsim$ 1\,g\,cm$^{-2}$ 
and its luminosity-to-mass ratio, L/M, $\gtrsim$ 10 \Lsun/\Msun\, \citep{Krumholz08}. 
Under these conditions, further fragmentation is suppressed by accretion onto 
nearby lower-mass protostars and the subsequent  heating of the surrounding 
material. It is thought  that an increase in temperature to $\sim$ 100~K is sufficient 
to increase the Jeans mass and, thus, halt the fragmentation process. Figure~\ref{LM-sigma} 
plots L/M versus $\Sigma$ for our sample of cores, with these thresholds overlaid. 

We find that 6 red cores and 3 active cores have L/M $>$ 10\,\Lsun/\Msun\, and 
$\Sigma$ $>$ 1\,g\,\cms\, and, thus, have sufficient luminosity and mass to form a 
high-mass protostar (according to the \citealp{Krumholz08} criteria). For the remaining 
cores, it appears that the fragmentation and the formation of lower-mass protostars is 
still possible. If these conditions hold, then these cores may be resolved into multiple 
protostellar condensations in high-angular resolution data. Because the size scales of 
these cores are $\sim$ 0.5 pc and are comparable to the size scales over which a star 
cluster may form rather than the individual stars, it is unclear if these criteria are 
applicable. However, their low $\Sigma$ is consistent with the fact that they will likely 
fragment further. 

\subsection{Evolutionary sequence}

The combination of data sets spanning wavelengths from the millimeter through the 
sub-millimeter and far-IR allows us to investigate the differences between the cores that
show evidence for high-mass star formation activity from those that are more quiescent. 
Because we can derive both masses and temperatures from the data, we can separate 
their effects.  One hypothetical possibility is that the core properties merely reflect an 
evolutionary sequence.  If this idea is correct, then the colder `quiescent'
cores are pre-protostellar and will evolve into the warmer, `active' protostellar cores as 
protostars form within them 
and begin to heat them internally. In this case, only the temperature of the cores should 
vary, since the mass for the entire cluster-forming core is unlikely to change very much 
as it evolves.  Thus, one would expect the active cores to be warmer than the 
quiescent cores, but that their mass distributions should be the same.

An alternative possibility is that the cores that show evidence for star-formation simply
contain more luminous, high-mass protostars, whereas the more quiescent cores contain
only lower mass protostars.  Thus, the lack of obvious star-formation activity in the quiescent  
cores would then result from the difficulty to detect the much fainter lower-mass protostars.  
Since low-mass stars form from lower-mass cores, in this 
scenario, one would expect the quiescent cores to have both lower temperatures {\it {and}} 
smaller masses on average than the active cores. Thus, the data allow us to distinguish between 
these two competing ideas:  evolution and mass.

While the derived dust temperatures and luminosities are higher for the 
cores that show evidence for star-formation compared to those that are more quiescent, 
the mass distributions for the two populations are similar.  In fact, the KS test shows that the probability 
that the active and quiescent cores are derived from the same parent population is $\sim$40\%. 
The similarity in their mass distributions suggests that the underlying populations are nearly the 
same, and that the differences in their temperature and luminosity reflect different 
evolutionary stages. If this is the case, then the quiescent cores have enough mass to form 
a high-mass protostar, however, they simply have not yet begun the process. Such cores 
should still be cold, and since no high-mass protostars have formed yet, also have low 
bolometric luminosities.  In contrast, the active cores contain an internal protostellar heating source(s)
which results in their higher dust  temperatures and luminosities. 

\begin{figure}
\begin{center}
\includegraphics[width=0.7\textwidth,clip=true]{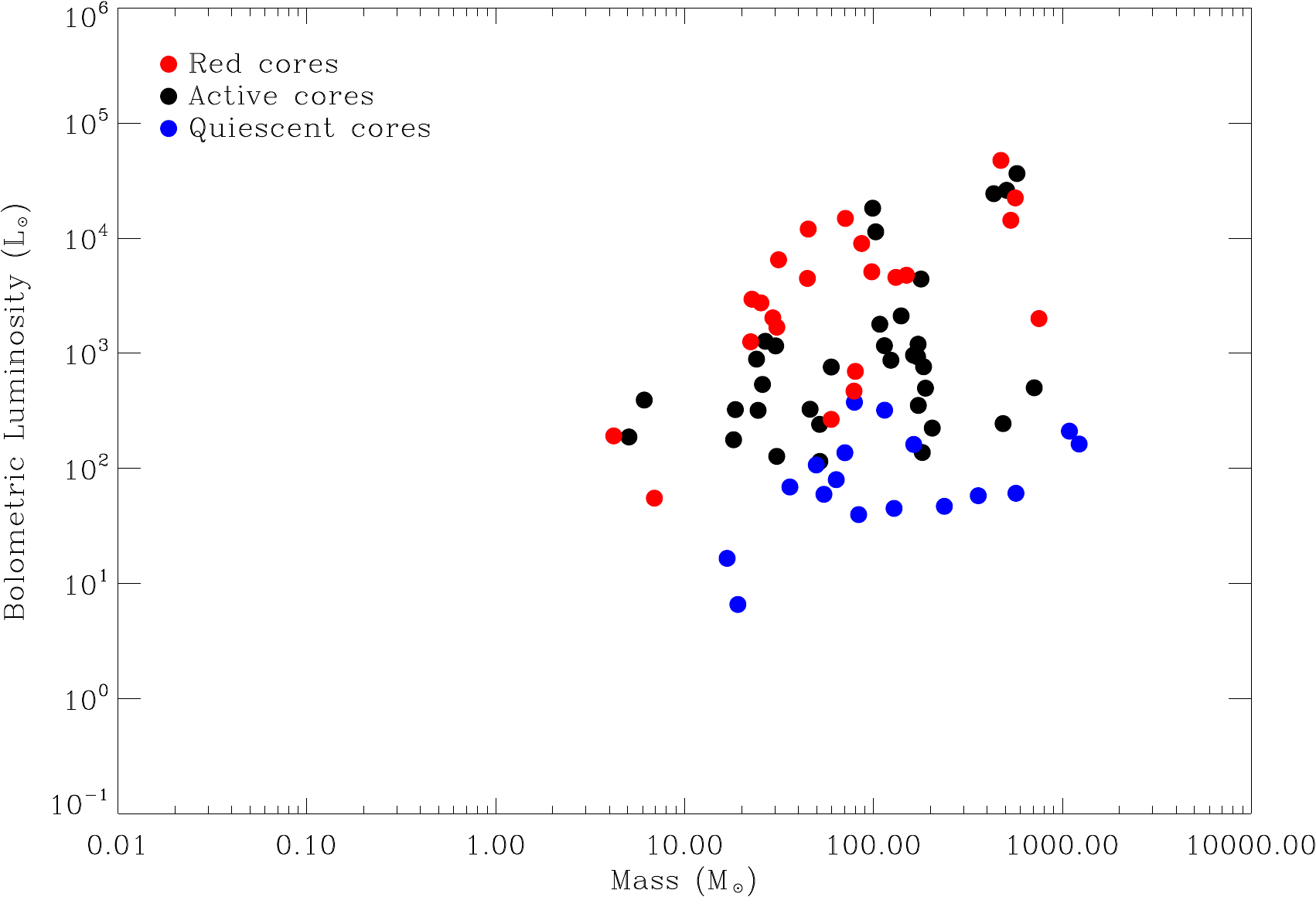}
\end{center}
\caption{\label{LM}Plot of bolometric luminosity versus mass for the cores. We find  
that for a given mass the luminosity increases from the quiescent, to the 
active, to the red cores. This is consistent with the idea that the quiescent cores are 
in an earlier evolutionary phase compared to the active and red cores.}
\end{figure}

The evolutionary differences between the samples of cores can also be seen in a 
bolometric luminosity versus core dust mass diagram (e.g., Fig.~\ref{LM}; \citealp{Sridharan02,Molinari08}). 
For both the low- and high-mass regimes, sources in different evolutionary 
phases have been shown to lie in distinct regions within this diagram. As the protostellar activity
increases, the luminosity also increases such that the different stages within the star formation
process occupy regions that overlap in mass but are offset to higher luminosities. 
Similar to the results of \cite{Molinari08}, we find that these IRDC cores lie in the high-mass regime and that,
for a given mass, the bolometric luminosities increase from the quiescent, to the active cores, to the red cores. 
This is consistent with the idea that the quiescent cores are in an earlier evolutionary phase 
compared to the active and red cores. 


\section{Conclusions}

To characterize the physical properties of cores within IRDCs we have 
obtained new 24\,\um, 60--100\,\um, and sub-millimeter continuum data toward a 
sample of 38 IRDCs. These IRDCs contain 190 compact cores, 140 of which are 
dark at 8\,\um\, and are cold, compact and dense. The \Spitzer/MIPS 24\,\um\, 
images reveal that while the IRDCs remain dark, many of their cores are 
associated with bright 24\,\um\, emission sources. The sub-millimeter 
continuum data elucidate both the large- and small-scale structure of the IRDCs. 
Because emission at millimeter/sub-millimeter wavelengths can trace either 
temperature and/or density enhancements one needs an additional method to 
distinguish between these two parameters. Using the presence or absence of 24\,\um\, point
source emission, in combination with their \Spitzer/IRAC 
3--8\,\um\, colors,  we have classified the cores into five groups: red, active, intermediate,
quiescent, and blue. We find that, of the 190 cores, 35 can be classified as red, 38 as active, 32 as intermediate, 79 as quiescent, and  6
as blue.

From gray-body fits to their spectral energy distributions (SEDs) we have determined 
the dust temperatures, emissivities, opacities,  bolometric luminosities, and masses 
for a large sample of the IRDC cores.  The derived distributions of the dust 
temperatures, luminosities, and masses for the different groups of cores
reveals that the dust temperatures and luminosities are higher for those cores 
that show active, high-mass star formation compared to those cores that are more
quiescent. Lower dust temperatures and luminosities are expected for the quiescent cores because they 
presumably have no high-mass internal source to significantly heat the 
dust. Comparing the derived masses for the core samples, however, we find that their 
mass distributions are similar. We interpret this similarity to be a result of evolutionary 
differences: the cooler quiescent cores may be the pre-protostellar precursors to the warmer, more
active protostellar cores. 

Using their derived bolometric luminosities, we estimate that $\sim$10\%  of the 
cores that show evidence for star formation may contain high-mass protostars. 
If the quiescent cores are indeed devoid of star formation, then the 
most massive of these are excellent candidates for the `high-mass starless core' phase, 
a very early phase in the  formation of a high-mass star. Because of their distances, we 
cannot yet rule out the possibility that many of these cores may contain low-mass stars.
Observations with ALMA will be crucial to address this issue.  Nevertheless, 
our study supports the idea that IRDCs harbor the very earliest evolutionary stages in the formation of 
high-mass stars and, thus,  clusters.

\acknowledgments
The authors gratefully acknowledge funding support through NASA grant NNG04GGC92G and
NSF grant AST0808001.
This work is based in part on observations made with the {\it Spitzer Space Telescope}, 
which is operated by the Jet Propulsion Laboratory, California Institute of Technology 
under NASA contract 1407. Support for this work was provided by NASA through 
contract 1267945 issued by JPL/Caltech. The JCMT is operated by JAC, Hilo, 
on behalf of the parent organizations of the Particle Physics and Astronomy Research 
Council in the UK, the National Research Council in Canada, and the Scientific Research 
Organization of the Netherlands. IRAM is supported by INSU/CNRS (France), MPG (Germany), 
and IGN (Spain). The CSO telescope is operated by Caltech under a contract from the 
National Science Foundation (NSF). 

\appendix
\section*{Appendix}
\setcounter{section}{1}

Figures~\ref{seds-27}--\ref{seds-34} show the 24\,\um\, image toward the IRDCs overlaid 
with the 1.2\,mm continuum emission. Also included in these figures are the SEDs and gray-body fits
for each of the cores that had sufficient data. Table~\ref{cores} list the coordinates, distances, and fluxes measured toward all cores. Marked on the SEDs and listed in 
Table~\ref{core-properties-good} and Table~\ref{core-properties-limits} are the derived parameters.


\begin{table}
\centering
{\scriptsize{
\caption{\label{obs-summary}Summary of the observations.}
\begin{tabular}{ccccccc}
\tableline \tableline
Telescope & Instrument & Wavelength      & Date           & Angular              &1 $\sigma$  \\
          &            &                 &                & Resolution & sensitivity\\
\tableline
\Spitzer\, 0.85\,m & MIPS           & 24\,\um              & 2004 Oct, 2005 Apr                      & 6\,\arcsec                  & $\sim$150\,$\mu$Jy\\
\Spitzer\, 0.85\,m & SED-mode  & 60--100\,\um     & 2006 Oct, 2007 May                    & 13--24\,\arcsec               & $\sim$100\,mJy\\
CSO 10\,m            & SHARC-2    & 350, 450\,\um    & 2005 Sept, 2005 Apr, 2006 Apr  & 8\,\arcsec                  & $\sim$\,200 mJy\\
JCMT 15 m            & SCUBA        & 450, 850\,\um    & 2004 Sept                                      & 8, 15\,\arcsec            & $\sim$\,60 mJy\\
IRAM 30\,m\tablenotemark{a} & MAMBO-II  & 1.2\,mm & 2004 Feb                                & 11\,\arcsec                & $\sim$\,10 mJy\\
\tableline
\end{tabular}\vspace{-0.5cm}
\tablenotetext{a}{These data were presented in \cite{Rathborne06}.}}}
\end{table}

\begin{table}
\centering
{\scriptsize{
\caption{\label{summary}Summary of the derived core properties}
\begin{tabular}{lcccccccc}
\tableline \tableline
Property & \multicolumn{2}{c}{Red} &\multicolumn{2}{c}{Active}&\multicolumn{2}{c}{Intermediate} &\multicolumn{2}{c}{Quiescent} \\
         & Median & Stddev & Median & Stddev & Median & Stddev & Median & Stddev \\
\tableline
T$_{D}$ (K)  &    40.4  &     5.7  &    34.5  &     6.5  &    30.4  &     8.1  &    23.7  &     5.3   \\
Log[$\tau_{250}$]  &  -1.894  &   0.575  &  -2.081  &   0.550  &  -2.456  &   0.628  &  -2.404  &   0.501   \\
$\beta$  &     1.7  &     0.4  &     1.6  &     0.4  &     1.4  &     0.5  &     1.2  &     0.5   \\
Log[Mass] (\Msun)  &    1.85  &    0.59  &    2.06  &    0.54  &    1.97  &    0.61  &    1.92  &    0.55   \\
Log[Lum] (\Lsun)  &    3.65  &    0.72  &    2.88  &    0.69  &    2.27  &    0.79  &    1.84  &    0.44   \\
$\Sigma$ (g\,\cms)  &   -0.39  &    0.39  &   -0.15  &    0.34  &   -0.29  &    0.42  &   -0.25  &    0.40   \\
Log[N(\hh)] (\cms)  &   22.05  &    0.42  &   22.10  &    0.43  &   22.03  &    0.36  &   22.01  &    0.29   \\
Log[n(\hh)] (\cmc)  &    5.82  &    0.38  &    5.98  &    0.32  &    5.81  &    0.42  &    6.06  &    0.39   \\
\tableline
\end{tabular}}}
\end{table}


\clearpage 
\clearpage 
\clearpage 
\clearpage 
\clearpage 
\clearpage 
\clearpage 
\clearpage 
\clearpage 
\clearpage 

\renewcommand{\thefigure}{A-\arabic{figure}}
\setcounter{figure}{0}  

\clearpage 
\begin{figure}
\begin{center}
\includegraphics[angle=0,width=0.5\textwidth]{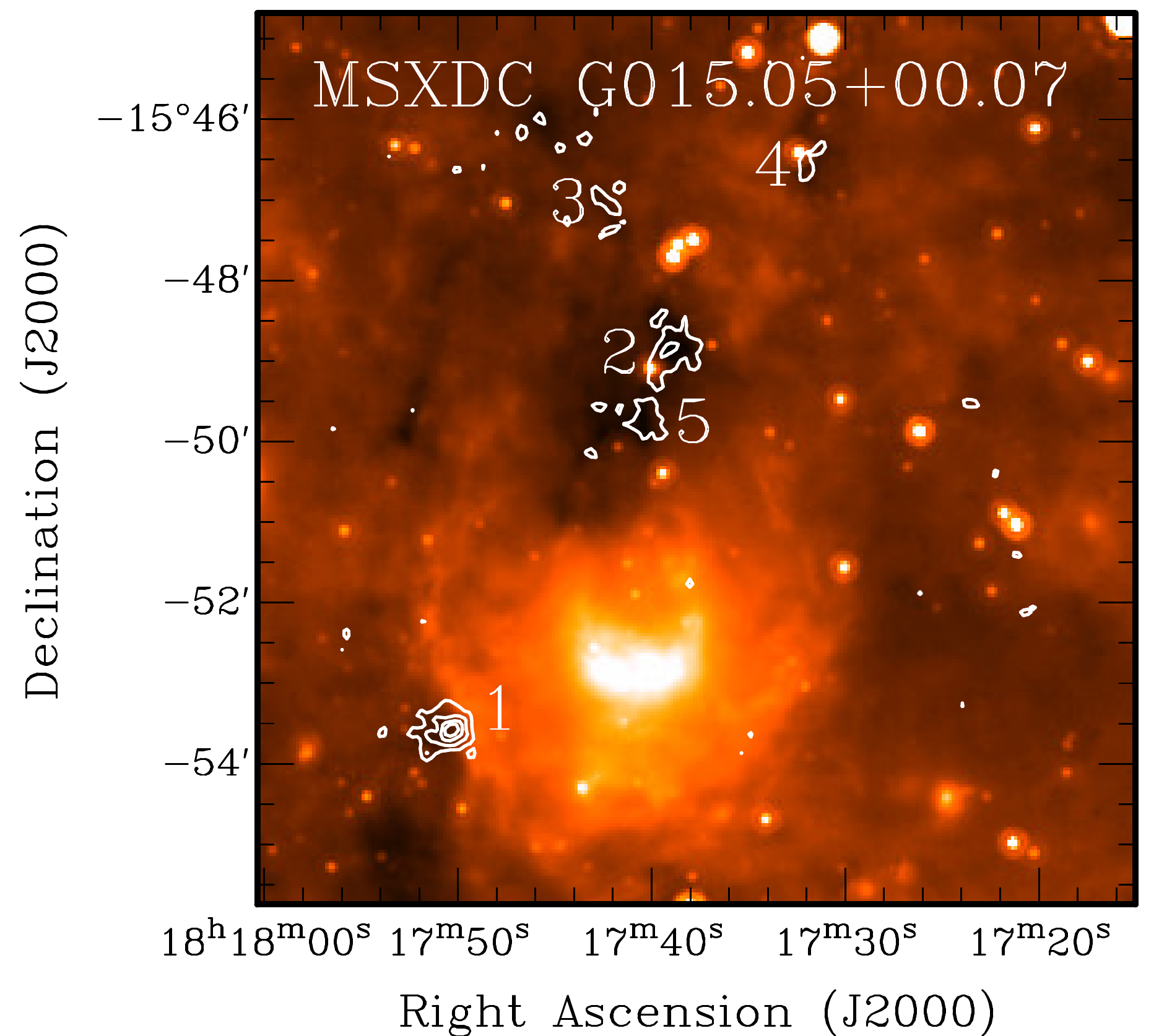}\\
\end{center}
\includegraphics[angle=90,width=0.5\textwidth]{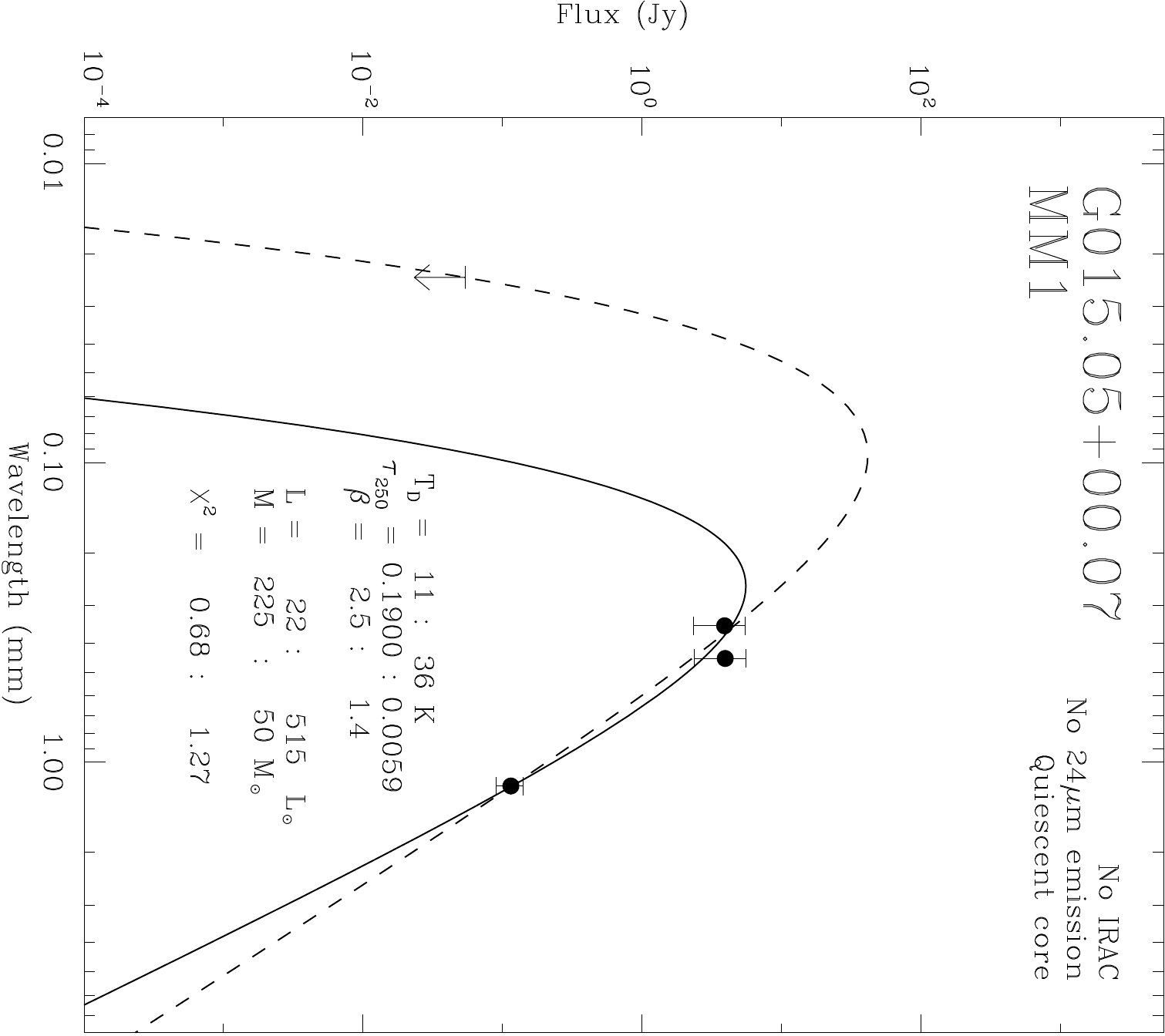}
\includegraphics[angle=90,width=0.5\textwidth]{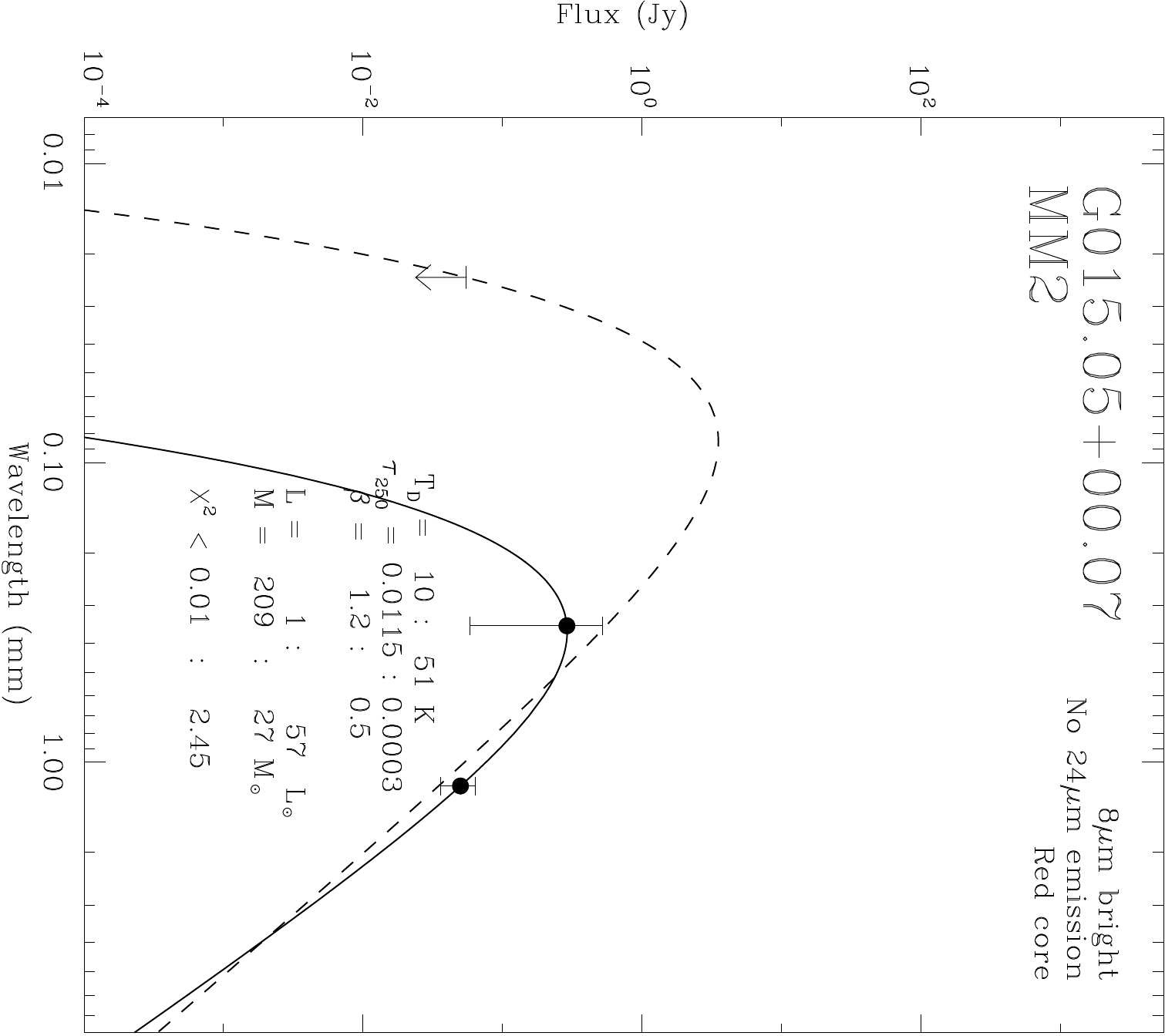}\\
\end{figure}
\clearpage 
\begin{figure}
\includegraphics[angle=90,width=0.5\textwidth]{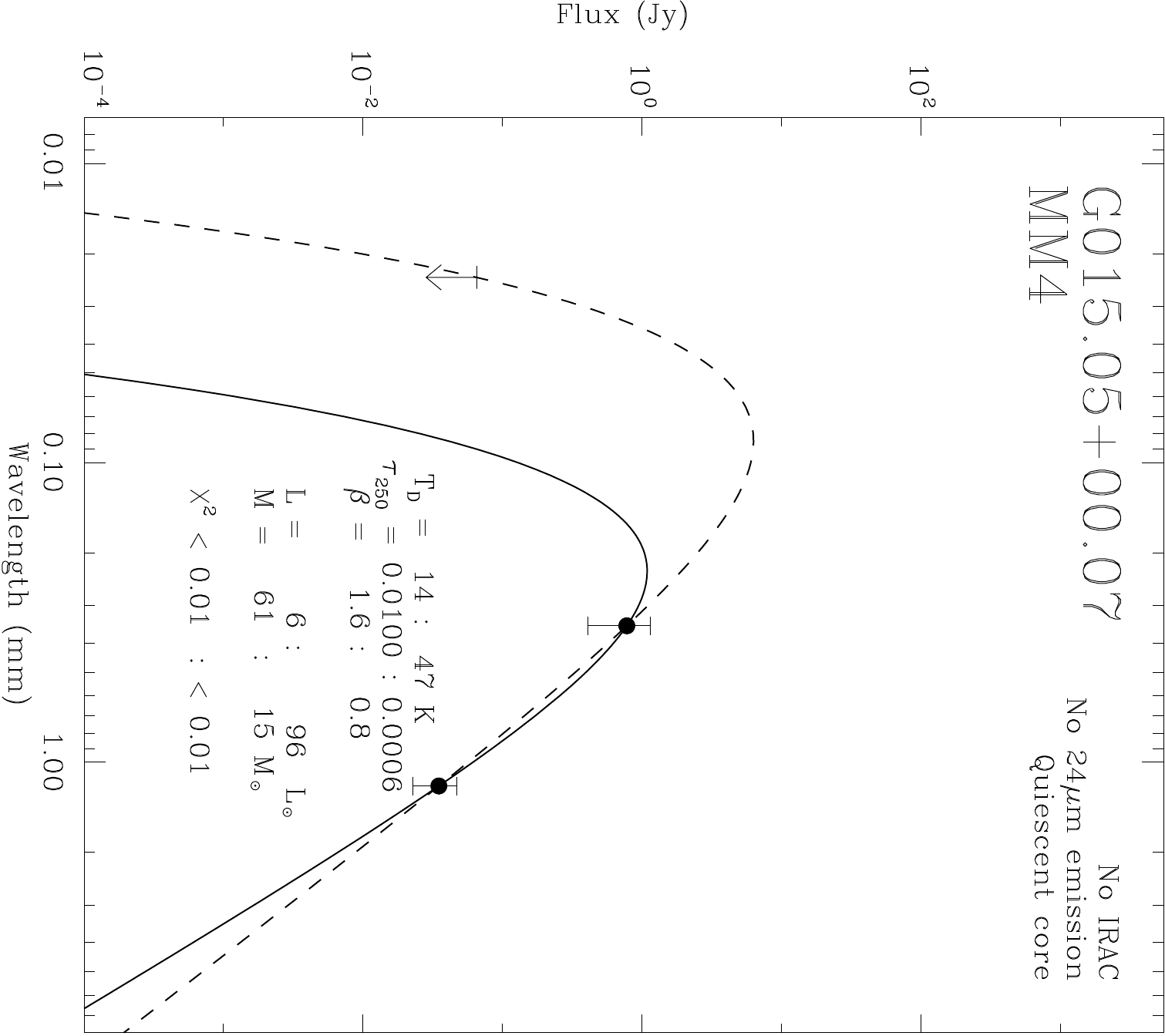}
\includegraphics[angle=90,width=0.5\textwidth]{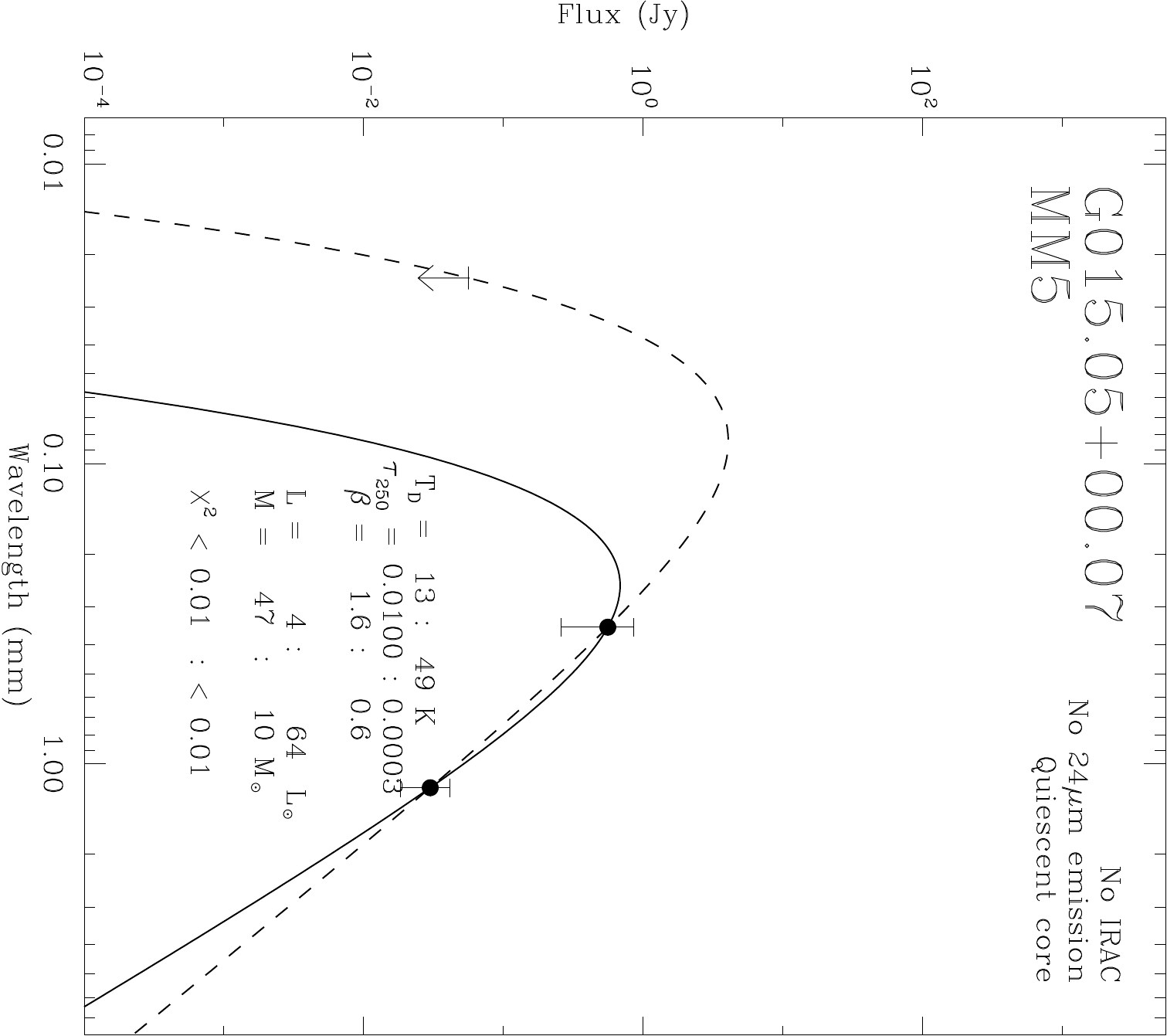}\\
\caption{\label{seds-27} \Spitzer\, 24\,\um\, image overlaid  
   with 1.2\,mm continuum emission for \irdctwentyseven\, (contour
   levels are 30 and 60\,mJy beam$^{-1}$). The lower panels show the broadband
   SEDs for cores within this IRDC.  The fluxes derived from the
   millimeter, sub-millimeter, and far-IR  continuum data are shown as filled
   circles (with the corresponding error bars), while the 24\,\um\, fluxes are shown as  either a filled circle (when included within the fit), an open circle (when excluded from the fit),  or as an upper limit arrow. For cores that have measured fluxes only in the millimeter/sub-millimeter regime (i.e.\, a limit at 24\,\um), we show the results from two fits: one using only the measured fluxes (solid line; lower limit), while the other includes the 24\,\um\, limit as a real data (dashed line; upper limit). In all other cases, the solid line is the best fit gray-body, while the dotted lines correspond to the functions determined using the errors for the T$_{D}$, $\tau$, and $\beta$ output from the fitting.  Labeled on each plot is the IRDC and core name,  classification, and the derived parameters.}
\end{figure}
\clearpage 
\begin{figure}
\begin{center}
\includegraphics[angle=0,width=0.6\textwidth]{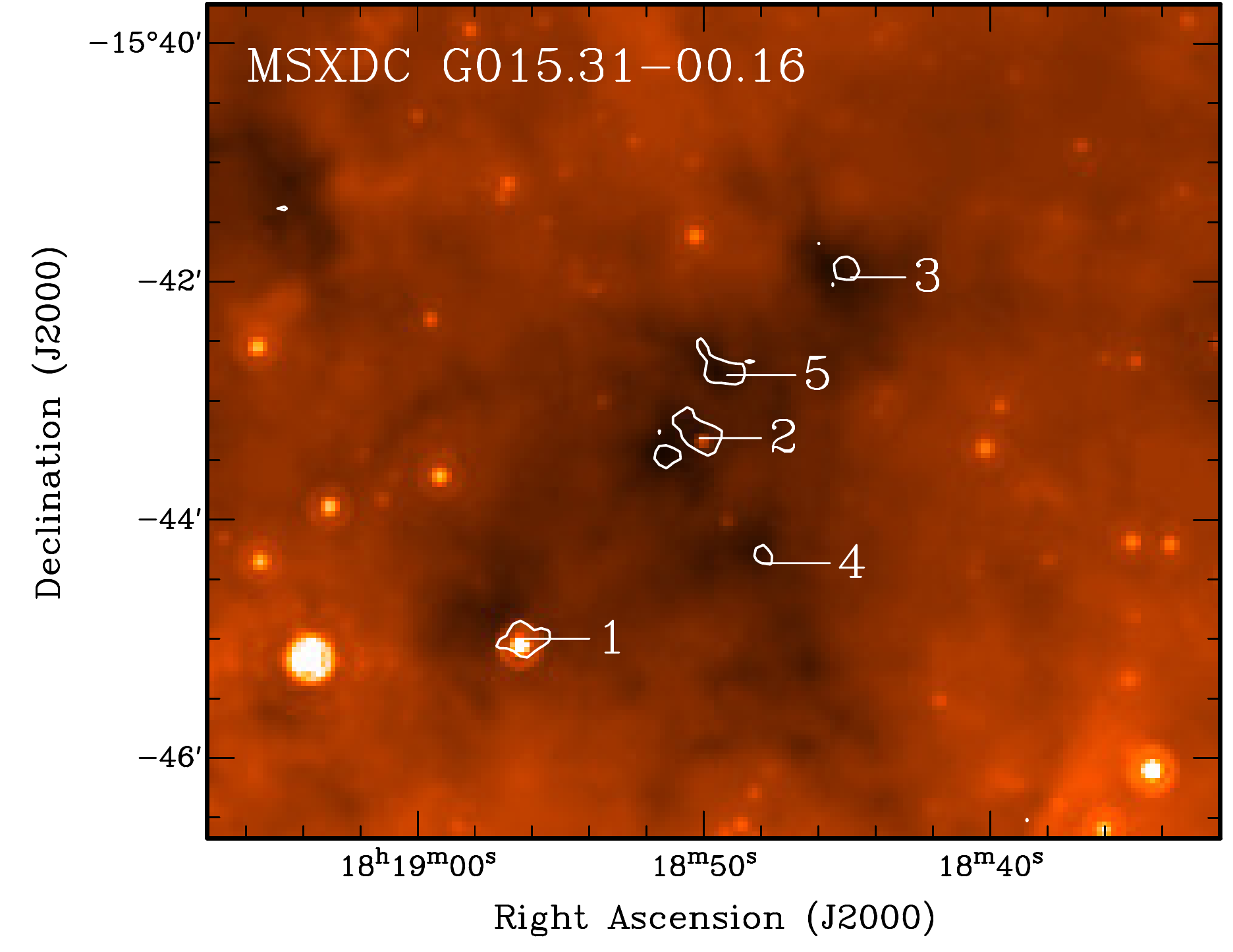}\\
\end{center}
\includegraphics[angle=90,width=0.5\textwidth]{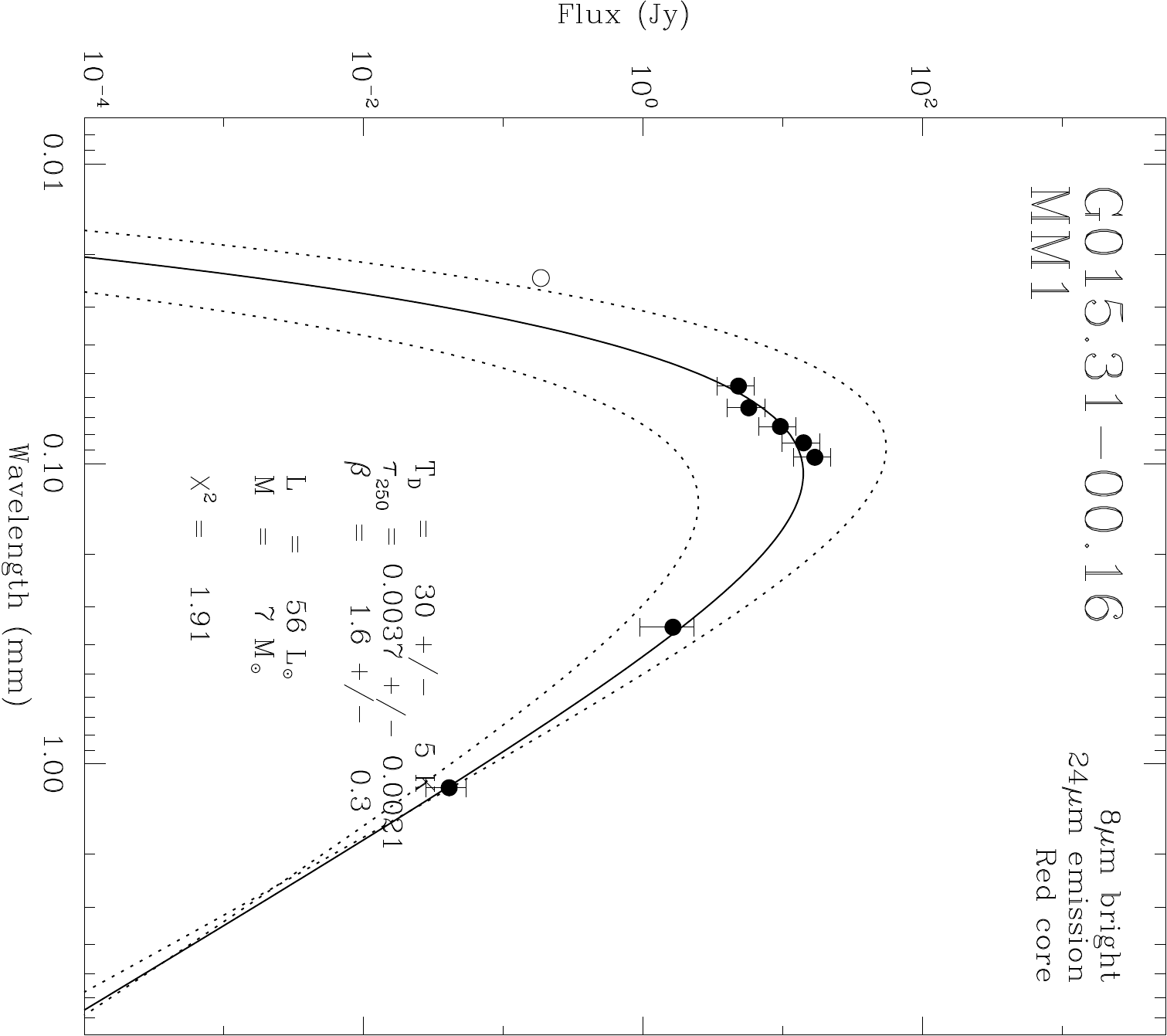}
\includegraphics[angle=90,width=0.5\textwidth]{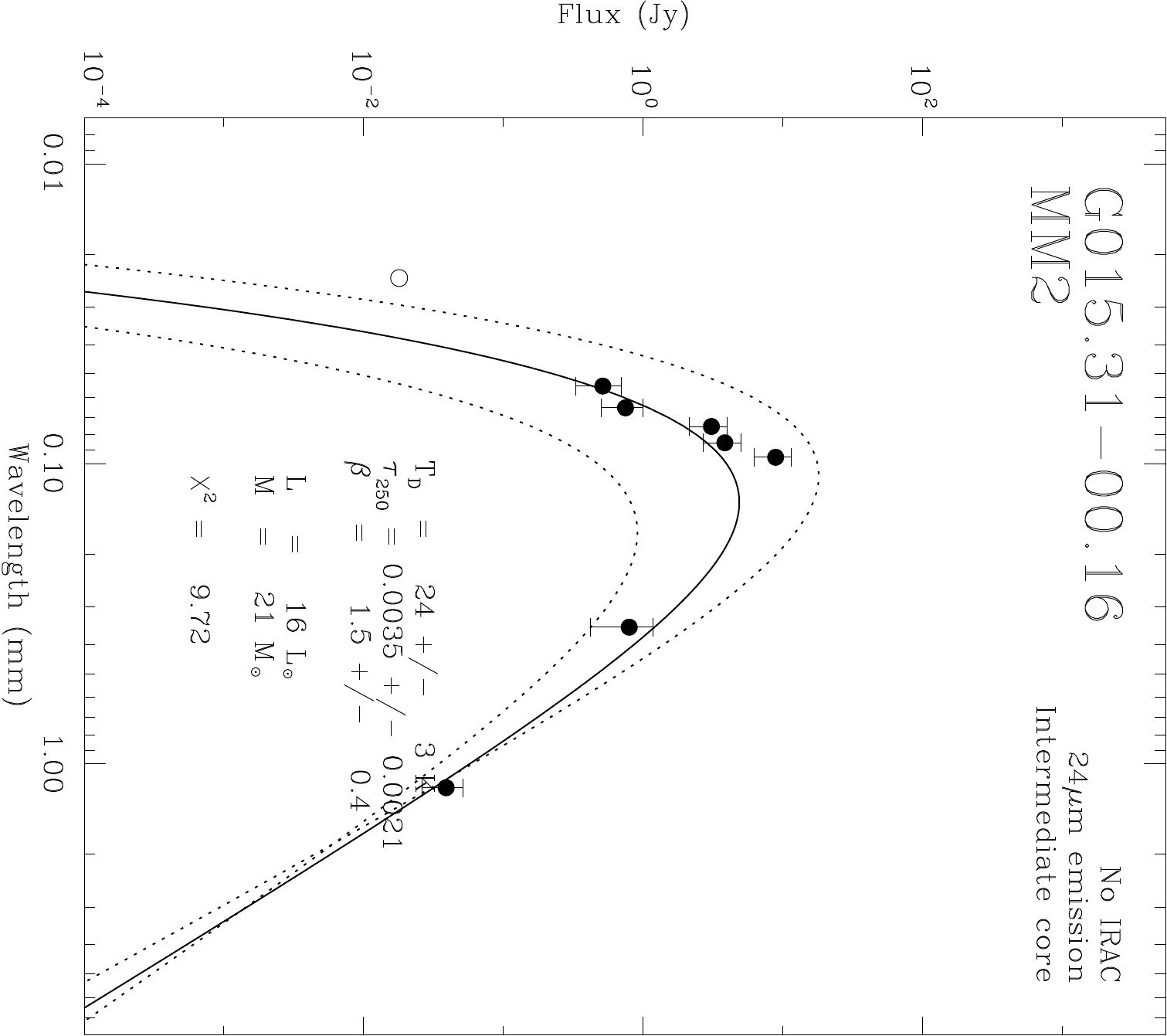}\\
\end{figure}
\clearpage 
\begin{figure}
\includegraphics[angle=90,width=0.5\textwidth]{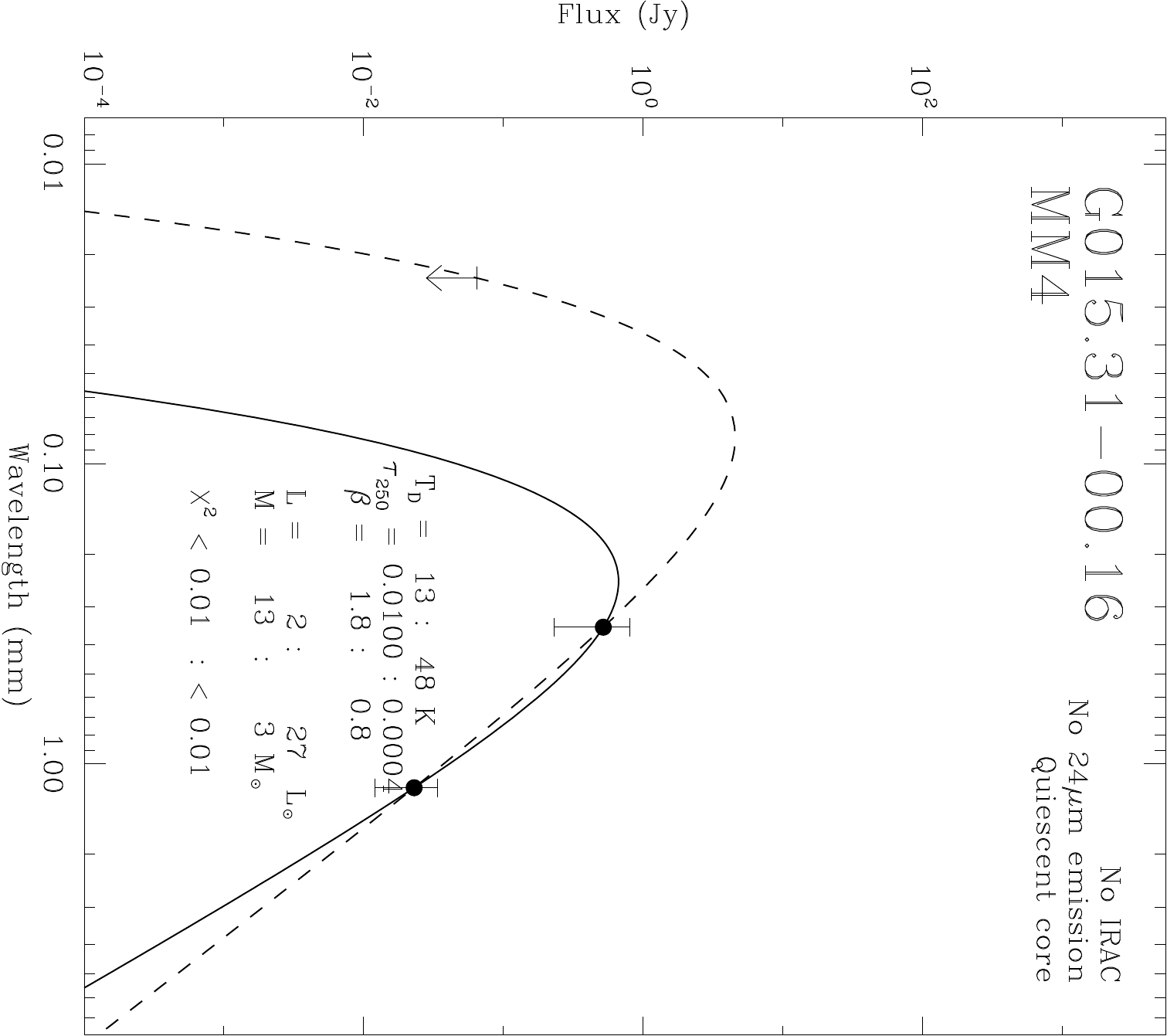}
\includegraphics[angle=90,width=0.5\textwidth]{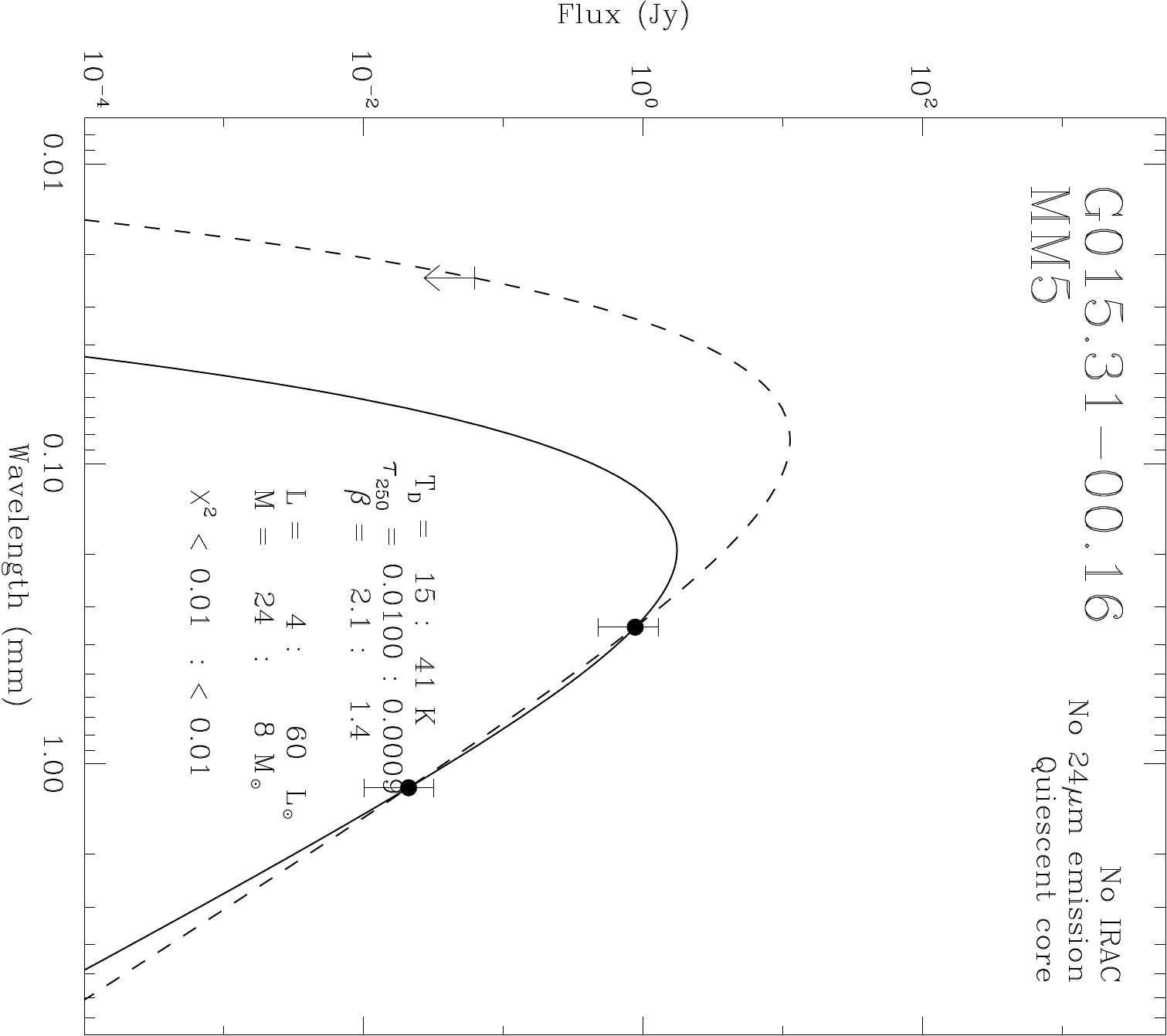}\\
\caption{\label{seds-15} \Spitzer\, 24\,\um\, image overlaid  
   with 1.2\,mm continuum emission for \irdcfifteen\, (contour
   level is 30\,mJy beam$^{-1}$).   The lower panels show the broadband
   SEDs for cores within this IRDC.  The fluxes derived from the
   millimeter, sub-millimeter, and far-IR  continuum data are shown as filled
   circles (with the corresponding error bars), while the 24\,\um\, fluxes are shown as  either a filled circle (when included within the fit), an open circle (when excluded from the fit),  or as an upper limit arrow. For cores that have measured fluxes only in the millimeter/sub-millimeter regime (i.e.\, a limit at 24\,\um), we show the results from two fits: one using only the measured fluxes (solid line; lower limit), while the other includes the 24\,\um\, limit as a real data (dashed line; upper limit). In all other cases, the solid line is the best fit gray-body, while the dotted lines correspond to the functions determined using the errors for the T$_{D}$, $\tau$, and $\beta$ output from the fitting.  Labeled on each plot is the IRDC and core name,  classification, and the derived parameters.}
\end{figure}
\clearpage 
\begin{figure}
\begin{center}
\includegraphics[angle=0,width=0.6\textwidth]{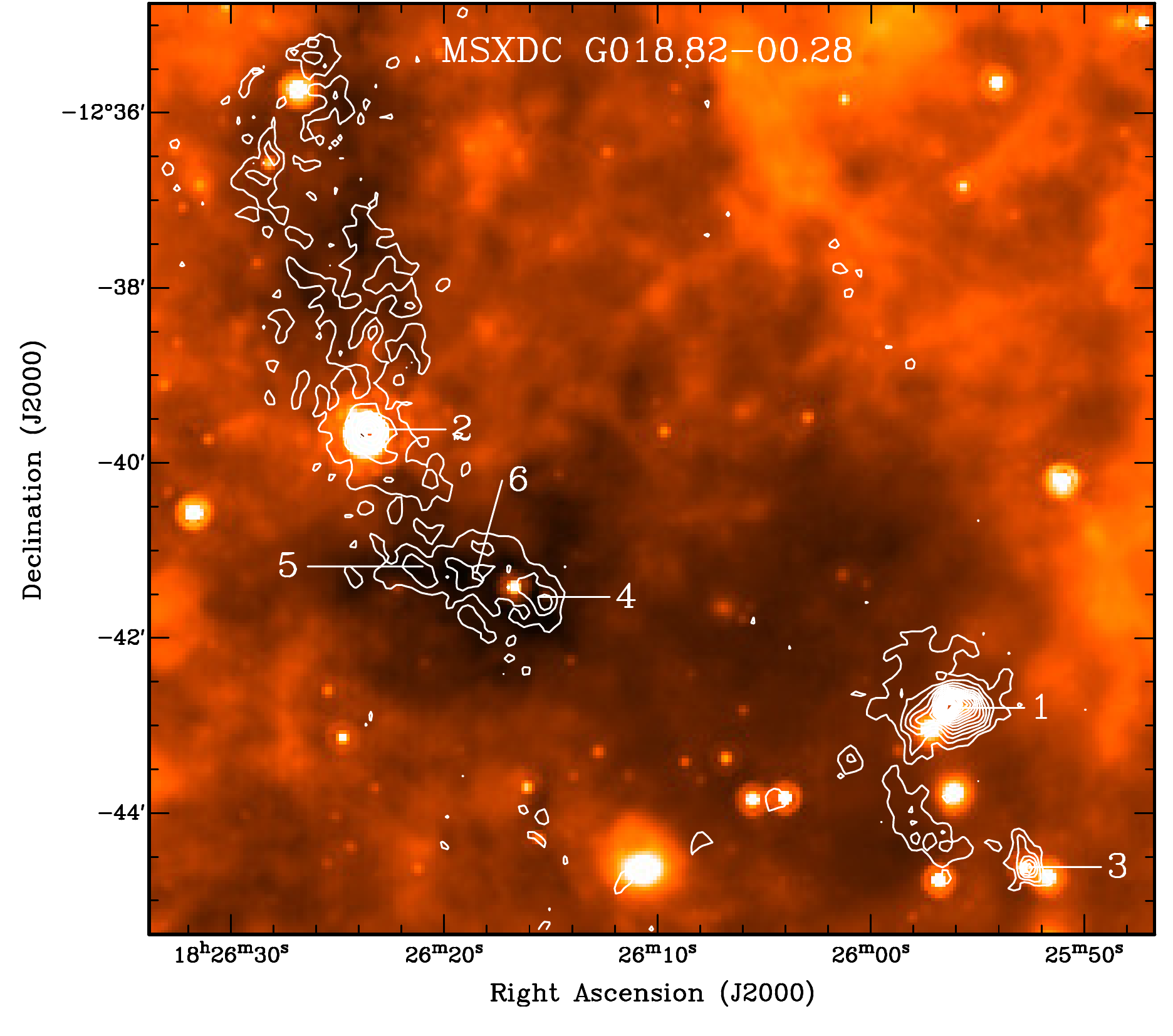}\\
\end{center}
\includegraphics[angle=90,width=0.5\textwidth]{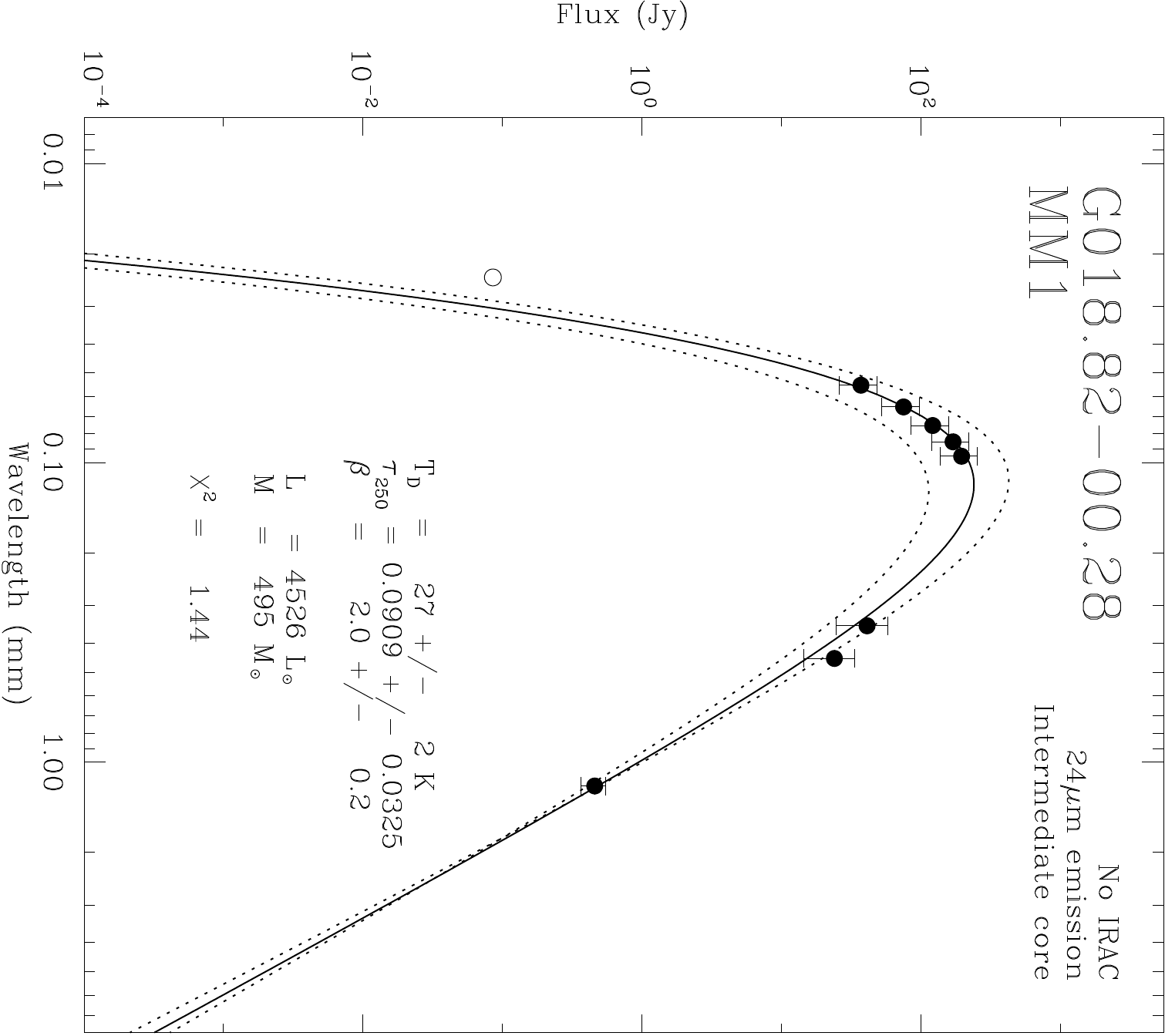}
\includegraphics[angle=90,width=0.5\textwidth]{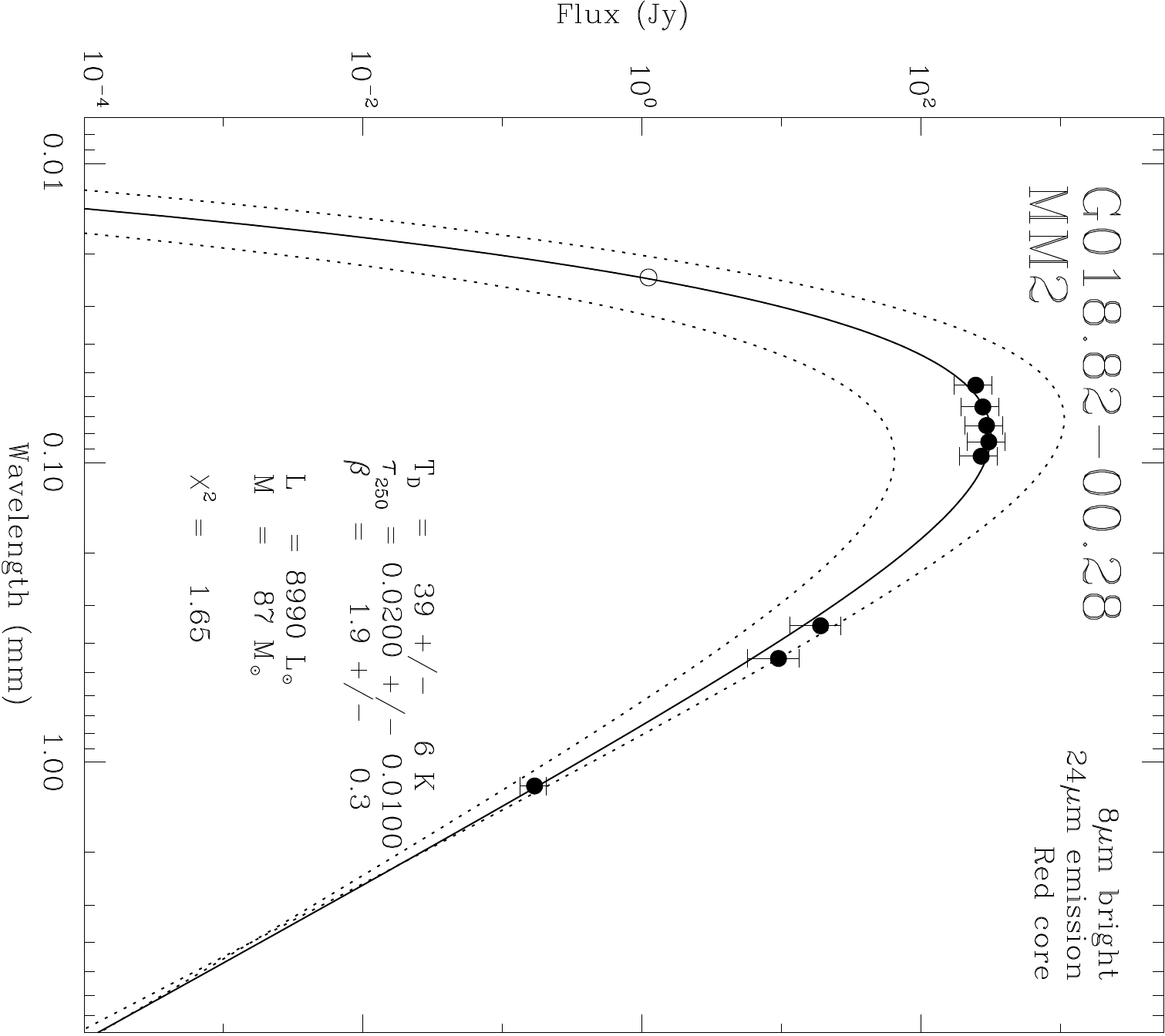}\\
\end{figure}
\clearpage 
\begin{figure}
\includegraphics[angle=90,width=0.5\textwidth]{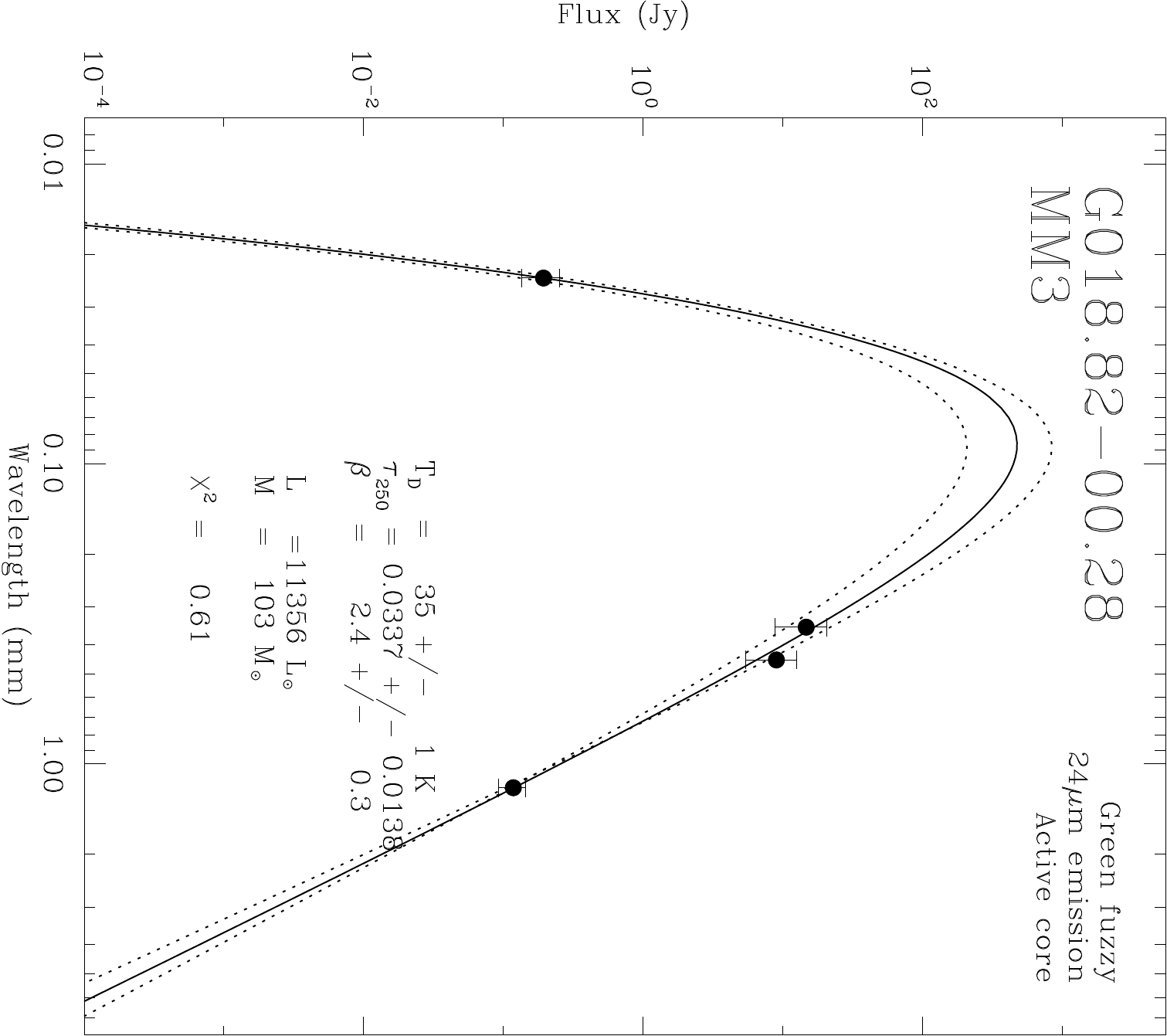}
\includegraphics[angle=90,width=0.5\textwidth]{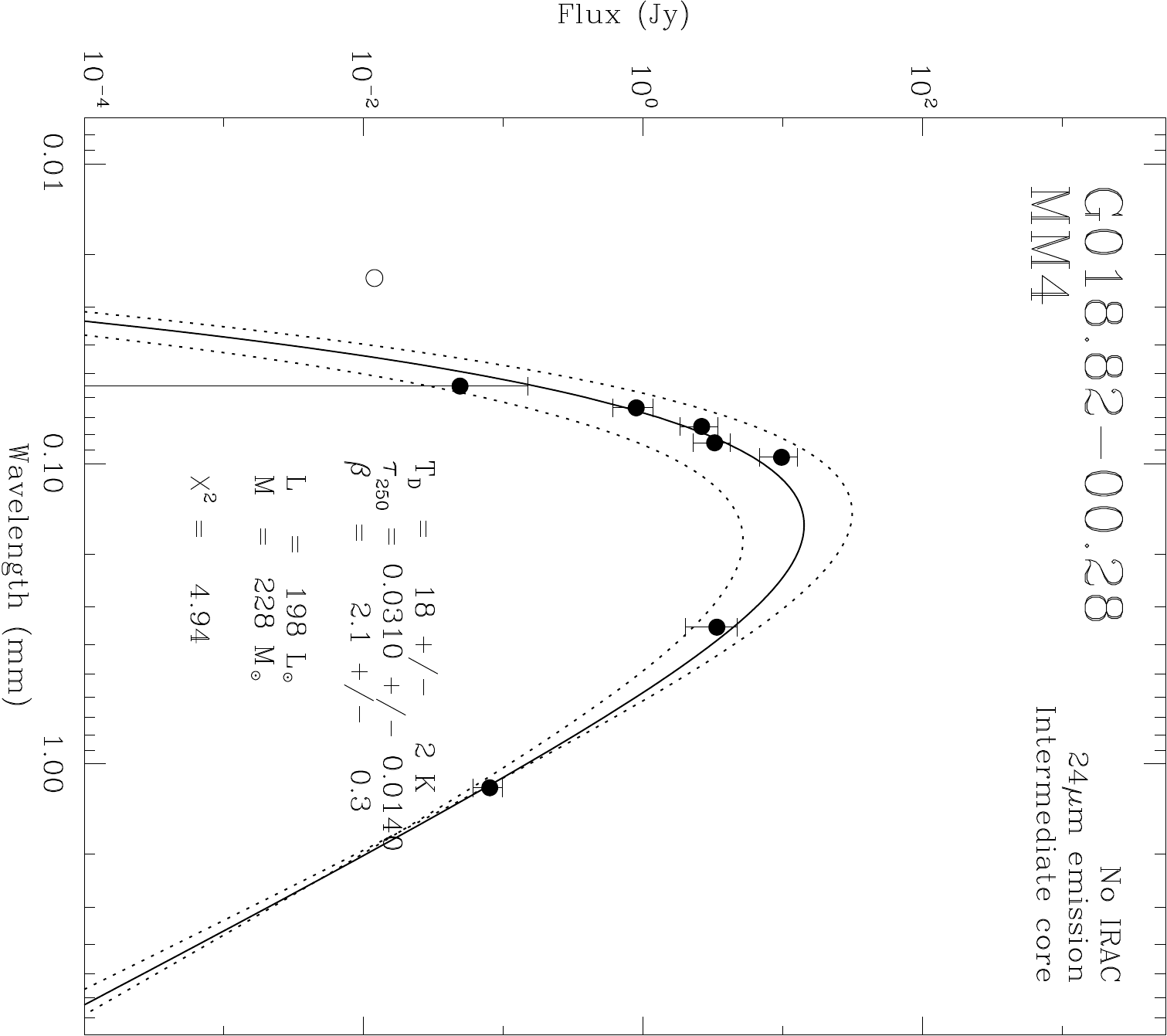}\\
\includegraphics[angle=90,width=0.5\textwidth]{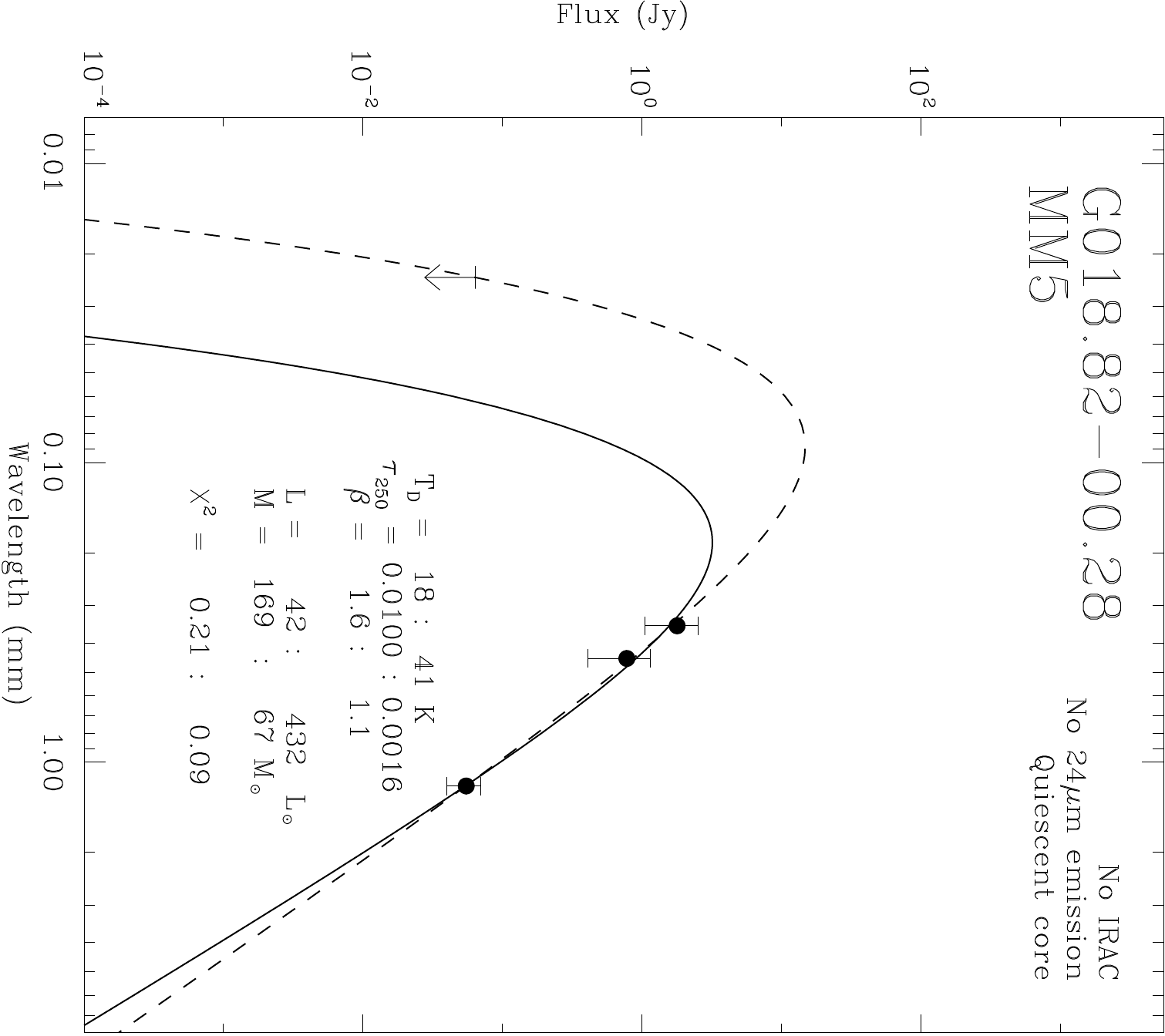}\\
\caption{\label{seds-16} \Spitzer\, 24\,\um\, image overlaid  
   with 1.2\,mm continuum emission for \irdcsixteen\, (contour levels
   are 30, 60, 90, 120, 240\,mJy beam$^{-1}$). The lower panels show the broadband
   SEDs for cores within this IRDC.  The fluxes derived from the
   millimeter, sub-millimeter, and far-IR  continuum data are shown as filled
   circles (with the corresponding error bars), while the 24\,\um\, fluxes are shown as  either a filled circle (when included within the fit), an open circle (when excluded from the fit),  or as an upper limit arrow. For cores that have measured fluxes only in the millimeter/sub-millimeter regime (i.e.\, a limit at 24\,\um), we show the results from two fits: one using only the measured fluxes (solid line; lower limit), while the other includes the 24\,\um\, limit as a real data (dashed line; upper limit). In all other cases, the solid line is the best fit gray-body, while the dotted lines correspond to the functions determined using the errors for the T$_{D}$, $\tau$, and $\beta$ output from the fitting.  Labeled on each plot is the IRDC and core name,  classification, and the derived parameters.}
\end{figure}
\clearpage 
\begin{figure}
\begin{center}
\includegraphics[angle=0,width=0.6\textwidth]{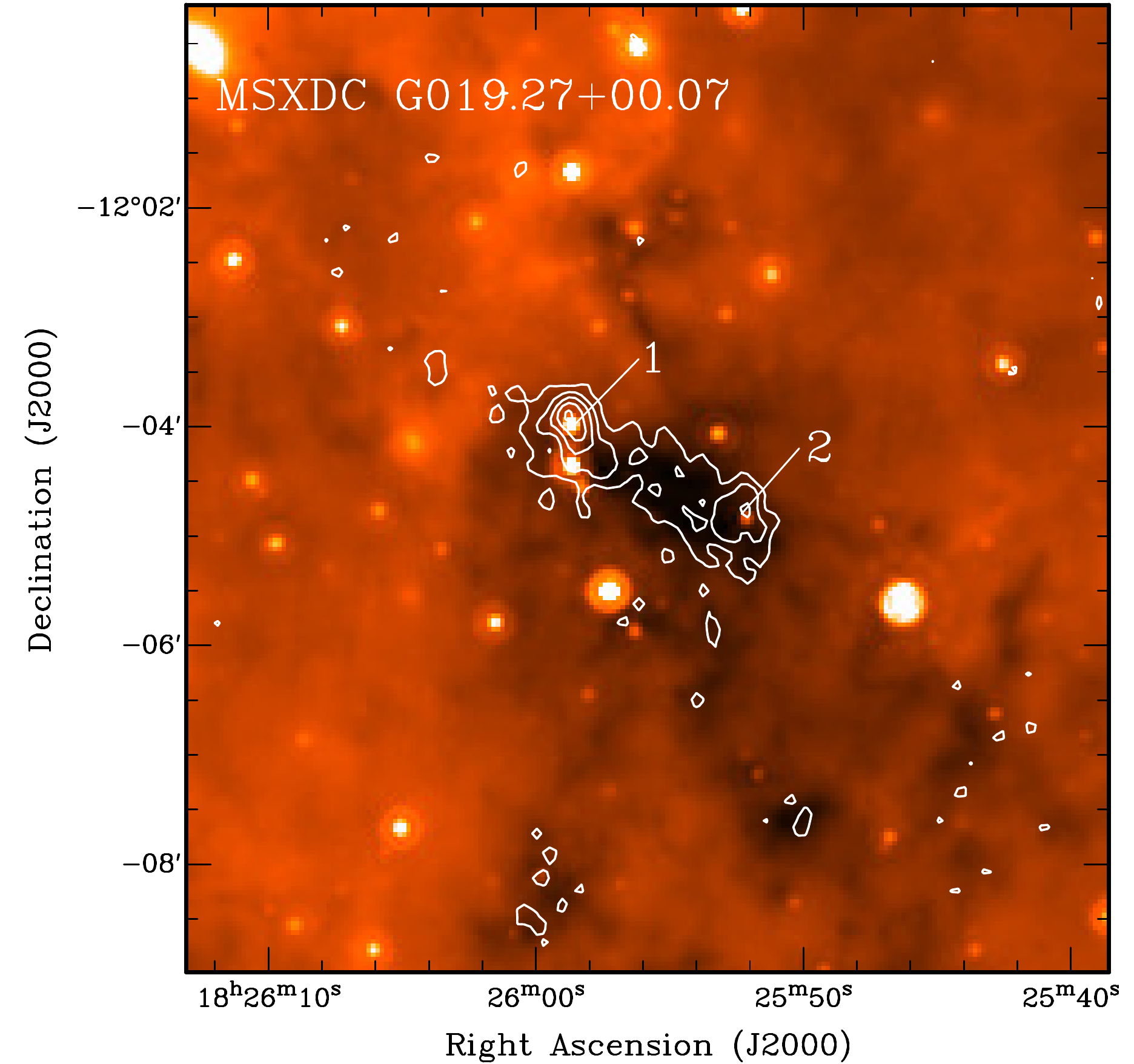}\\
\end{center}
\includegraphics[angle=90,width=0.5\textwidth]{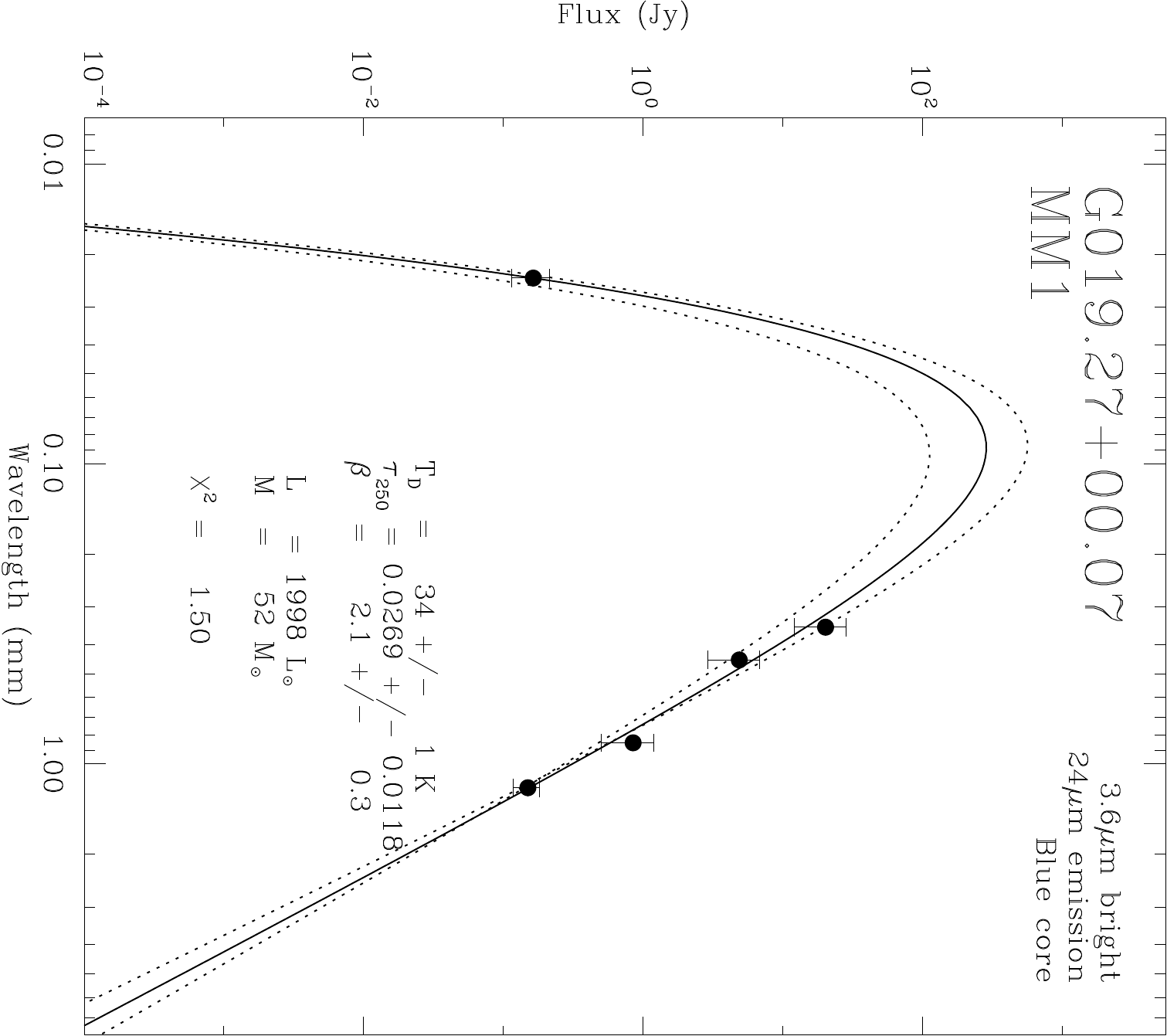}
\includegraphics[angle=90,width=0.5\textwidth]{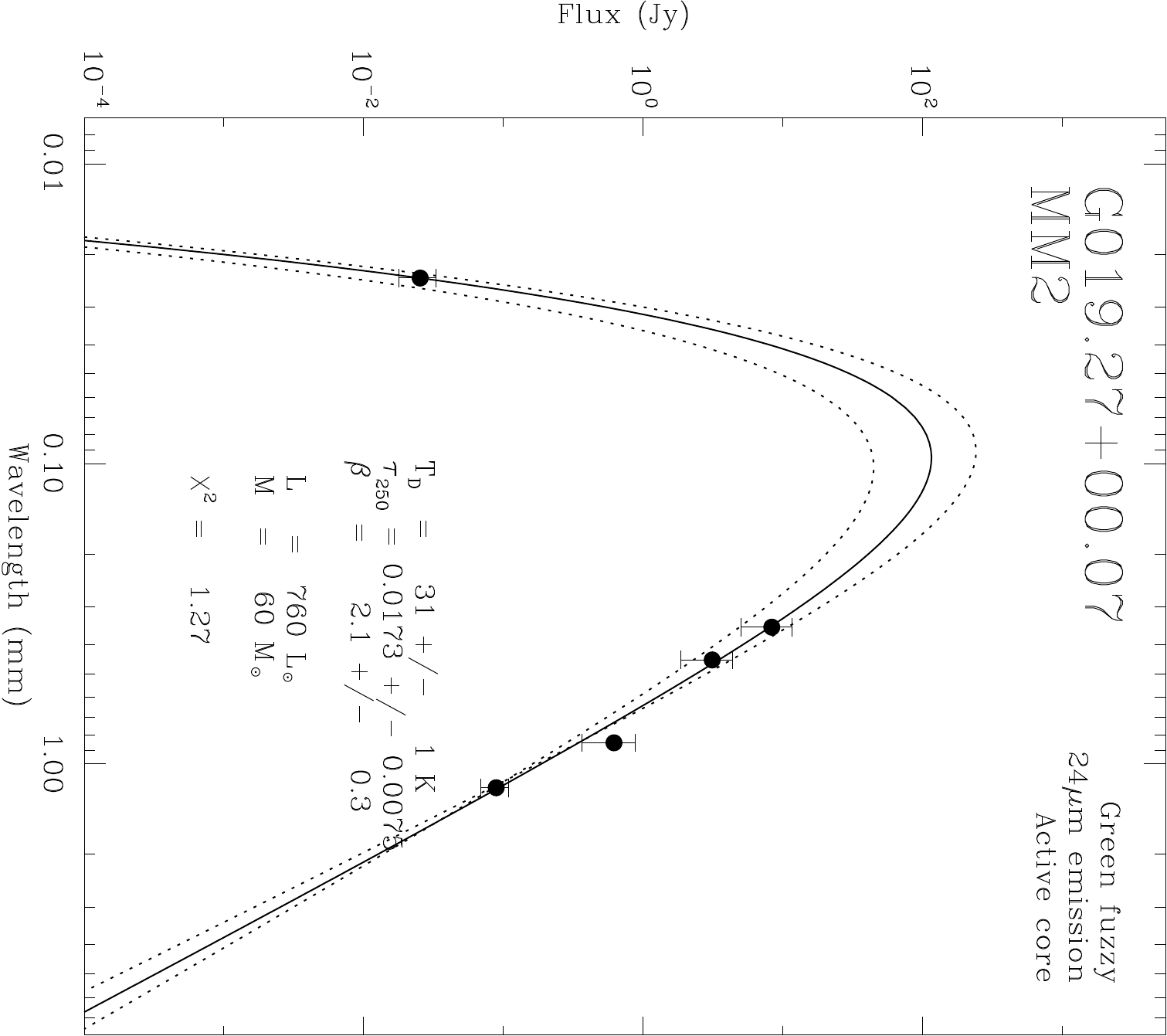}\\
\caption{\label{seds-7} \Spitzer\, 24\,\um\, image overlaid  
   with 1.2\,mm continuum emission for \irdcseven\, (contour levels
   are 30, 60, 90, 120, 240\,mJy beam$^{-1}$). The lower panels show the broadband
   SEDs for cores within this IRDC.  The fluxes derived from the
   millimeter, sub-millimeter, and far-IR  continuum data are shown as filled
   circles (with the corresponding error bars), while the 24\,\um\, fluxes are shown as  either a filled circle (when included within the fit), an open circle (when excluded from the fit),  or as an upper limit arrow. For cores that have measured fluxes only in the millimeter/sub-millimeter regime (i.e.\, a limit at 24\,\um), we show the results from two fits: one using only the measured fluxes (solid line; lower limit), while the other includes the 24\,\um\, limit as a real data (dashed line; upper limit). In all other cases, the solid line is the best fit gray-body, while the dotted lines correspond to the functions determined using the errors for the T$_{D}$, $\tau$, and $\beta$ output from the fitting.  Labeled on each plot is the IRDC and core name,  classification, and the derived parameters.}
\end{figure}
\clearpage 
\begin{figure}
\begin{center}
\includegraphics[angle=0,width=0.6\textwidth]{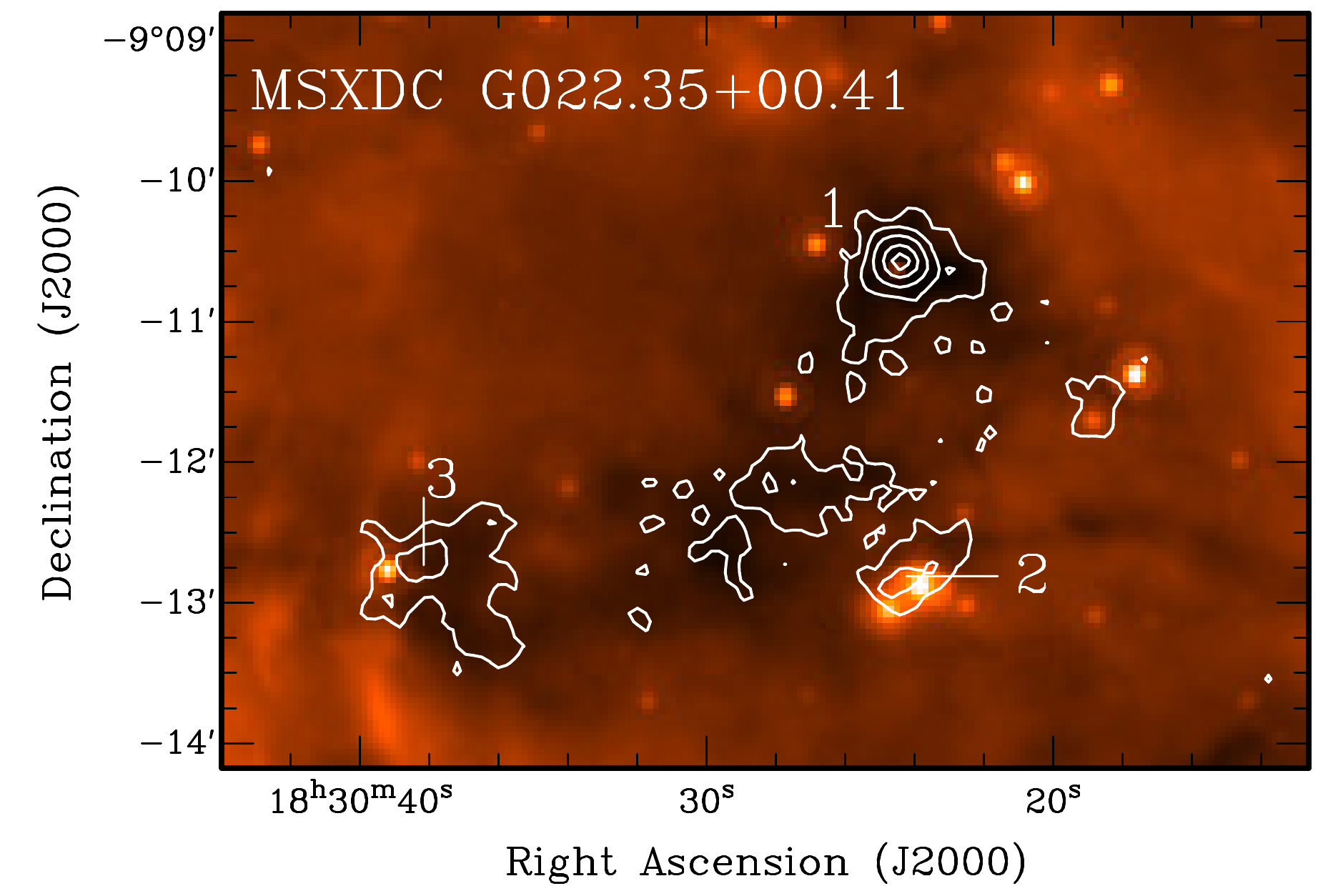}\\
\end{center}
\includegraphics[angle=90,width=0.5\textwidth]{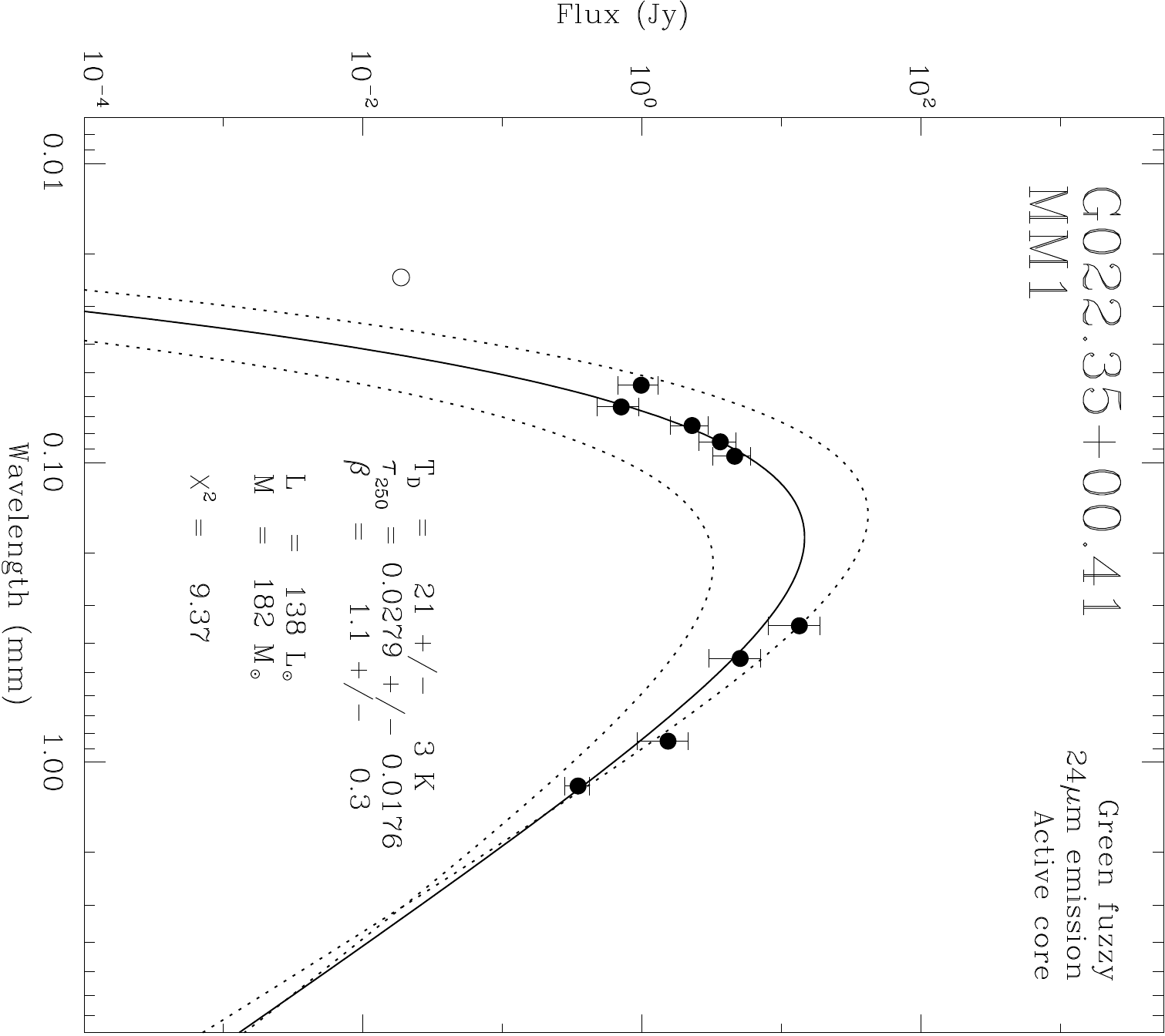}
\includegraphics[angle=90,width=0.5\textwidth]{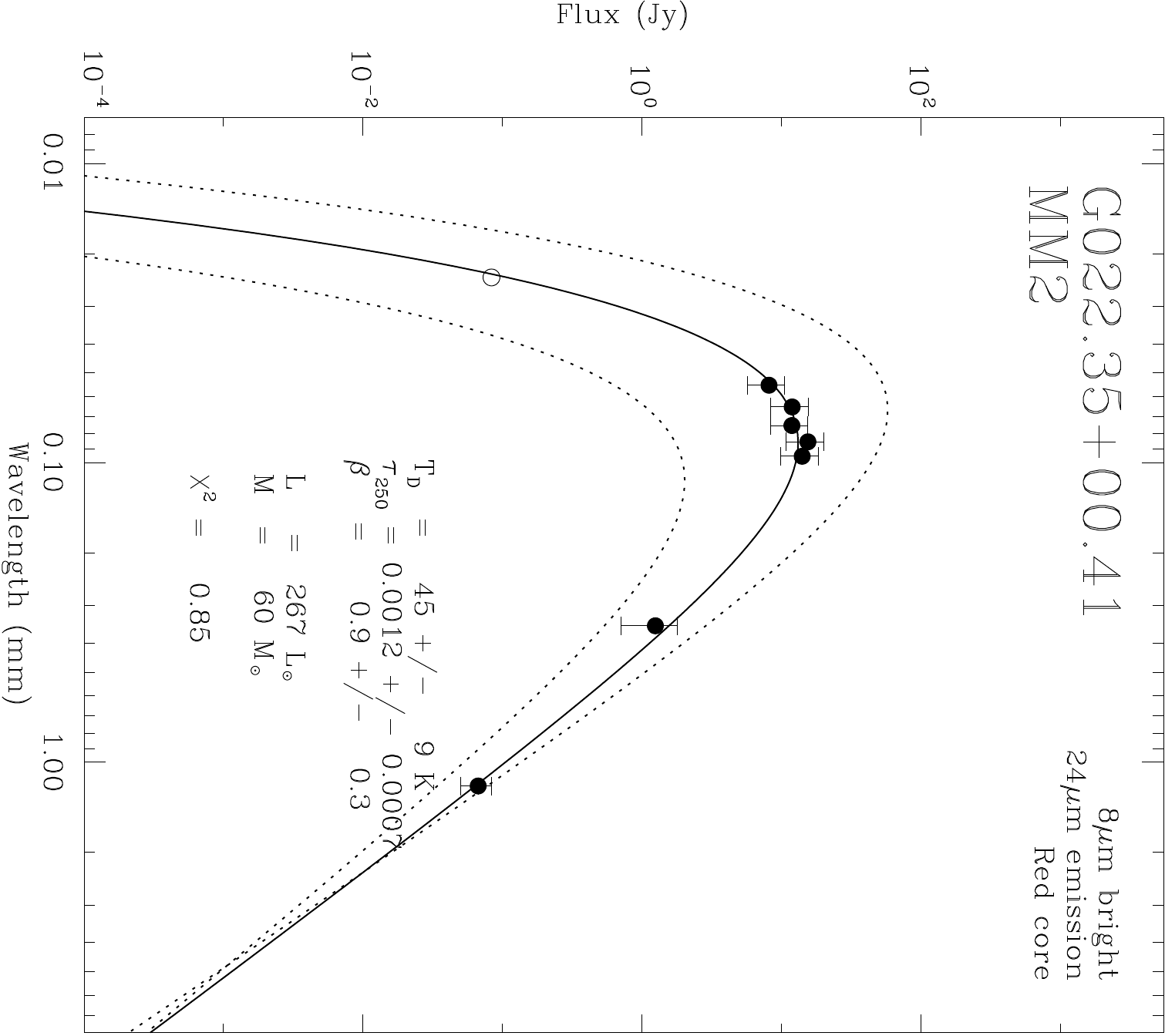}\\
\end{figure}
\clearpage 
\begin{figure}
\includegraphics[angle=90,width=0.5\textwidth]{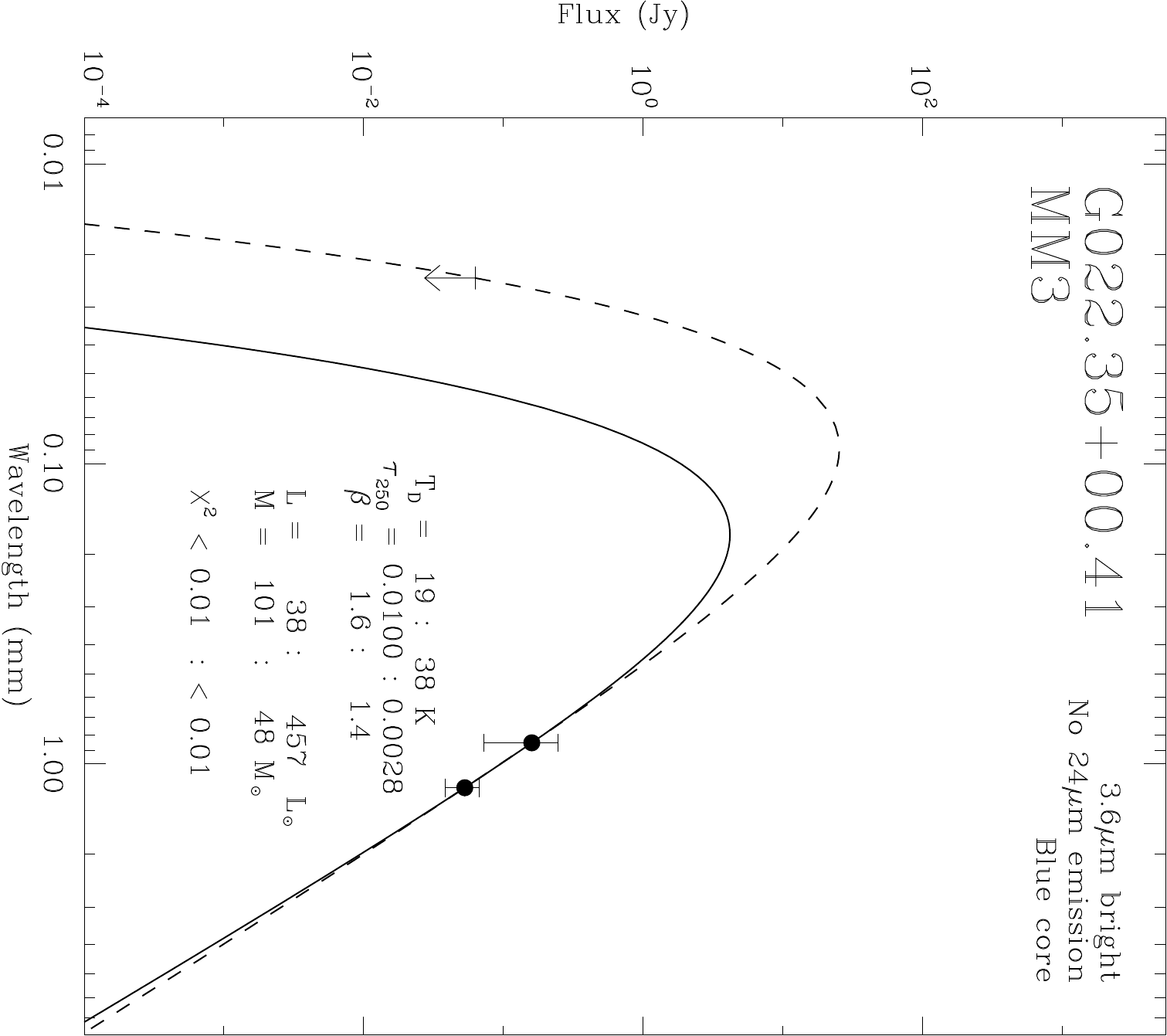}
\caption{\label{seds-6} \Spitzer\, 24\,\um\, image overlaid  
   with 1.2\,mm continuum emission for \irdcsix\, (contour levels are
   30, 60, 90, 120, 240\,mJy beam$^{-1}$). The lower panels show the broadband
   SEDs for cores within this IRDC.  The fluxes derived from the
   millimeter, sub-millimeter, and far-IR  continuum data are shown as filled
   circles (with the corresponding error bars), while the 24\,\um\, fluxes are shown as  either a filled circle (when included within the fit), an open circle (when excluded from the fit),  or as an upper limit arrow. For cores that have measured fluxes only in the millimeter/sub-millimeter regime (i.e.\, a limit at 24\,\um), we show the results from two fits: one using only the measured fluxes (solid line; lower limit), while the other includes the 24\,\um\, limit as a real data (dashed line; upper limit). In all other cases, the solid line is the best fit gray-body, while the dotted lines correspond to the functions determined using the errors for the T$_{D}$, $\tau$, and $\beta$ output from the fitting.  Labeled on each plot is the IRDC and core name,  classification, and the derived parameters.}
\end{figure}
\clearpage 
\begin{figure}
\begin{center}
\includegraphics[angle=0,width=0.6\textwidth]{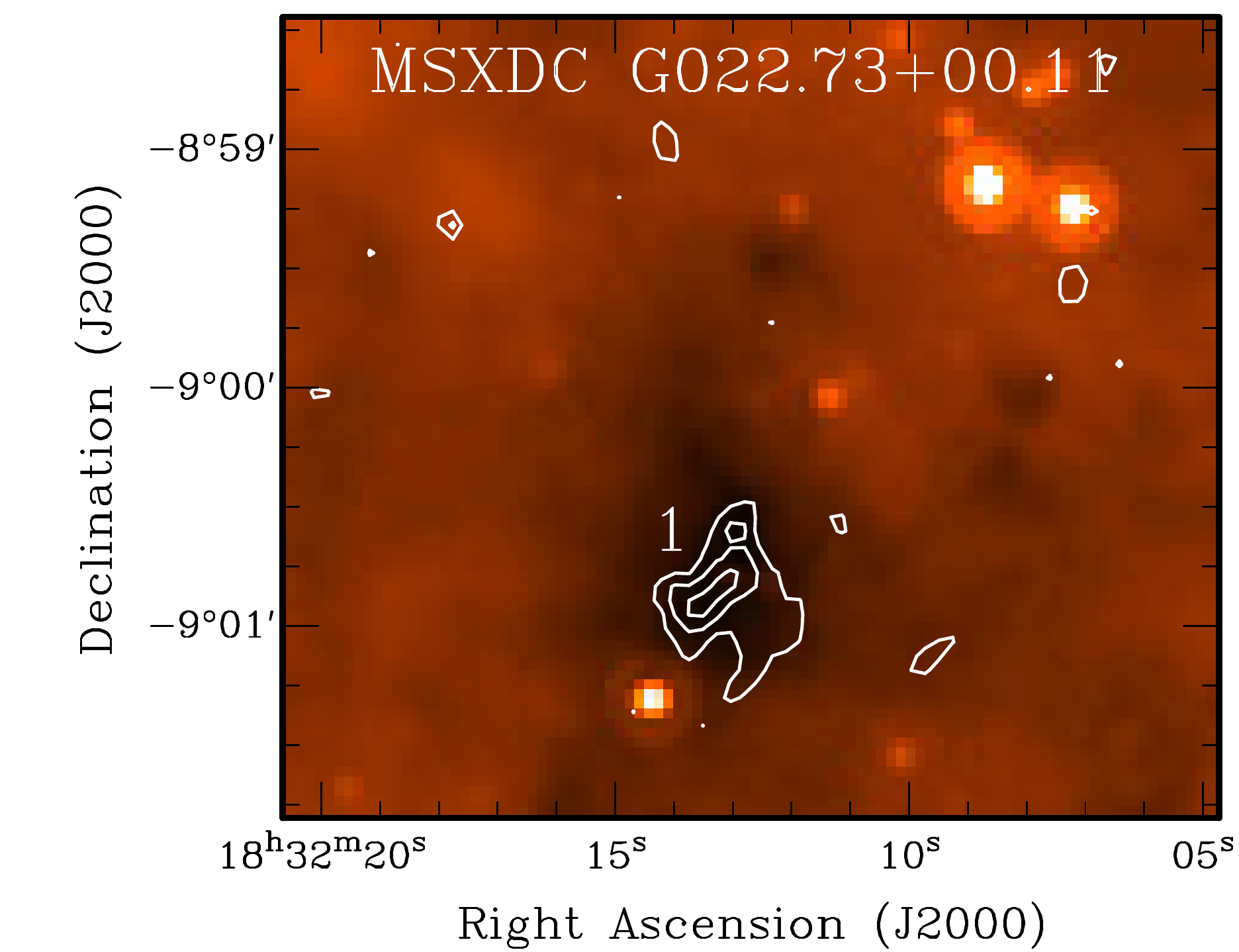}\\
\end{center}
\includegraphics[angle=90,width=0.5\textwidth]{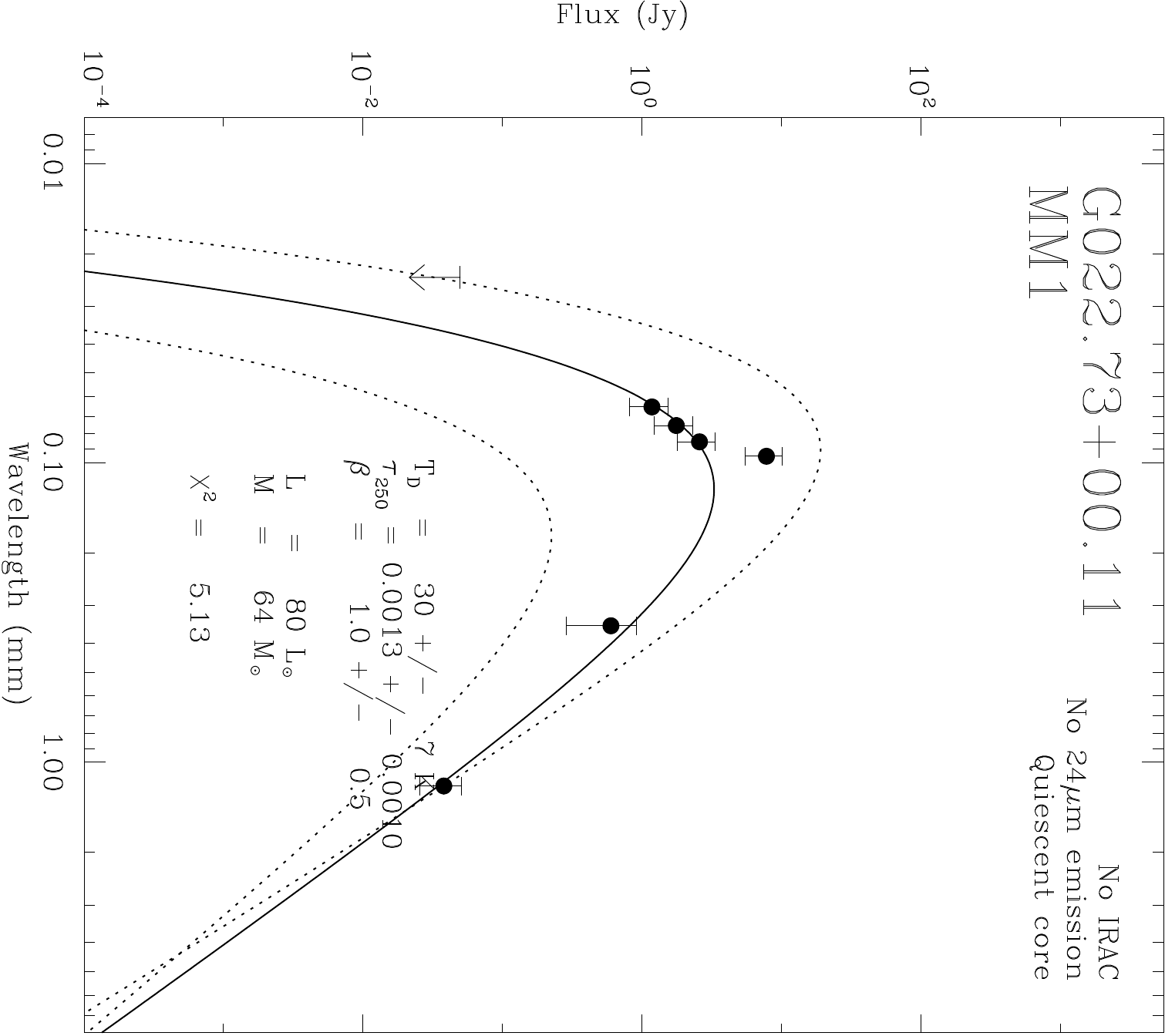}\\
\caption{\label{seds-49} \Spitzer\, 24\,\um\, image overlaid  
   with 1.2\,mm continuum emission for \irdcfortynine\, (contour
   levels are 20, 30 and 45\,mJy beam$^{-1}$). The lower panels show the broadband
   SEDs for cores within this IRDC.  The fluxes derived from the
   millimeter, sub-millimeter, and far-IR  continuum data are shown as filled
   circles (with the corresponding error bars), while the 24\,\um\, fluxes are shown as  either a filled circle (when included within the fit), an open circle (when excluded from the fit),  or as an upper limit arrow. For cores that have measured fluxes only in the millimeter/sub-millimeter regime (i.e.\, a limit at 24\,\um), we show the results from two fits: one using only the measured fluxes (solid line; lower limit), while the other includes the 24\,\um\, limit as a real data (dashed line; upper limit). In all other cases, the solid line is the best fit gray-body, while the dotted lines correspond to the functions determined using the errors for the T$_{D}$, $\tau$, and $\beta$ output from the fitting.  Labeled on each plot is the IRDC and core name,  classification, and the derived parameters.}
   \end{figure}
\clearpage 
\begin{figure}
\begin{center}
\includegraphics[angle=0,width=0.6\textwidth]{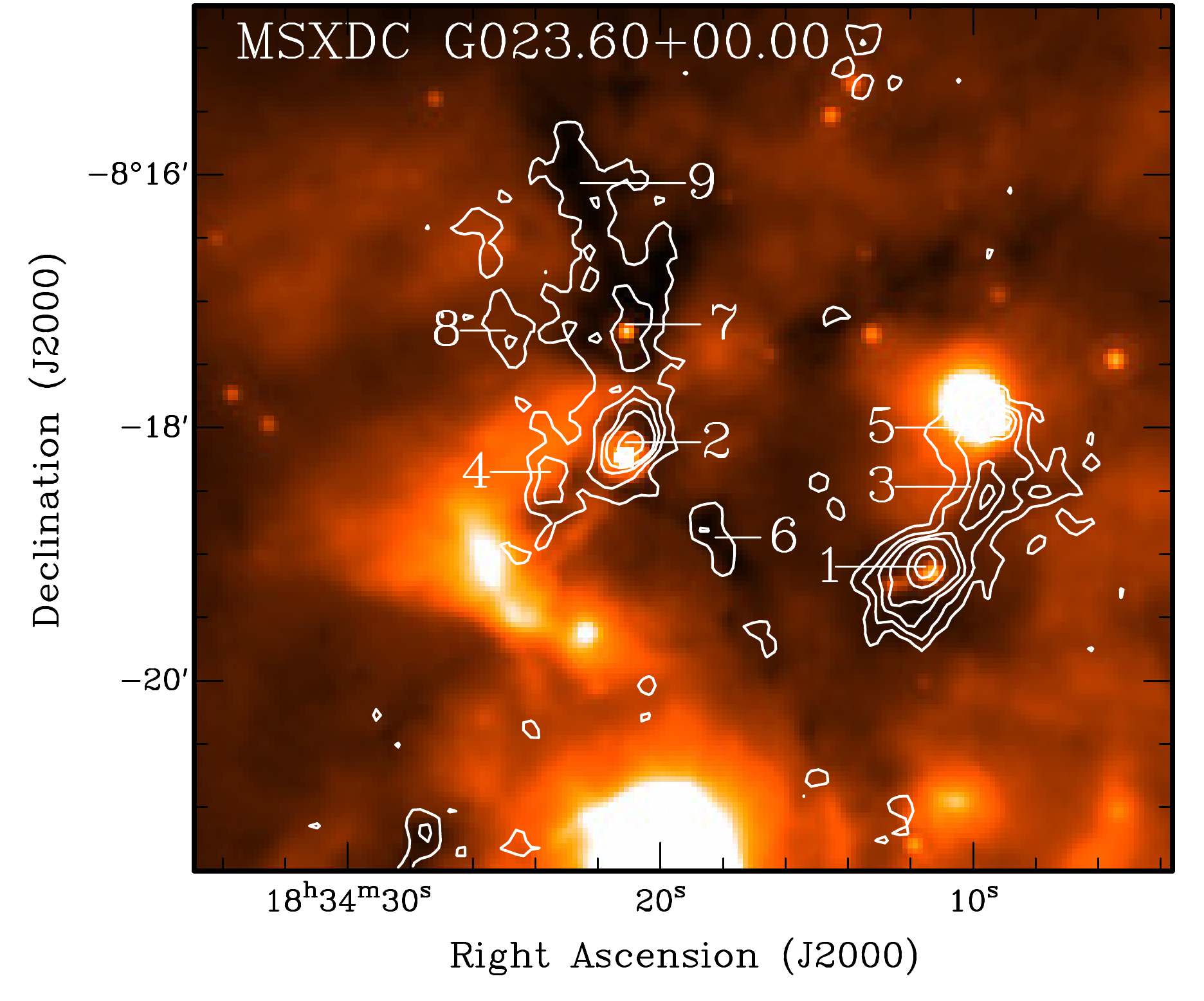}\\
\end{center}
\includegraphics[angle=90,width=0.5\textwidth]{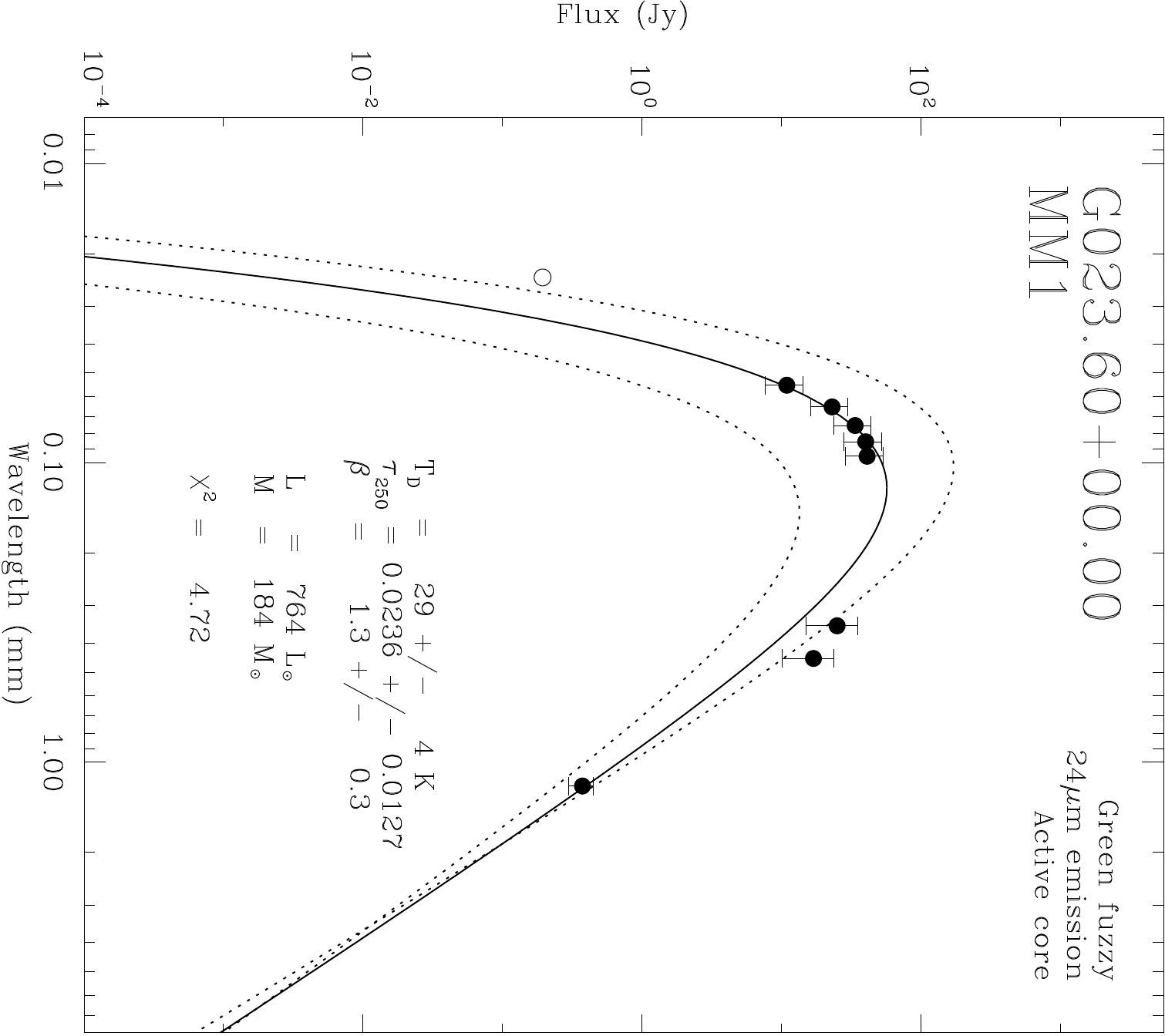}
\includegraphics[angle=90,width=0.5\textwidth]{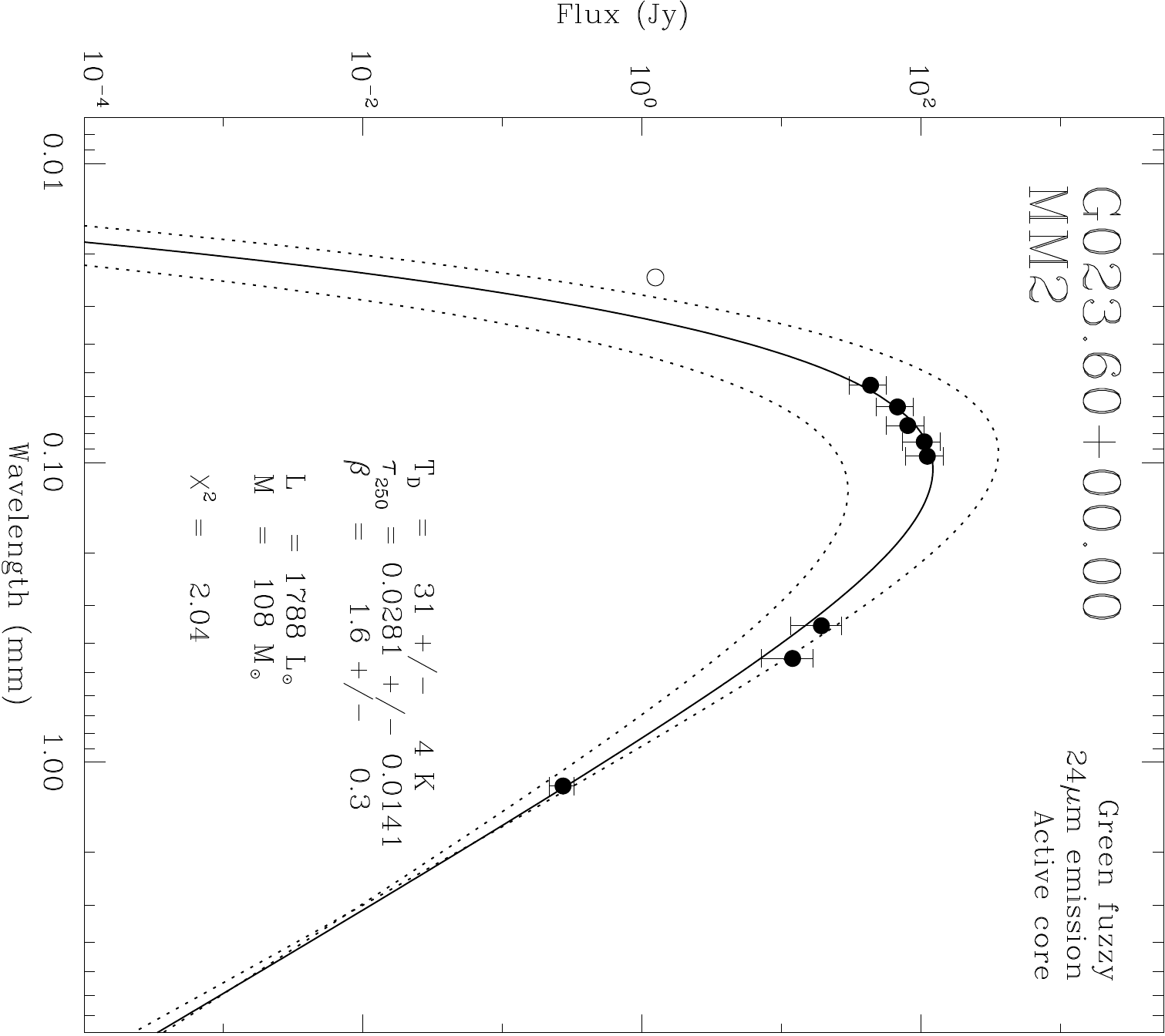}\\
\end{figure}
\clearpage 
\begin{figure}
\includegraphics[angle=90,width=0.5\textwidth]{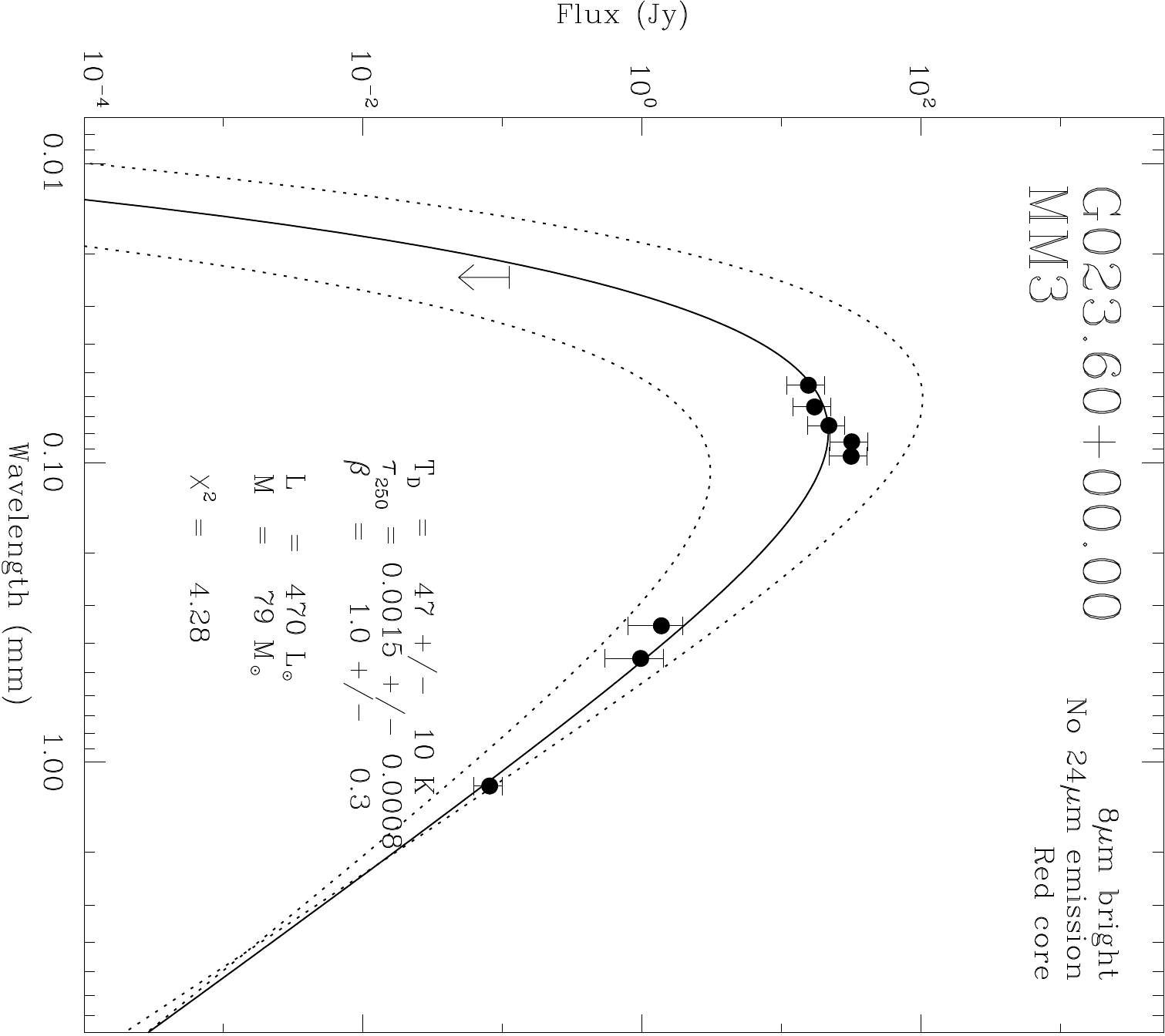}
\includegraphics[angle=90,width=0.5\textwidth]{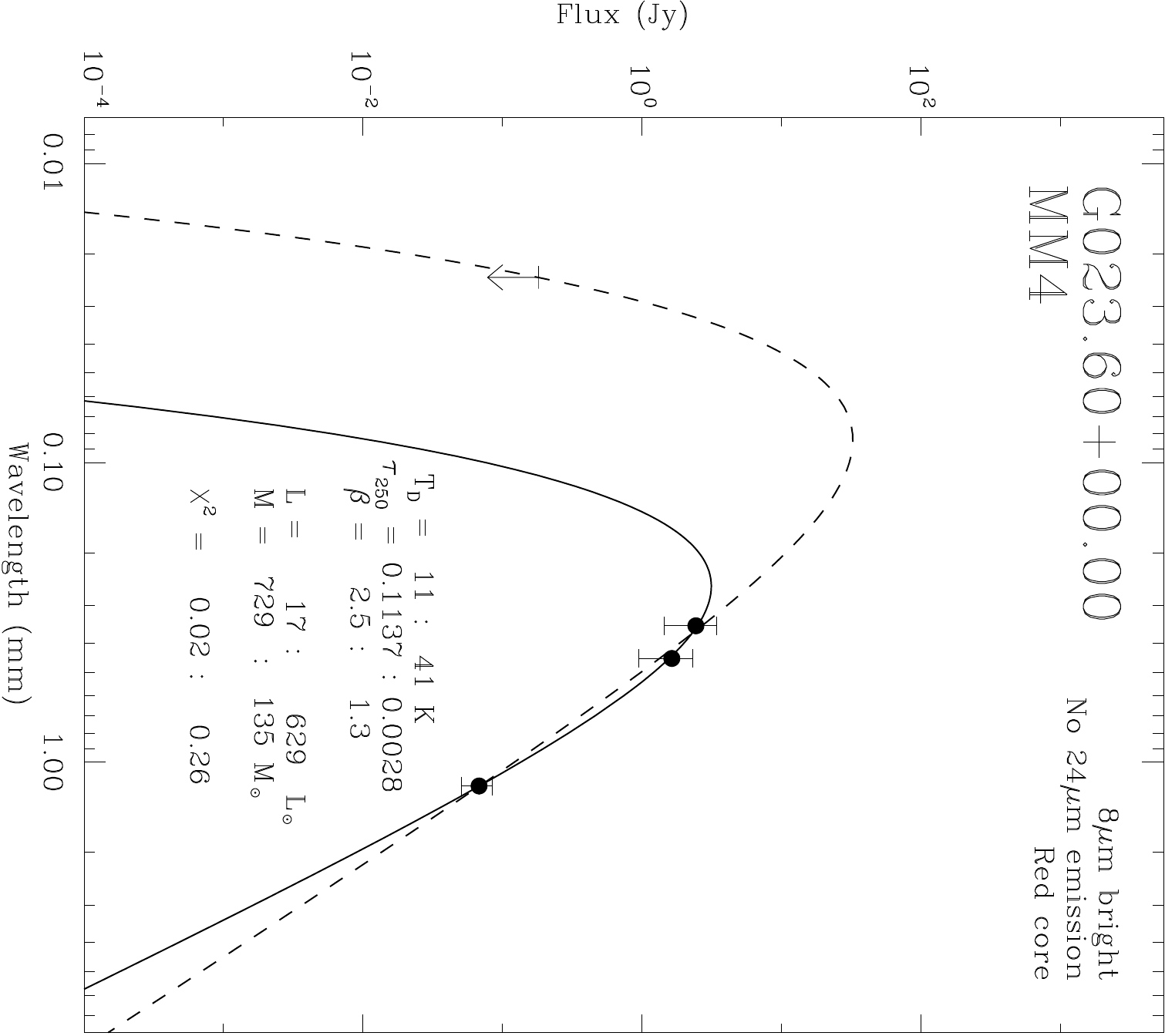}\\
\includegraphics[angle=90,width=0.5\textwidth]{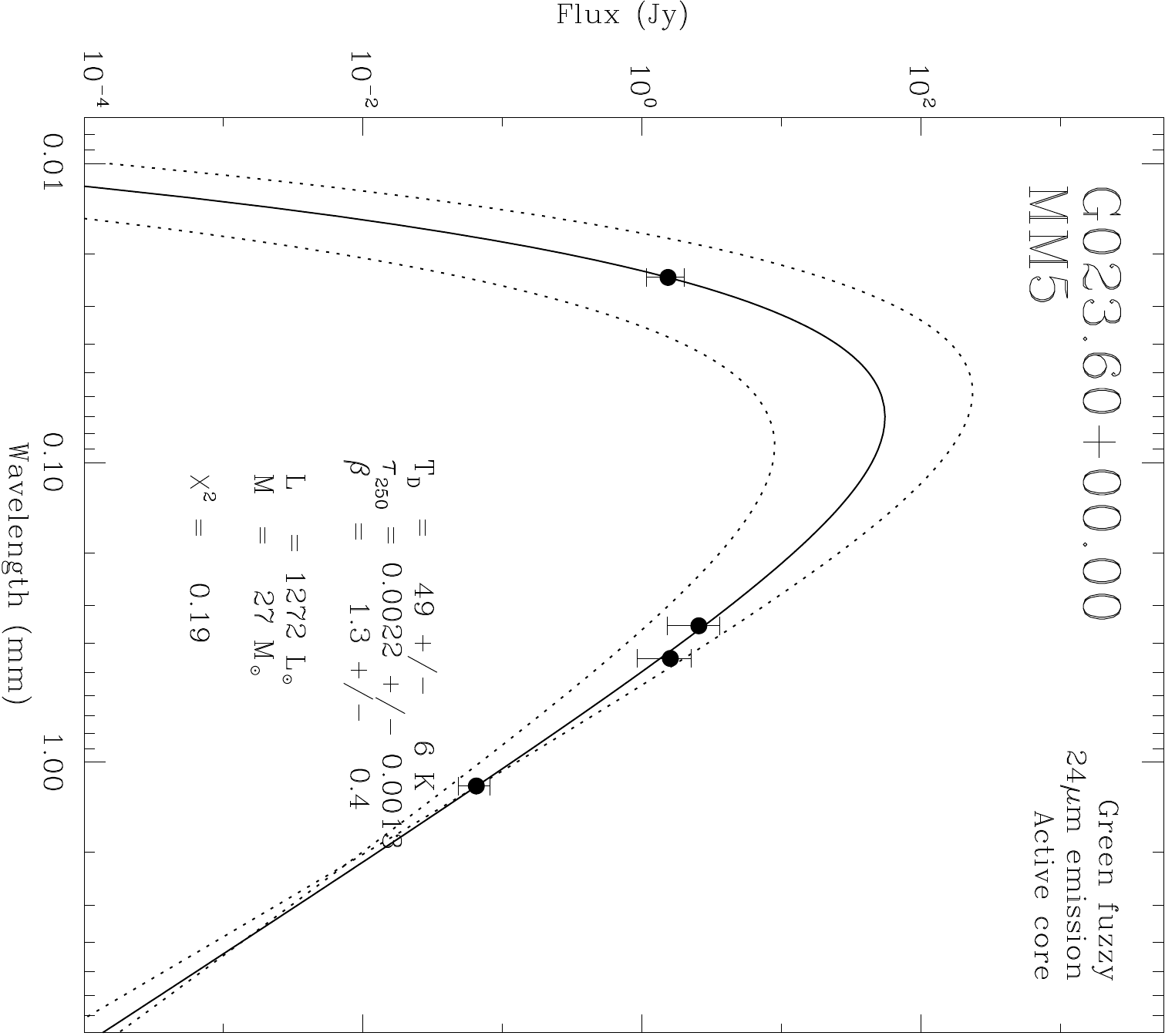}
\includegraphics[angle=90,width=0.5\textwidth]{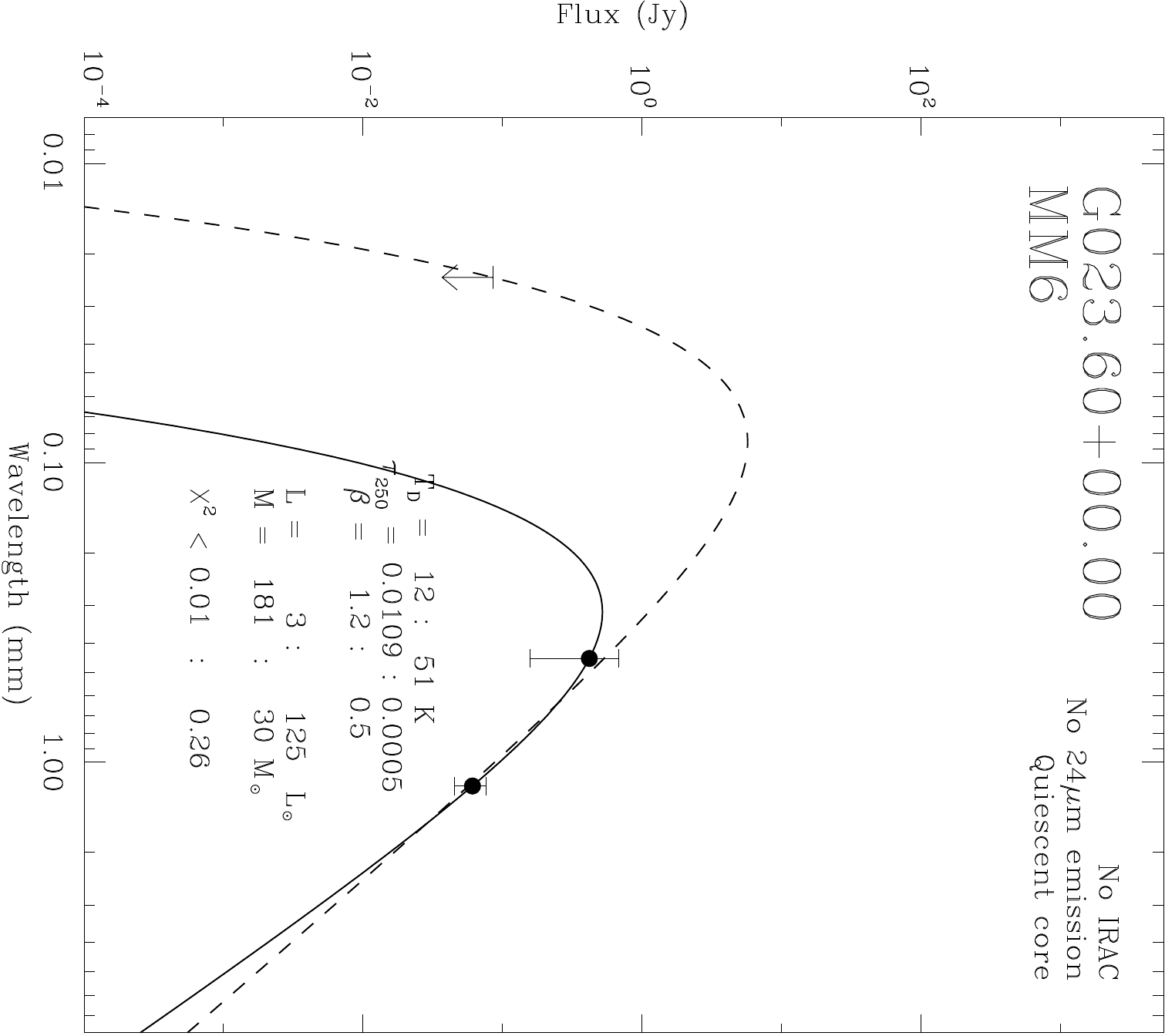}\\
\end{figure}
\clearpage 
\begin{figure}
\includegraphics[angle=90,width=0.5\textwidth]{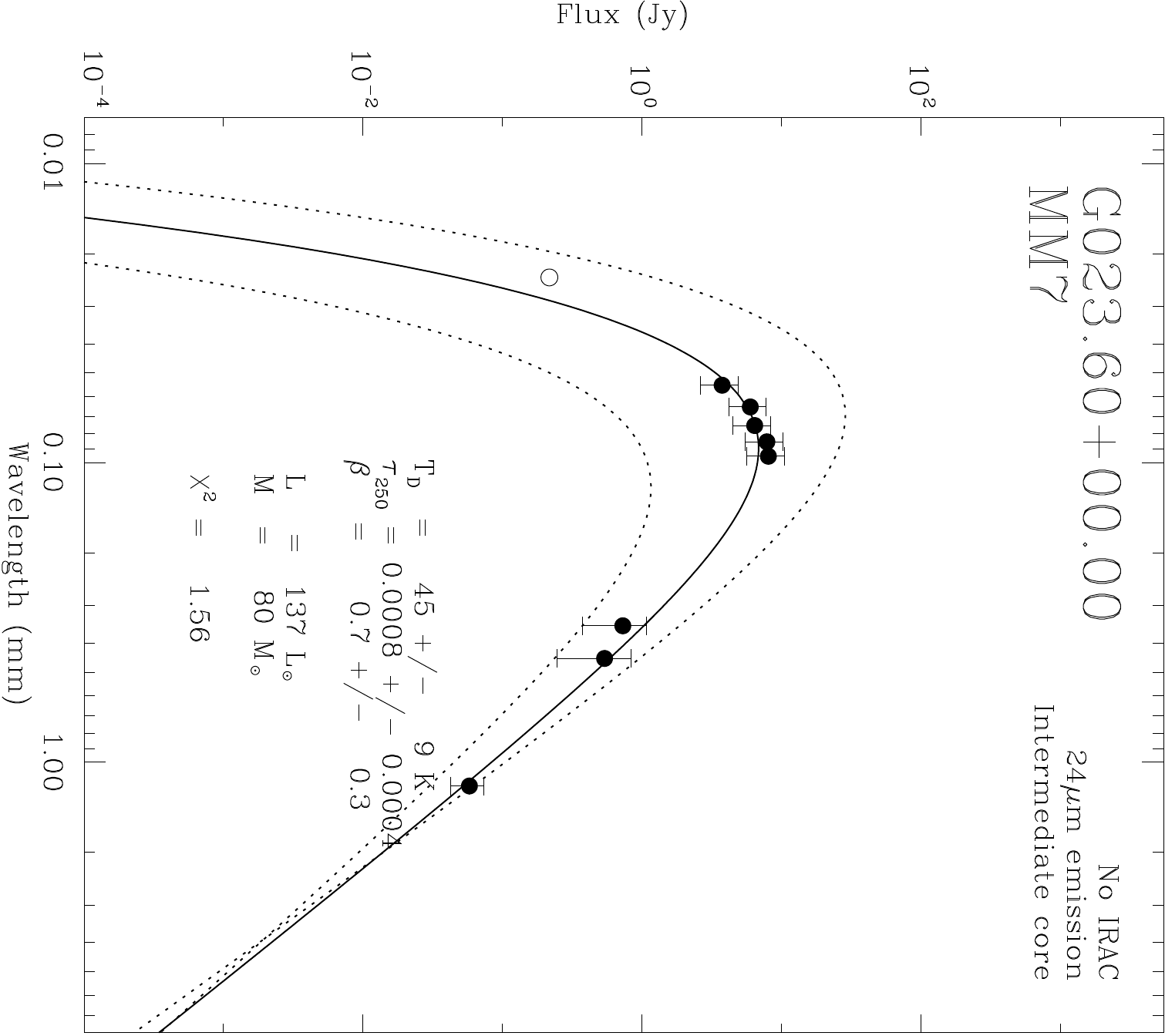}
\includegraphics[angle=90,width=0.5\textwidth]{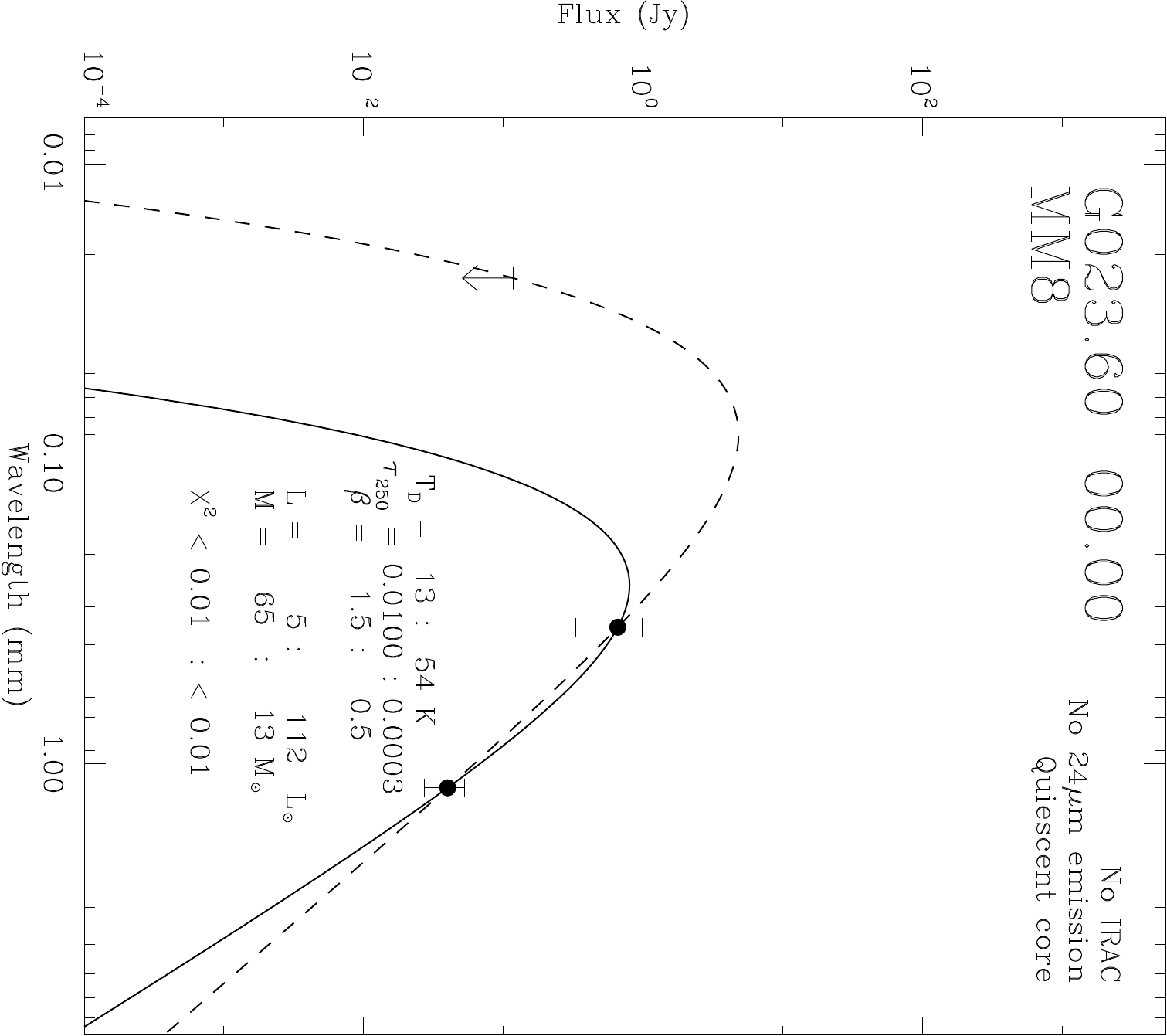}\\
\includegraphics[angle=90,width=0.5\textwidth]{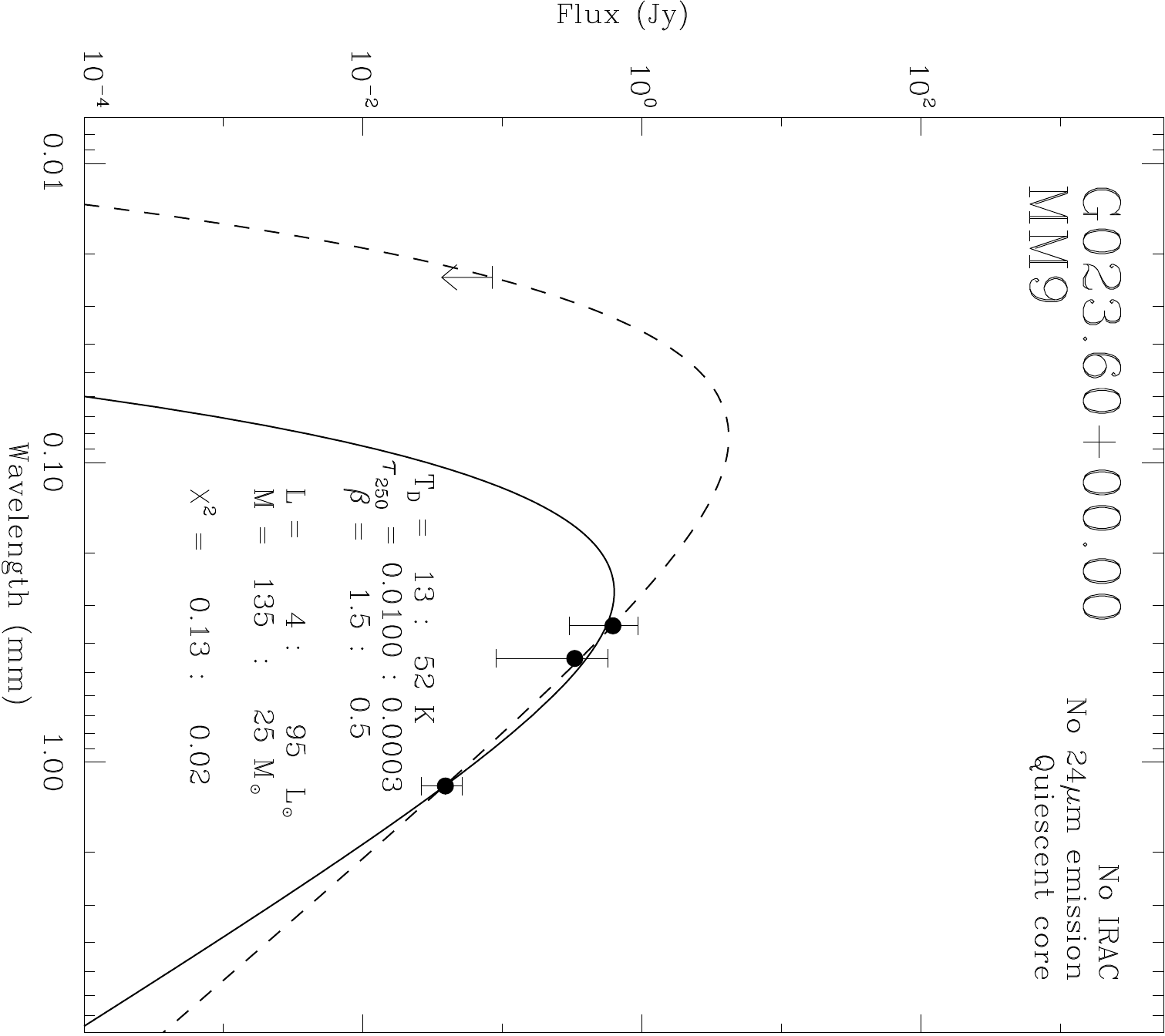}
\caption{\label{seds-33} \Spitzer\, 24\,\um\, image overlaid with 
    1.2\,mm continuum emission for \irdcthirtythree\, (contour levels
    are 30, 60, 90, 120, 240, 360, 480\,mJy beam$^{-1}$). The lower panels show the broadband
   SEDs for cores within this IRDC.  The fluxes derived from the
   millimeter, sub-millimeter, and far-IR  continuum data are shown as filled
   circles (with the corresponding error bars), while the 24\,\um\, fluxes are shown as  either a filled circle (when included within the fit), an open circle (when excluded from the fit),  or as an upper limit arrow. For cores that have measured fluxes only in the millimeter/sub-millimeter regime (i.e.\, a limit at 24\,\um), we show the results from two fits: one using only the measured fluxes (solid line; lower limit), while the other includes the 24\,\um\, limit as a real data (dashed line; upper limit). In all other cases, the solid line is the best fit gray-body, while the dotted lines correspond to the functions determined using the errors for the T$_{D}$, $\tau$, and $\beta$ output from the fitting.  Labeled on each plot is the IRDC and core name,  classification, and the derived parameters.}
\end{figure}
\clearpage 
\begin{figure}
\begin{center}
\includegraphics[angle=0,width=0.6\textwidth]{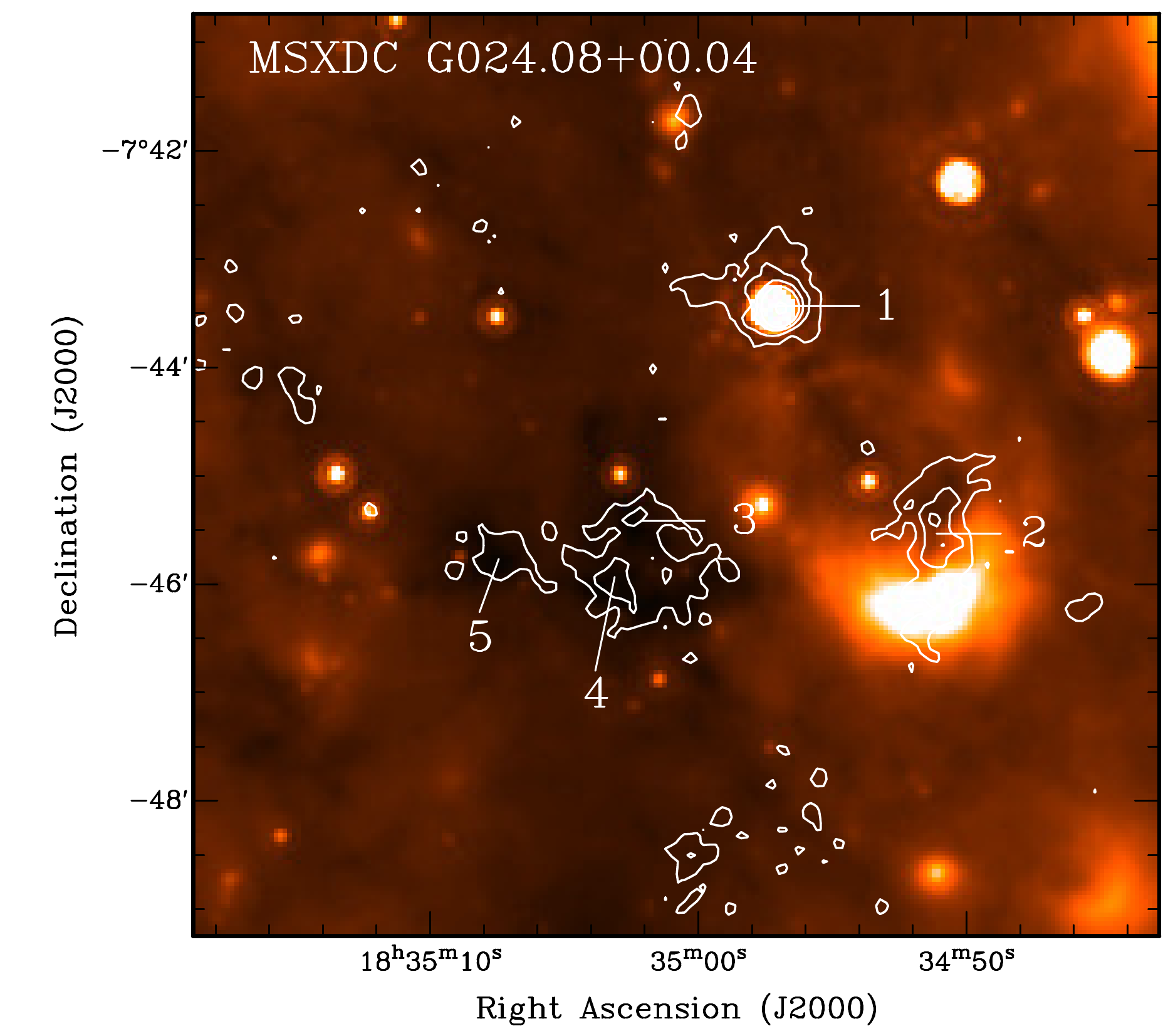}\\
\end{center}
\includegraphics[angle=90,width=0.5\textwidth]{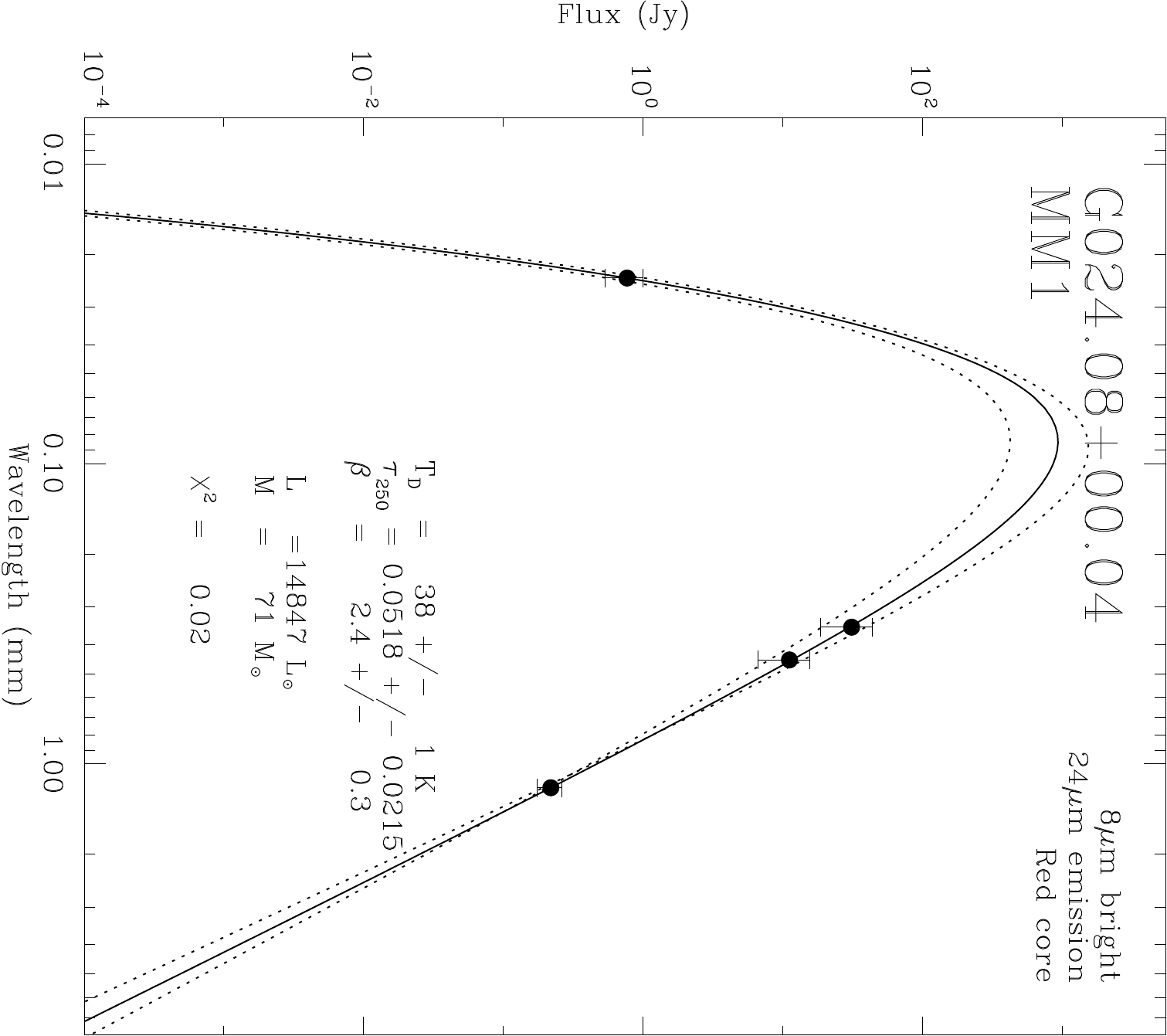}
\includegraphics[angle=90,width=0.5\textwidth]{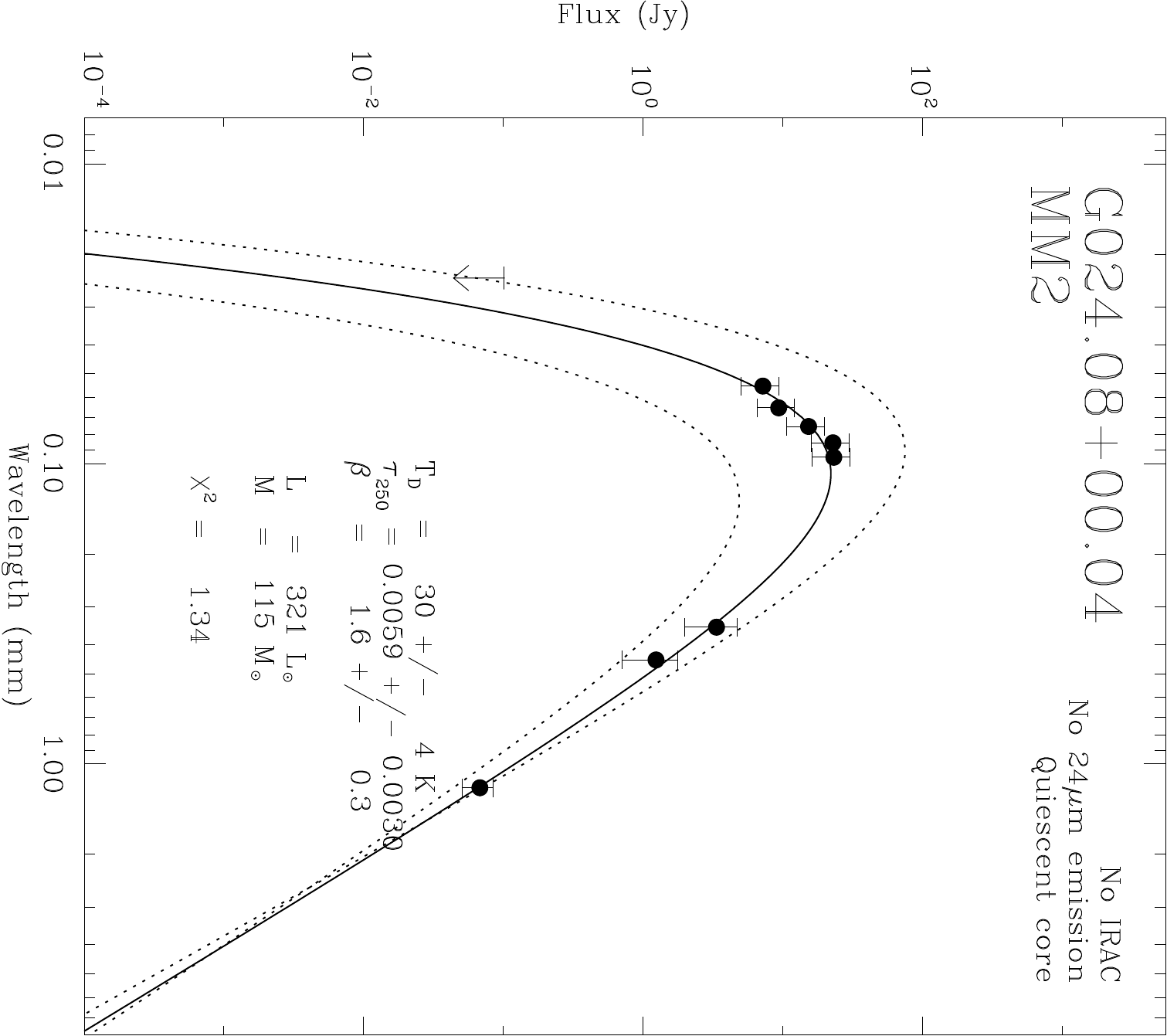}\\
\end{figure}
\clearpage 
\begin{figure}
\includegraphics[angle=90,width=0.5\textwidth]{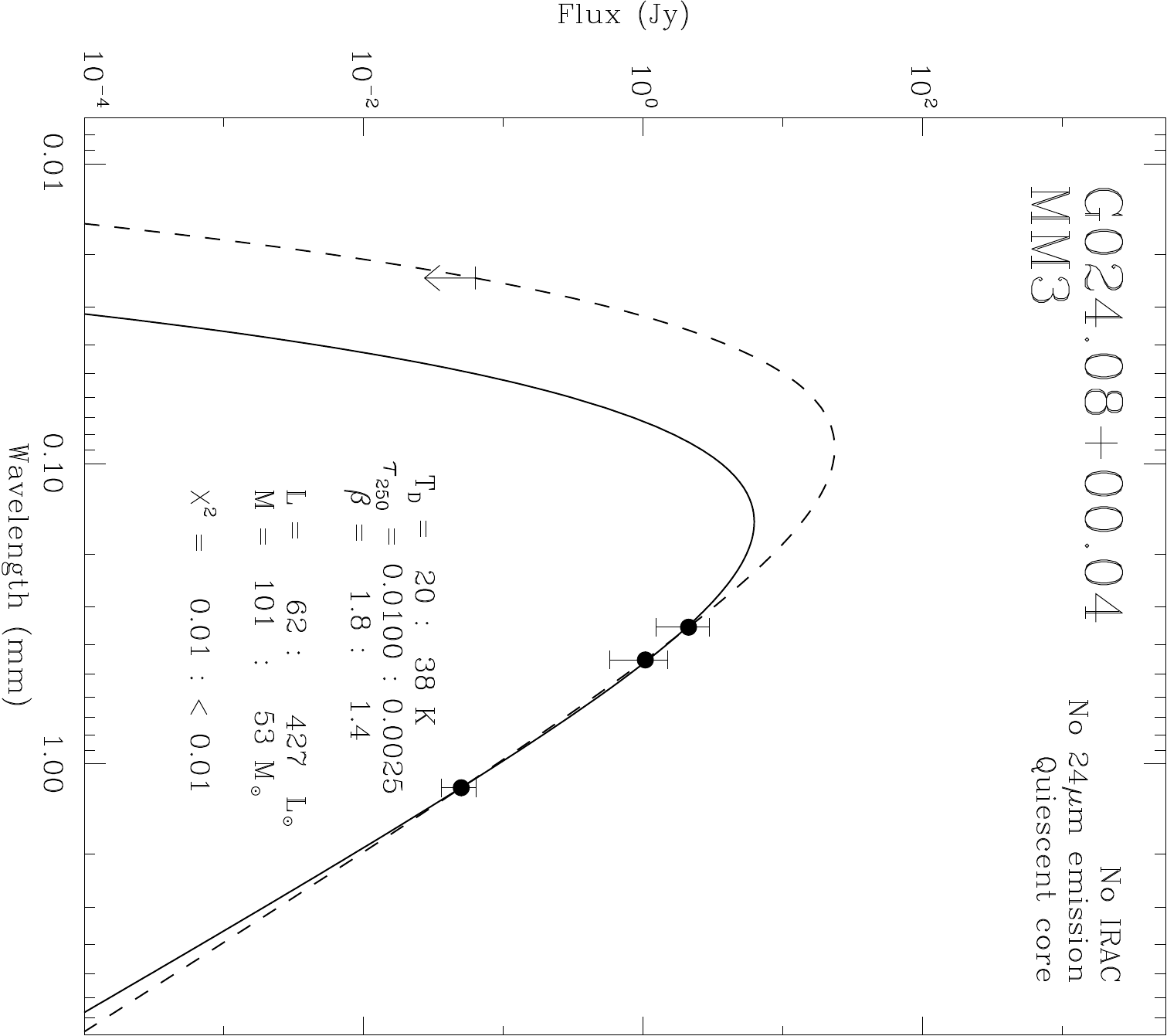}
\includegraphics[angle=90,width=0.5\textwidth]{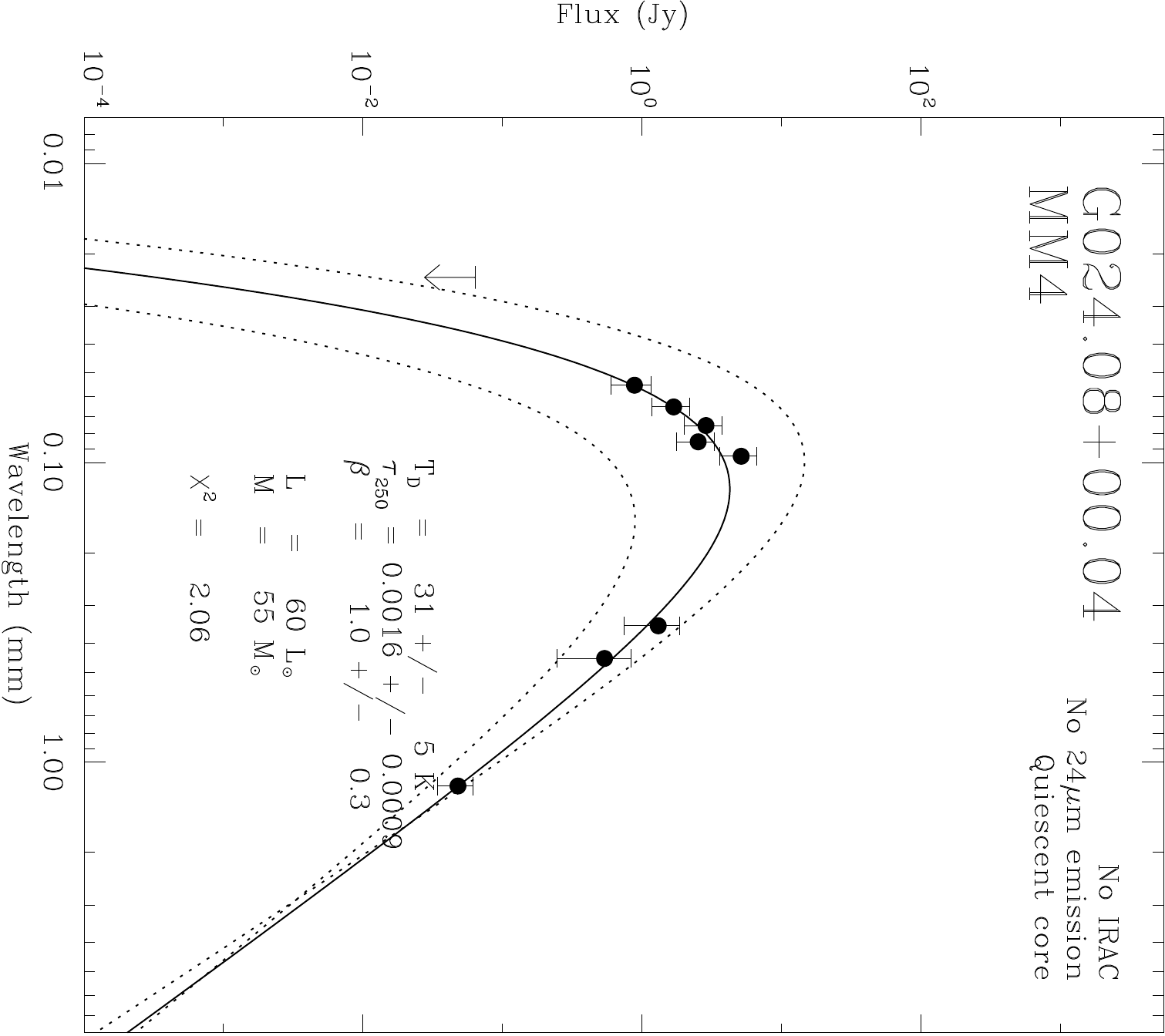}\\
\includegraphics[angle=90,width=0.5\textwidth]{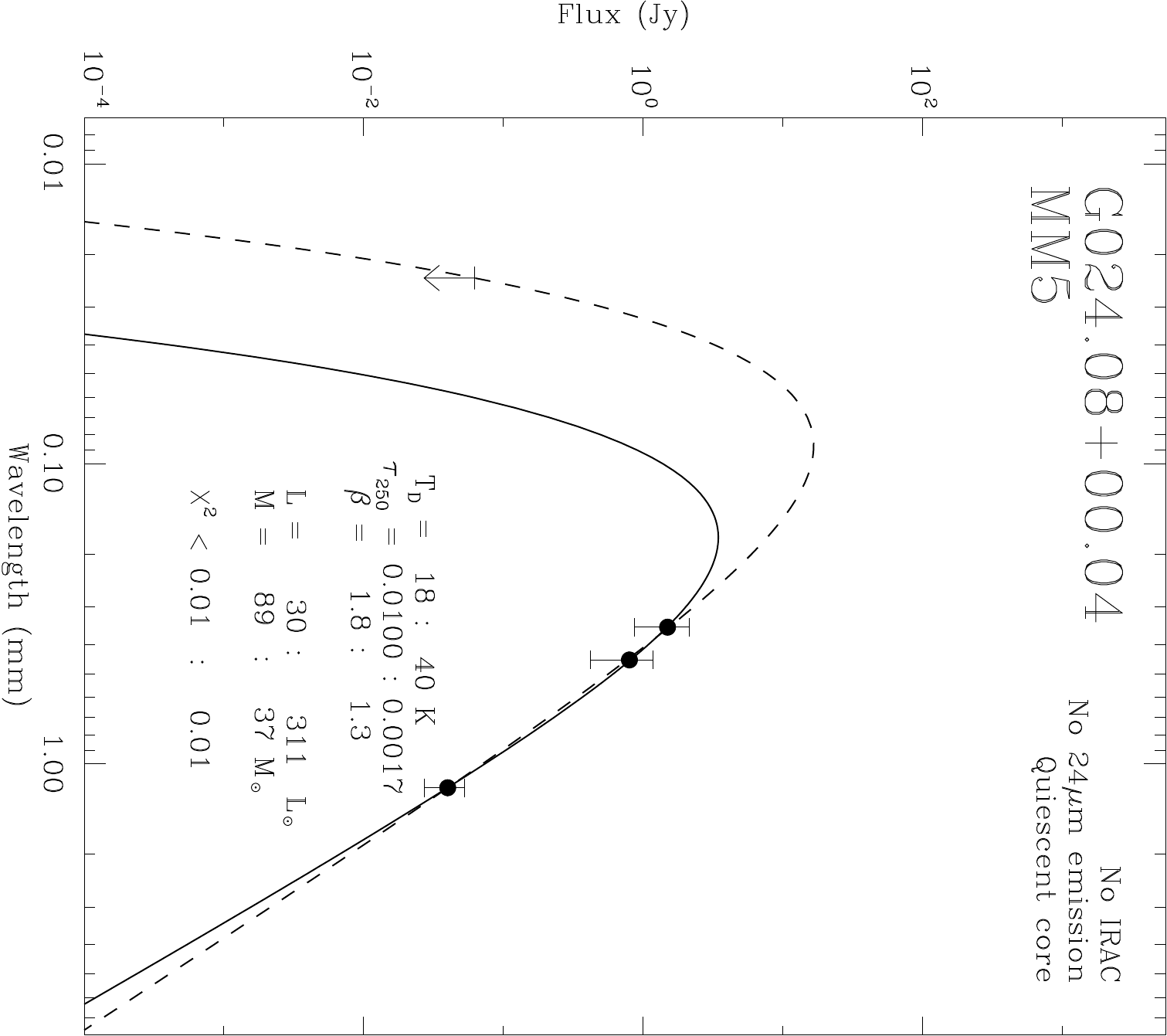}\\
\caption{\label{seds-45} \Spitzer\, 24\,\um\, image overlaid  
   with 1.2\,mm continuum emission for \irdcfortyfive\, (contour levels are
   30, 60, 90, 120, 240\,mJy beam$^{-1}$). The lower panels show the broadband
   SEDs for cores within this IRDC.  The fluxes derived from the
   millimeter, sub-millimeter, and far-IR  continuum data are shown as filled
   circles (with the corresponding error bars), while the 24\,\um\, fluxes are shown as  either a filled circle (when included within the fit), an open circle (when excluded from the fit),  or as an upper limit arrow. For cores that have measured fluxes only in the millimeter/sub-millimeter regime (i.e.\, a limit at 24\,\um), we show the results from two fits: one using only the measured fluxes (solid line; lower limit), while the other includes the 24\,\um\, limit as a real data (dashed line; upper limit). In all other cases, the solid line is the best fit gray-body, while the dotted lines correspond to the functions determined using the errors for the T$_{D}$, $\tau$, and $\beta$ output from the fitting.  Labeled on each plot is the IRDC and core name,  classification, and the derived parameters.}
\end{figure}
\clearpage 
\begin{figure}
\begin{center}
\includegraphics[angle=0,width=0.6\textwidth]{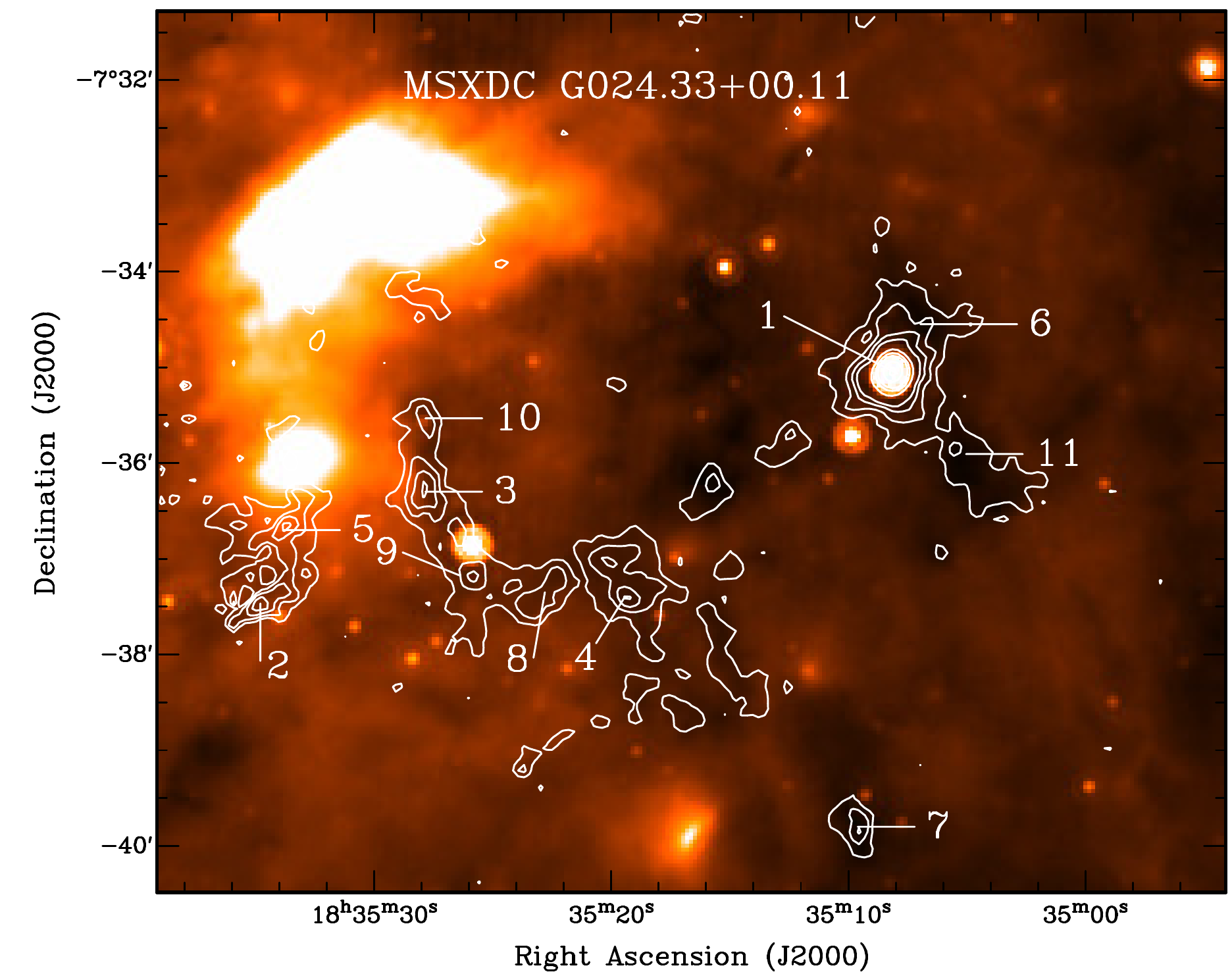}\\
\end{center}
\includegraphics[angle=90,width=0.5\textwidth]{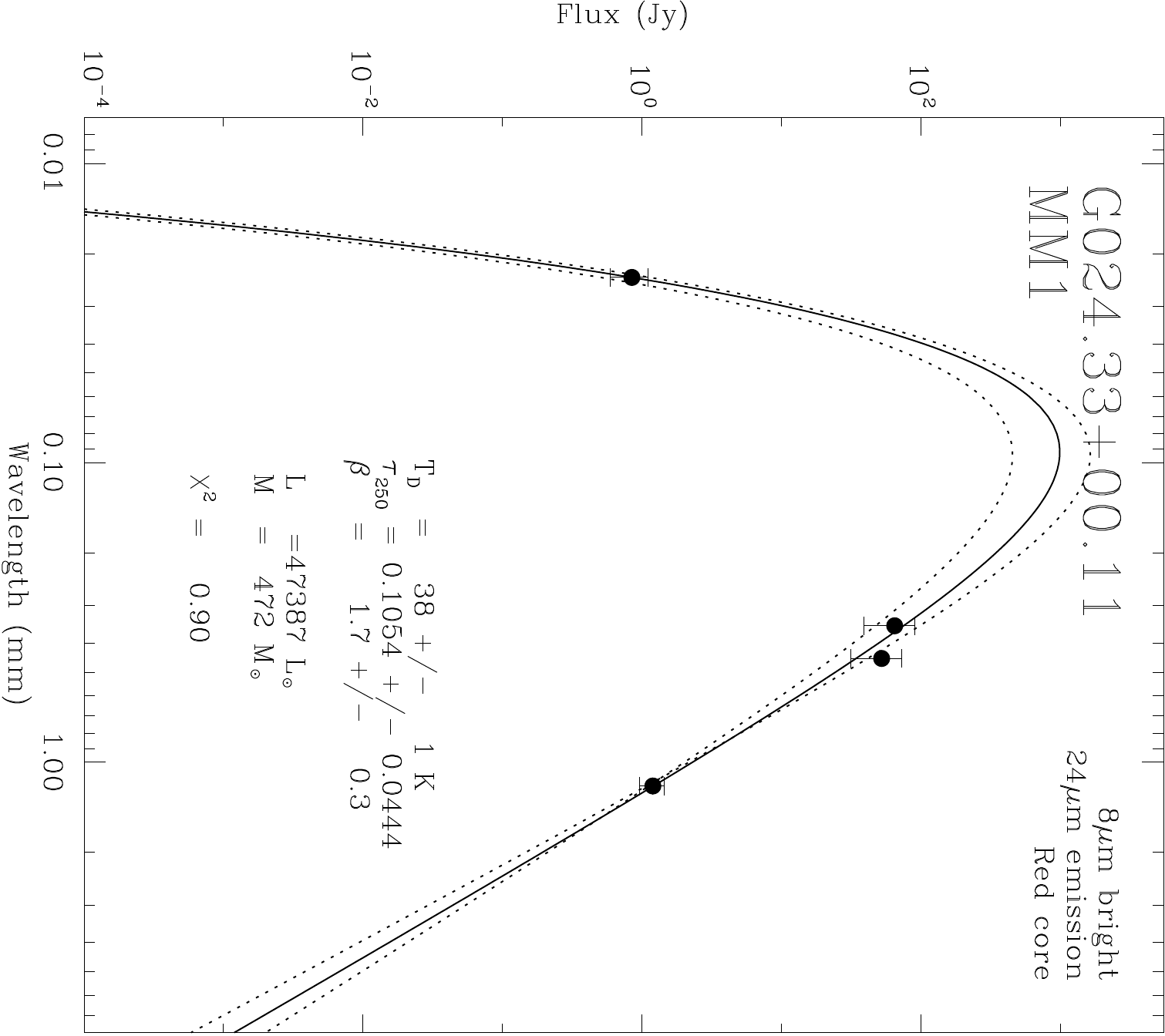}
\includegraphics[angle=90,width=0.5\textwidth]{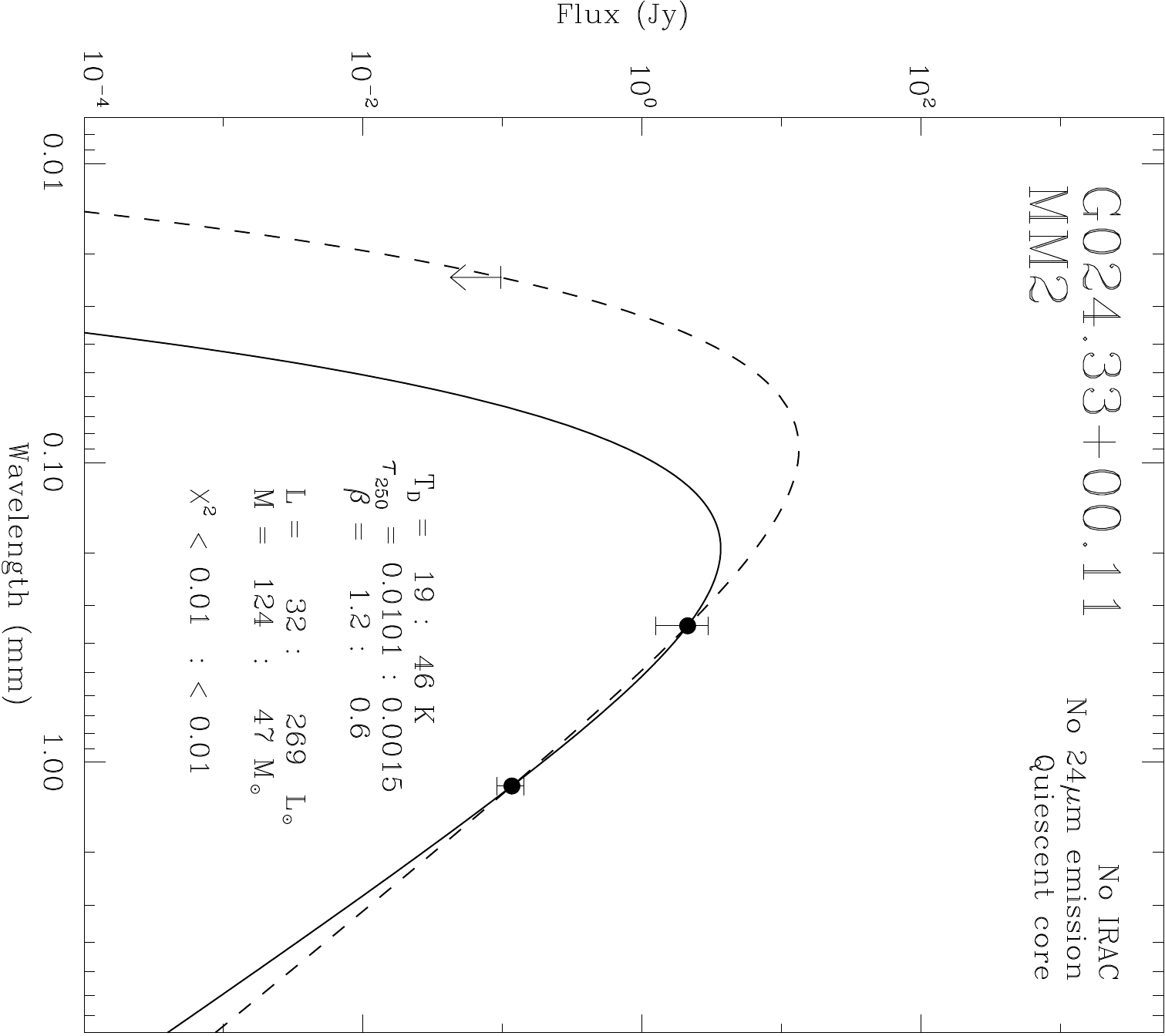}\\
\end{figure}
\clearpage 
\begin{figure}
\includegraphics[angle=90,width=0.5\textwidth]{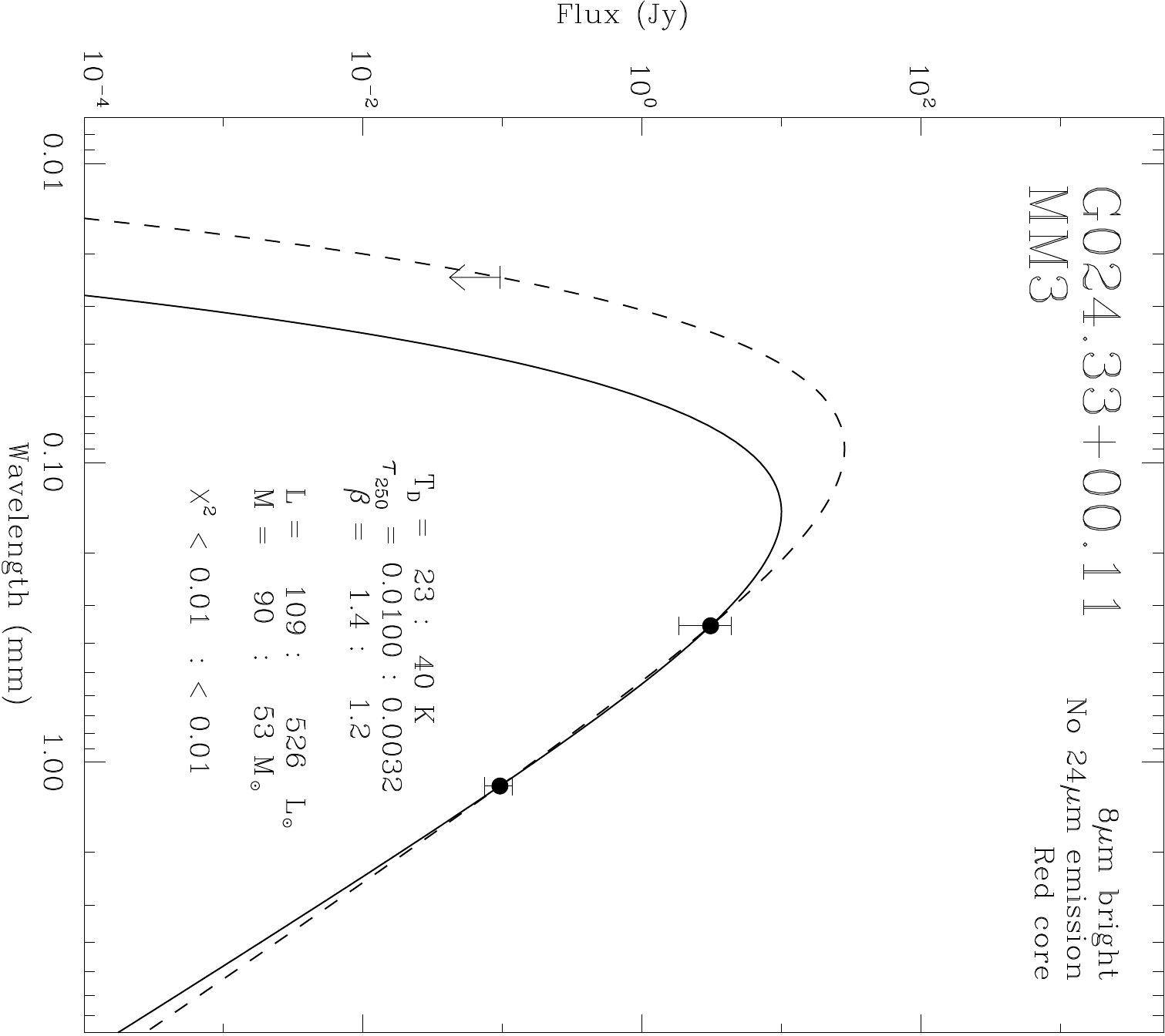}
\includegraphics[angle=90,width=0.5\textwidth]{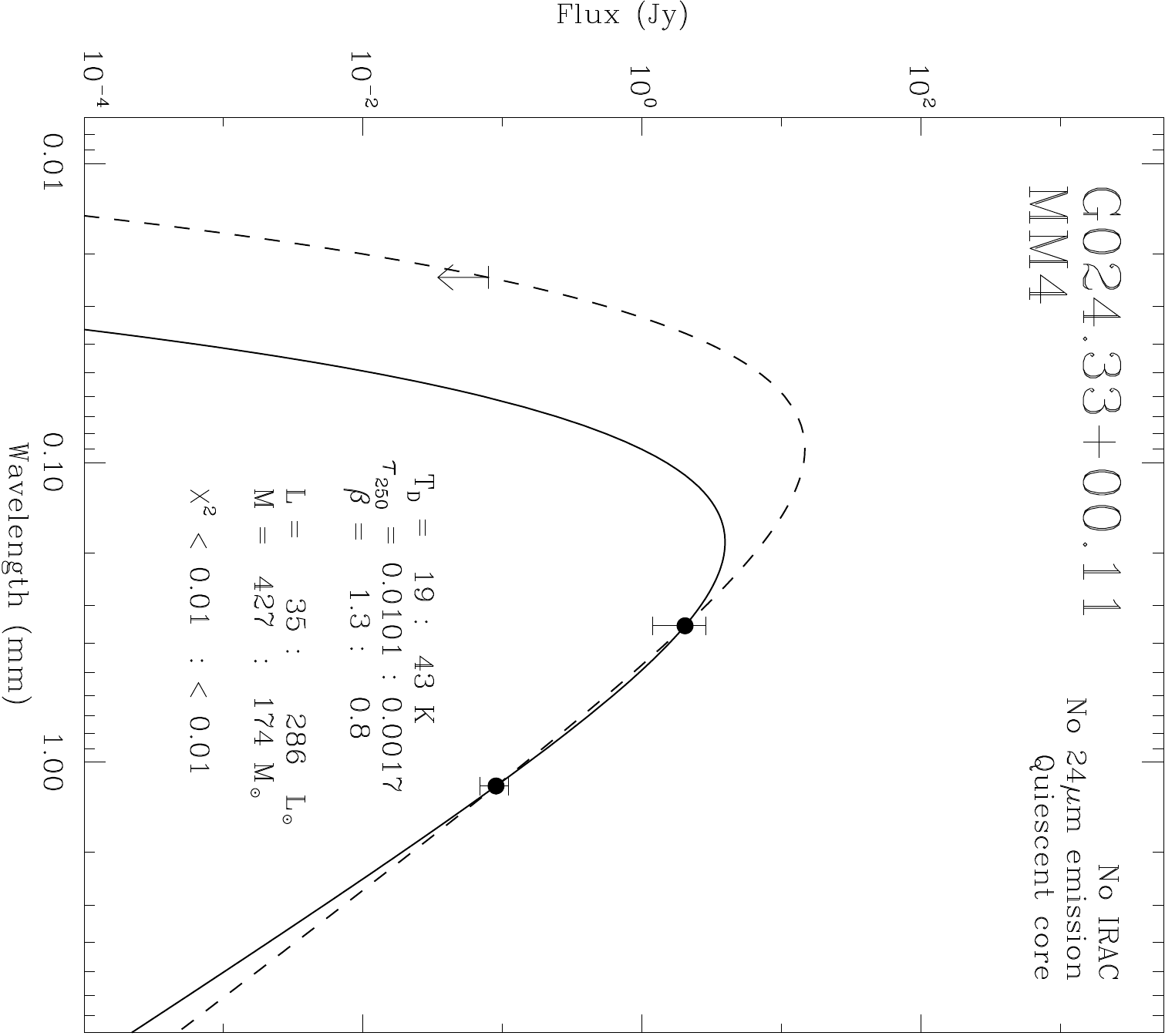}\\
\includegraphics[angle=90,width=0.5\textwidth]{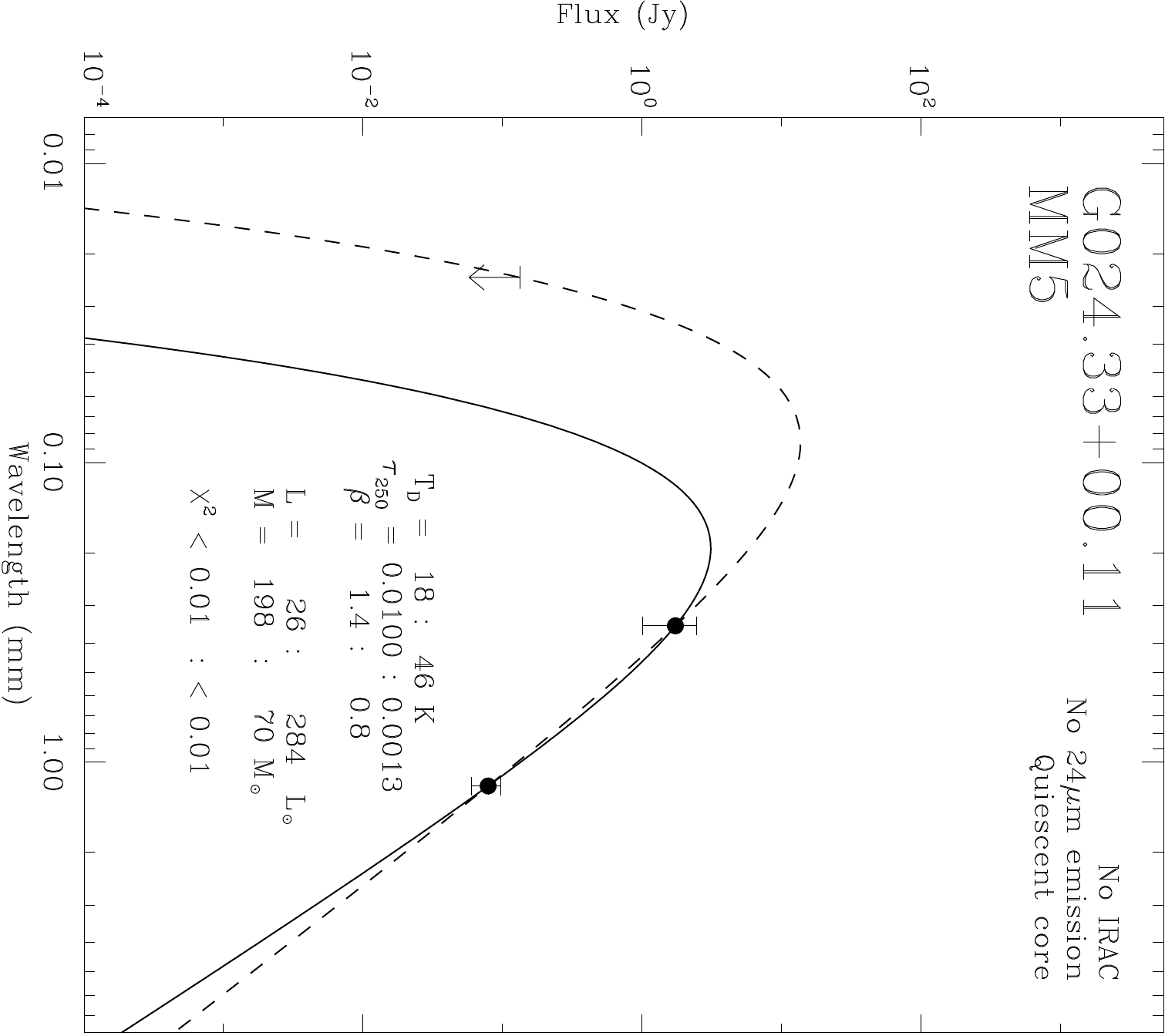}
\includegraphics[angle=90,width=0.5\textwidth]{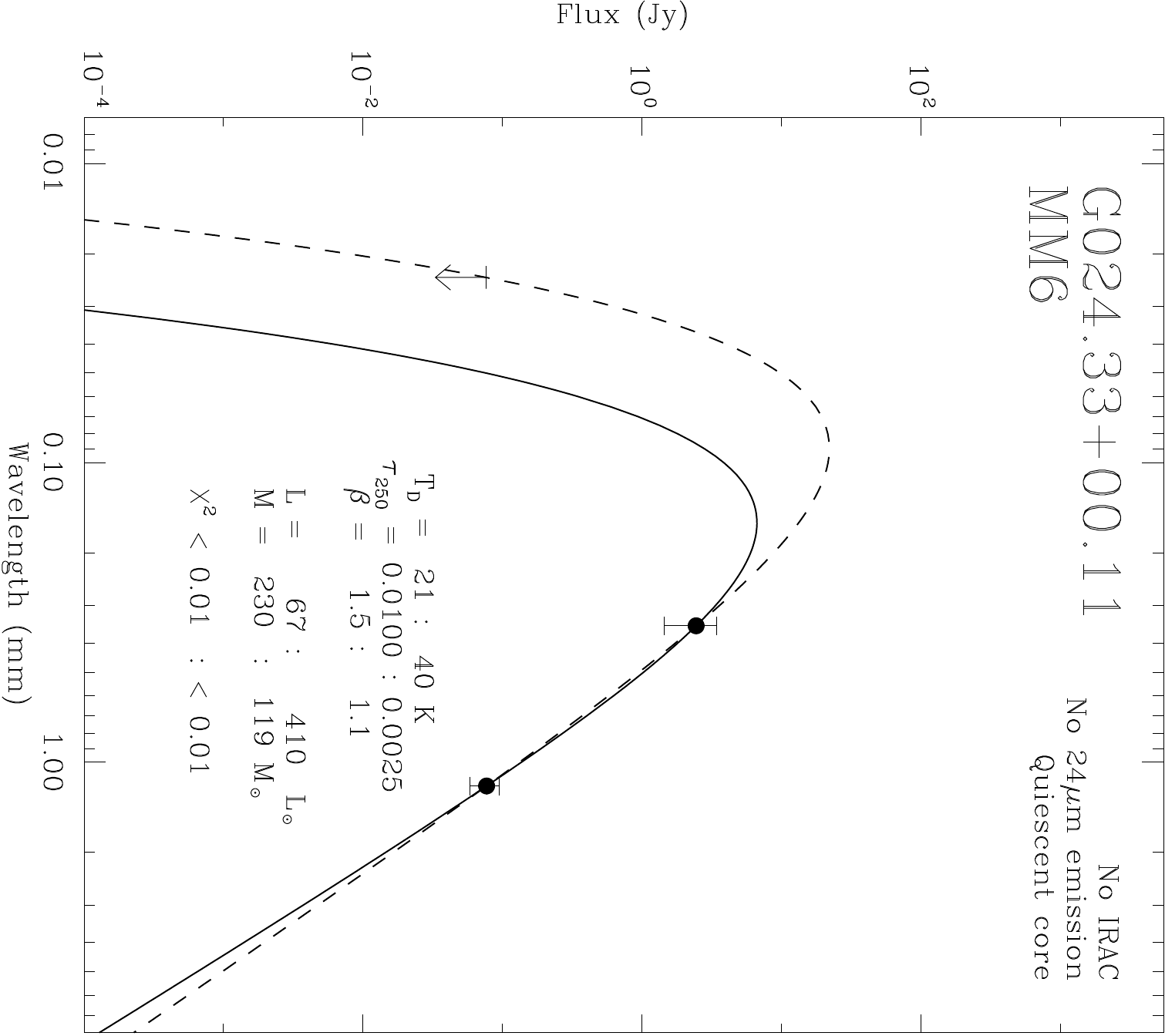}\\
\end{figure}
\clearpage 
\begin{figure}
\includegraphics[angle=90,width=0.5\textwidth]{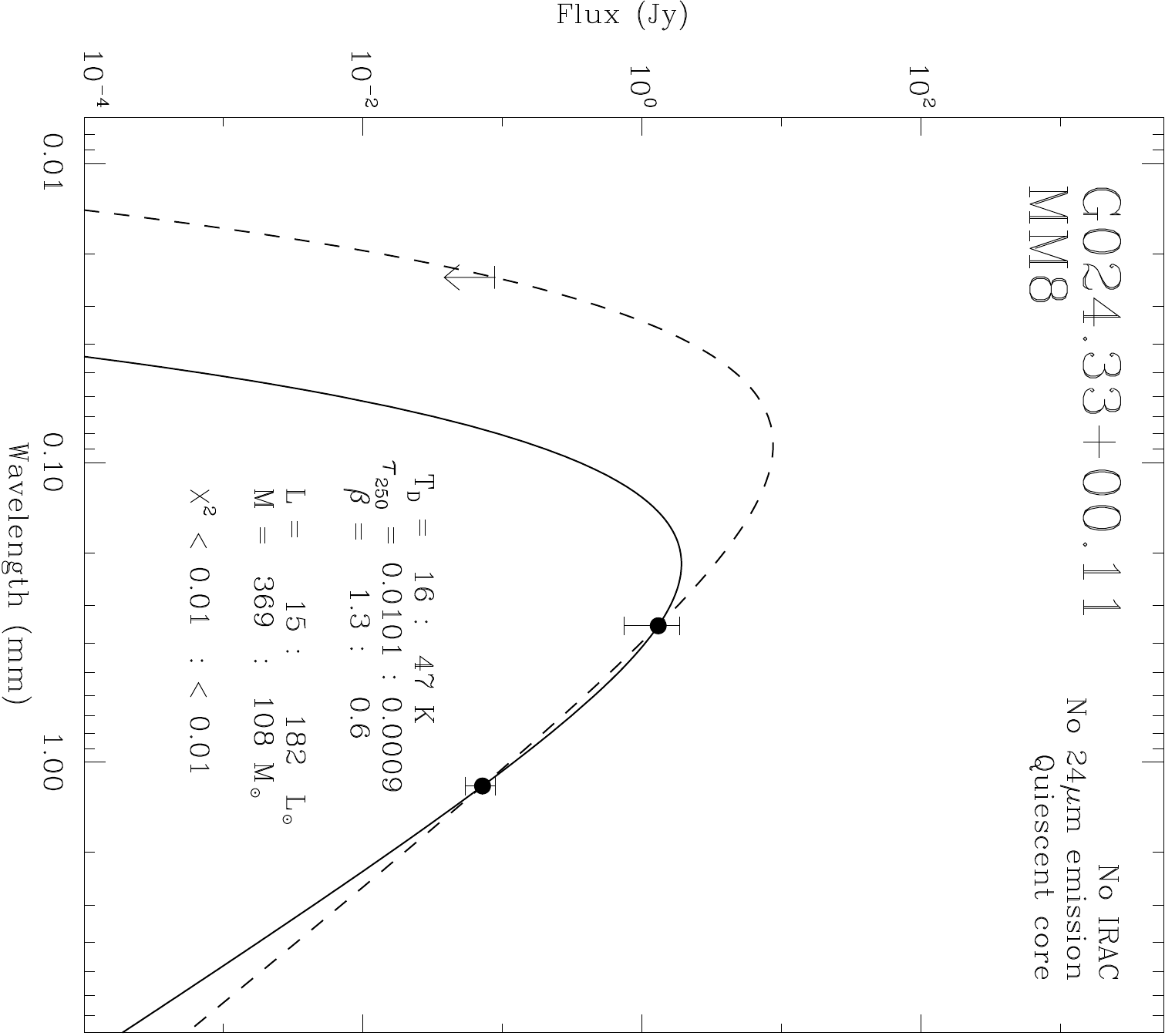}
\includegraphics[angle=90,width=0.5\textwidth]{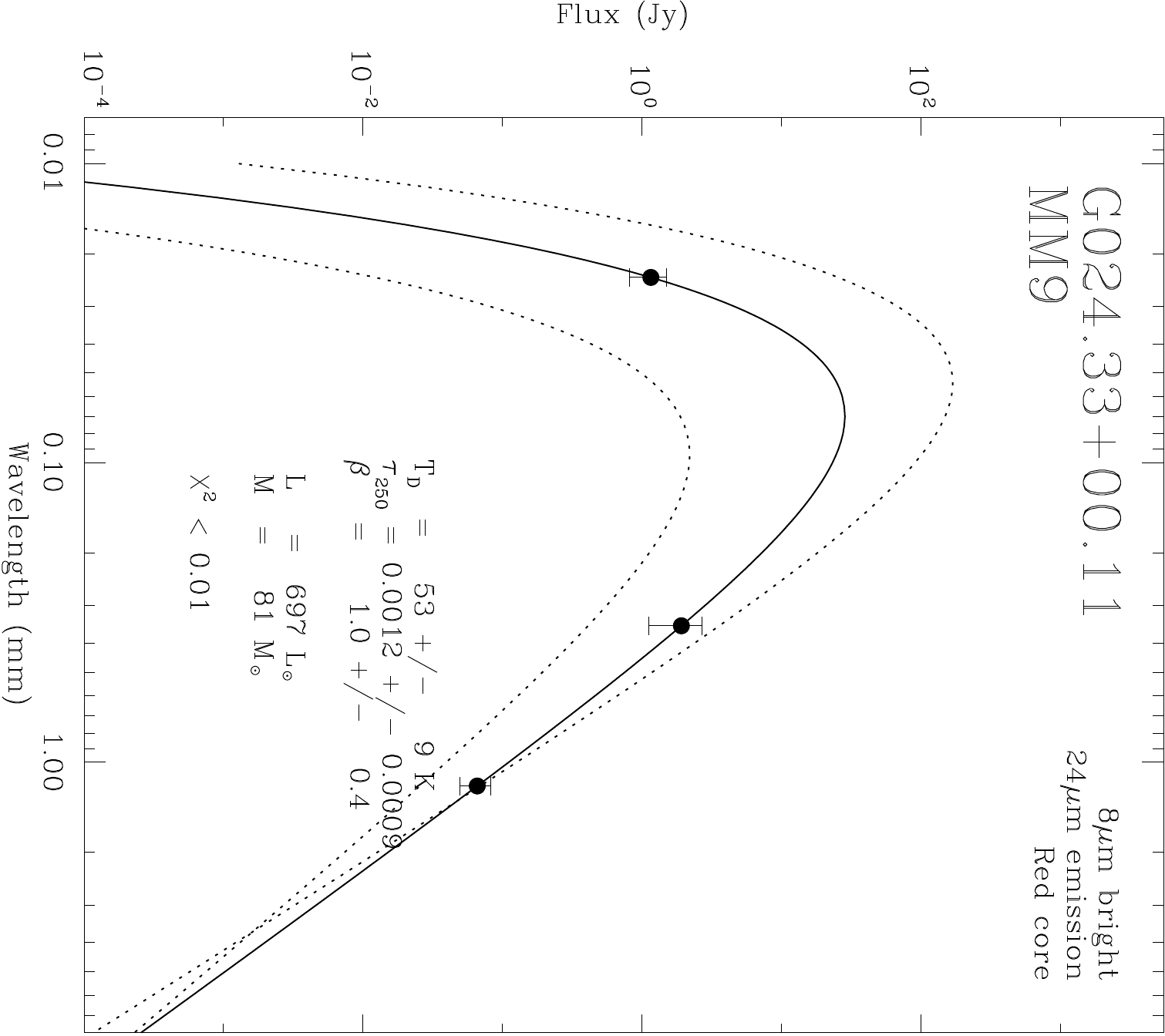}\\
\includegraphics[angle=90,width=0.5\textwidth]{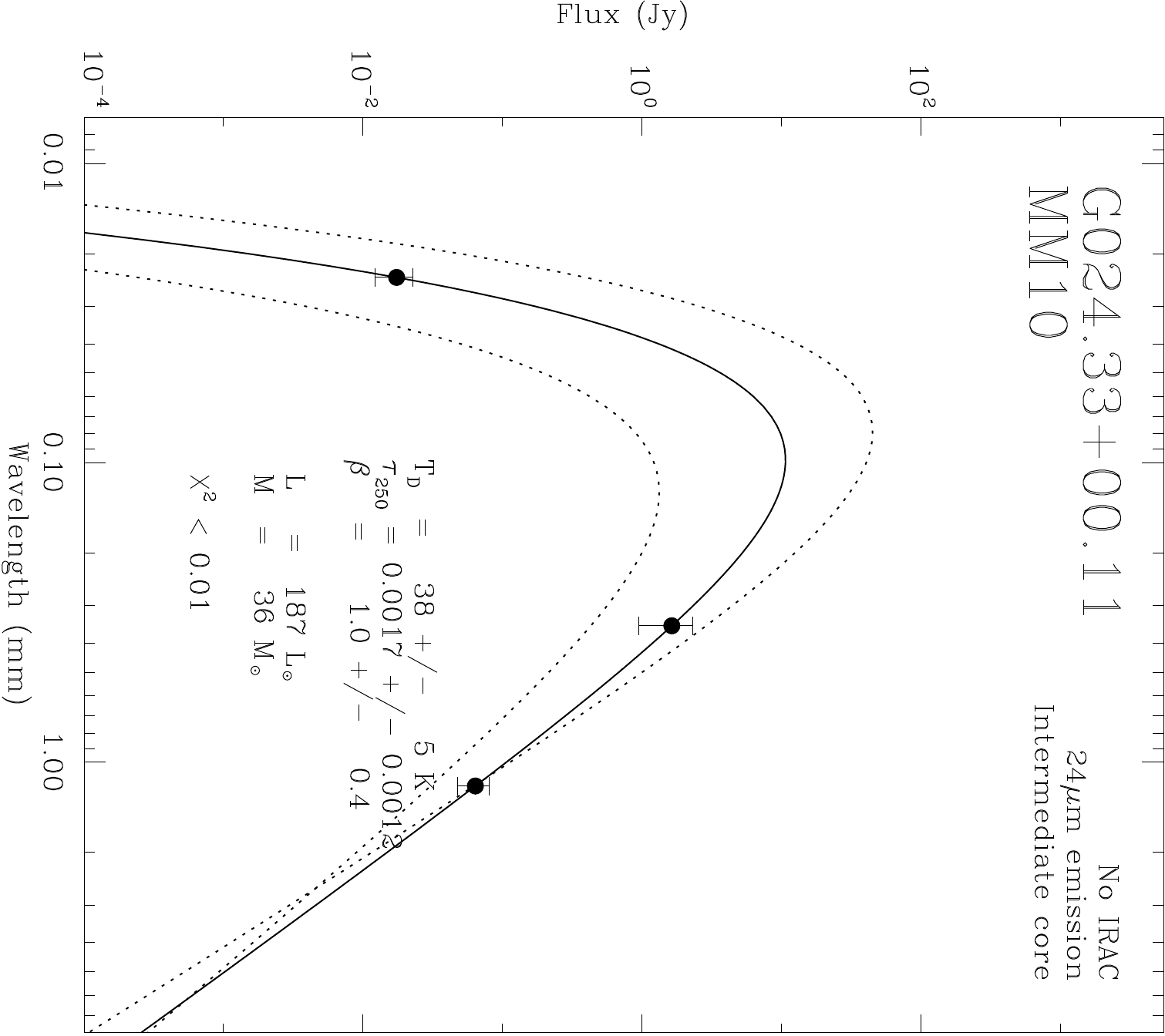}
\includegraphics[angle=90,width=0.5\textwidth]{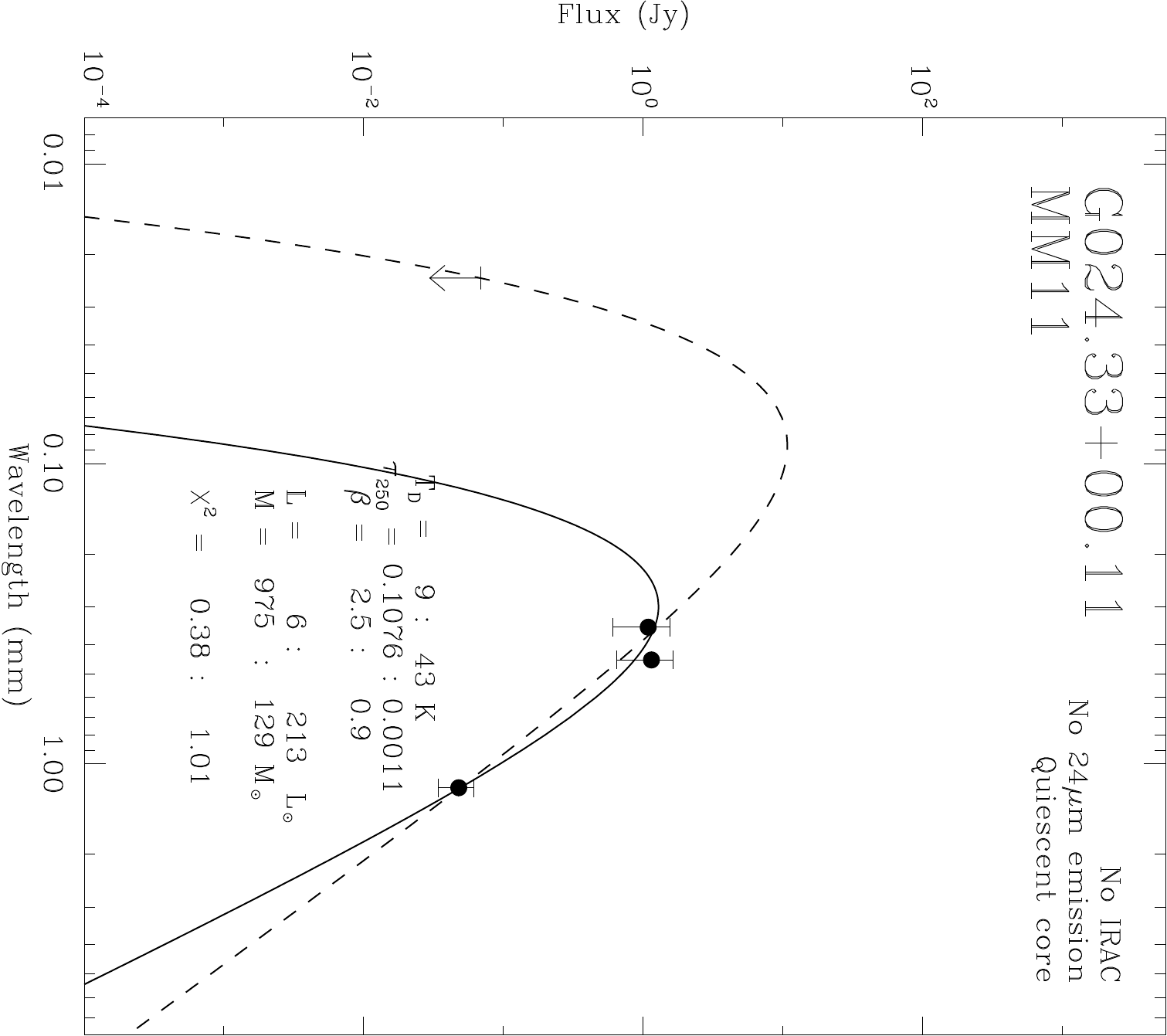}\\
\caption{\label{seds-51} \Spitzer\, 24\,\um\, image overlaid  
   with 1.2\,mm continuum emission for \irdcfiftyone\, (contour levels
   are 30, 60, 90, 120, 240, 360, 480, 840, 1200\,mJy
   beam$^{-1}$). The lower panels show the broadband
   SEDs for cores within this IRDC.  The fluxes derived from the
   millimeter, sub-millimeter, and far-IR  continuum data are shown as filled
   circles (with the corresponding error bars), while the 24\,\um\, fluxes are shown as  either a filled circle (when included within the fit), an open circle (when excluded from the fit),  or as an upper limit arrow. For cores that have measured fluxes only in the millimeter/sub-millimeter regime (i.e.\, a limit at 24\,\um), we show the results from two fits: one using only the measured fluxes (solid line; lower limit), while the other includes the 24\,\um\, limit as a real data (dashed line; upper limit). In all other cases, the solid line is the best fit gray-body, while the dotted lines correspond to the functions determined using the errors for the T$_{D}$, $\tau$, and $\beta$ output from the fitting.  Labeled on each plot is the IRDC and core name,  classification, and the derived parameters.}
\end{figure}
\clearpage 
\begin{figure}
\begin{center}
\includegraphics[angle=0,width=0.6\textwidth]{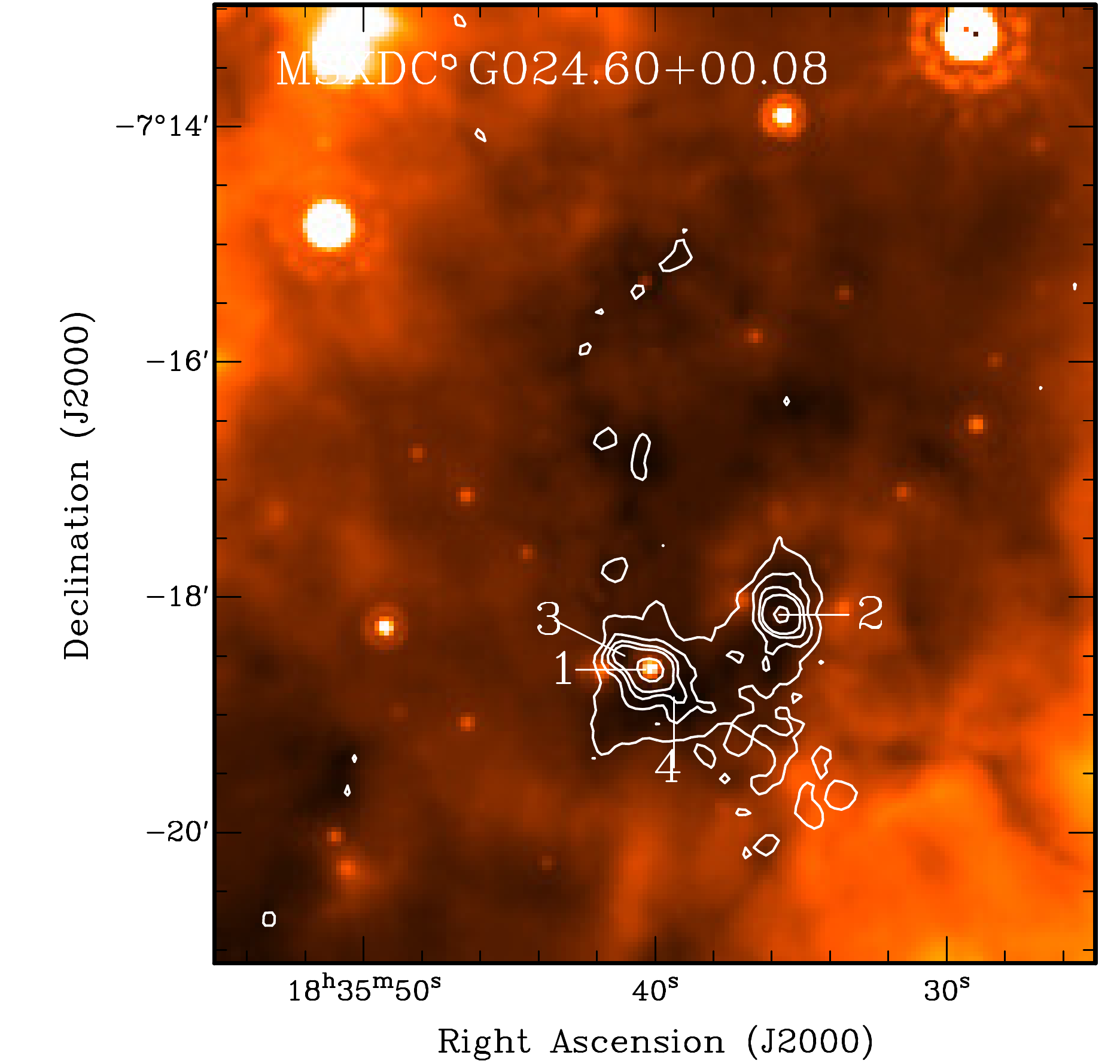}\\
\end{center}
\includegraphics[angle=90,width=0.5\textwidth,clip=true]{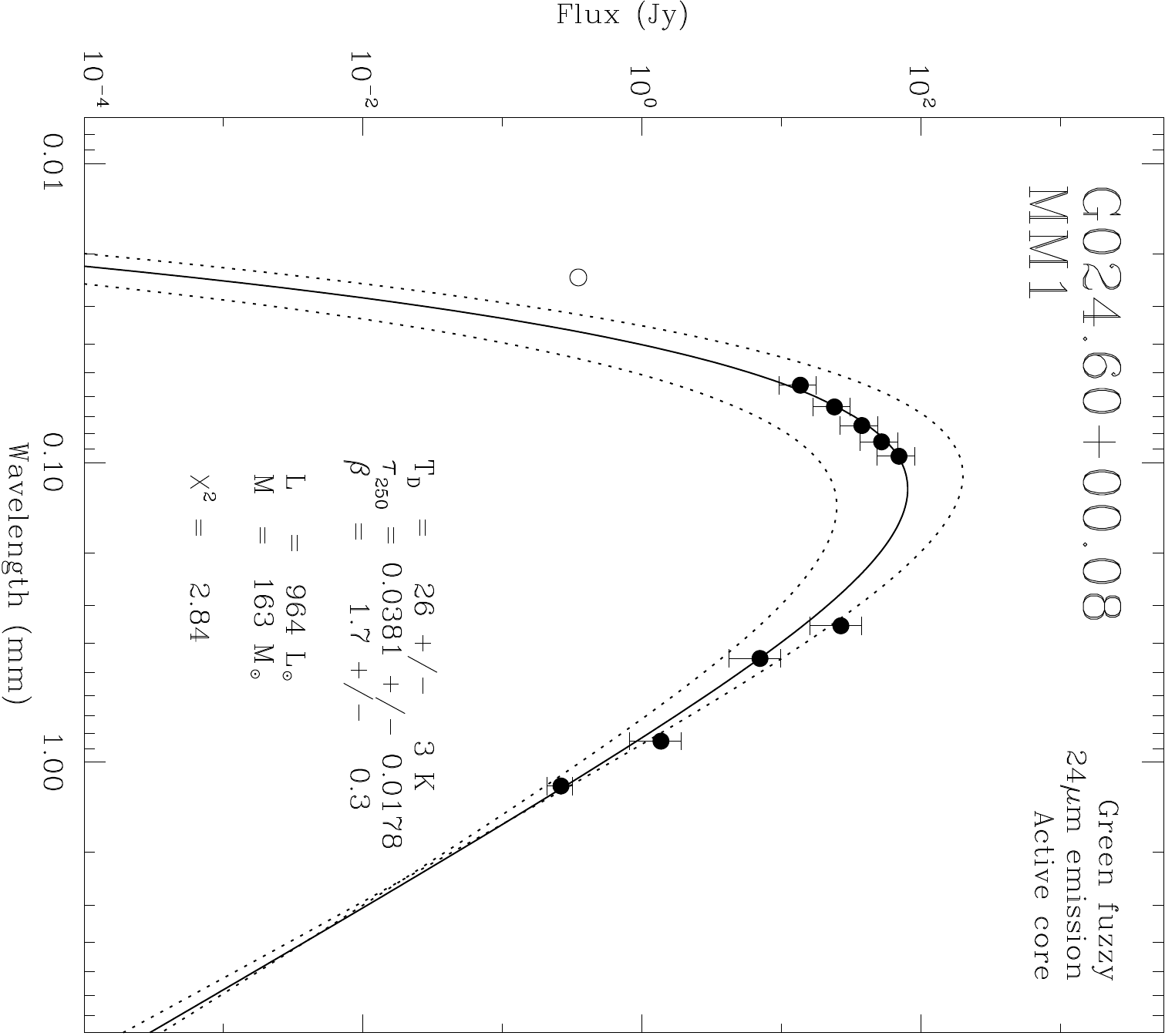}
\includegraphics[angle=90,width=0.5\textwidth,clip=true]{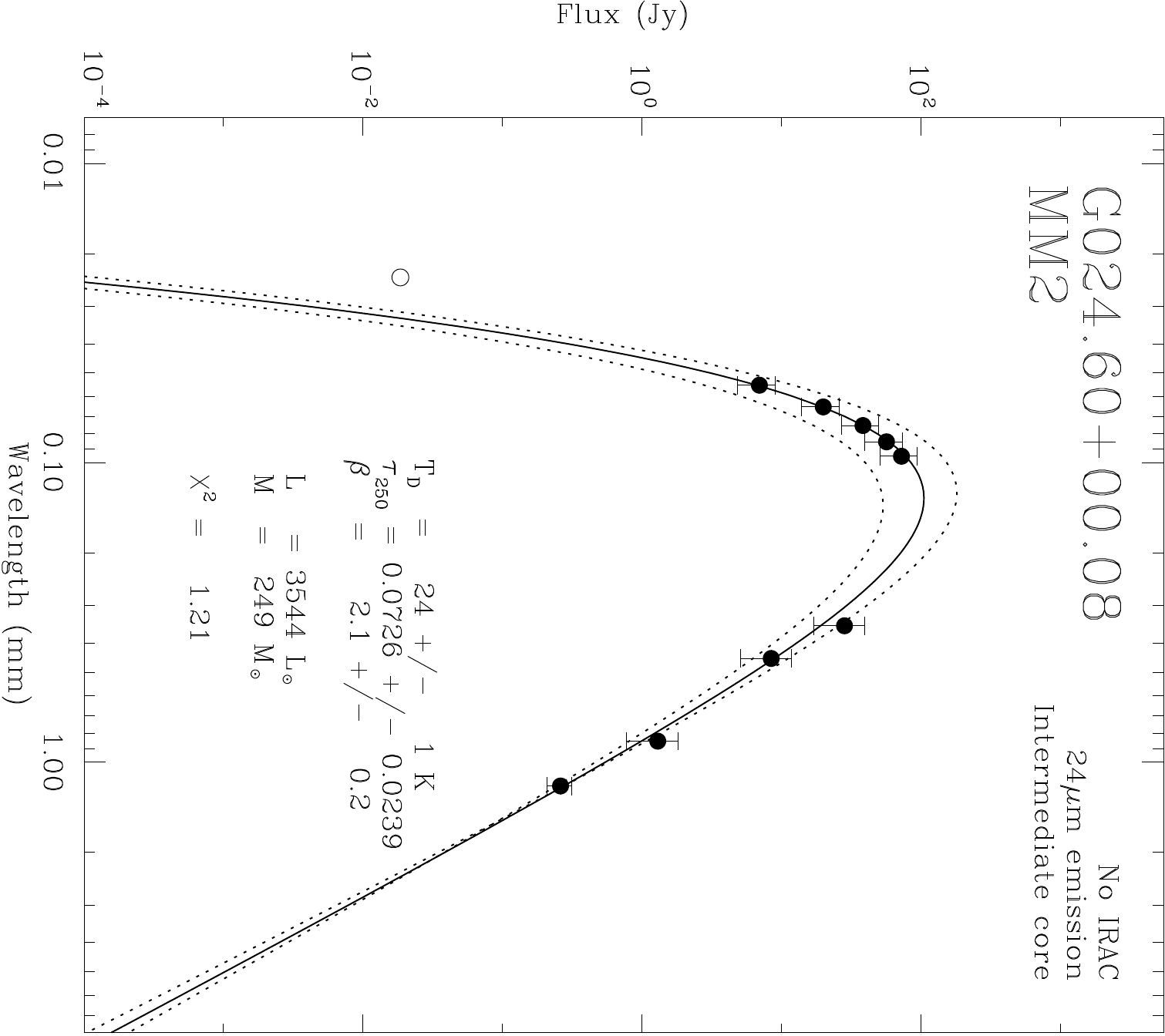}\\
\end{figure}
\clearpage 
\begin{figure}
\includegraphics[angle=90,width=0.5\textwidth,clip=true]{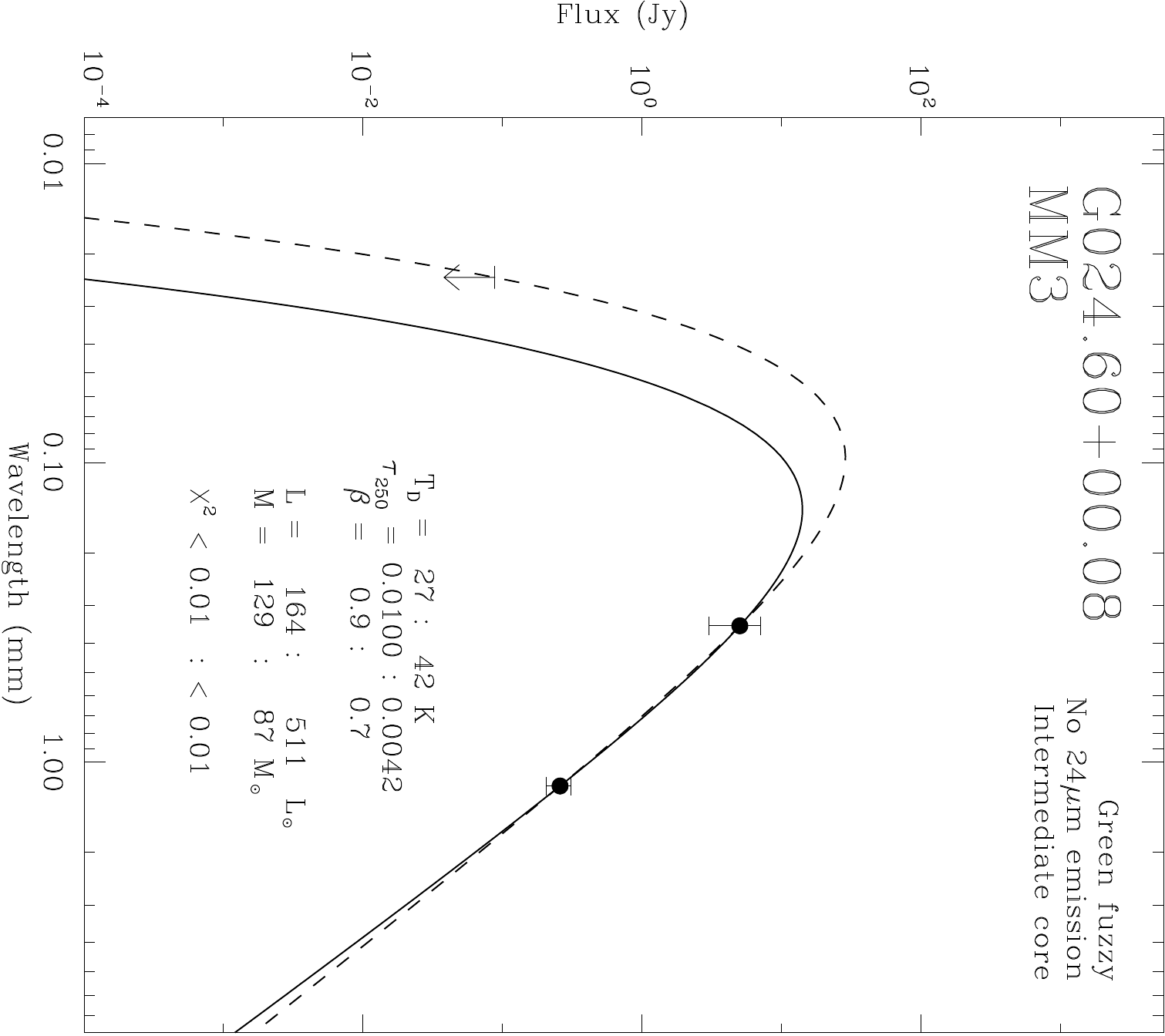}
\caption{\label{seds-9} \Spitzer\, 24\,\um\, image overlaid  
   with 1.2\,mm continuum emission for \irdcnine\, (contour levels are
   30, 60, 90, 120, 240, 360, 480, 840, 1200\,mJy beam$^{-1}$). The lower panels show the broadband
   SEDs for cores within this IRDC.  The fluxes derived from the
   millimeter, sub-millimeter, and far-IR  continuum data are shown as filled
   circles (with the corresponding error bars), while the 24\,\um\, fluxes are shown as  either a filled circle (when included within the fit), an open circle (when excluded from the fit),  or as an upper limit arrow. For cores that have measured fluxes only in the millimeter/sub-millimeter regime (i.e.\, a limit at 24\,\um), we show the results from two fits: one using only the measured fluxes (solid line; lower limit), while the other includes the 24\,\um\, limit as a real data (dashed line; upper limit). In all other cases, the solid line is the best fit gray-body, while the dotted lines correspond to the functions determined using the errors for the T$_{D}$, $\tau$, and $\beta$ output from the fitting.  Labeled on each plot is the IRDC and core name,  classification, and the derived parameters.}
\end{figure}
\clearpage 
\begin{figure}
\begin{center}
\includegraphics[angle=0,width=0.6\textwidth]{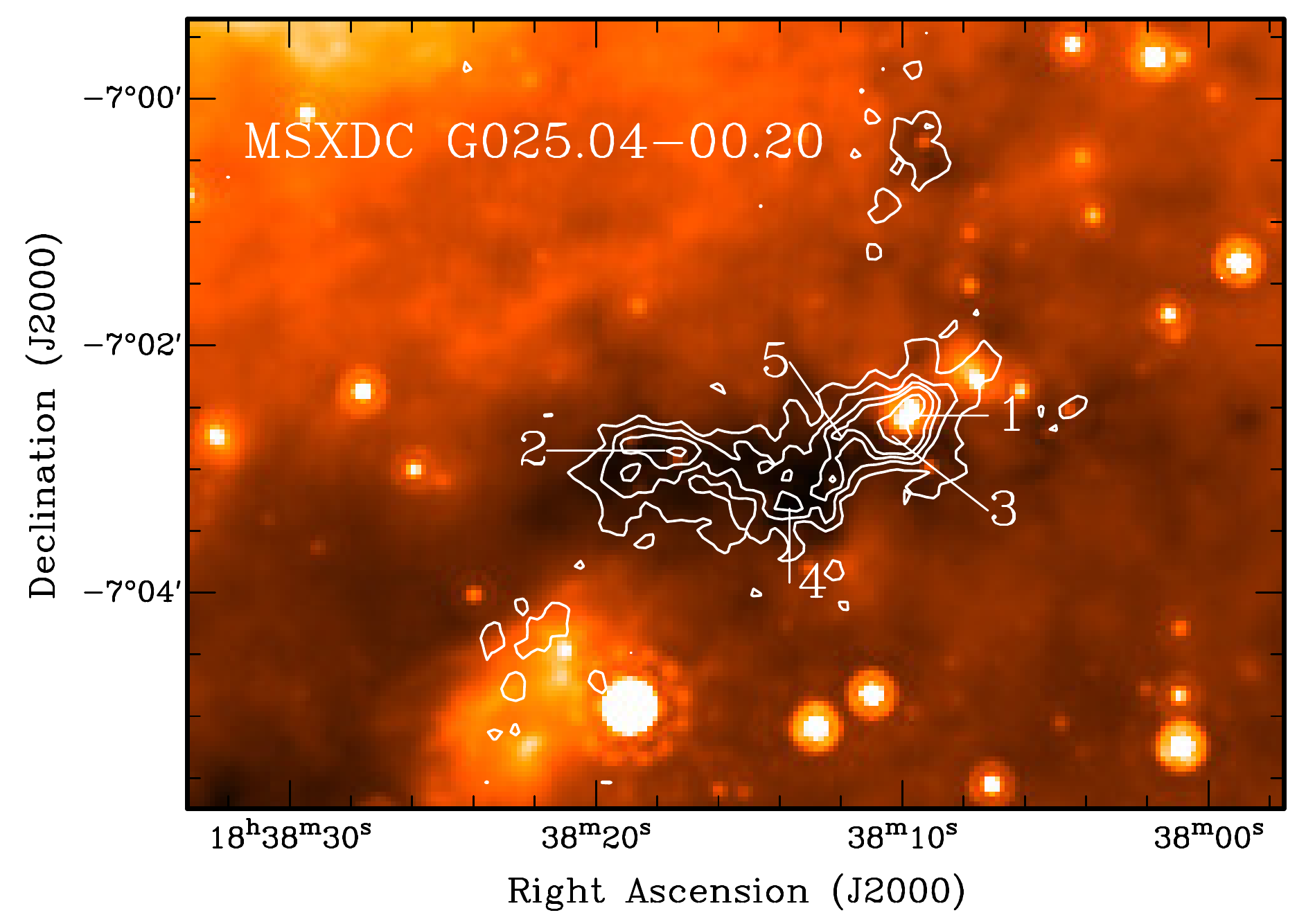}\\
\end{center}
\includegraphics[angle=90,width=0.5\textwidth]{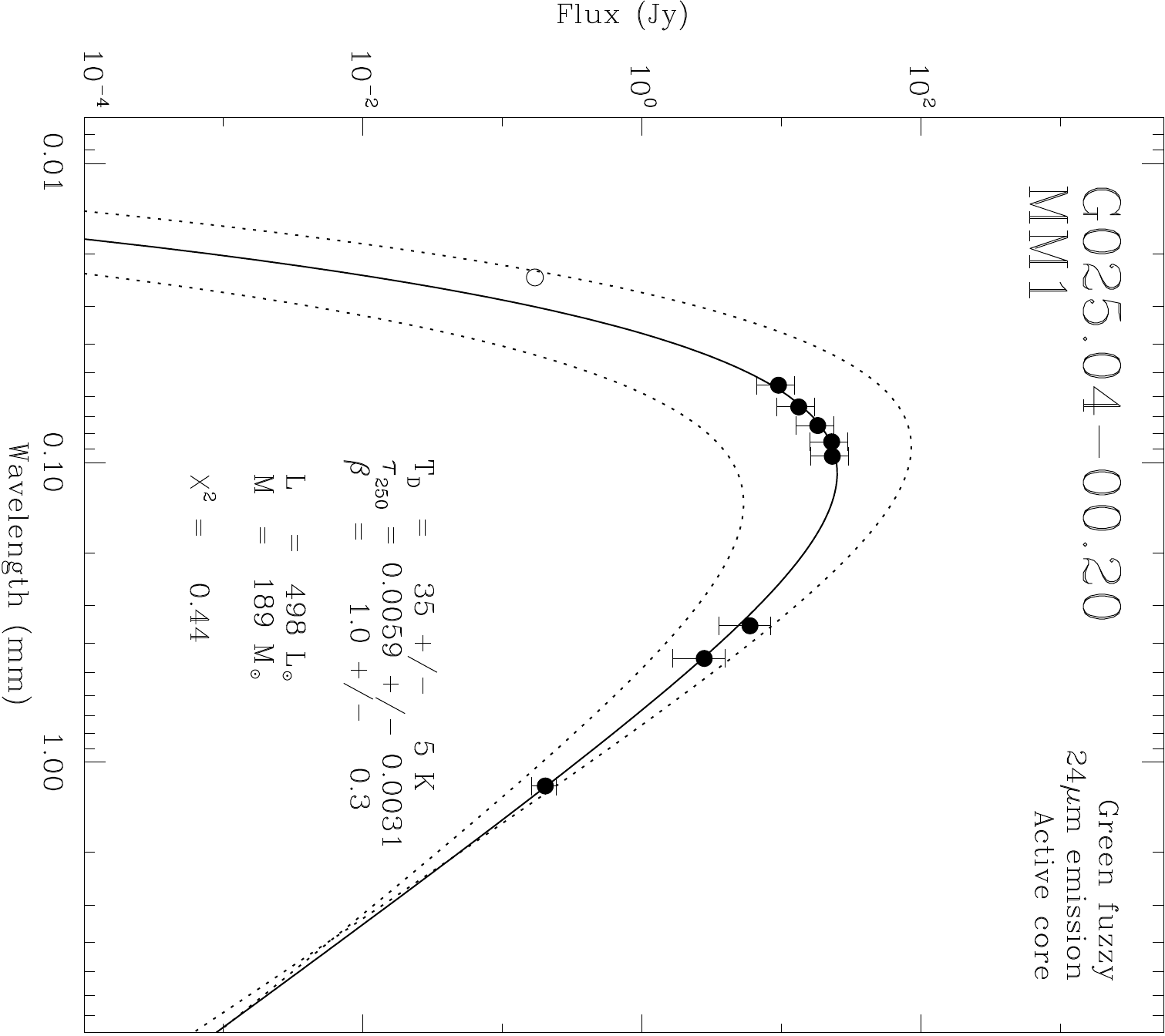}
\includegraphics[angle=90,width=0.5\textwidth]{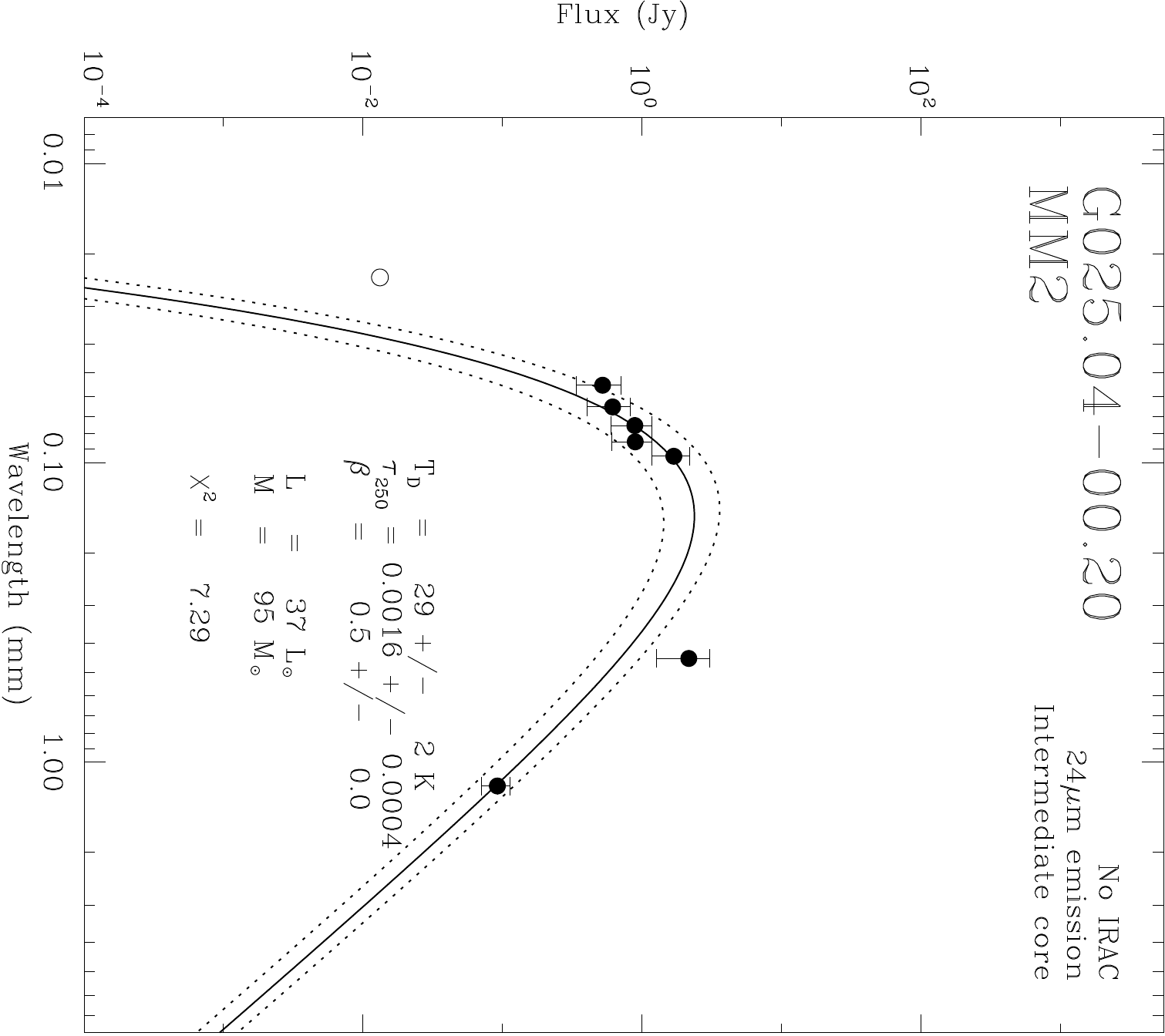}\\
\end{figure}
\clearpage 
\begin{figure}
\includegraphics[angle=90,width=0.5\textwidth]{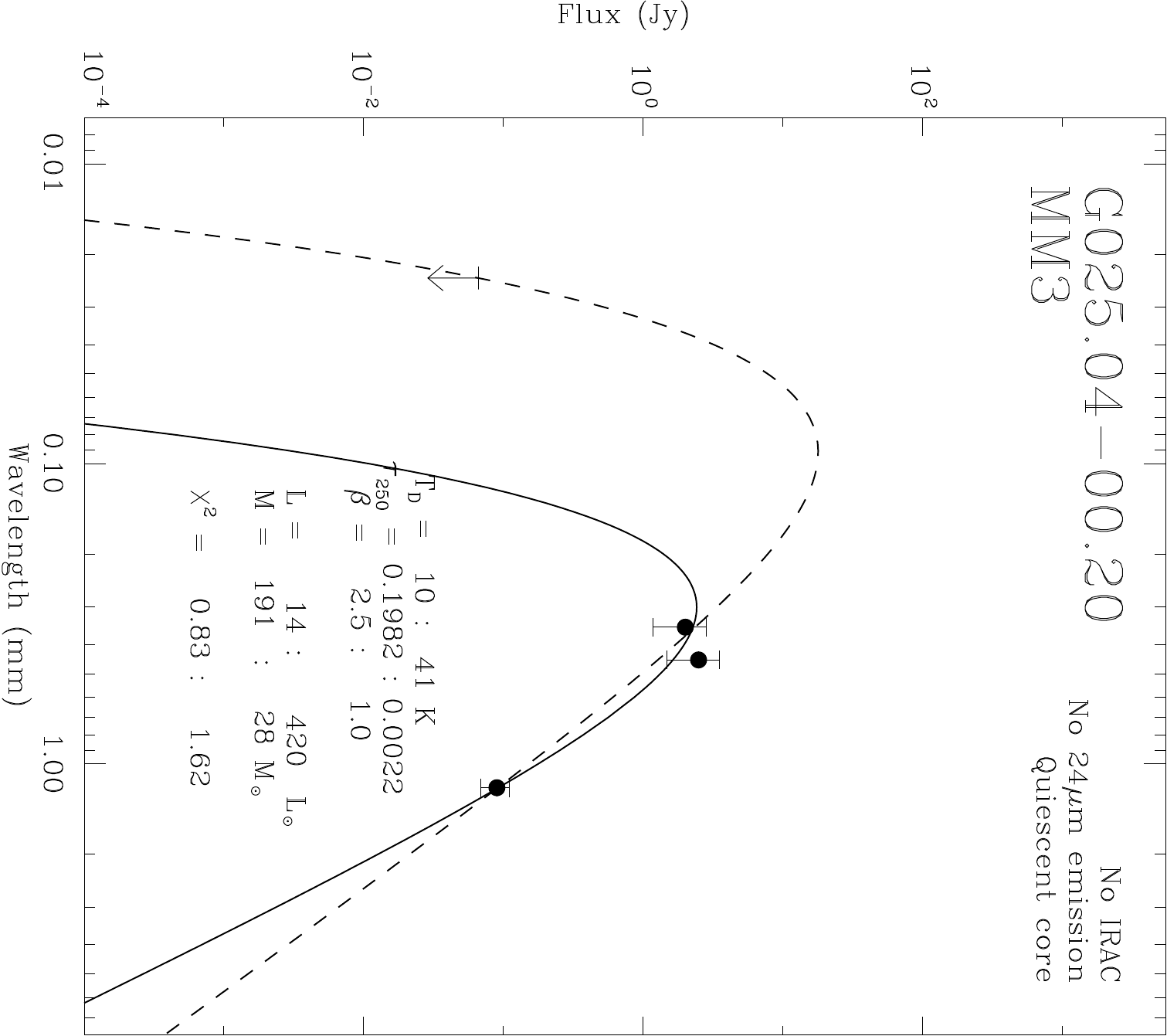}
\includegraphics[angle=90,width=0.5\textwidth]{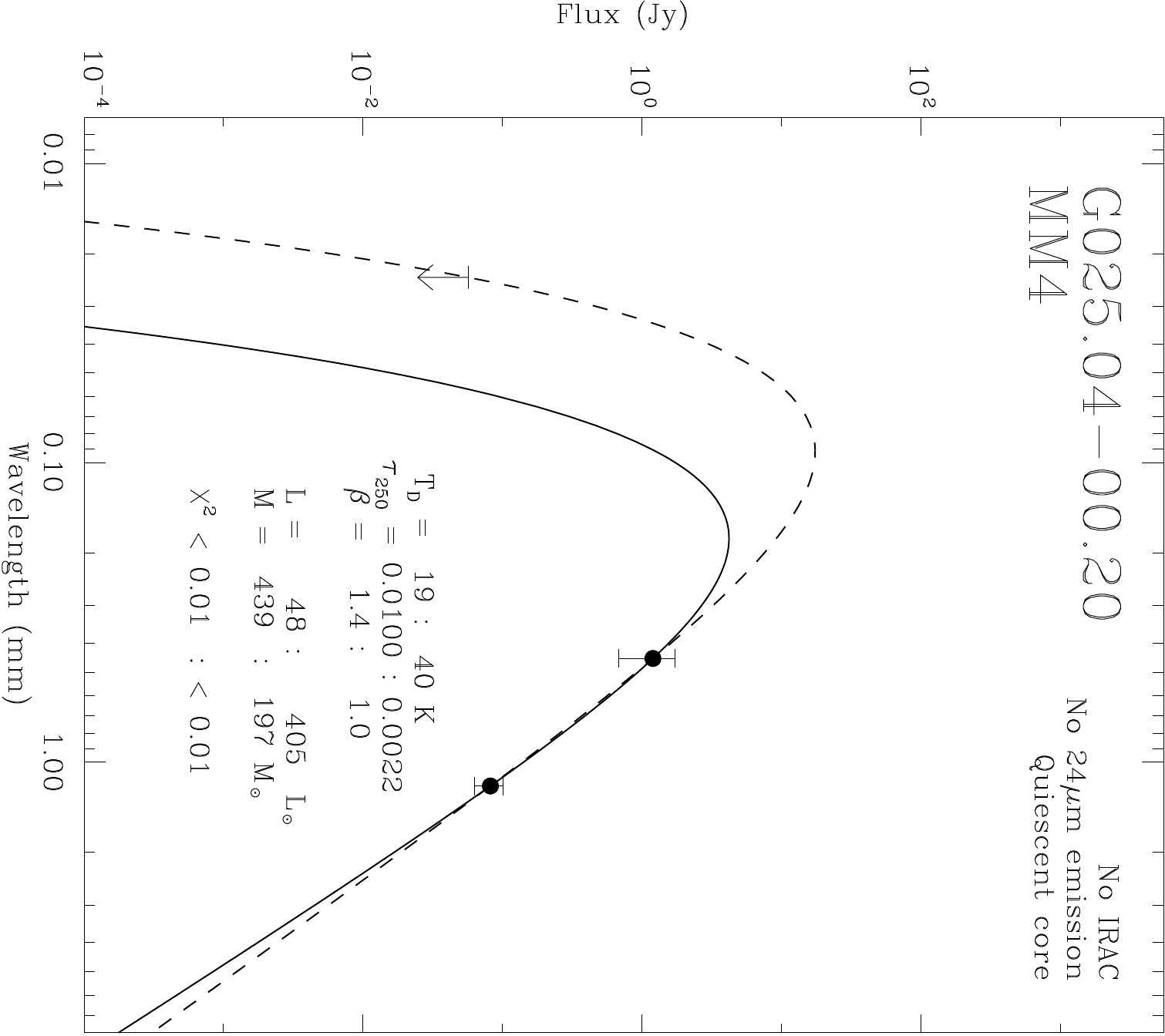}\\
\includegraphics[angle=90,width=0.5\textwidth]{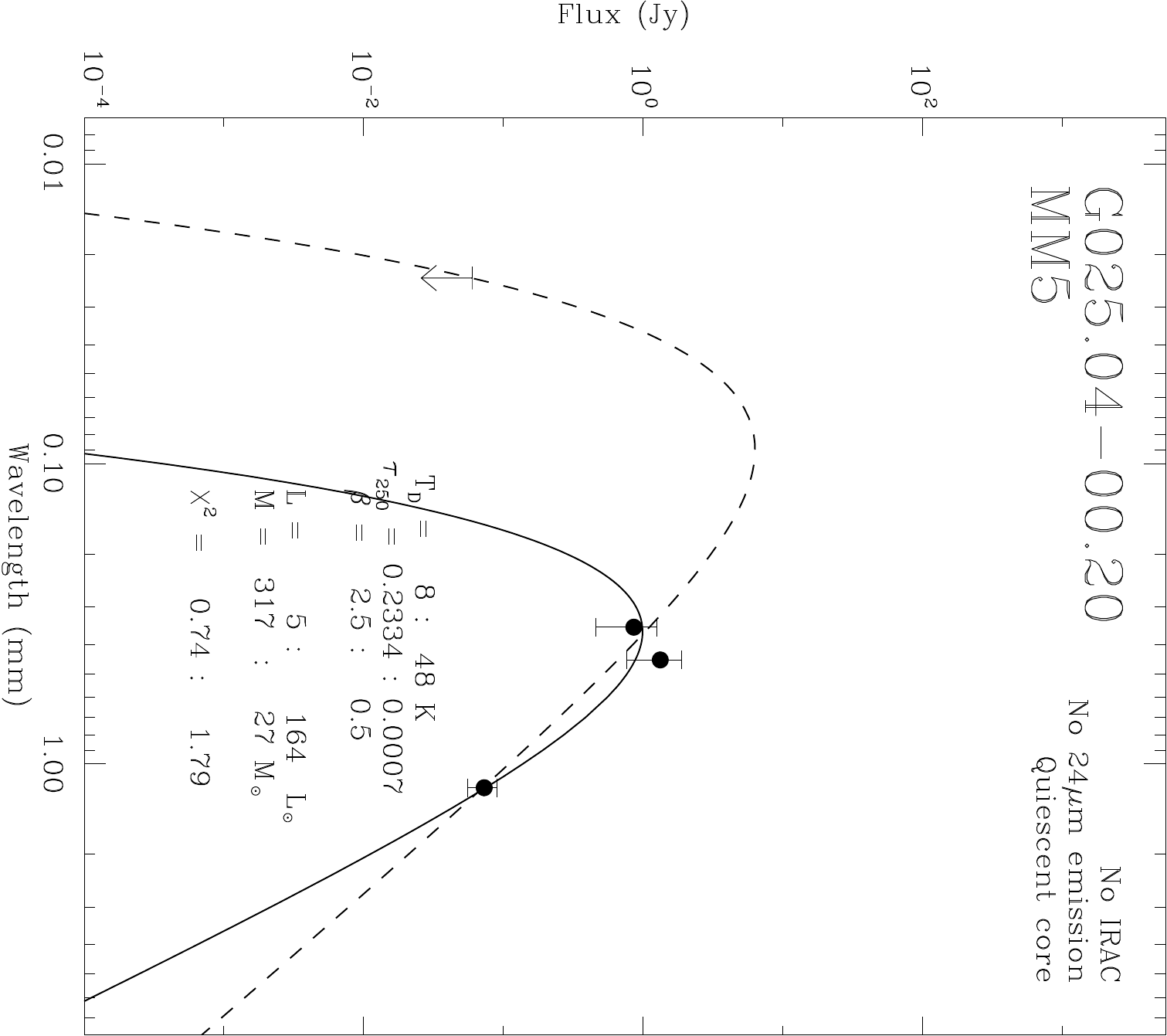}\\
\caption{\label{seds-14}3 \Spitzer\, 24\,\um\, image overlaid  
   with 1.2\,mm continuum emission for \irdcforteen\, (contour levels
   are 30, 60, 90, 120, 240, 360, 480, 840, 1200\,mJy
   beam$^{-1}$). The lower panels show the broadband
   SEDs for cores within this IRDC.  The fluxes derived from the
   millimeter, sub-millimeter, and far-IR  continuum data are shown as filled
   circles (with the corresponding error bars), while the 24\,\um\, fluxes are shown as  either a filled circle (when included within the fit), an open circle (when excluded from the fit),  or as an upper limit arrow. For cores that have measured fluxes only in the millimeter/sub-millimeter regime (i.e.\, a limit at 24\,\um), we show the results from two fits: one using only the measured fluxes (solid line; lower limit), while the other includes the 24\,\um\, limit as a real data (dashed line; upper limit). In all other cases, the solid line is the best fit gray-body, while the dotted lines correspond to the functions determined using the errors for the T$_{D}$, $\tau$, and $\beta$ output from the fitting.  Labeled on each plot is the IRDC and core name,  classification, and the derived parameters.}
\end{figure}
\clearpage 
\begin{figure}
\begin{center}
\includegraphics[angle=0,width=0.6\textwidth]{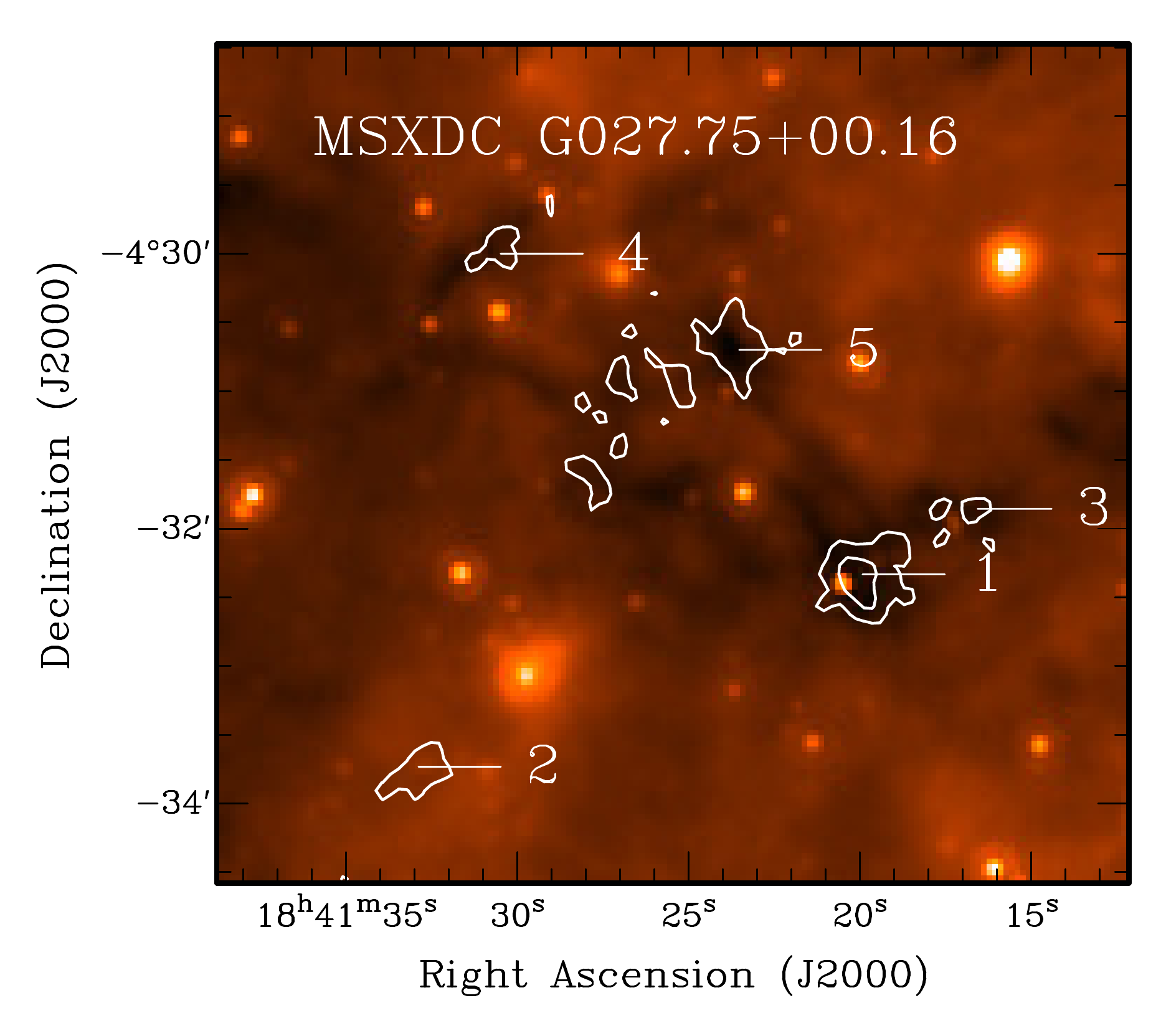}\\
\end{center}
\includegraphics[angle=90,width=0.5\textwidth]{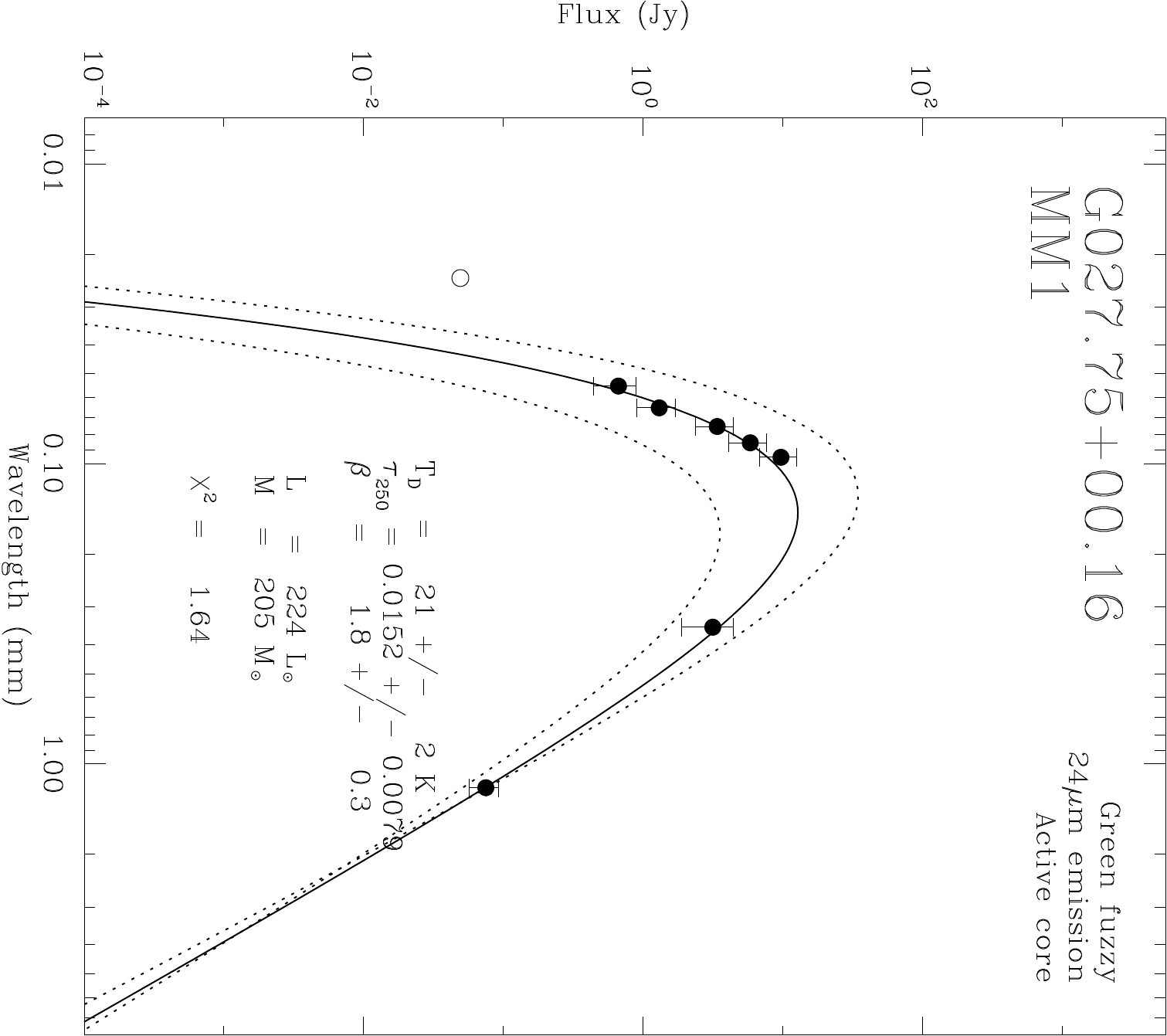}
\includegraphics[angle=90,width=0.5\textwidth]{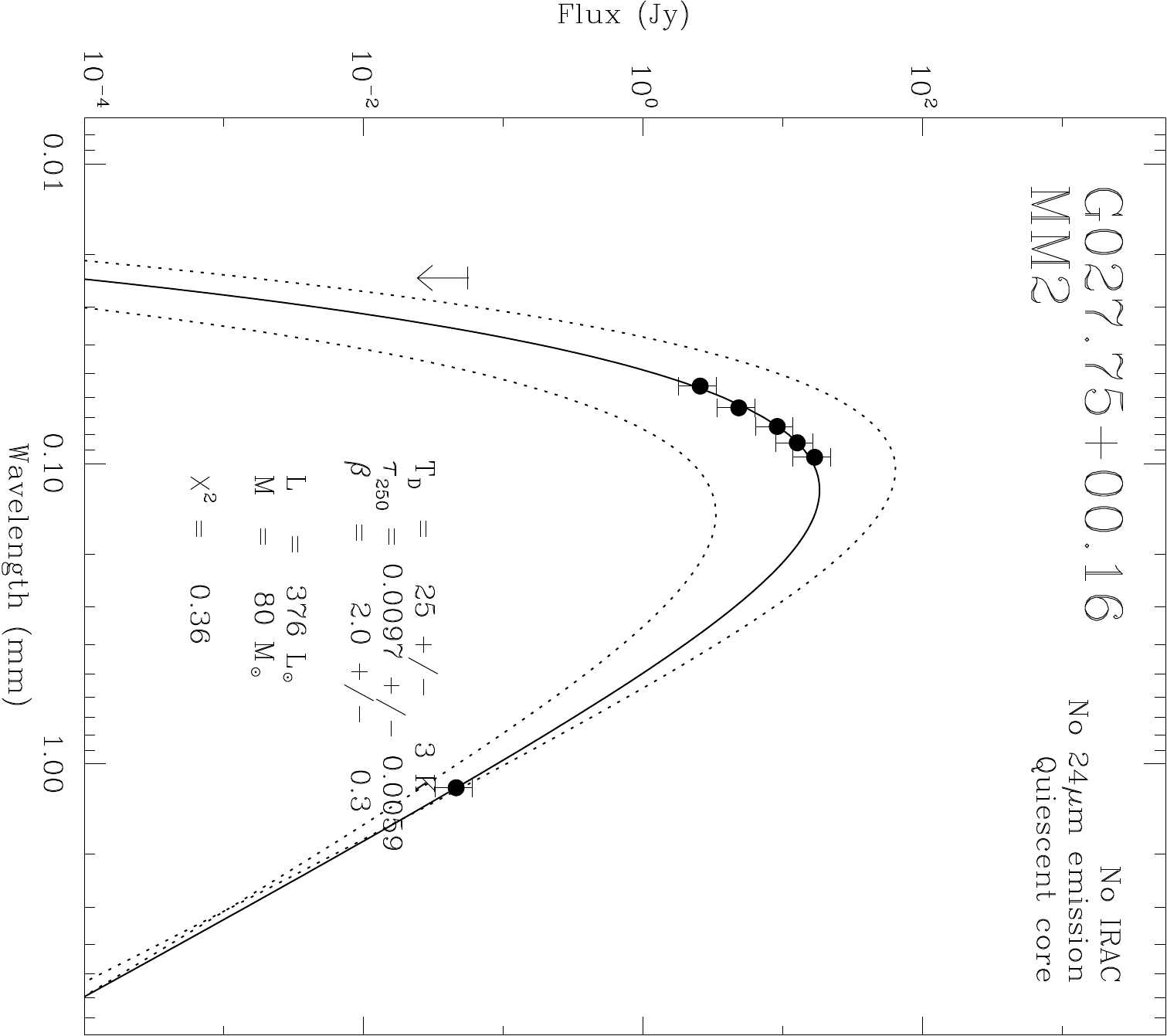}\\
\end{figure}
\clearpage 
\begin{figure}
\includegraphics[angle=90,width=0.5\textwidth]{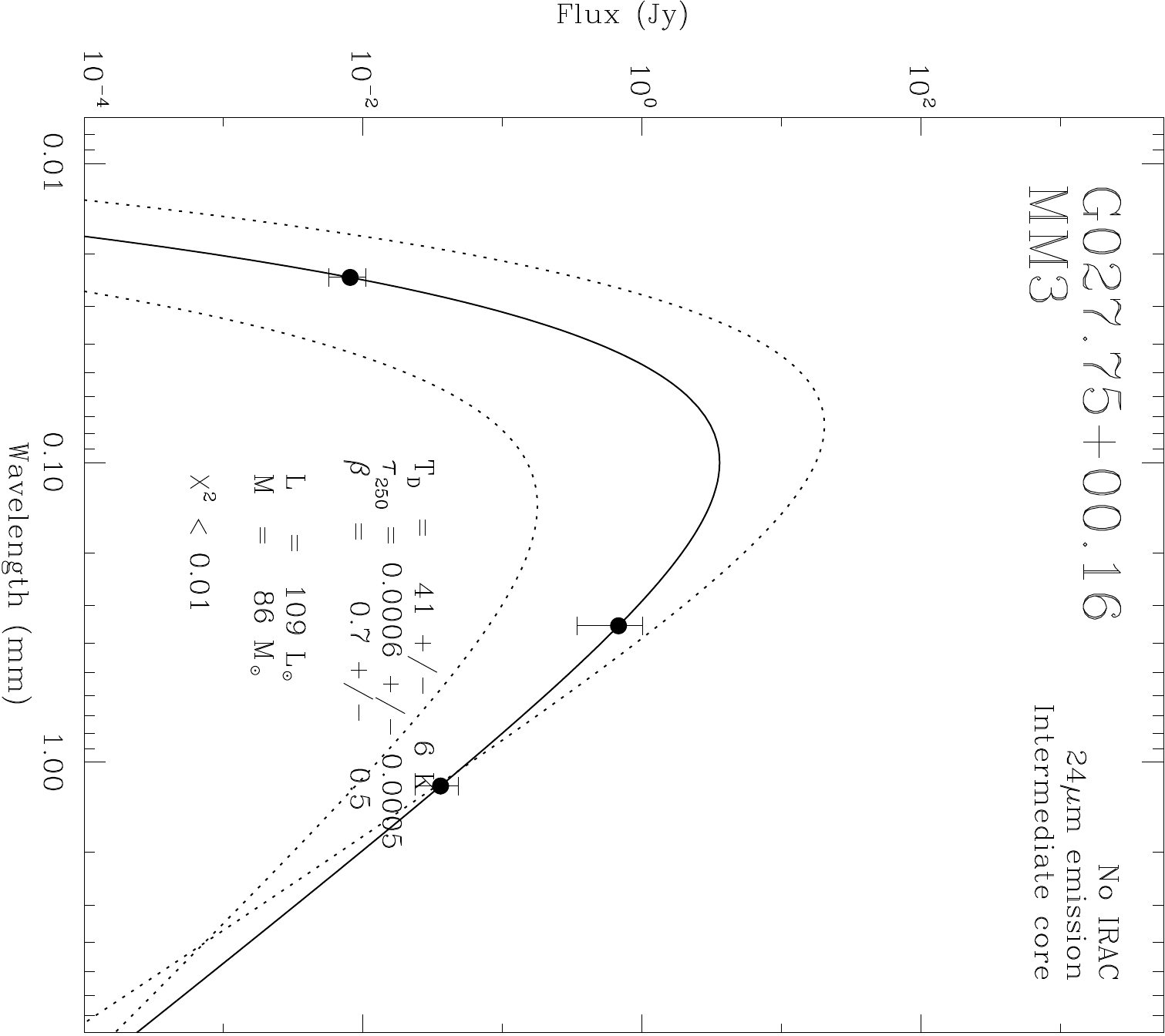}
\includegraphics[angle=90,width=0.5\textwidth]{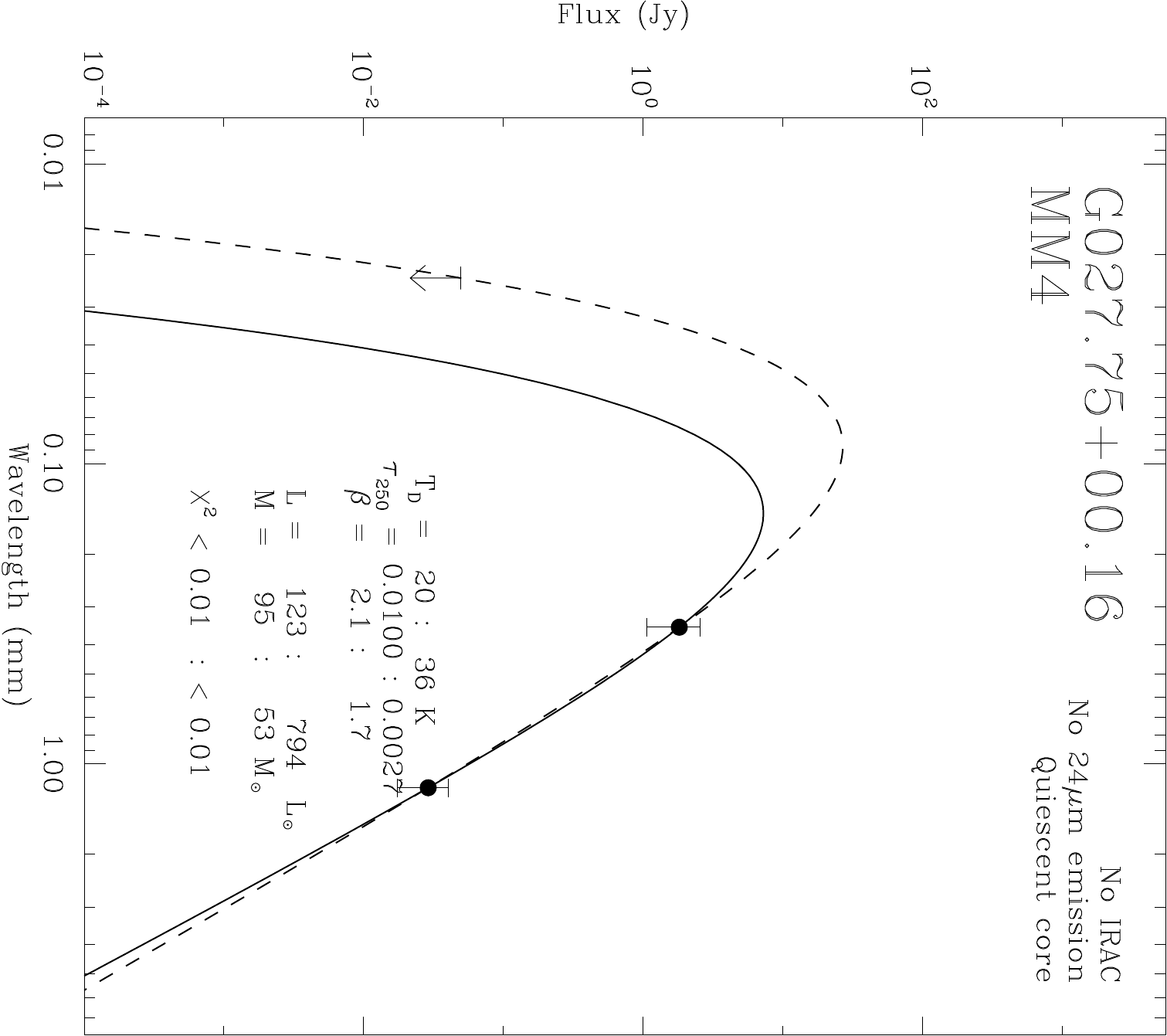}\\
\includegraphics[angle=90,width=0.5\textwidth]{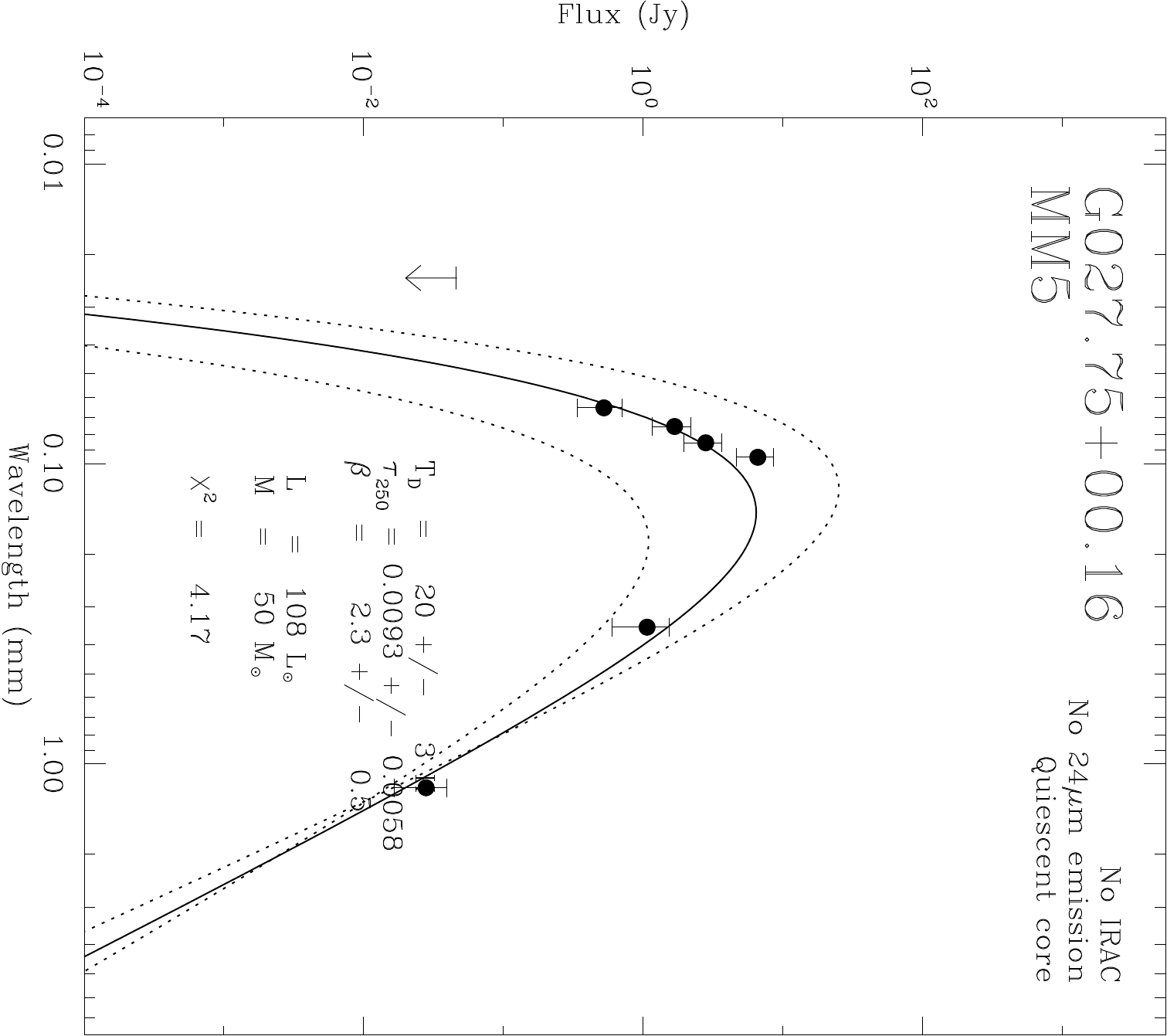}\\
\caption{\label{seds-53} \Spitzer\, 24\,\um\, image overlaid  
   with 1.2\,mm continuum emission for \irdcfiftythree\, (contour levels are
   30, 60, 90, 120, 180\,mJy beam$^{-1}$). The lower panels show the broadband
   SEDs for cores within this IRDC.  The fluxes derived from the
   millimeter, sub-millimeter, and far-IR  continuum data are shown as filled
   circles (with the corresponding error bars), while the 24\,\um\, fluxes are shown as  either a filled circle (when included within the fit), an open circle (when excluded from the fit),  or as an upper limit arrow. For cores that have measured fluxes only in the millimeter/sub-millimeter regime (i.e.\, a limit at 24\,\um), we show the results from two fits: one using only the measured fluxes (solid line; lower limit), while the other includes the 24\,\um\, limit as a real data (dashed line; upper limit). In all other cases, the solid line is the best fit gray-body, while the dotted lines correspond to the functions determined using the errors for the T$_{D}$, $\tau$, and $\beta$ output from the fitting.  Labeled on each plot is the IRDC and core name,  classification, and the derived parameters.}
\end{figure}
\clearpage 
\begin{figure}
\begin{center}
\includegraphics[angle=0,width=0.6\textwidth]{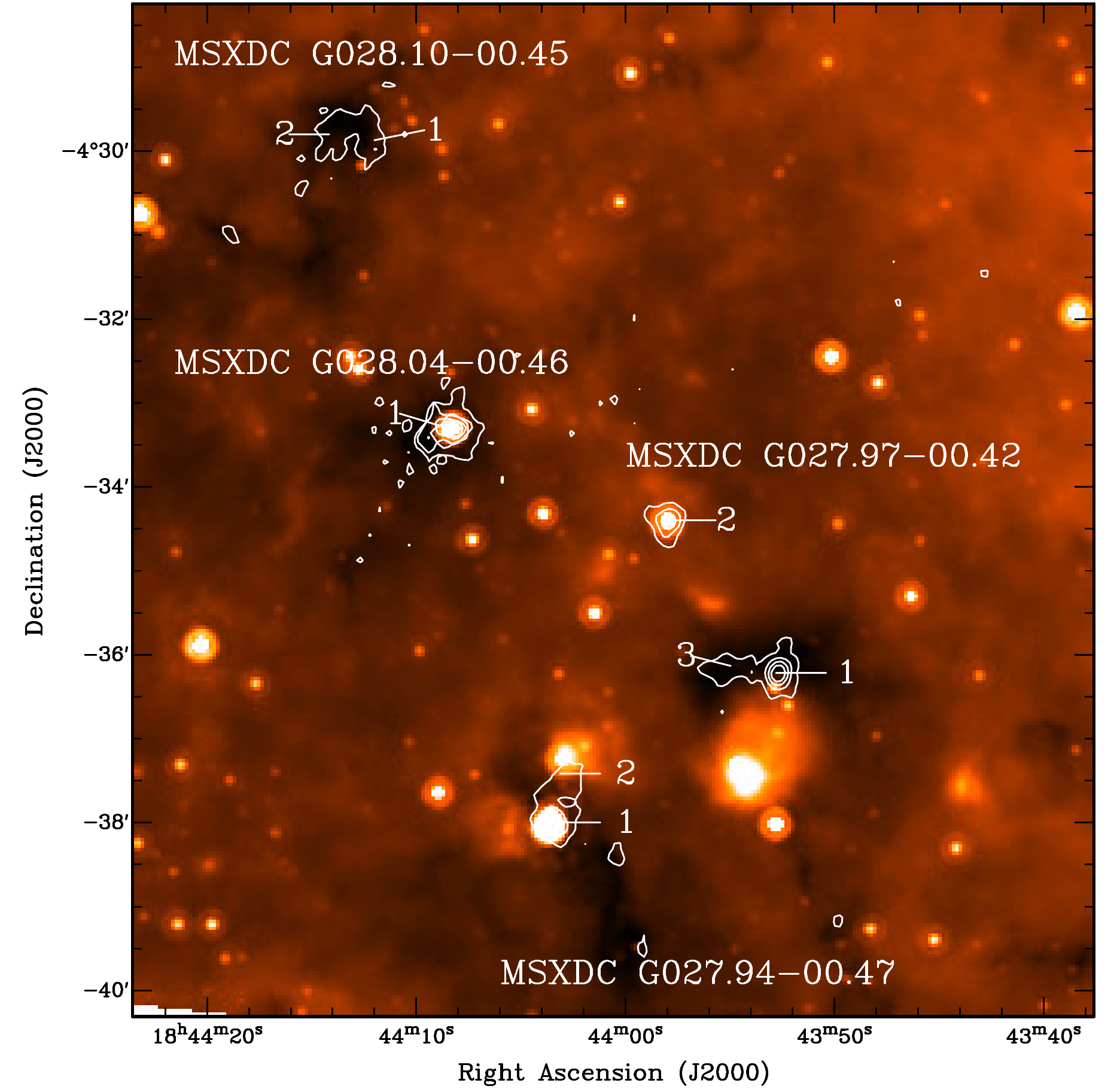}\\
\end{center}
\includegraphics[angle=90,width=0.5\textwidth]{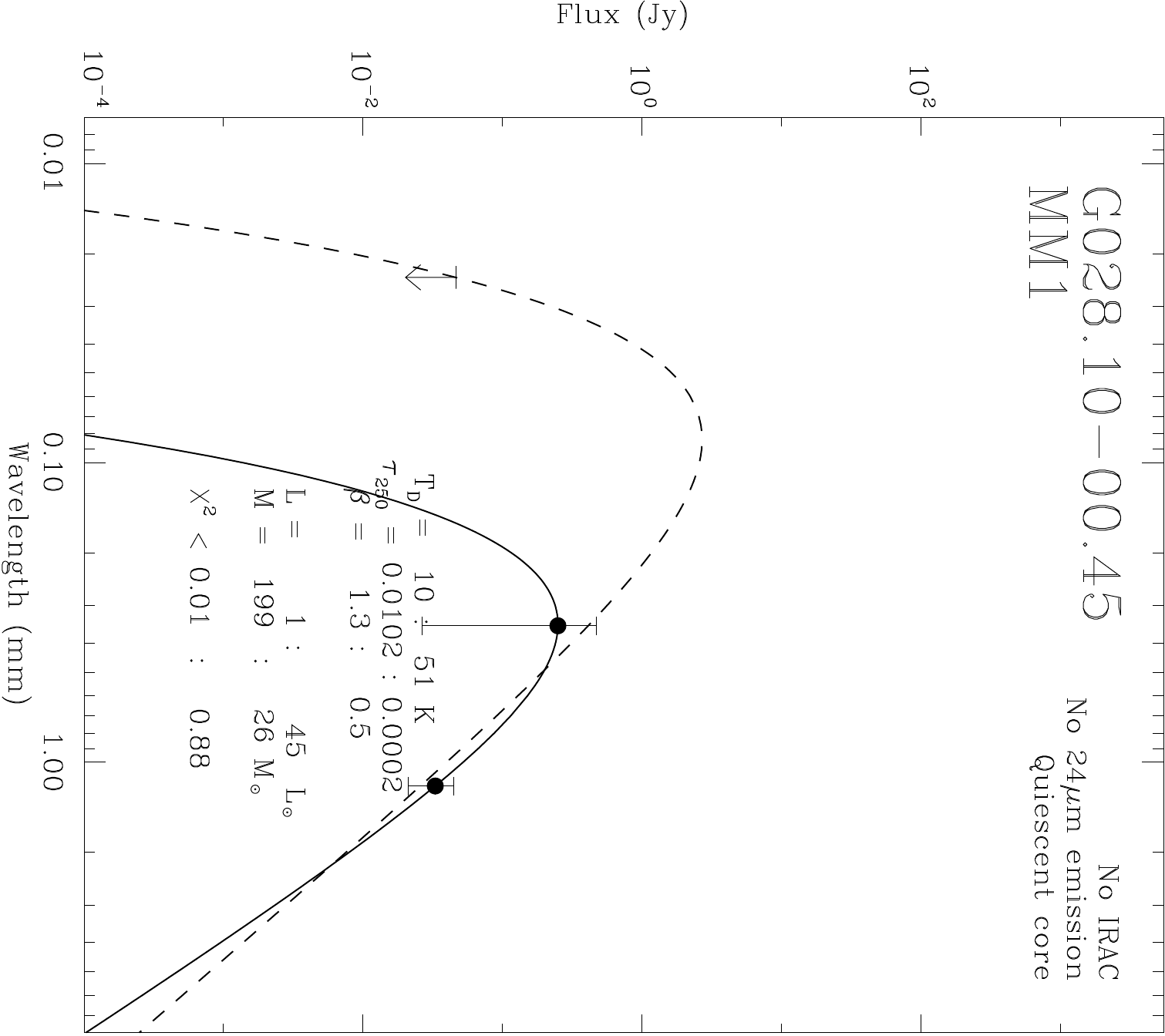}
\includegraphics[angle=90,width=0.5\textwidth]{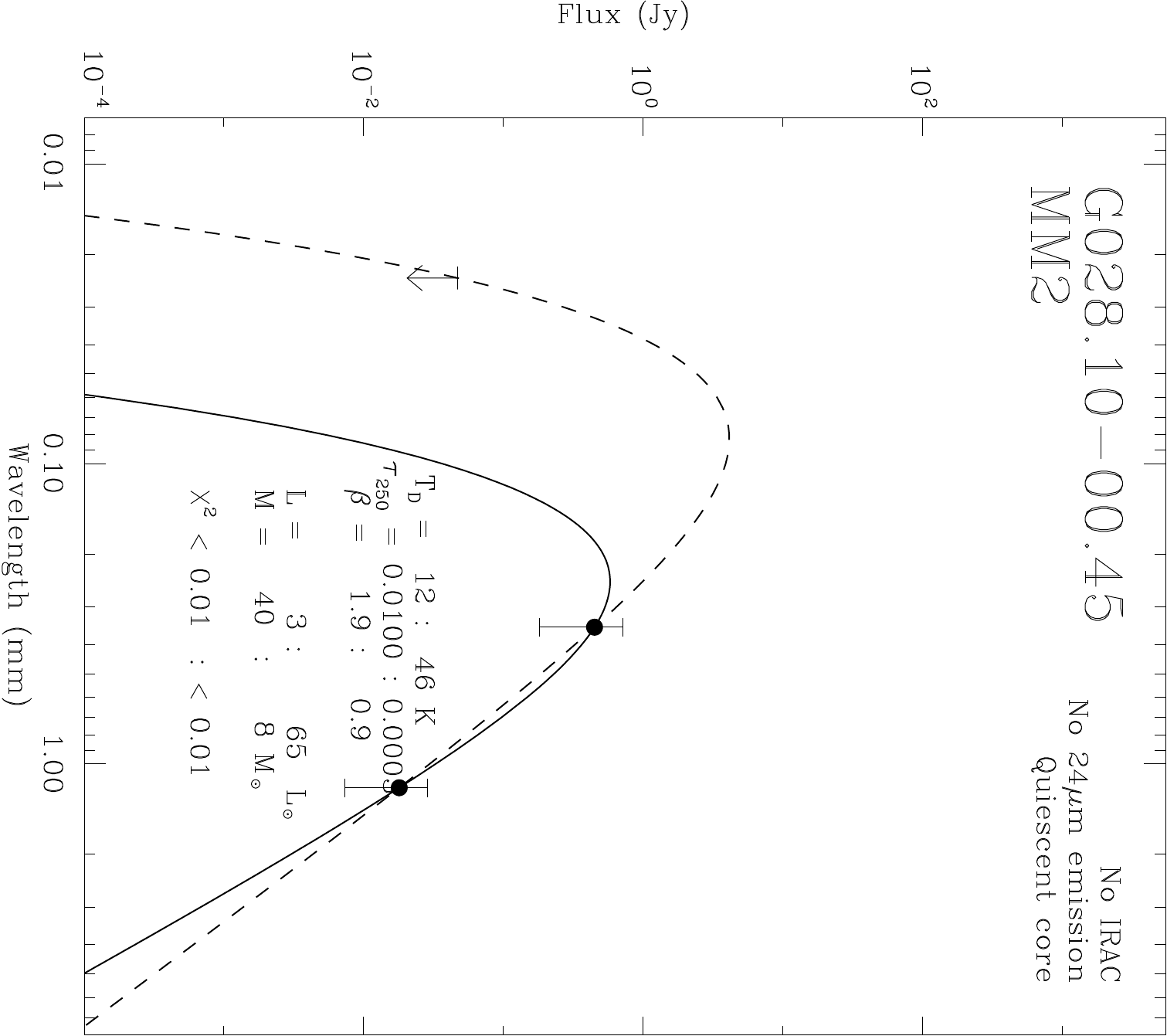}\\
\end{figure}
\clearpage 
\begin{figure}
\includegraphics[angle=90,width=0.5\textwidth]{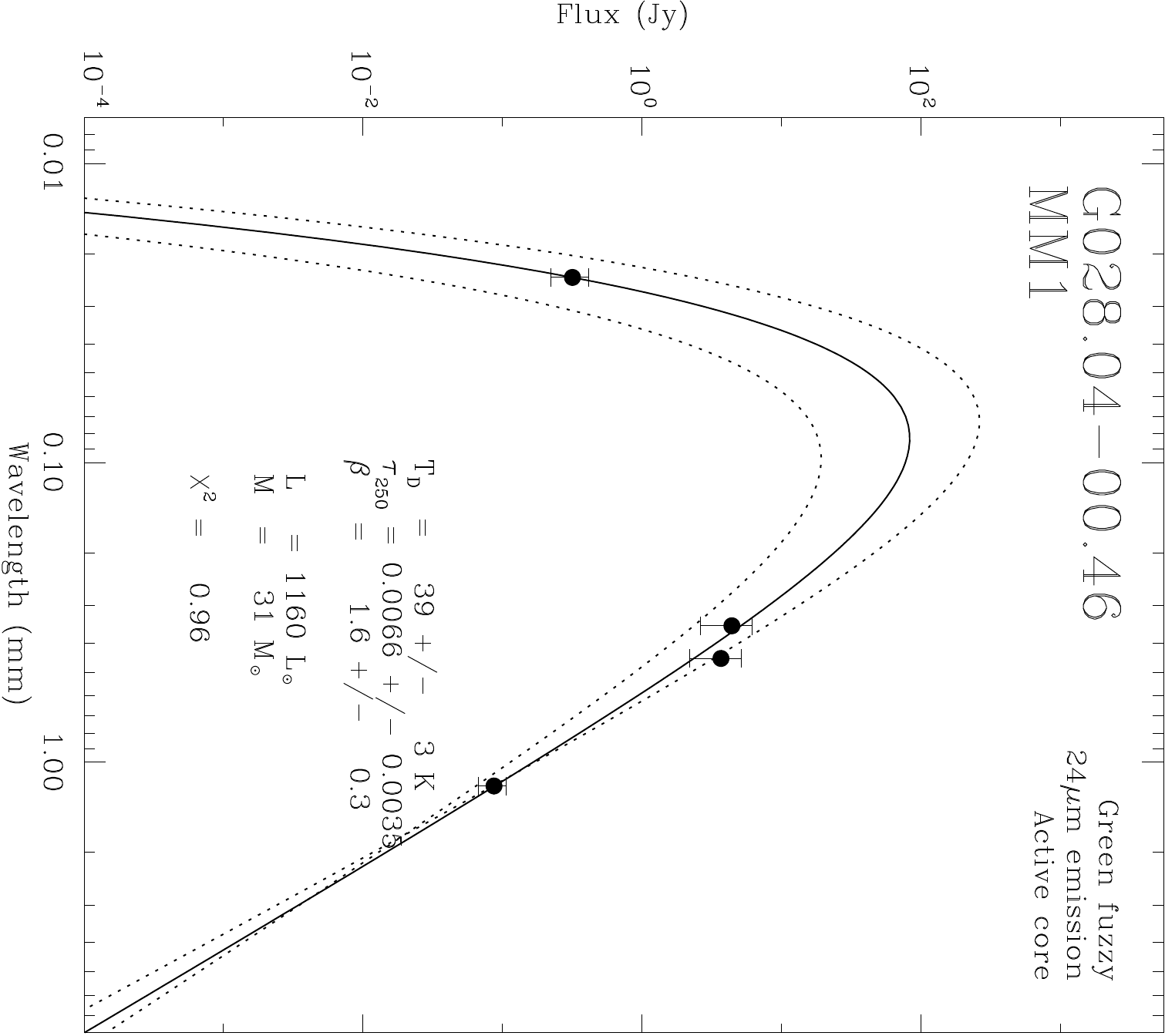}
\includegraphics[angle=90,width=0.5\textwidth]{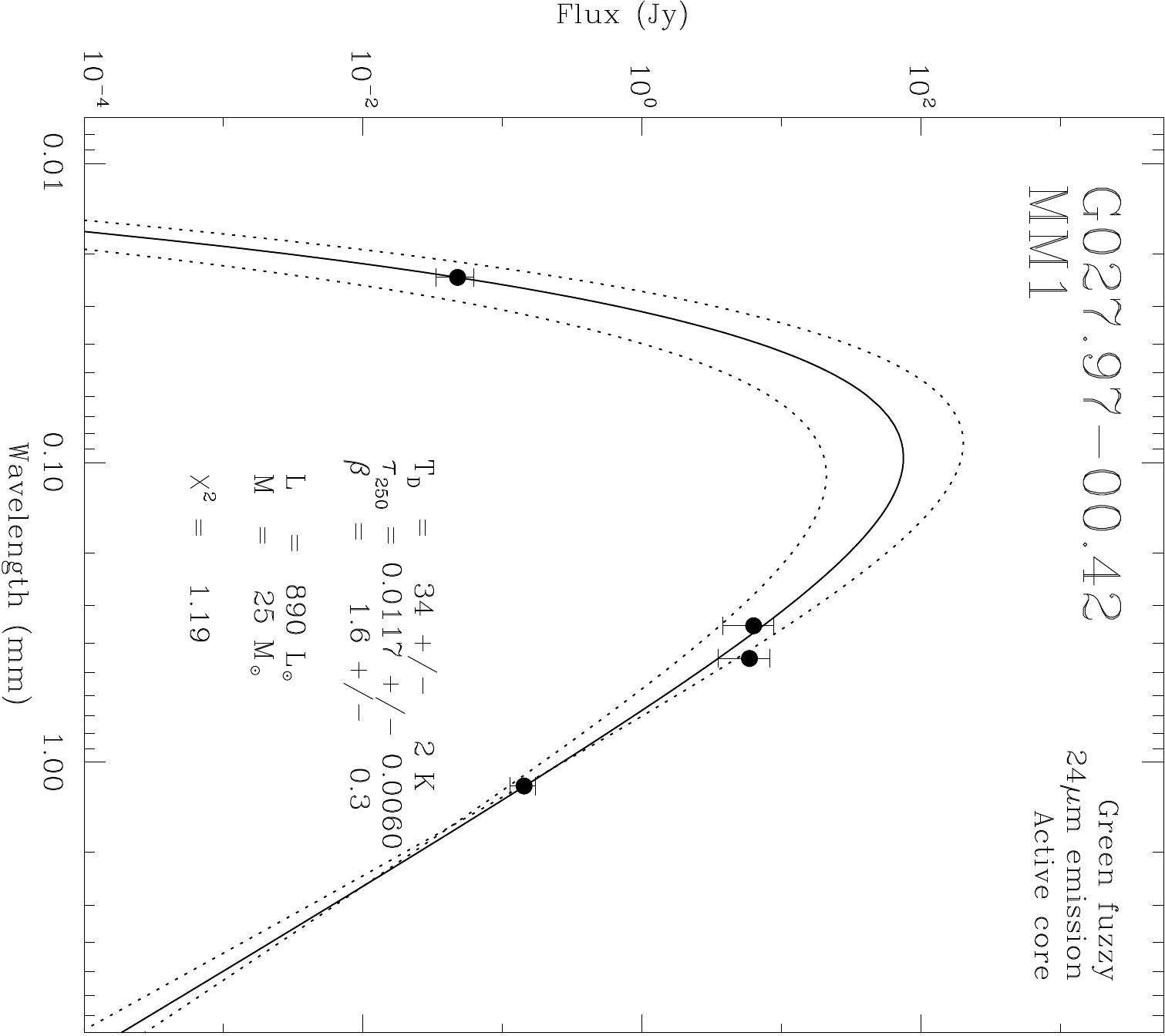}\\
\includegraphics[angle=90,width=0.5\textwidth]{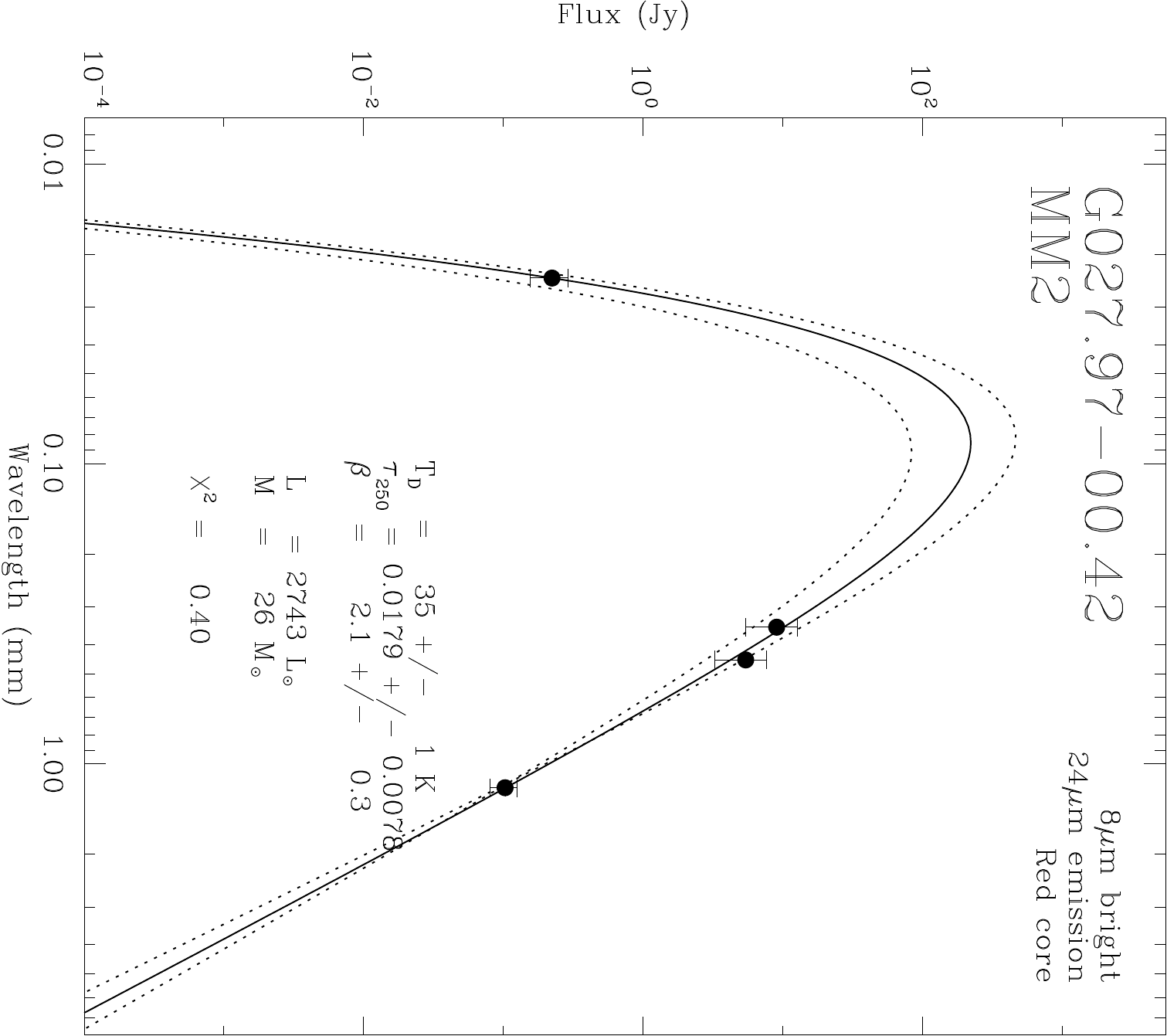}
\includegraphics[angle=90,width=0.5\textwidth]{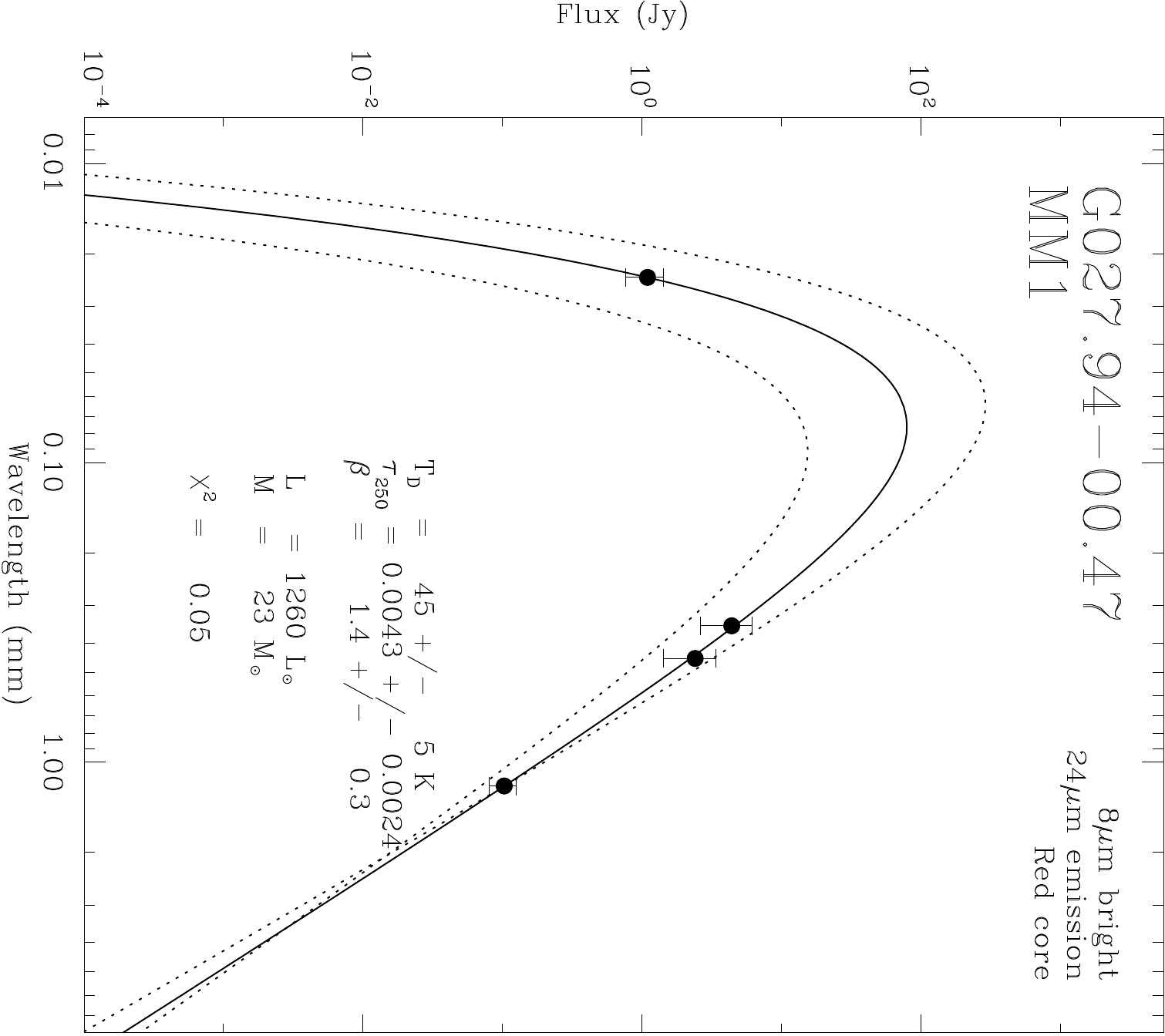}\\
\end{figure}
\clearpage 
\begin{figure}
\includegraphics[angle=90,width=0.5\textwidth]{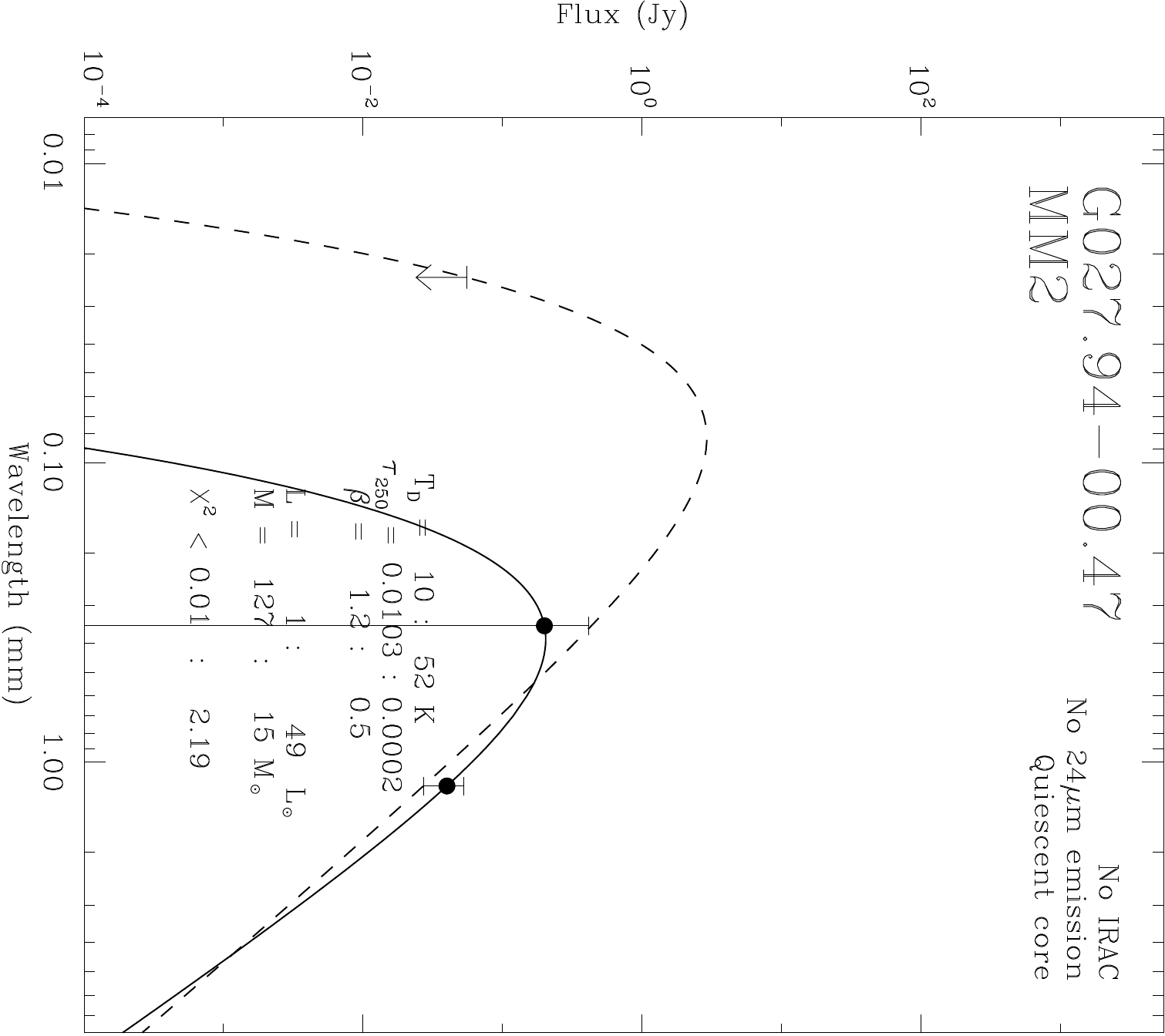}\\
\caption{\label{seds-10} \Spitzer\, 24\,\um\, image overlaid  
   with 1.2\,mm continuum emission for \irdcfortytwotwo, \irdcfortytwoone, \irdcten, and
   \irdcfortysix\, (contour levels are 30, 60, 90, 120\,mJy
   beam$^{-1}$). The lower panels show the broadband
   SEDs for cores within this IRDC.  The fluxes derived from the
   millimeter, sub-millimeter, and far-IR  continuum data are shown as filled
   circles (with the corresponding error bars), while the 24\,\um\, fluxes are shown as  either a filled circle (when included within the fit), an open circle (when excluded from the fit),  or as an upper limit arrow. For cores that have measured fluxes only in the millimeter/sub-millimeter regime (i.e.\, a limit at 24\,\um), we show the results from two fits: one using only the measured fluxes (solid line; lower limit), while the other includes the 24\,\um\, limit as a real data (dashed line; upper limit). In all other cases, the solid line is the best fit gray-body, while the dotted lines correspond to the functions determined using the errors for the T$_{D}$, $\tau$, and $\beta$ output from the fitting.  Labeled on each plot is the IRDC and core name,  classification, and the derived parameters.}
\end{figure}
\clearpage 
\begin{figure}
\begin{center}
\includegraphics[angle=0,width=0.6\textwidth]{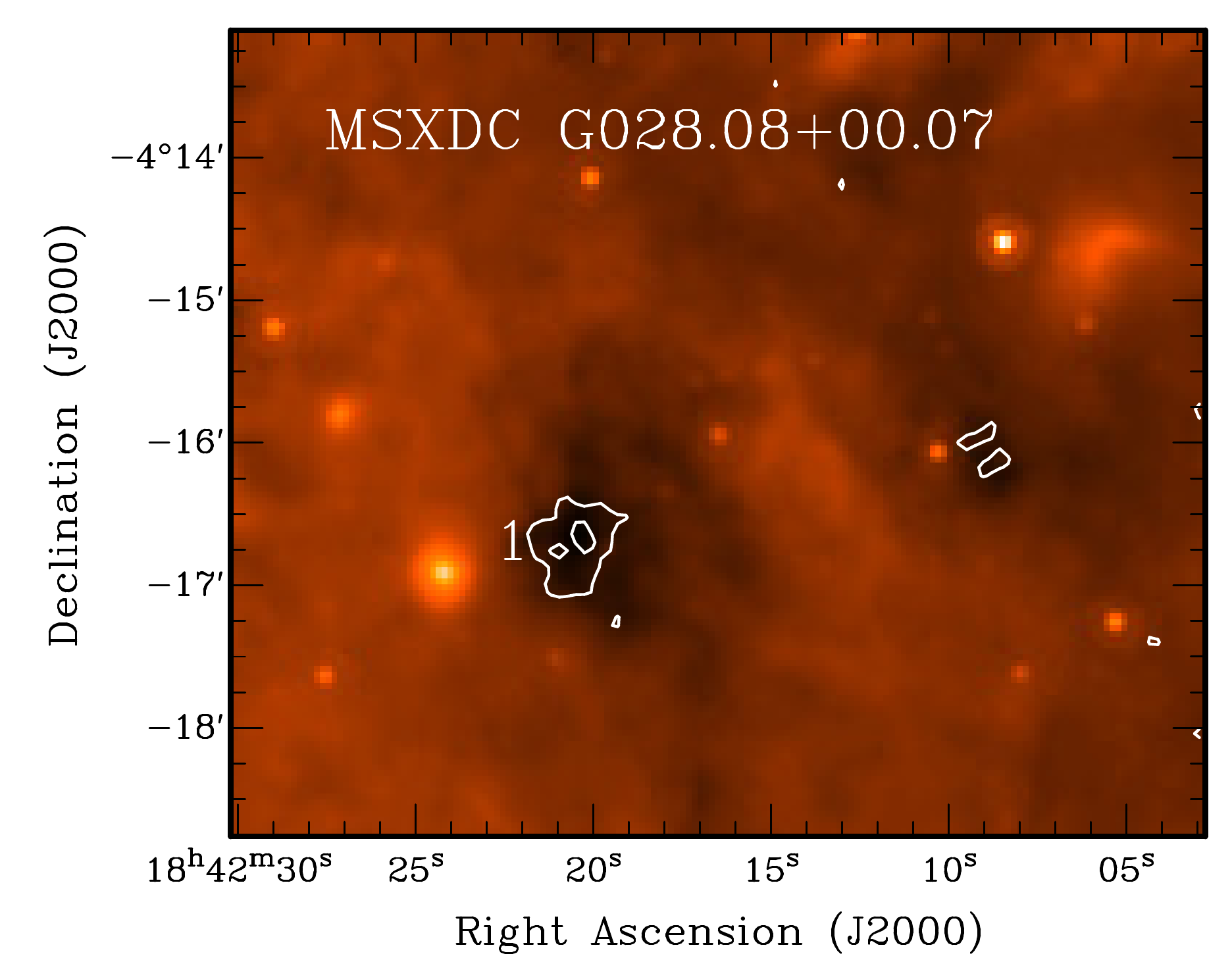}\\
\end{center}
\includegraphics[angle=90,width=0.5\textwidth]{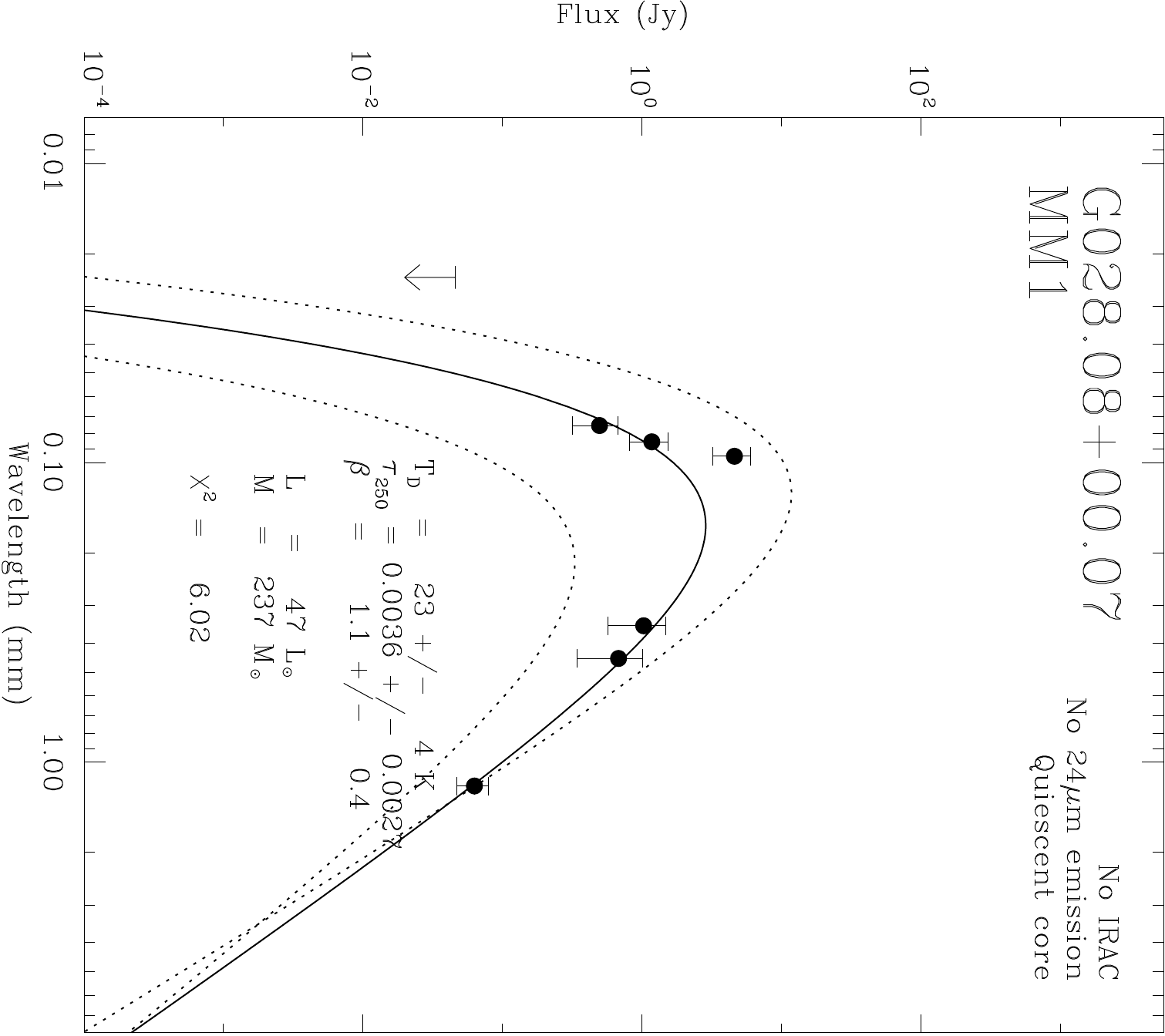}\\
\caption{\label{seds-12}\Spitzer\, 24\,\um\, image overlaid  
   with 1.2\,mm continuum emission for \irdctwelve\, (contour levels
   are 30, 60, 90, 120\,mJy beam$^{-1}$). The lower panels show the broadband
   SEDs for cores within this IRDC.  The fluxes derived from the
   millimeter, sub-millimeter, and far-IR  continuum data are shown as filled
   circles (with the corresponding error bars), while the 24\,\um\, fluxes are shown as  either a filled circle (when included within the fit), an open circle (when excluded from the fit),  or as an upper limit arrow. For cores that have measured fluxes only in the millimeter/sub-millimeter regime (i.e.\, a limit at 24\,\um), we show the results from two fits: one using only the measured fluxes (solid line; lower limit), while the other includes the 24\,\um\, limit as a real data (dashed line; upper limit). In all other cases, the solid line is the best fit gray-body, while the dotted lines correspond to the functions determined using the errors for the T$_{D}$, $\tau$, and $\beta$ output from the fitting.  Labeled on each plot is the IRDC and core name,  classification, and the derived parameters.}
\end{figure}
\clearpage 
\begin{figure}
\begin{center}
\includegraphics[angle=0,width=0.6\textwidth]{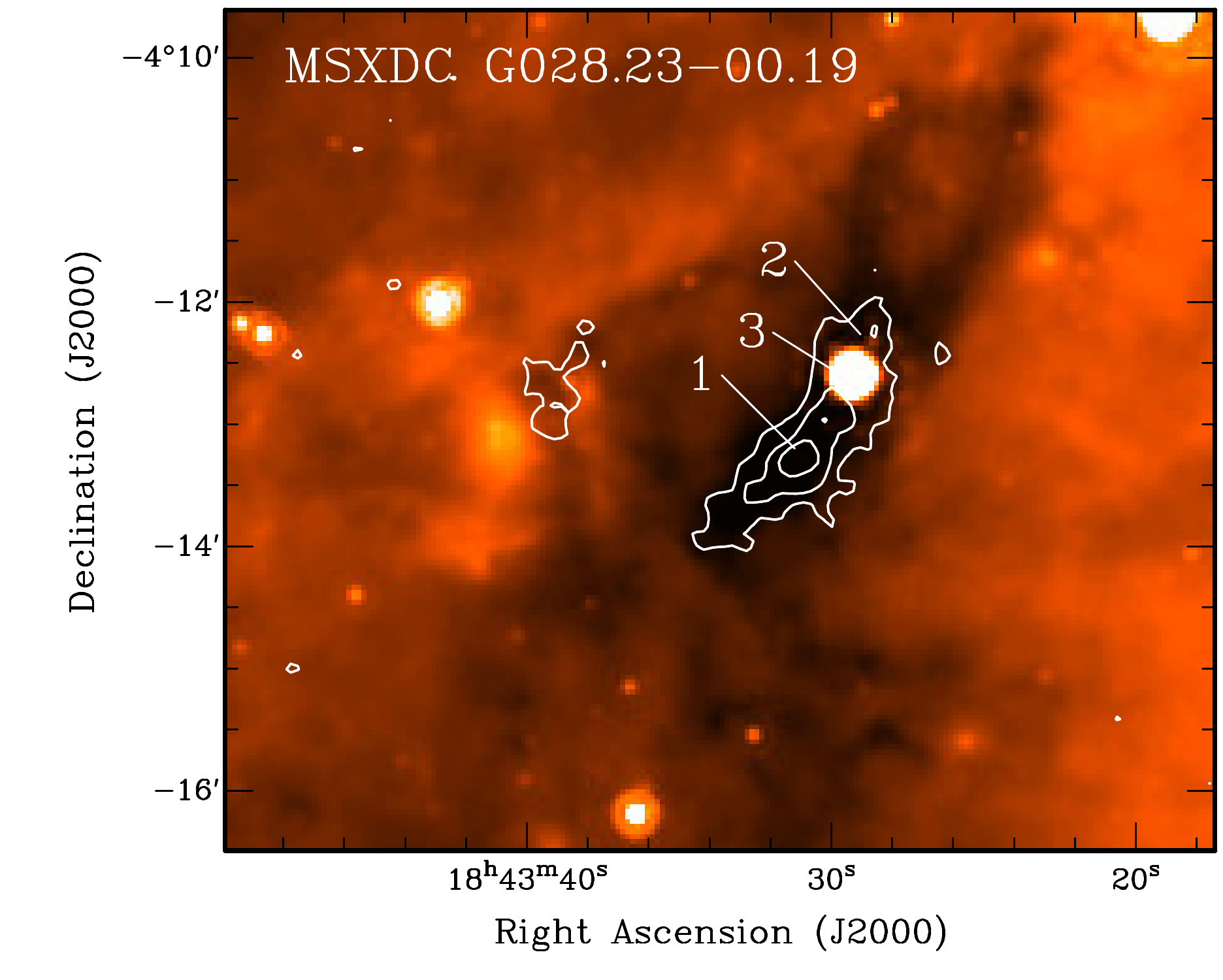}\\
\end{center}
\includegraphics[angle=90,width=0.5\textwidth]{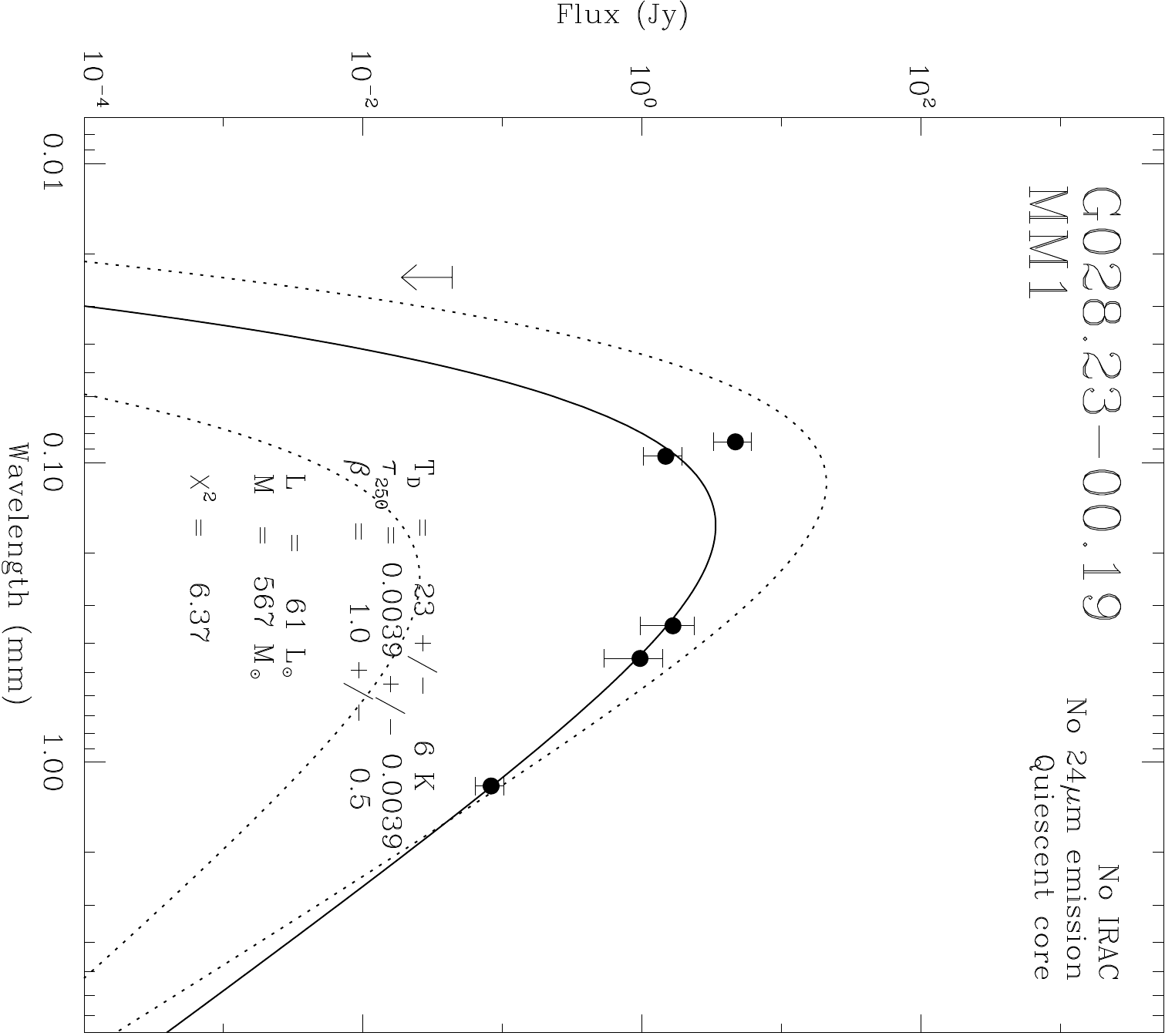}
\includegraphics[angle=90,width=0.5\textwidth]{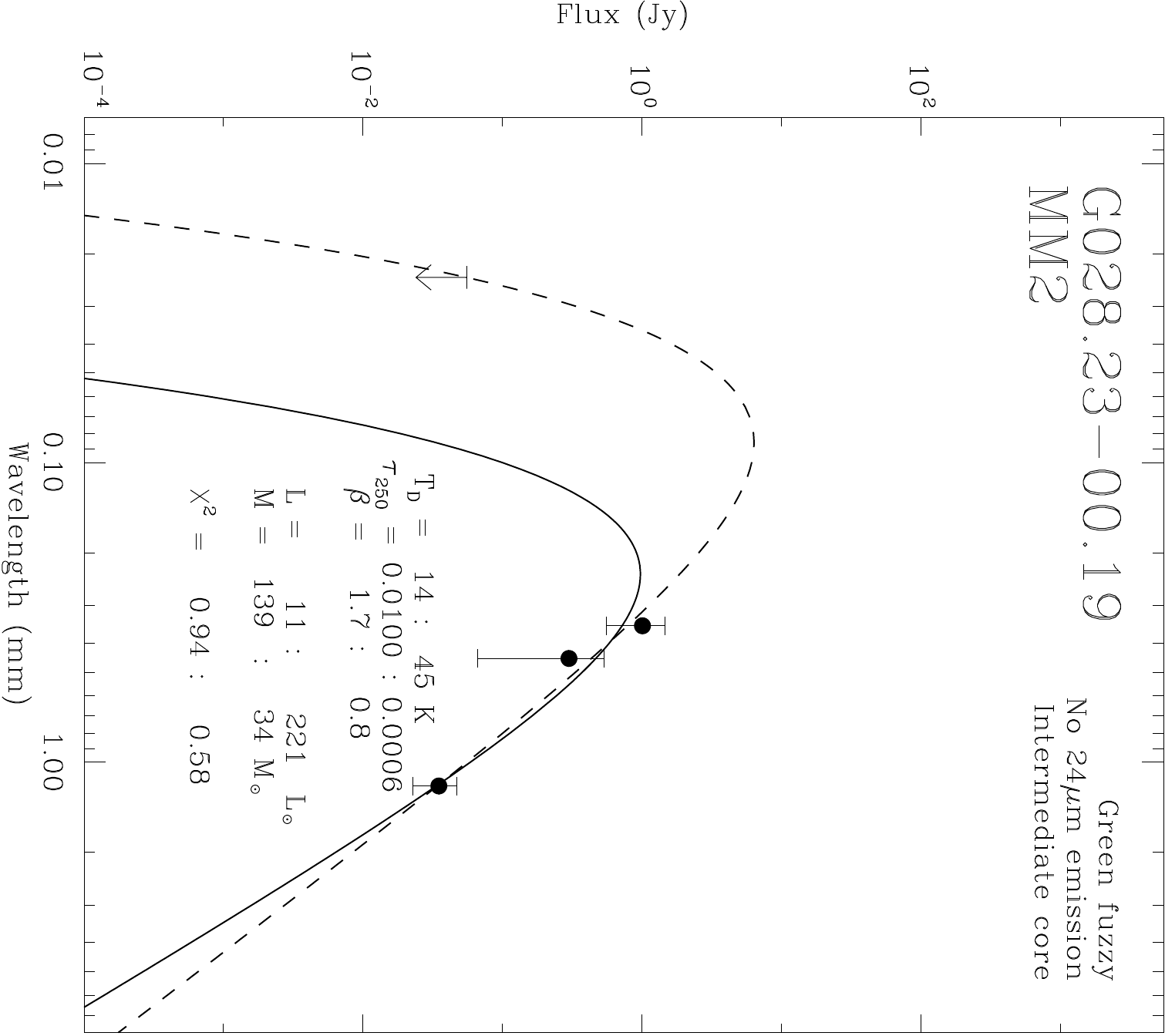}\\
\end{figure}
\clearpage 
\begin{figure}
\includegraphics[angle=90,width=0.5\textwidth]{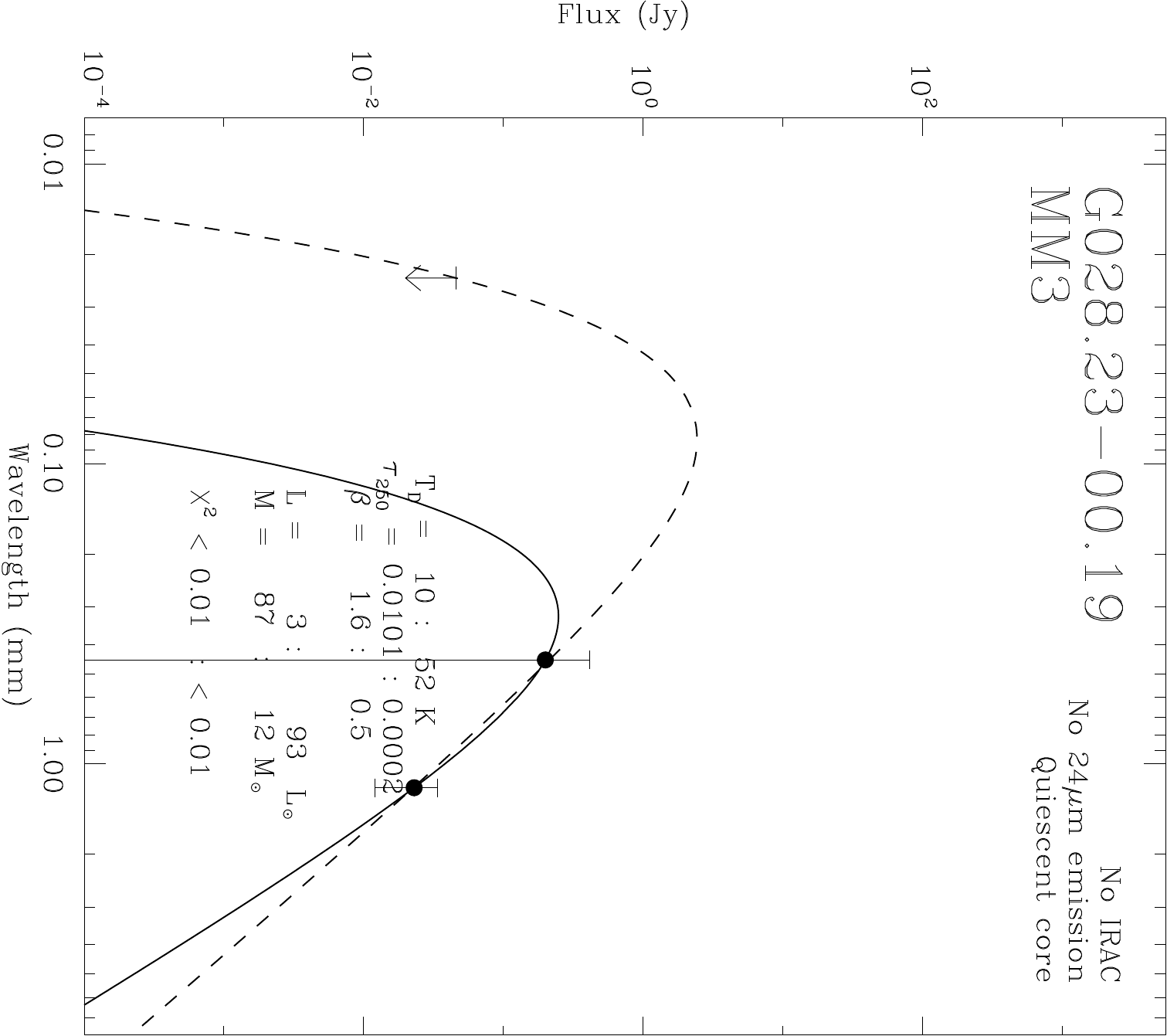}\\
\caption{\label{seds-2}\Spitzer\, 24\,\um\, image overlaid  
   with 1.2\,mm continuum emission for \irdctwo\, (contour levels are
   30, 60, 90\,mJy beam$^{-1}$). The lower panels show the broadband
   SEDs for cores within this IRDC.  The fluxes derived from the
   millimeter, sub-millimeter, and far-IR  continuum data are shown as filled
   circles (with the corresponding error bars), while the 24\,\um\, fluxes are shown as  either a filled circle (when included within the fit), an open circle (when excluded from the fit),  or as an upper limit arrow. For cores that have measured fluxes only in the millimeter/sub-millimeter regime (i.e.\, a limit at 24\,\um), we show the results from two fits: one using only the measured fluxes (solid line; lower limit), while the other includes the 24\,\um\, limit as a real data (dashed line; upper limit). In all other cases, the solid line is the best fit gray-body, while the dotted lines correspond to the functions determined using the errors for the T$_{D}$, $\tau$, and $\beta$ output from the fitting.  Labeled on each plot is the IRDC and core name,  classification, and the derived parameters.}
\end{figure}
\clearpage 
\begin{figure}
\begin{center}
\includegraphics[angle=0,width=0.6\textwidth]{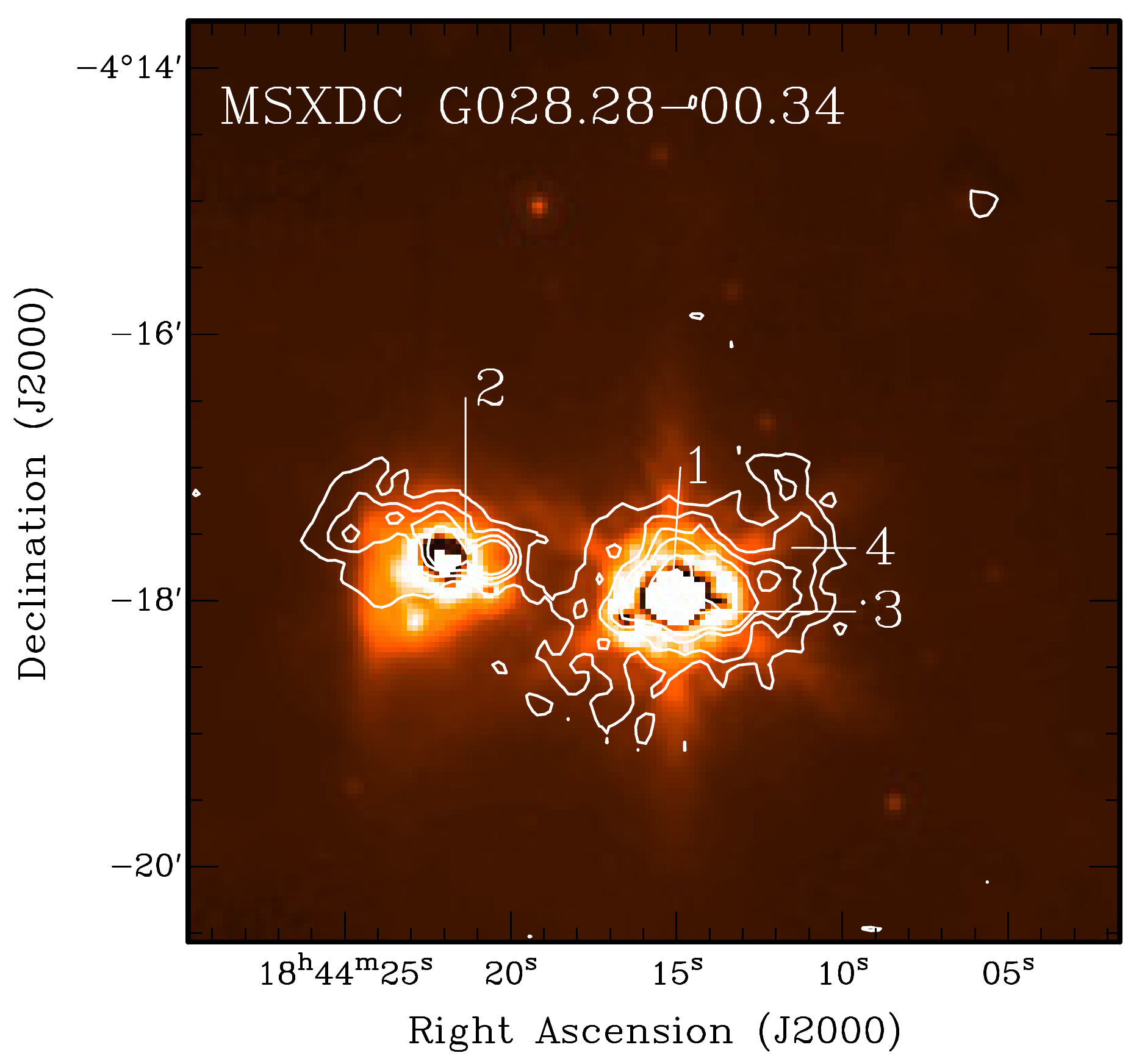}\\
\end{center}
\includegraphics[angle=90,width=0.5\textwidth]{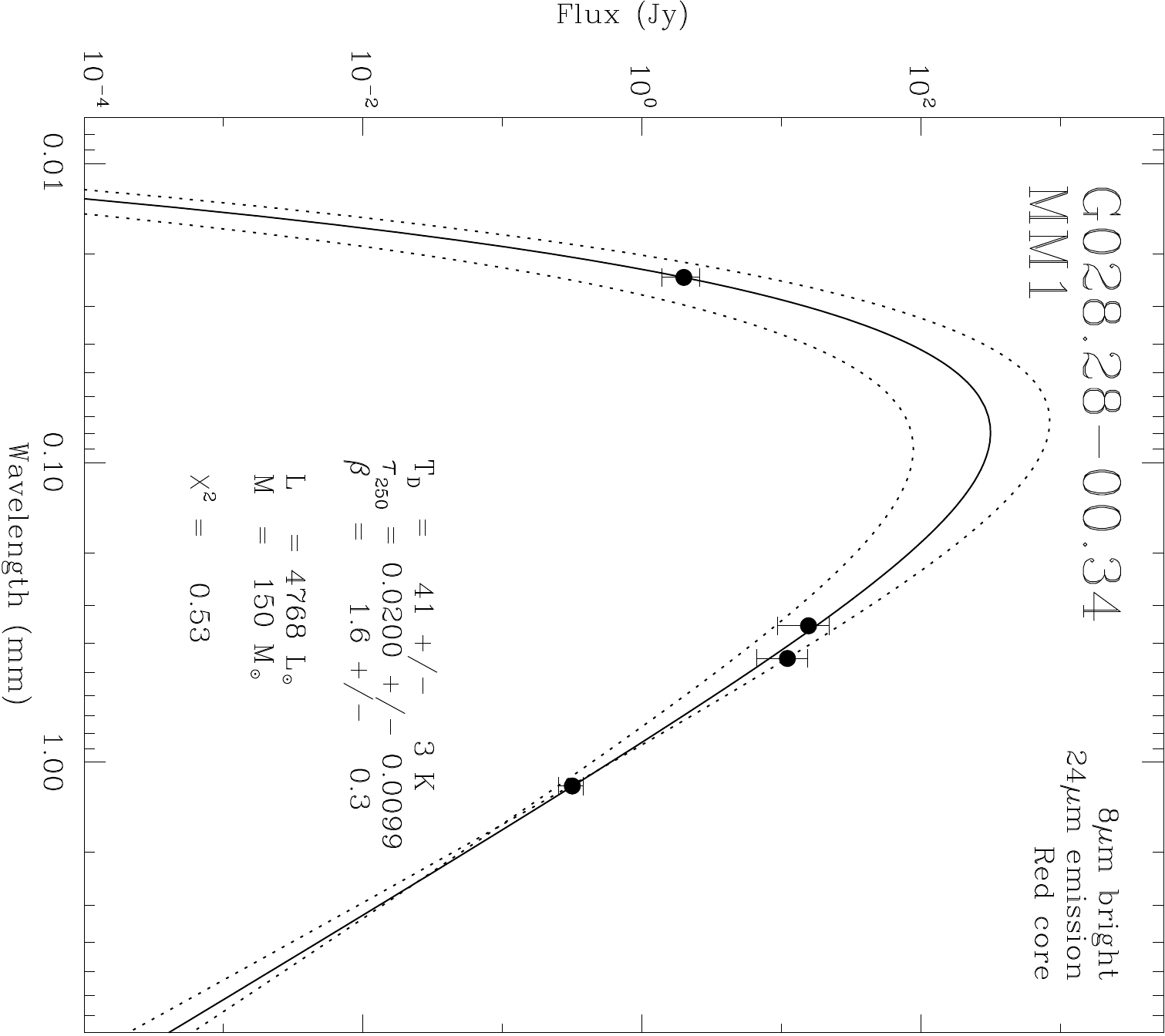}
\includegraphics[angle=90,width=0.5\textwidth]{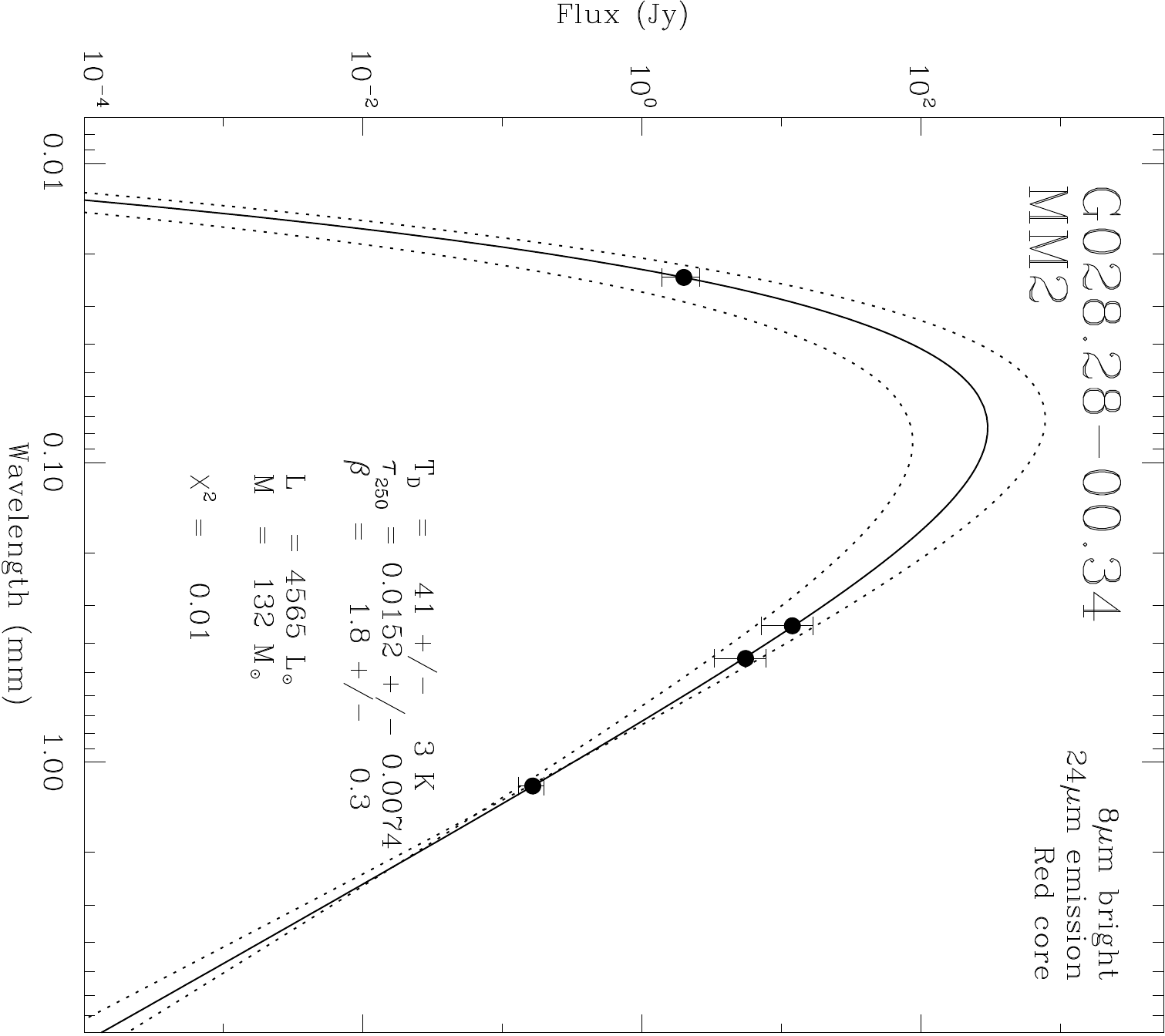}\\
\end{figure}
\clearpage 
\begin{figure}
\includegraphics[angle=90,width=0.5\textwidth]{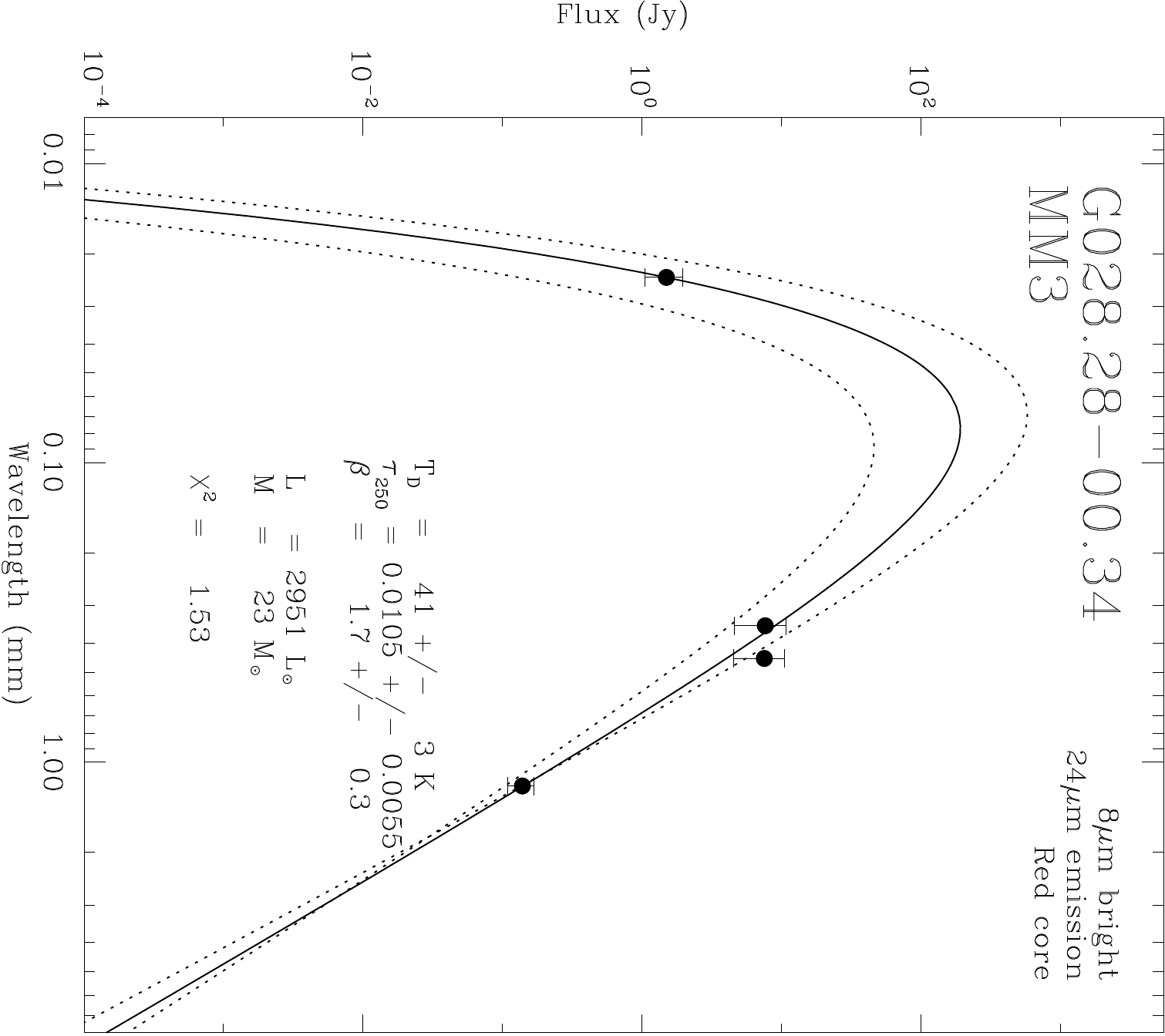}
\caption{\label{seds-26} \Spitzer\, 24\,\um\, image overlaid  
   with 1.2\,mm continuum emission for \irdctwentysix\, (contour levels
   are 30, 60, 90, 120, 240, 360, 480\,mJy beam$^{-1}$). The lower panels show the broadband
   SEDs for cores within this IRDC.  The fluxes derived from the
   millimeter, sub-millimeter, and far-IR  continuum data are shown as filled
   circles (with the corresponding error bars), while the 24\,\um\, fluxes are shown as  either a filled circle (when included within the fit), an open circle (when excluded from the fit),  or as an upper limit arrow. For cores that have measured fluxes only in the millimeter/sub-millimeter regime (i.e.\, a limit at 24\,\um), we show the results from two fits: one using only the measured fluxes (solid line; lower limit), while the other includes the 24\,\um\, limit as a real data (dashed line; upper limit). In all other cases, the solid line is the best fit gray-body, while the dotted lines correspond to the functions determined using the errors for the T$_{D}$, $\tau$, and $\beta$ output from the fitting.  Labeled on each plot is the IRDC and core name,  classification, and the derived parameters.}
\end{figure}
\clearpage 
\begin{figure}
\begin{center}
\includegraphics[angle=0,width=0.6\textwidth]{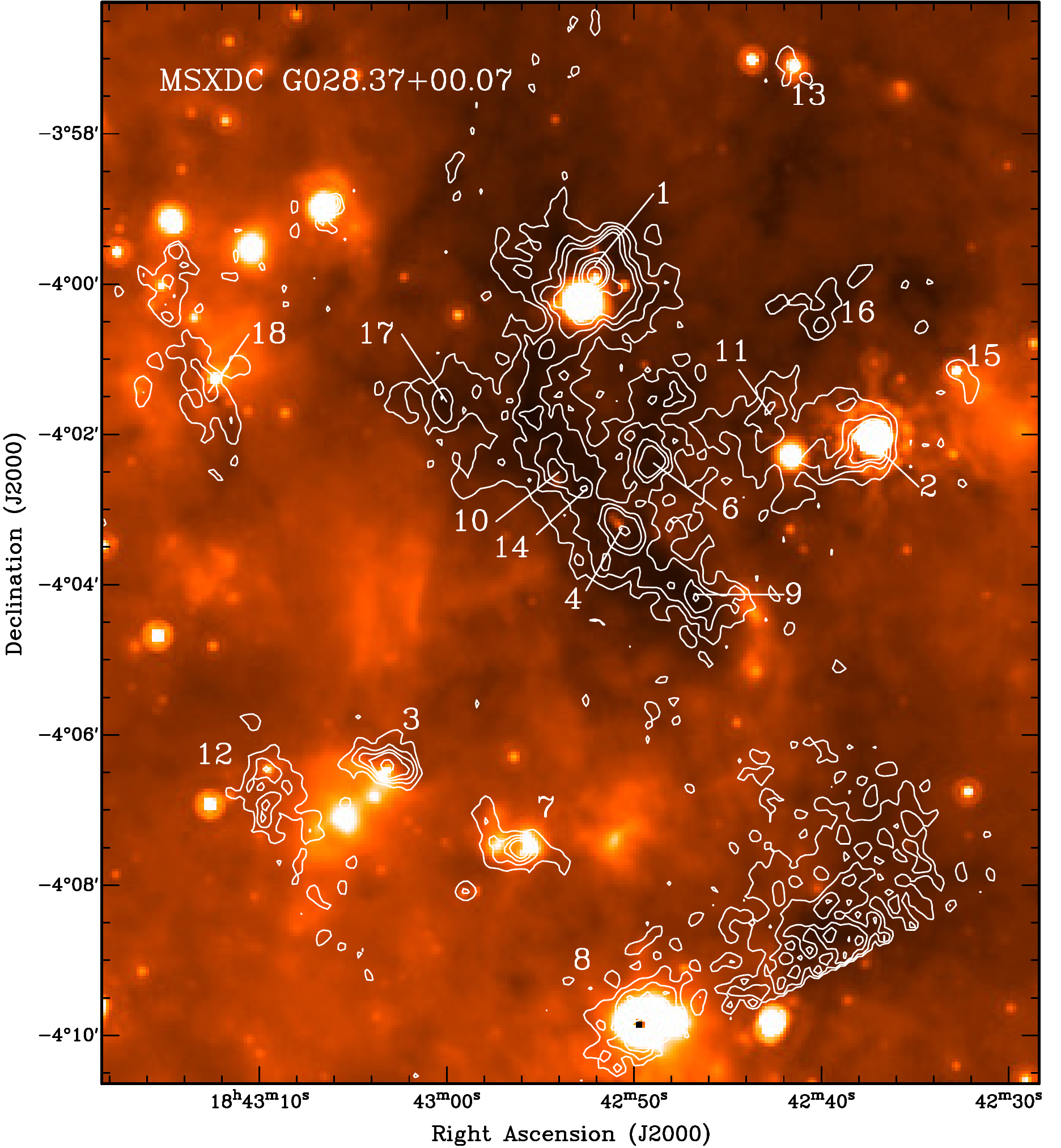}\\
\end{center}
\includegraphics[angle=90,width=0.5\textwidth]{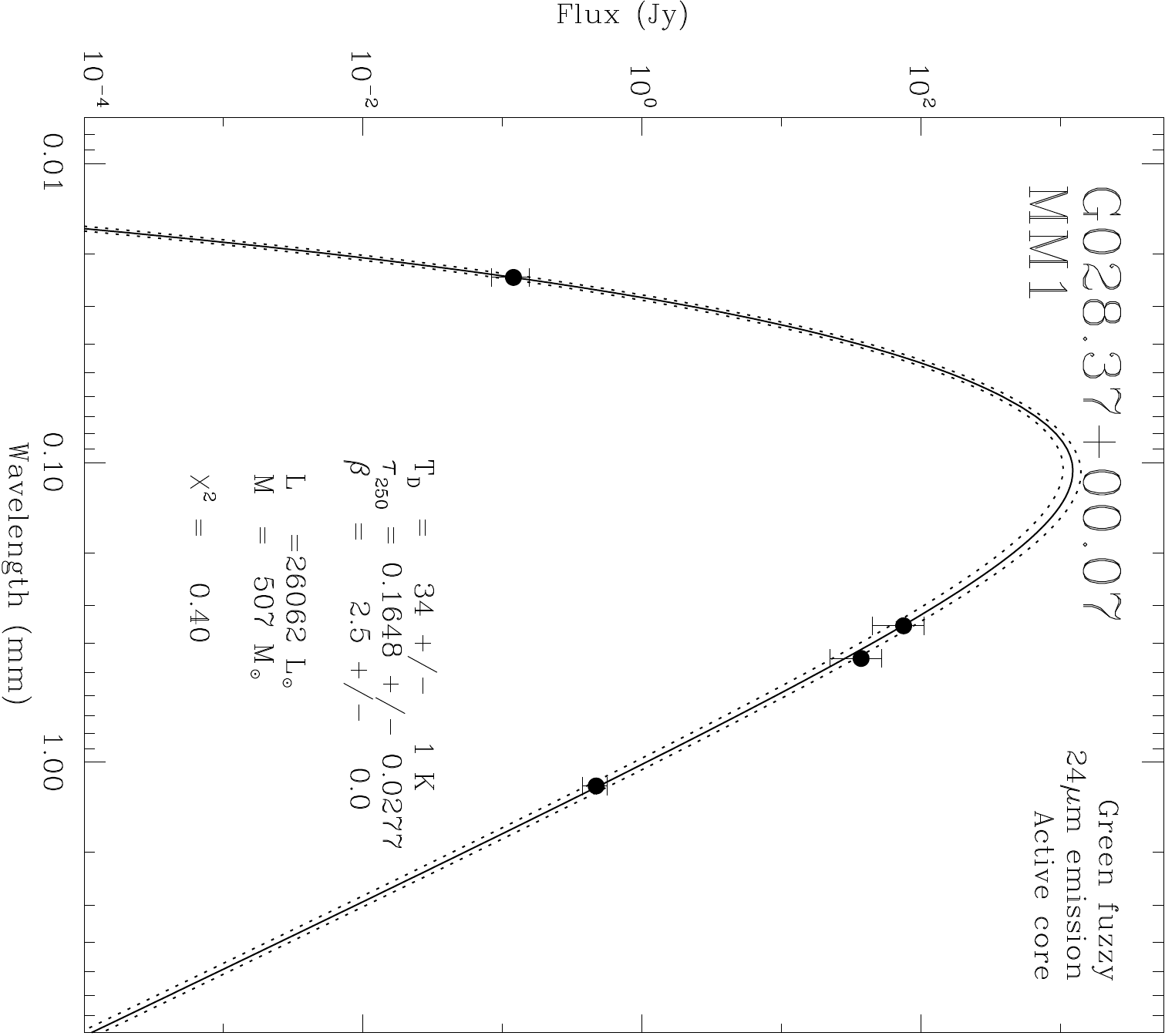}
\includegraphics[angle=90,width=0.5\textwidth]{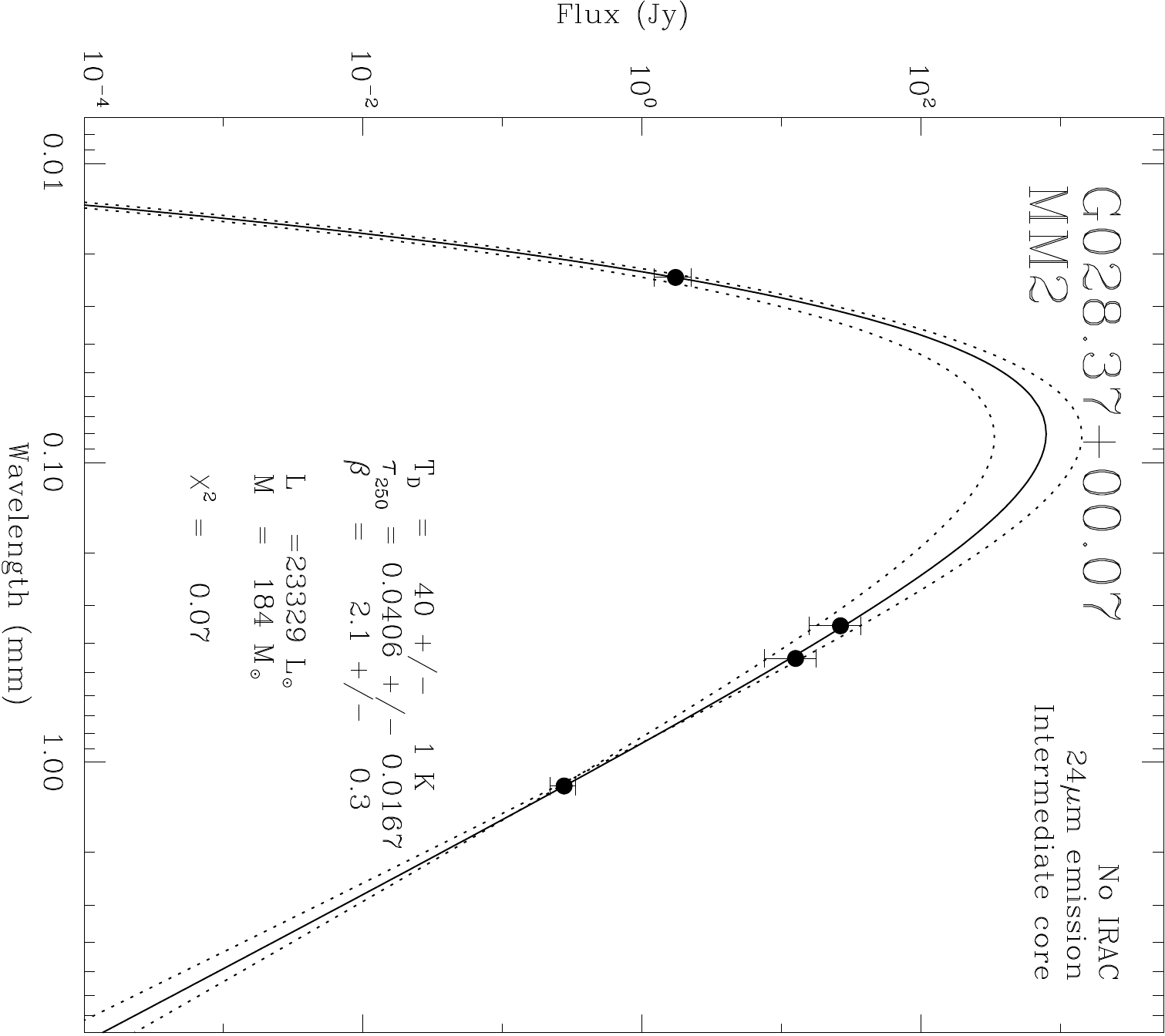}\\
\end{figure}
\clearpage 
\begin{figure}
\includegraphics[angle=90,width=0.5\textwidth]{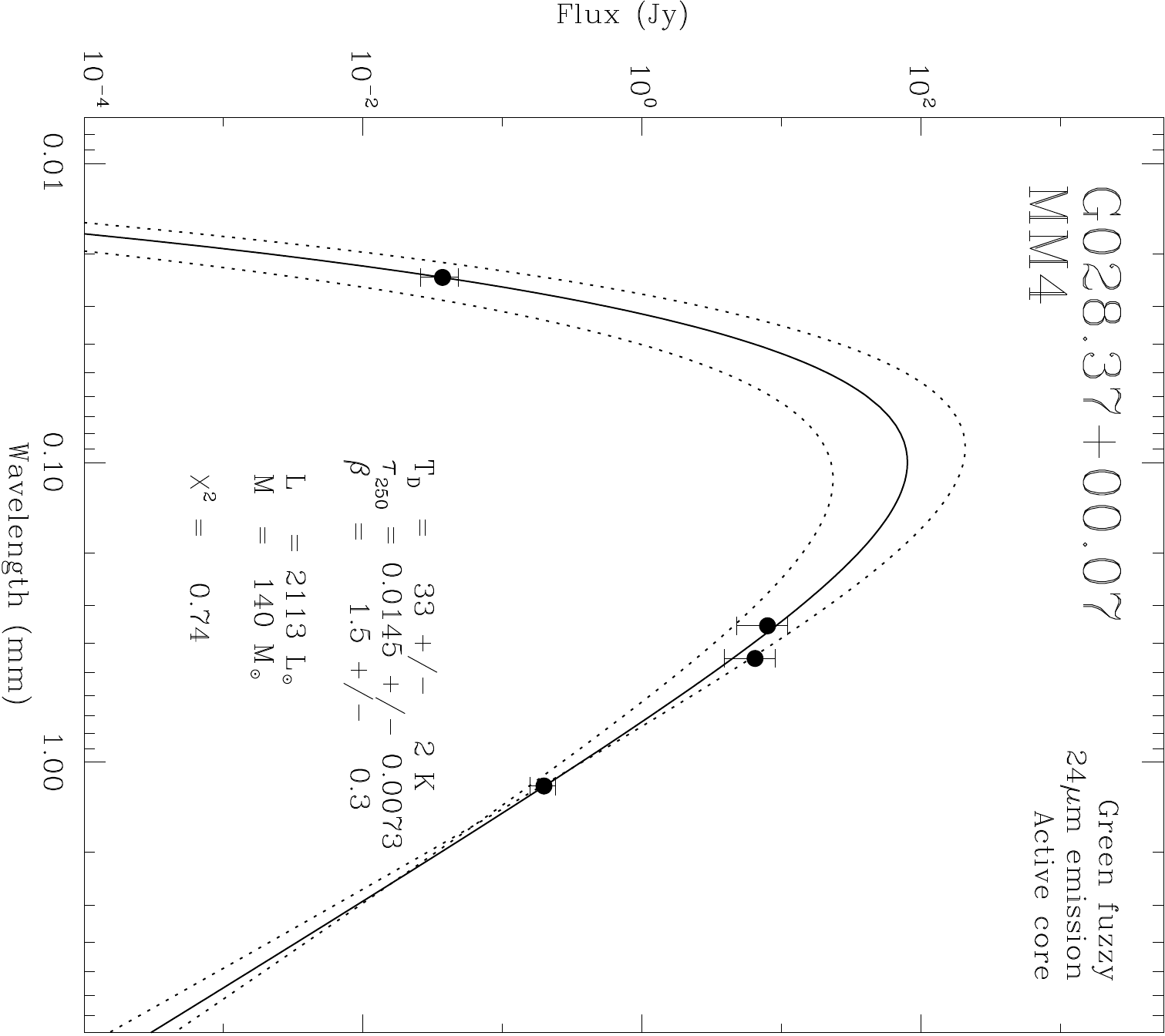}
\includegraphics[angle=90,width=0.5\textwidth]{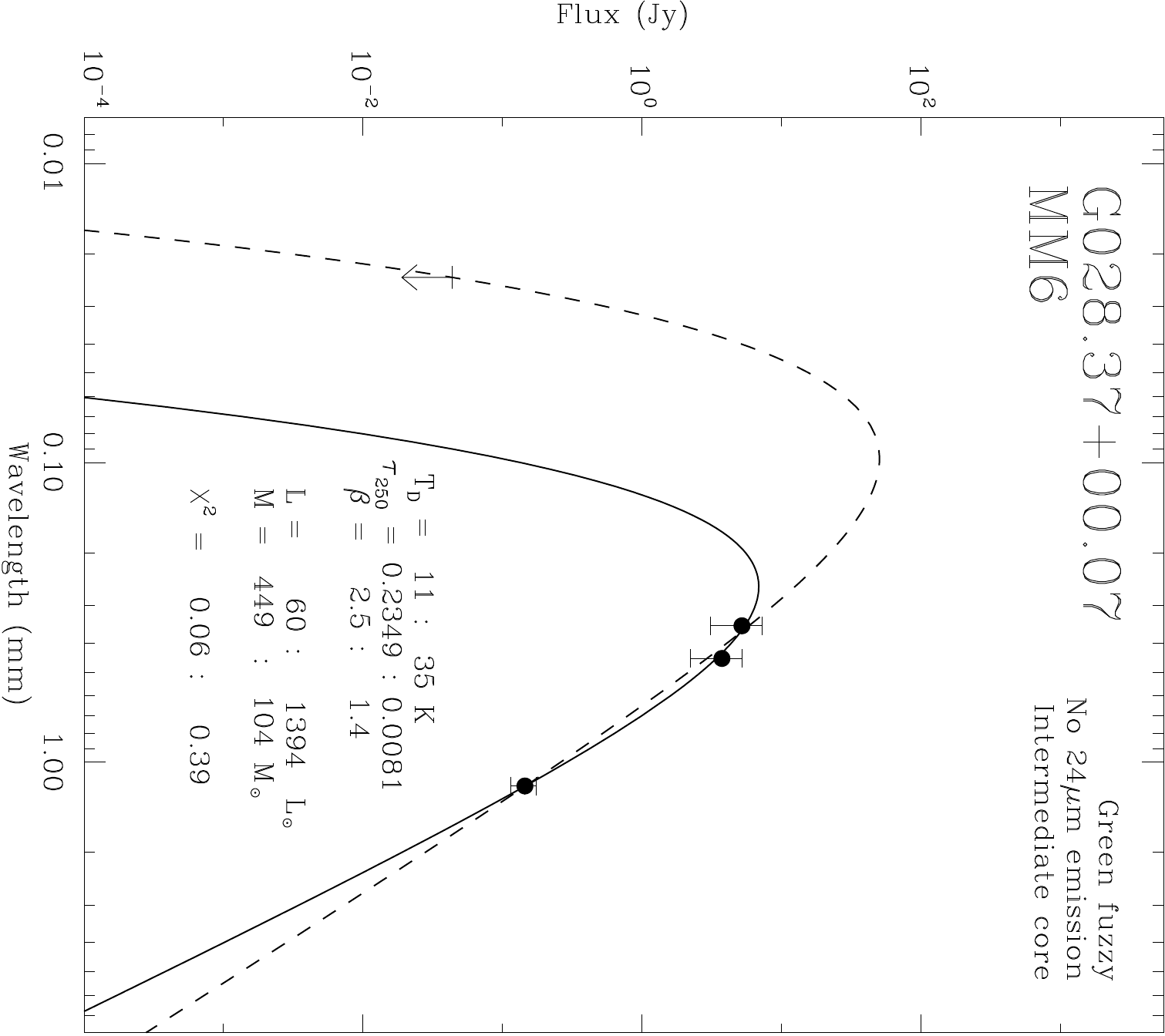}\\
\includegraphics[angle=90,width=0.5\textwidth]{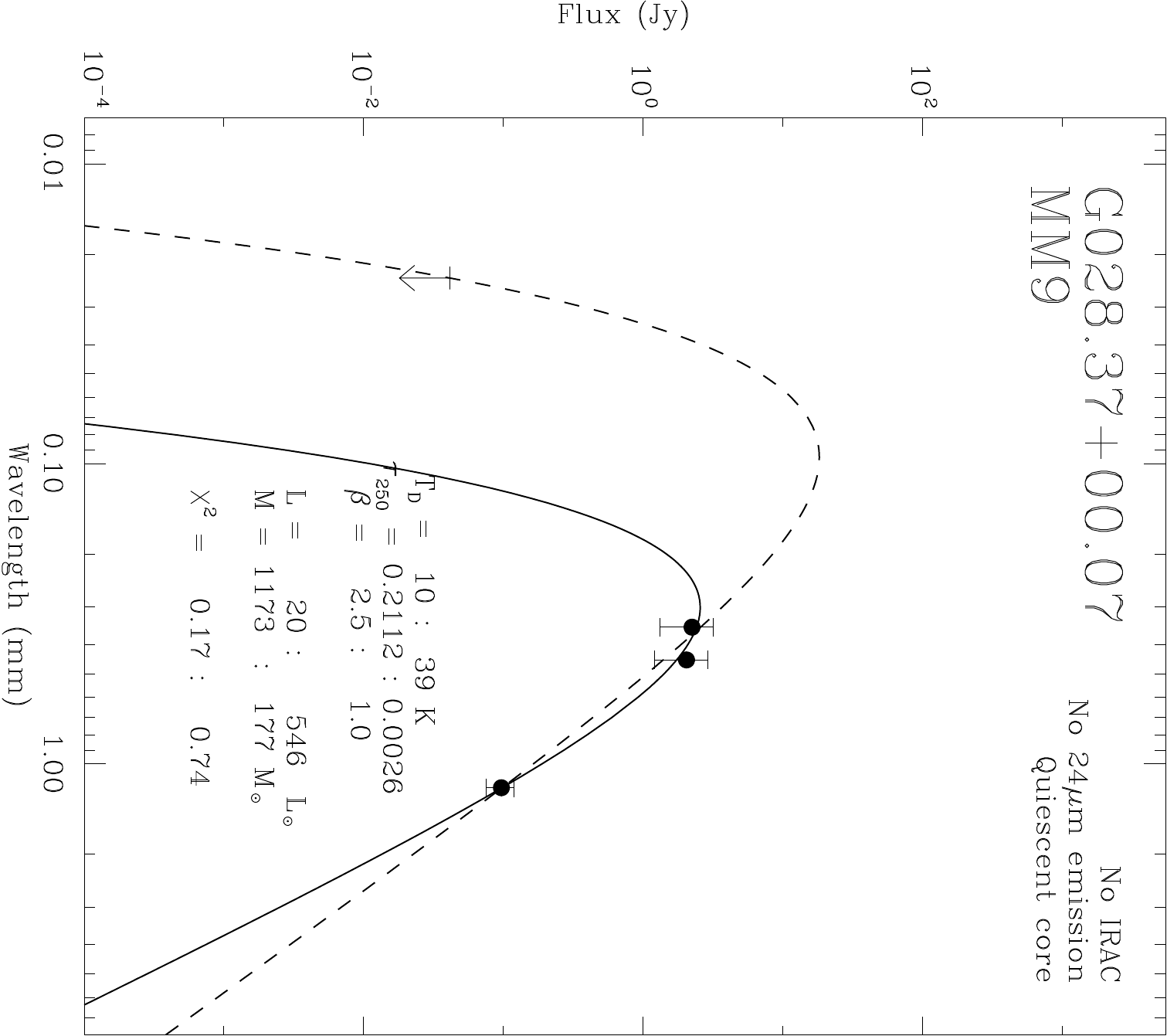}
\includegraphics[angle=90,width=0.5\textwidth]{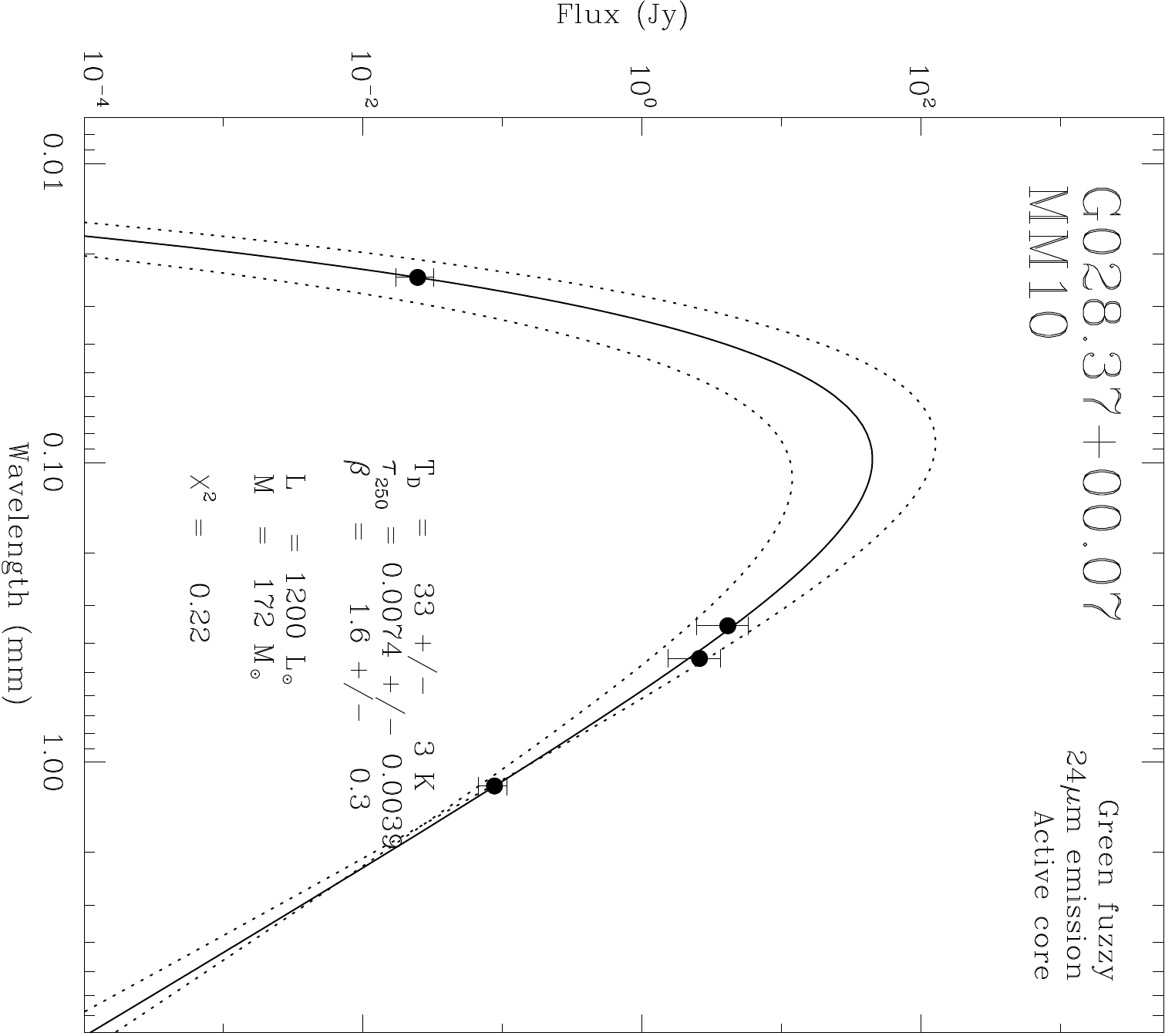}\\
\end{figure}
\clearpage 
\begin{figure}
\includegraphics[angle=90,width=0.5\textwidth]{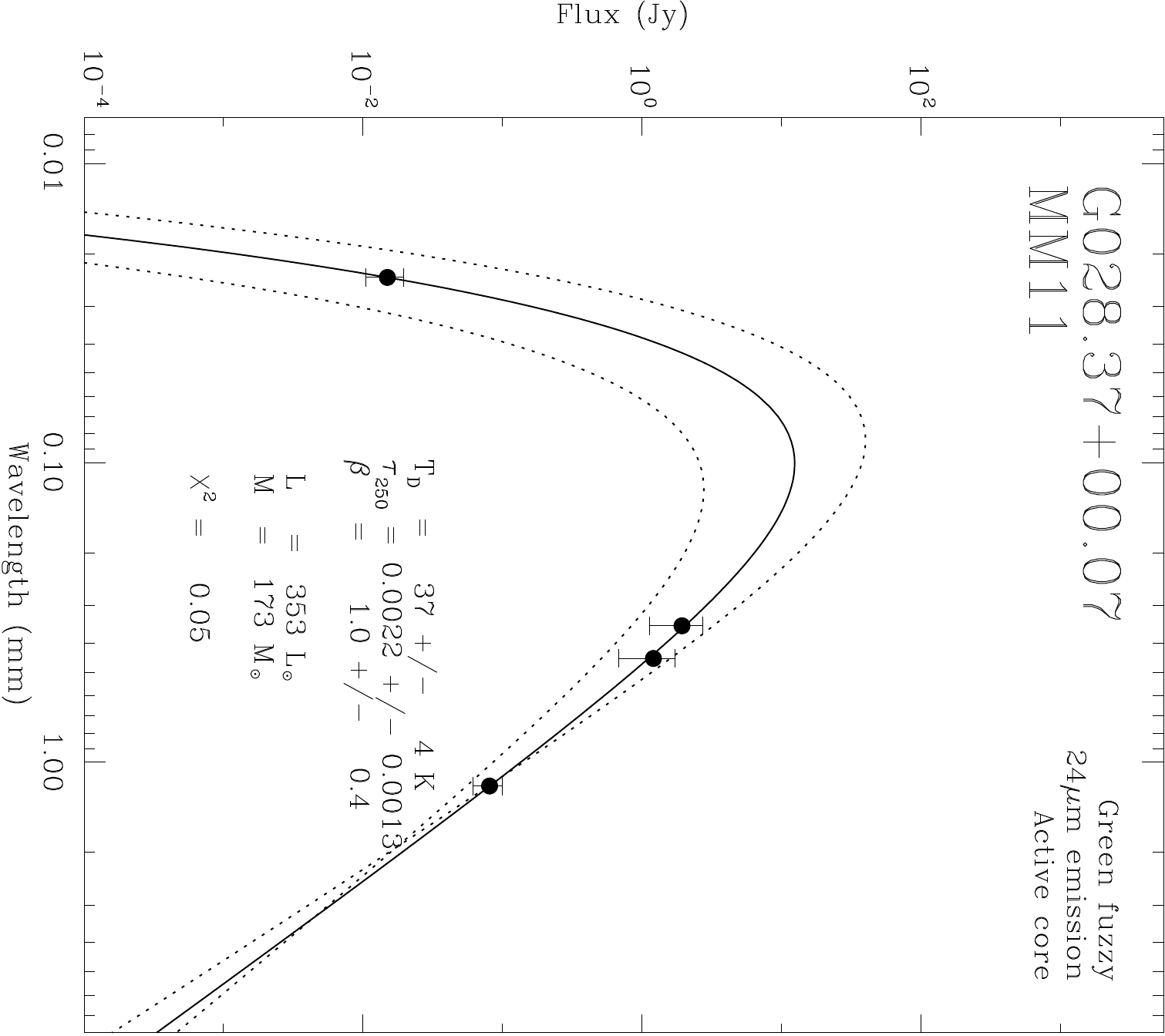}
\includegraphics[angle=90,width=0.5\textwidth]{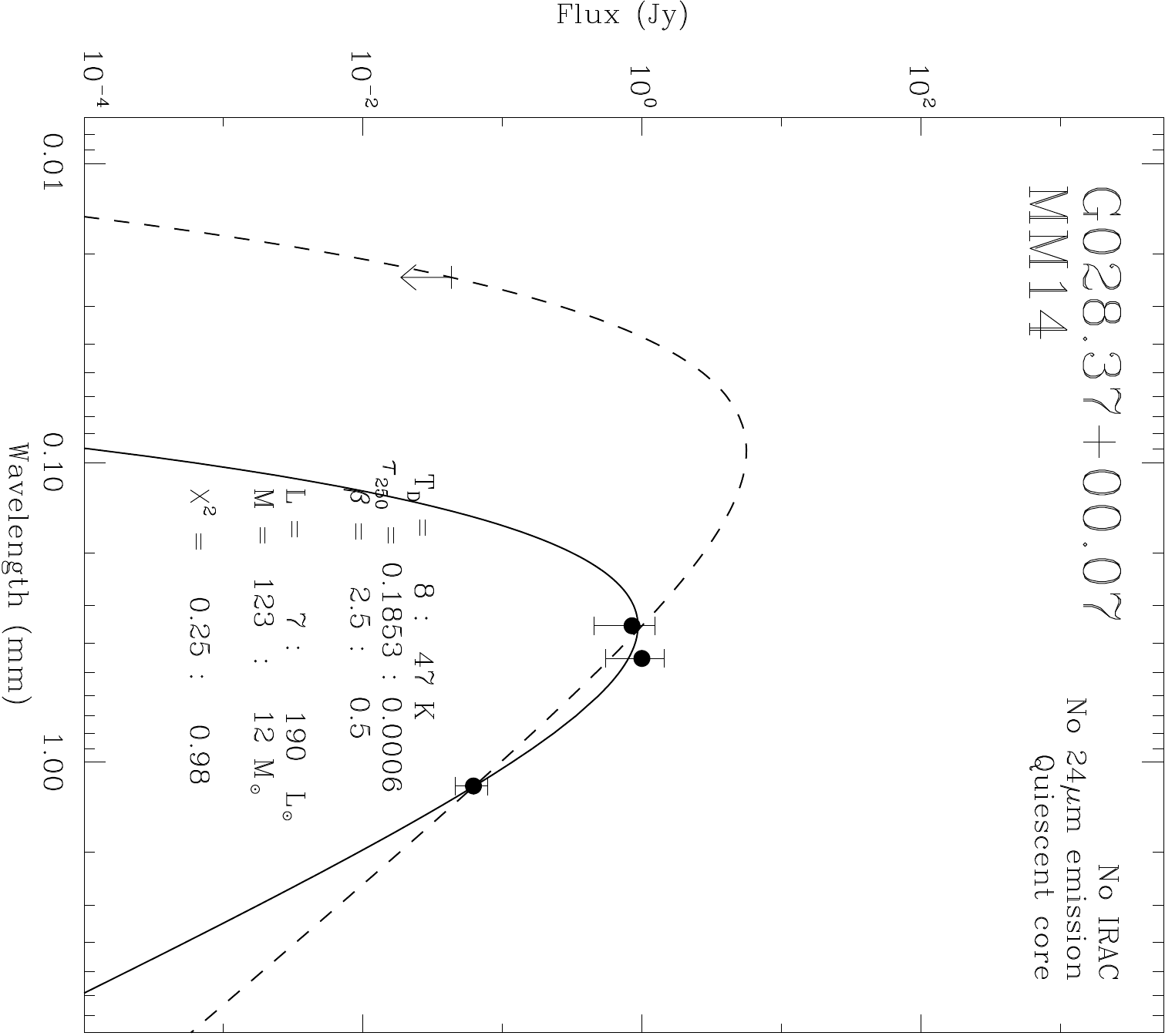}\\
\includegraphics[angle=90,width=0.5\textwidth]{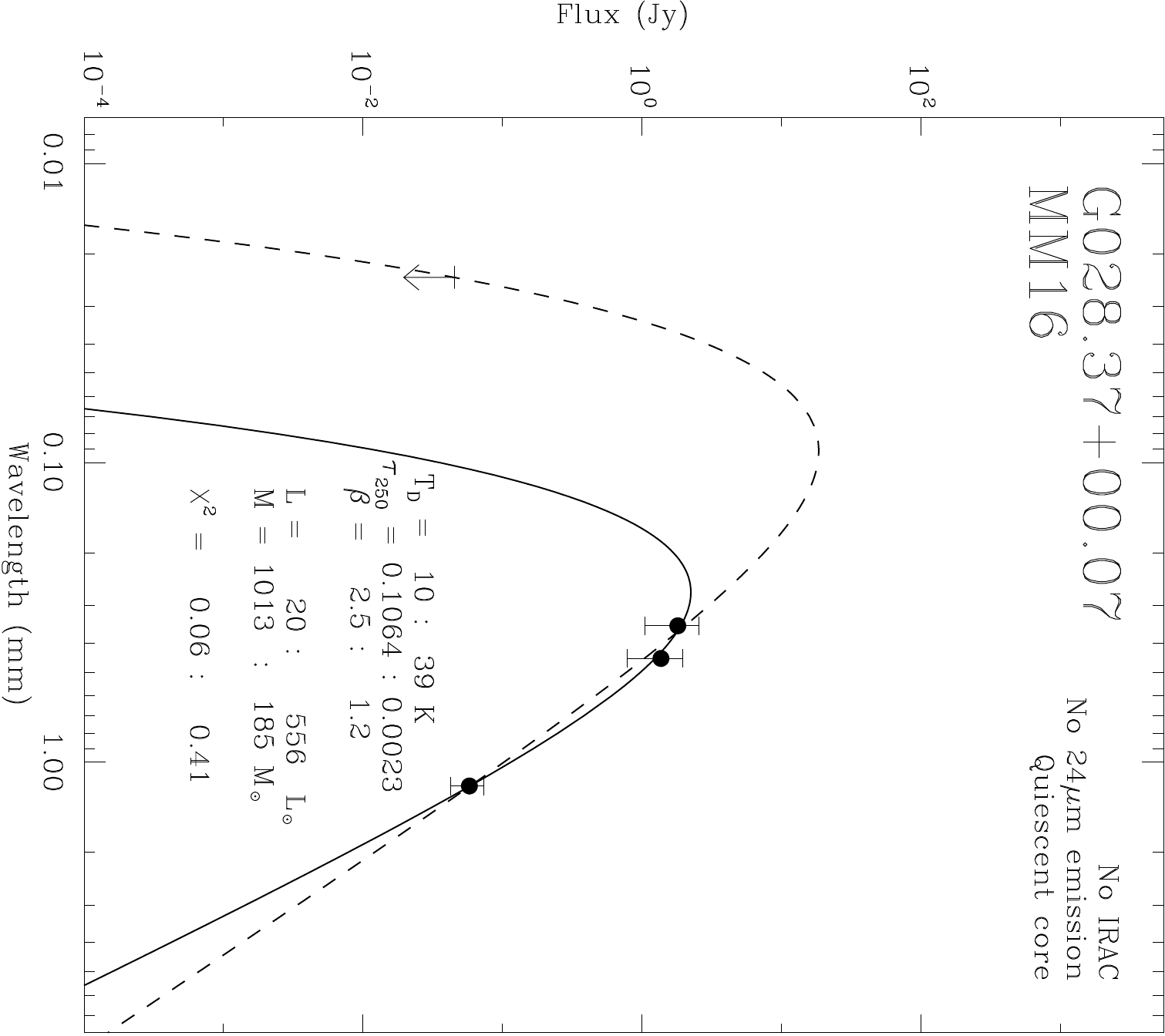}
\includegraphics[angle=90,width=0.5\textwidth]{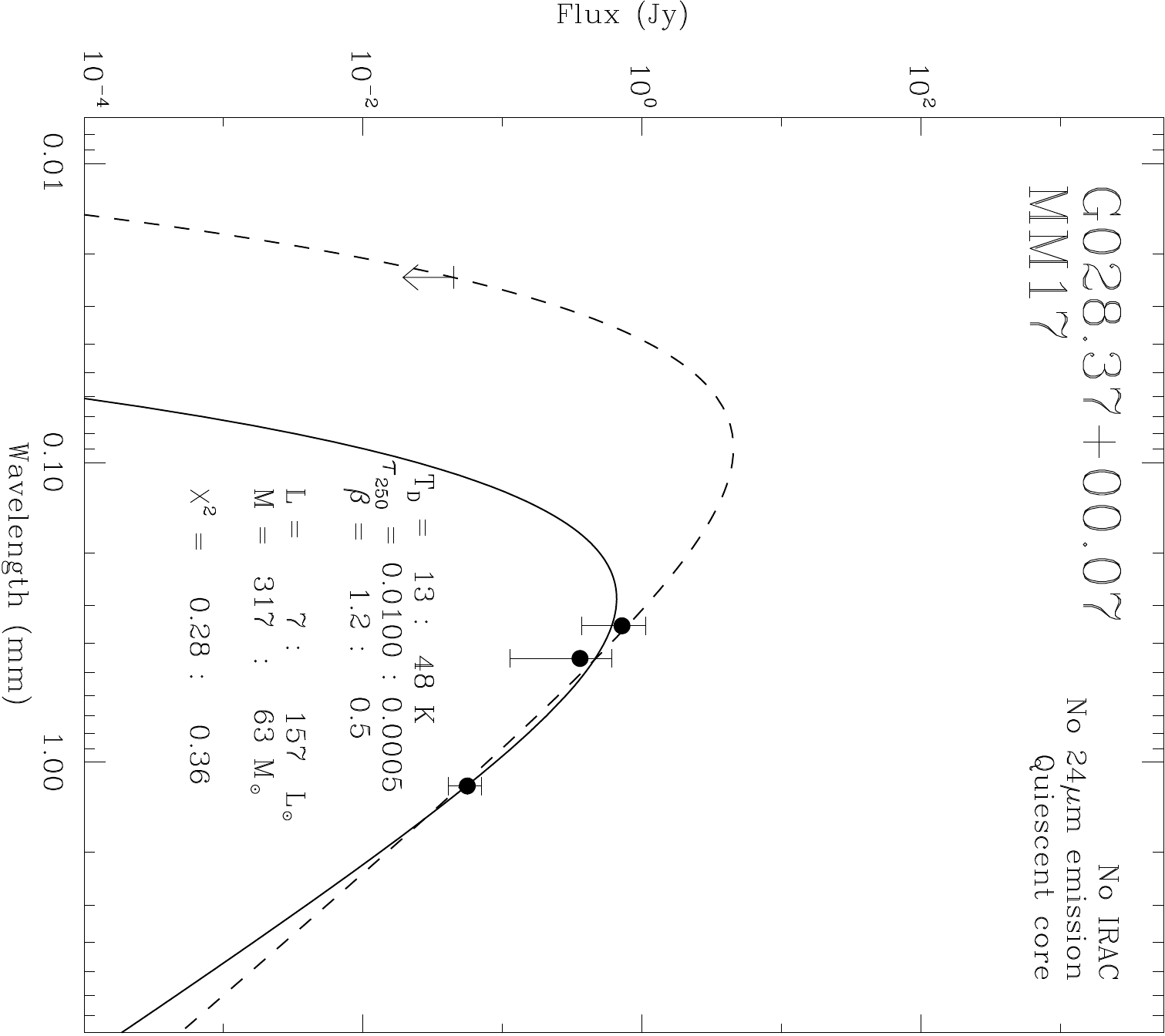}\\
\caption{\label{seds-1}\Spitzer\, 24\,\um\, image overlaid  
   with 1.2\,mm continuum emission for \irdcone\, (contour levels are
   30, 60, 90, 120, 240, 360, 480, 840, 1200\,mJy beam$^{-1}$). 
   The lower panels show the broadband
   SEDs for cores within this IRDC.  The fluxes derived from the
   millimeter, sub-millimeter, and far-IR  continuum data are shown as filled
   circles (with the corresponding error bars), while the 24\,\um\, fluxes are shown as  either a filled circle (when included within the fit), an open circle (when excluded from the fit),  or as an upper limit arrow. For cores that have measured fluxes only in the millimeter/sub-millimeter regime (i.e.\, a limit at 24\,\um), we show the results from two fits: one using only the measured fluxes (solid line; lower limit), while the other includes the 24\,\um\, limit as a real data (dashed line; upper limit). In all other cases, the solid line is the best fit gray-body, while the dotted lines correspond to the functions determined using the errors for the T$_{D}$, $\tau$, and $\beta$ output from the fitting.  Labeled on each plot is the IRDC and core name,  classification, and the derived parameters.}
\end{figure}
\clearpage 
\begin{figure}
\begin{center}
\includegraphics[angle=0,width=0.6\textwidth]{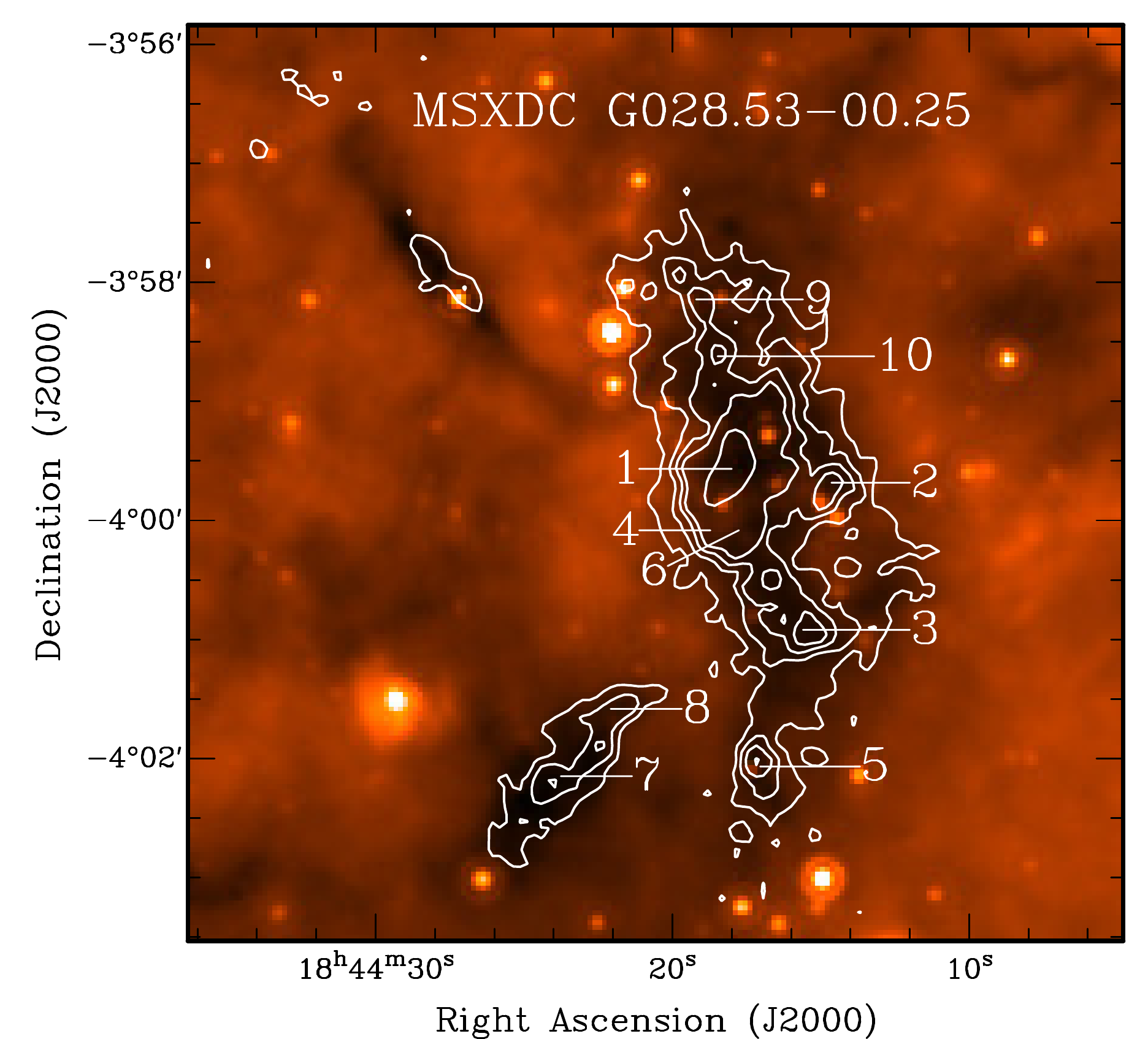}\\
\end{center}
\includegraphics[angle=90,width=0.5\textwidth]{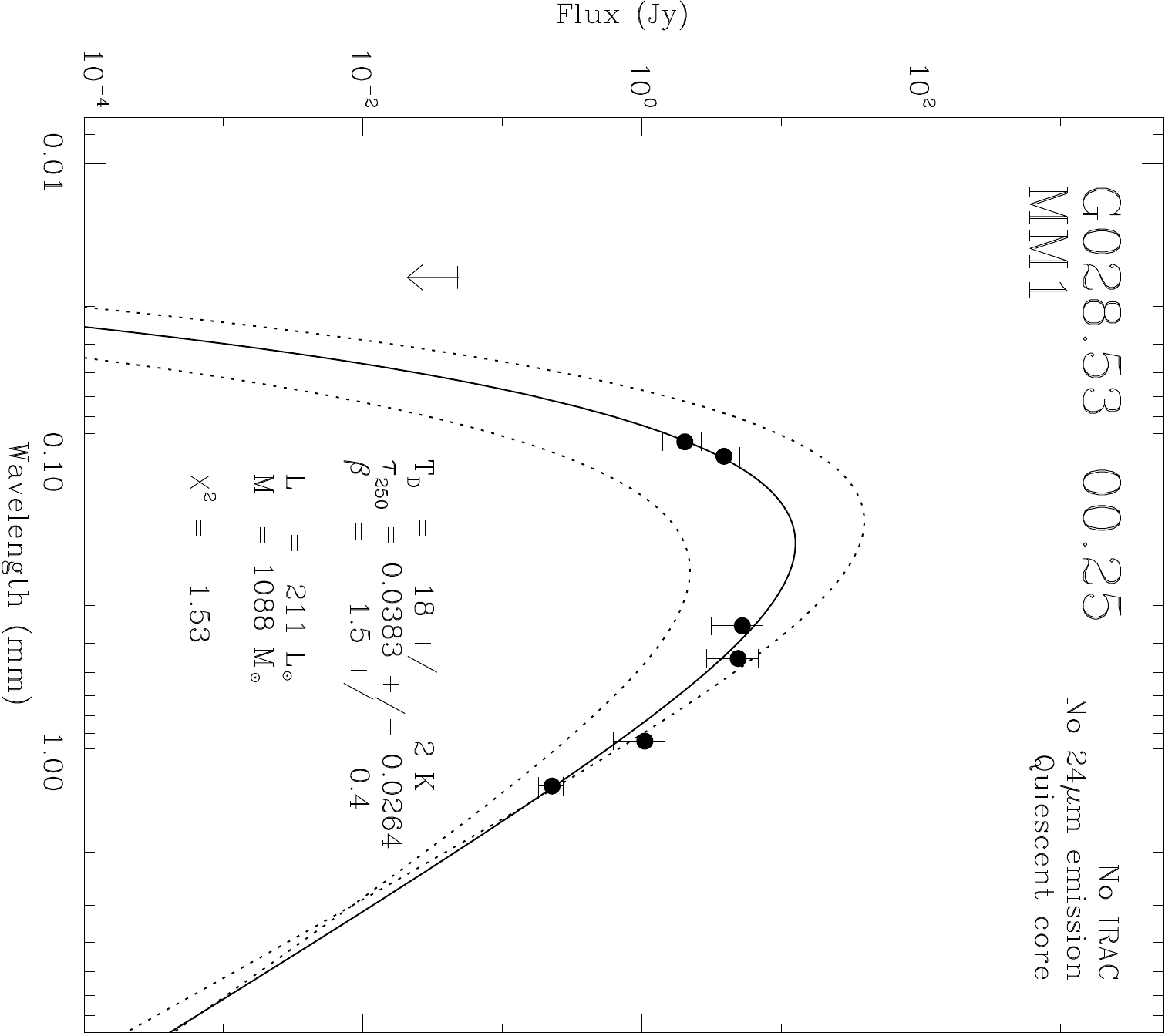}
\includegraphics[angle=90,width=0.5\textwidth]{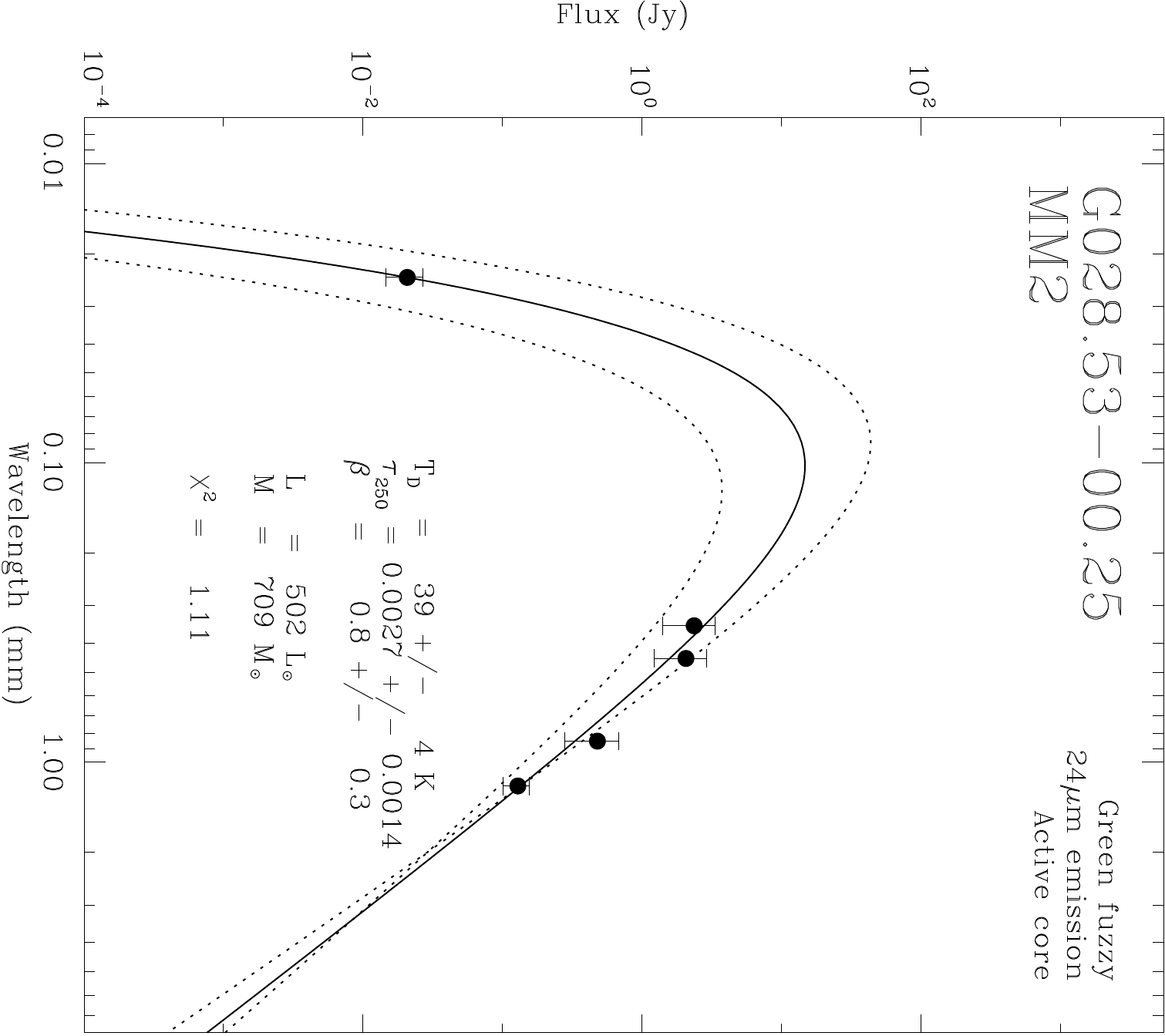}\\
\end{figure}
\clearpage 
\begin{figure}
\includegraphics[angle=90,width=0.5\textwidth]{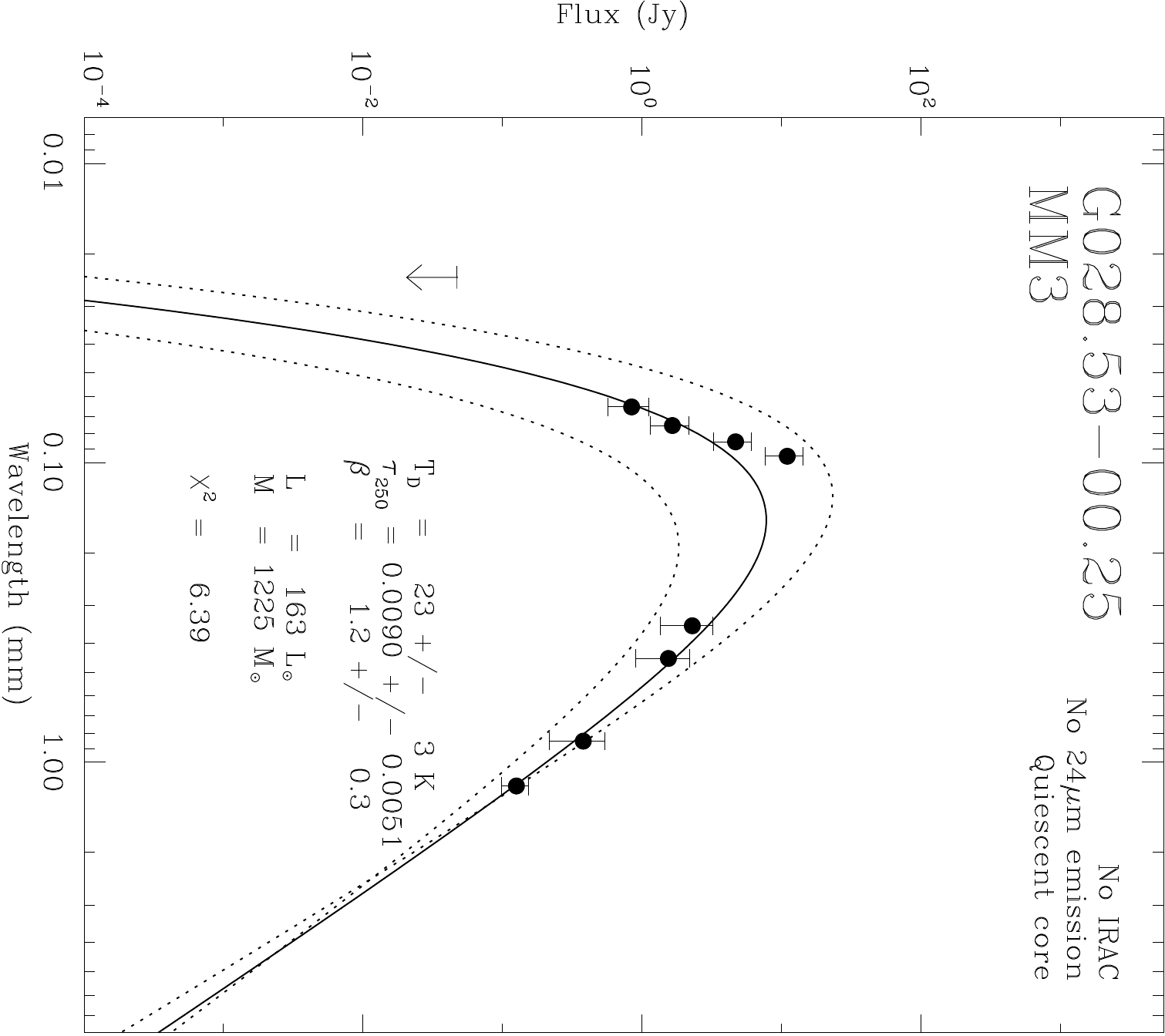}
\includegraphics[angle=90,width=0.5\textwidth]{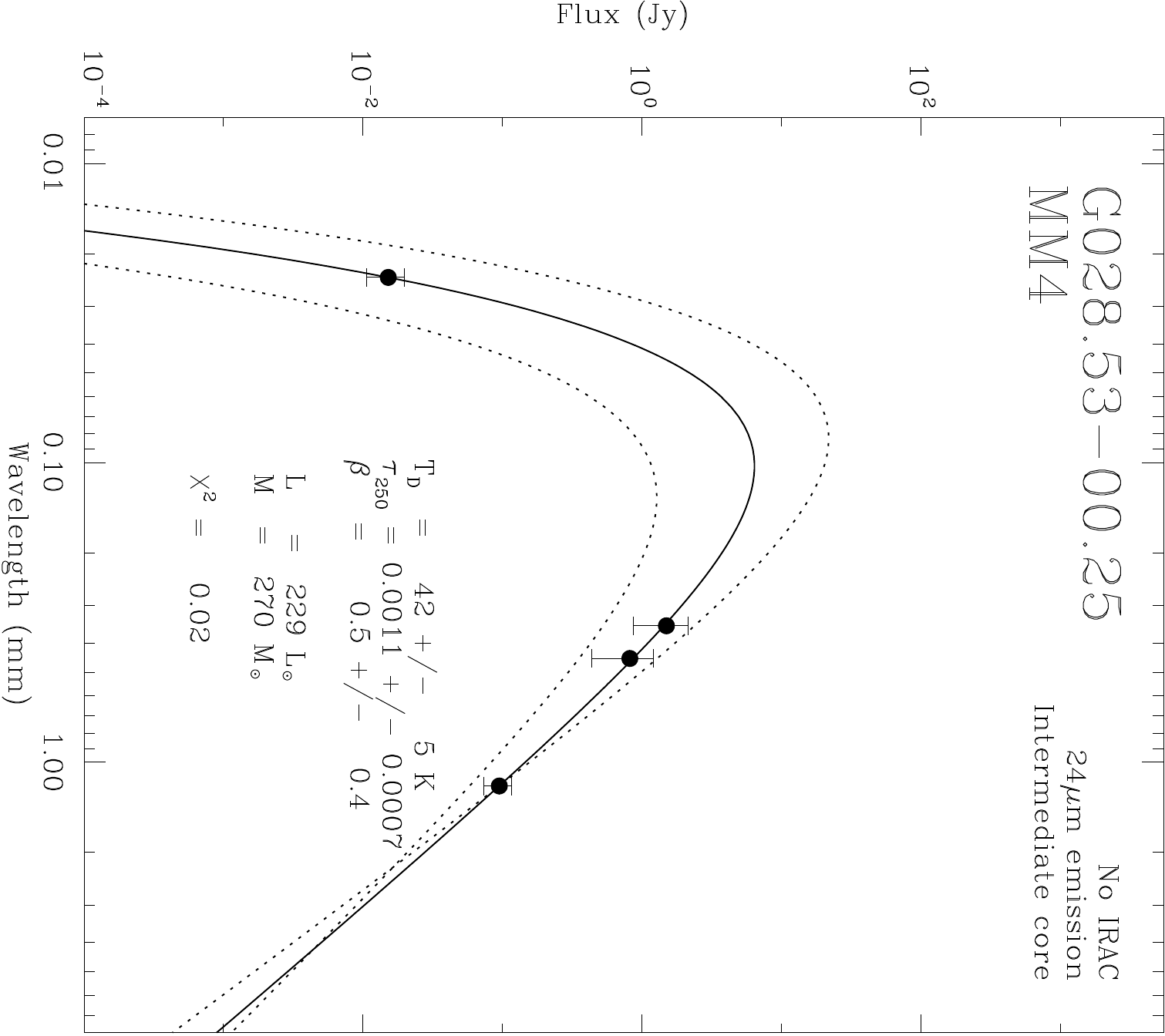}\\
\includegraphics[angle=90,width=0.5\textwidth]{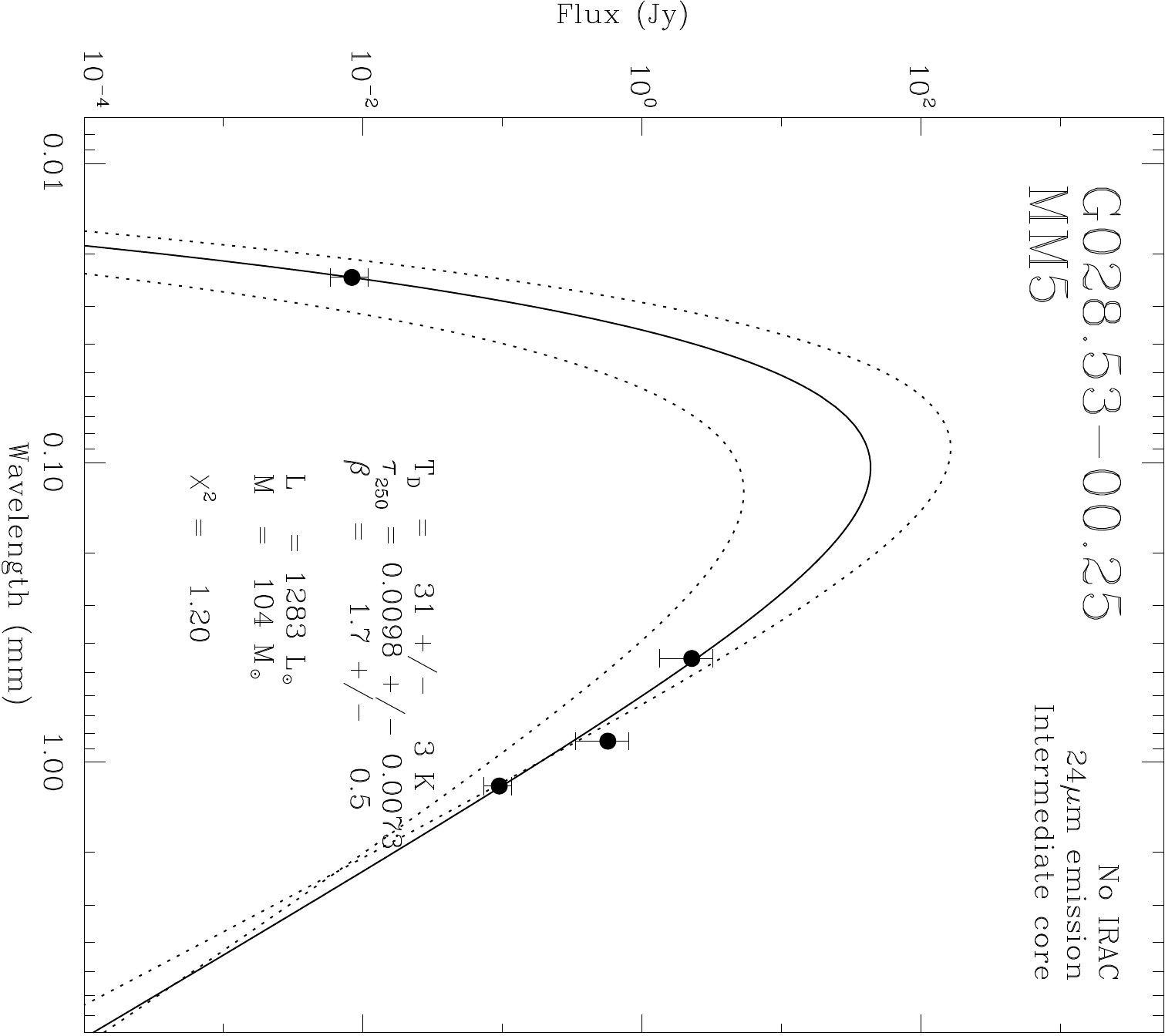}
\includegraphics[angle=90,width=0.5\textwidth]{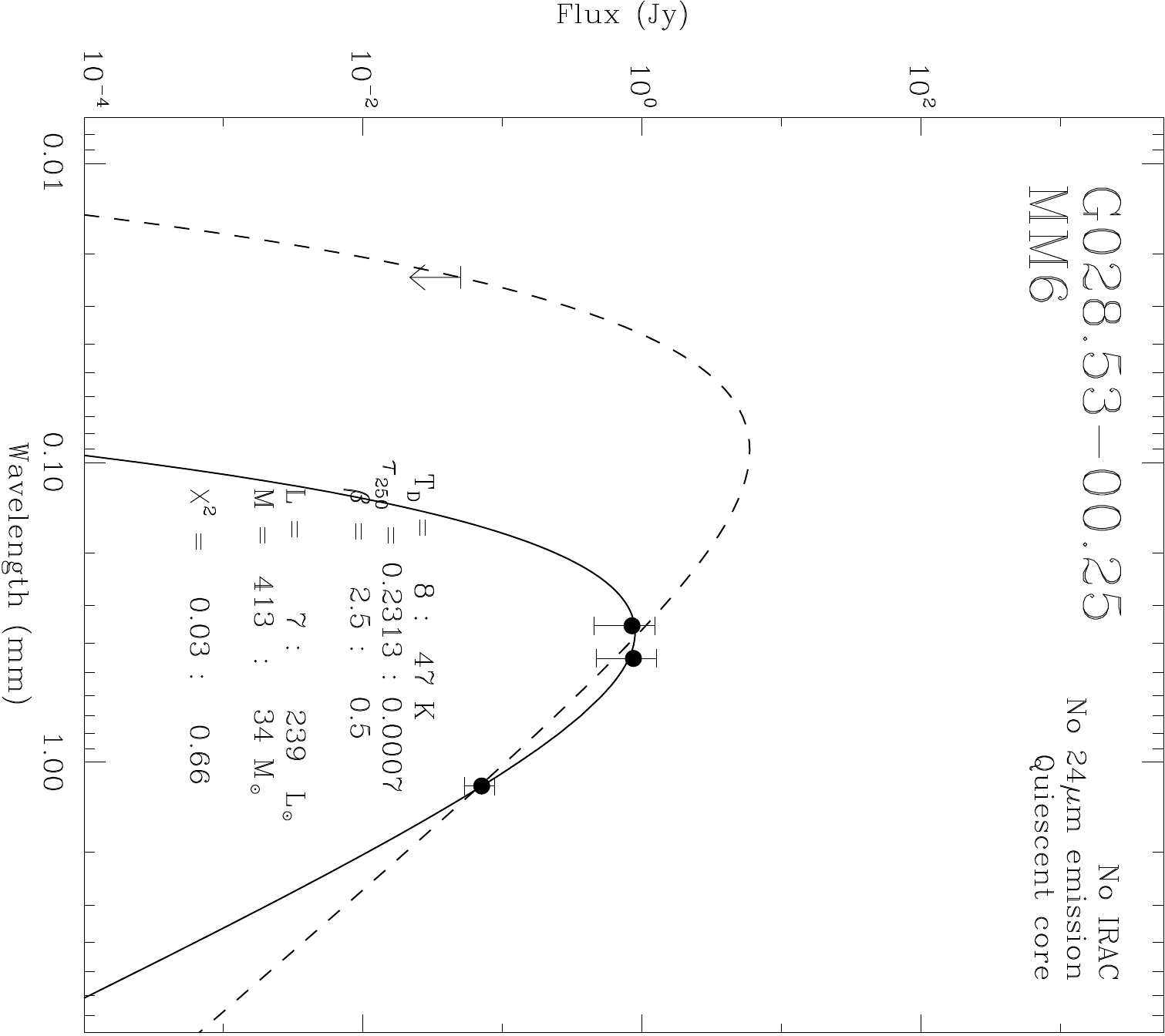}\\
\end{figure}
\clearpage 
\begin{figure}
\includegraphics[angle=90,width=0.5\textwidth]{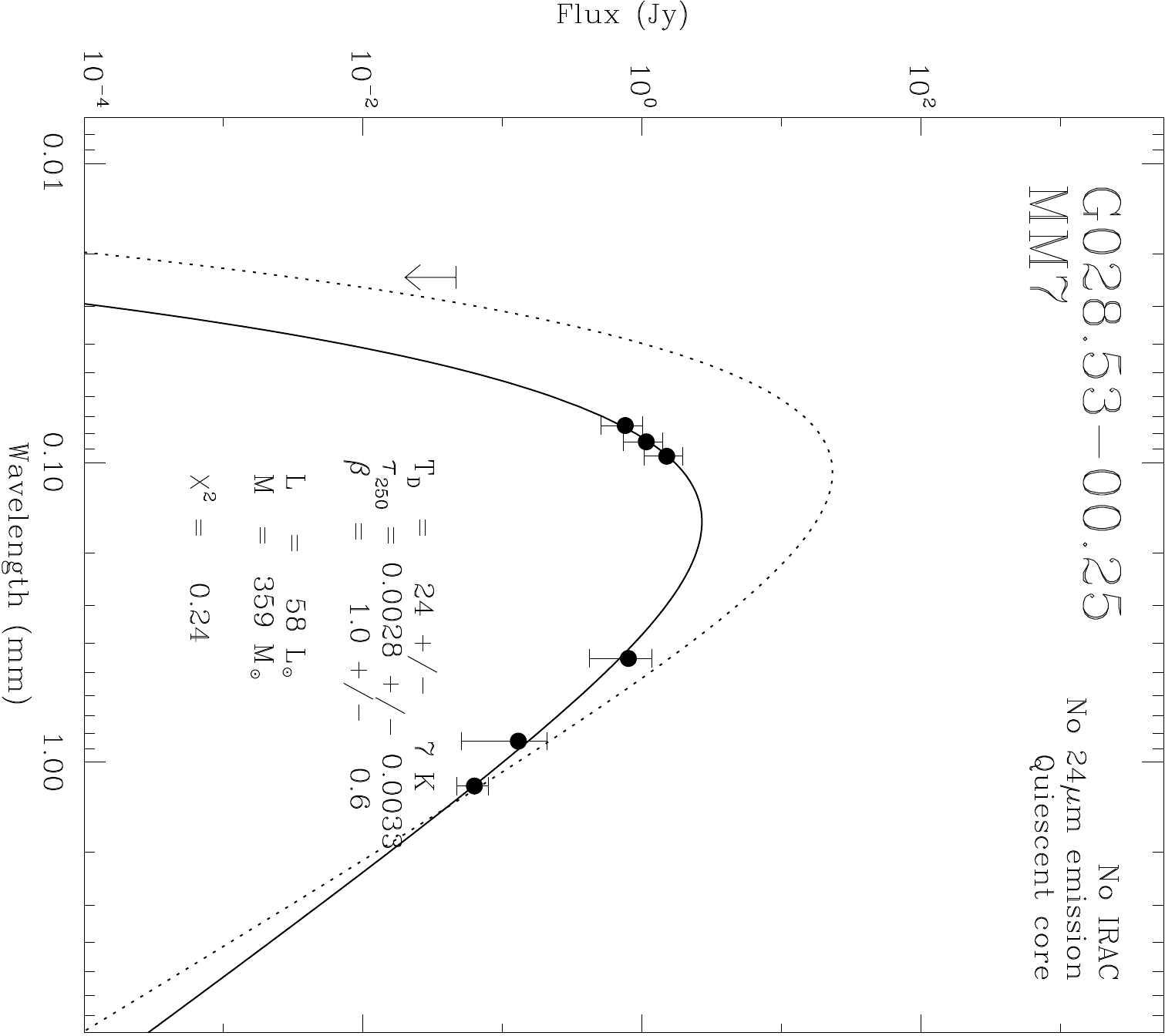}
\includegraphics[angle=90,width=0.5\textwidth]{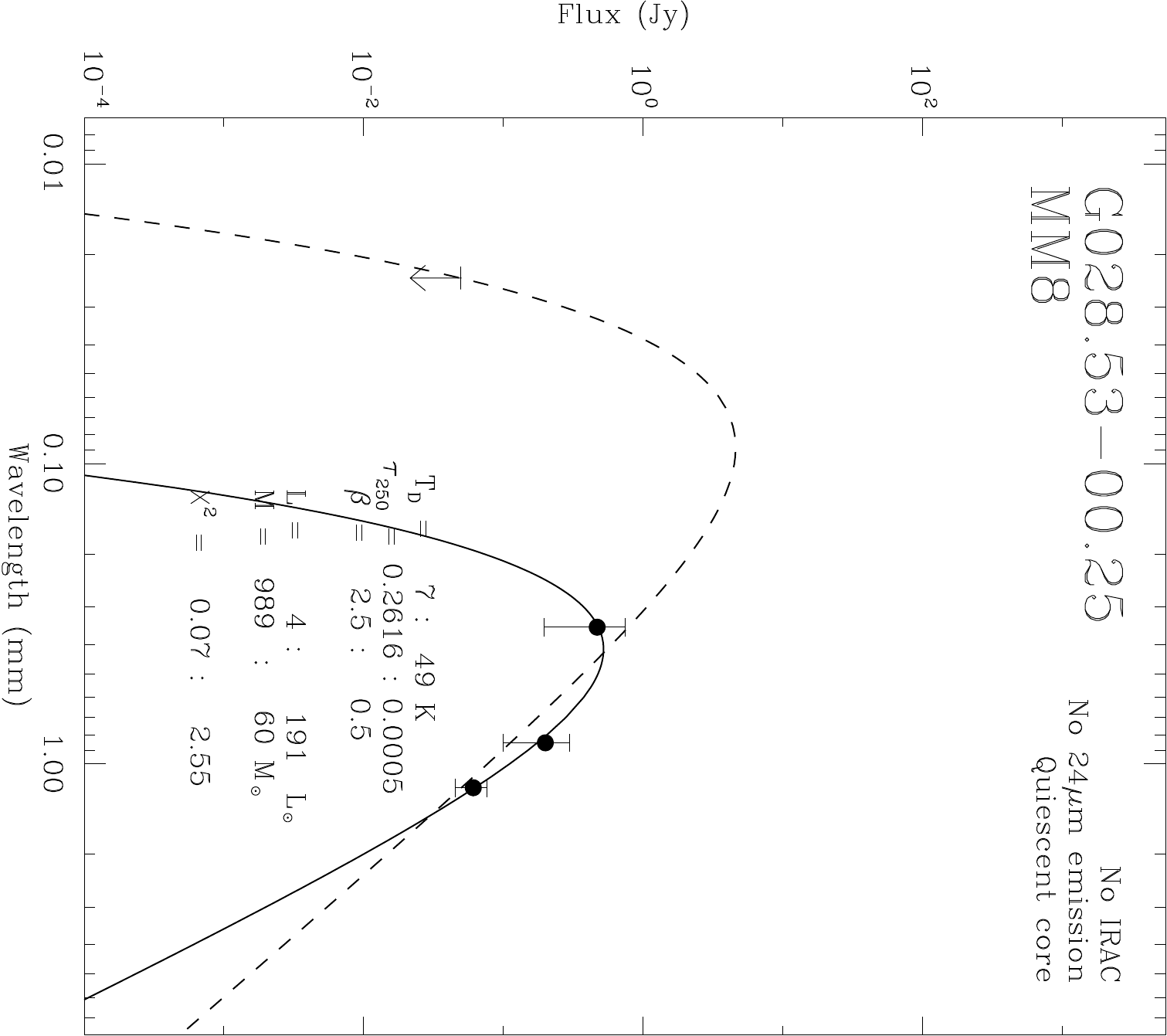}\\
\includegraphics[angle=90,width=0.5\textwidth]{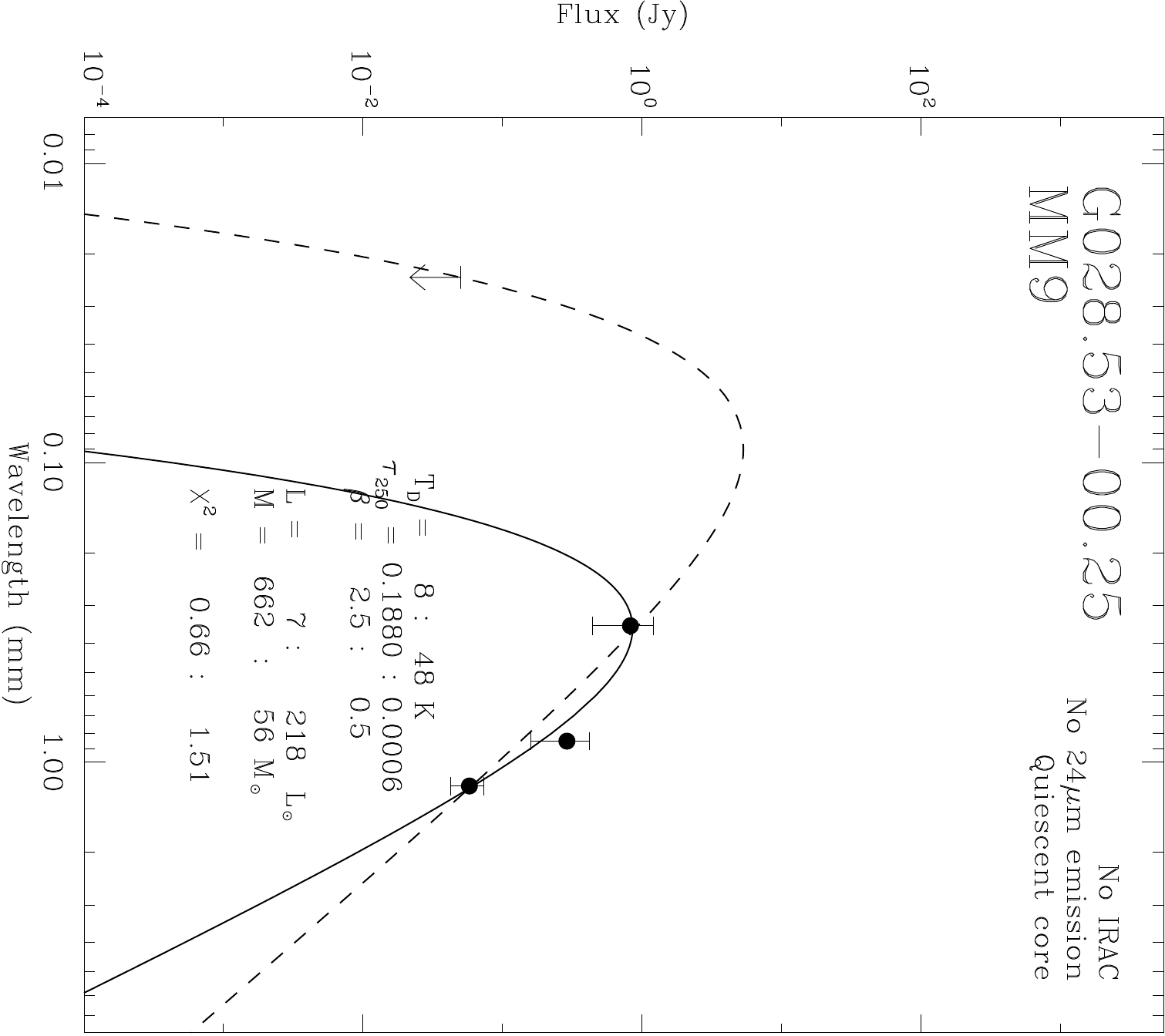}
\includegraphics[angle=90,width=0.5\textwidth]{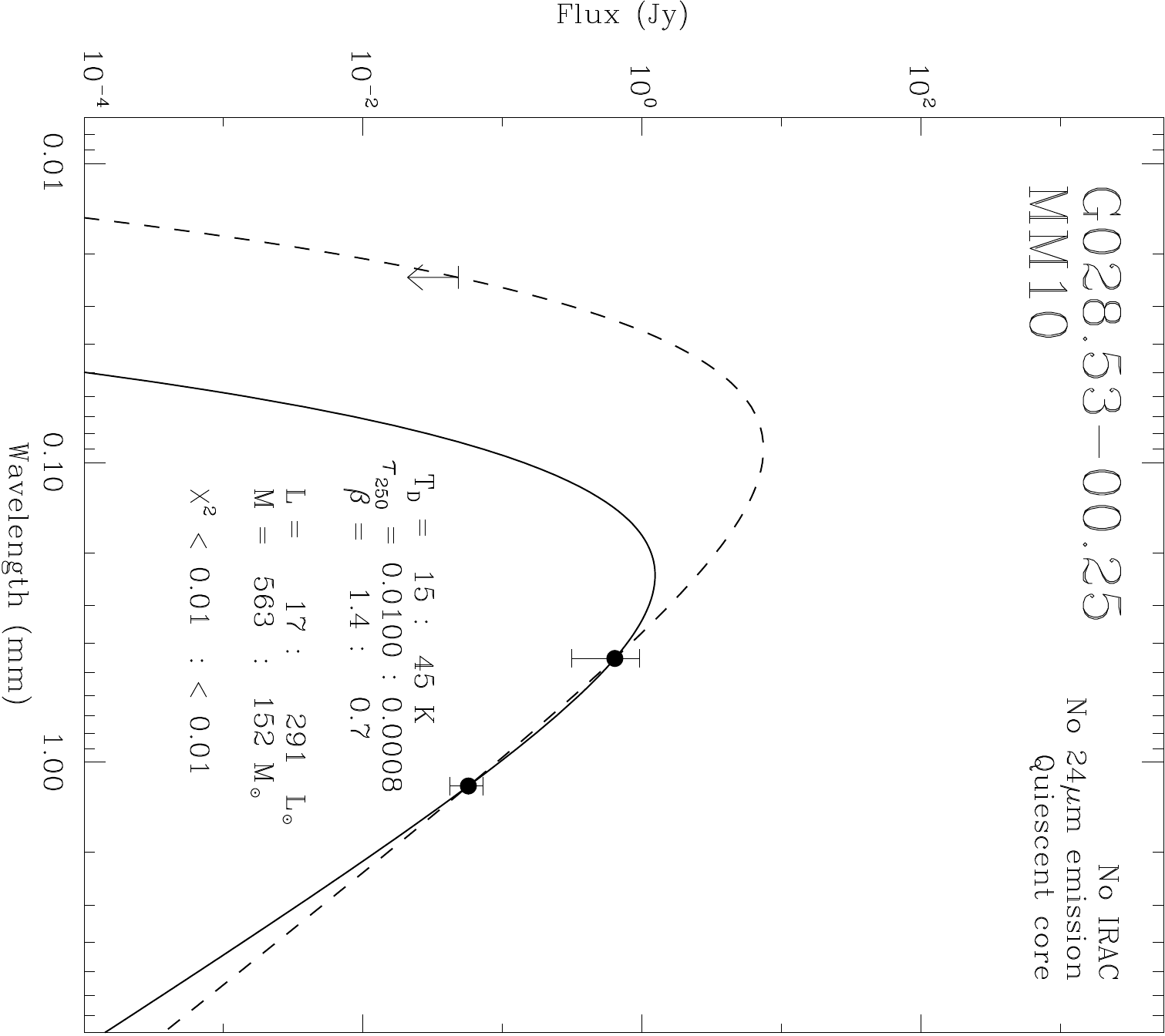}\\
\caption{\label{seds-30}\Spitzer\, 24\,\um\, image overlaid  
   with 1.2\,mm continuum emission for \irdcthirty\, (contour levels are
   30, 60, 90,120, 240\,mJy beam$^{-1}$). The lower panels show the broadband
   SEDs for cores within this IRDC.  The fluxes derived from the
   millimeter, sub-millimeter, and far-IR  continuum data are shown as filled
   circles (with the corresponding error bars), while the 24\,\um\, fluxes are shown as  either a filled circle (when included within the fit), an open circle (when excluded from the fit),  or as an upper limit arrow. For cores that have measured fluxes only in the millimeter/sub-millimeter regime (i.e.\, a limit at 24\,\um), we show the results from two fits: one using only the measured fluxes (solid line; lower limit), while the other includes the 24\,\um\, limit as a real data (dashed line; upper limit). In all other cases, the solid line is the best fit gray-body, while the dotted lines correspond to the functions determined using the errors for the T$_{D}$, $\tau$, and $\beta$ output from the fitting.  Labeled on each plot is the IRDC and core name,  classification, and the derived parameters.}
\end{figure}
\clearpage 
\begin{figure}
\begin{center}
\includegraphics[angle=0,width=0.6\textwidth]{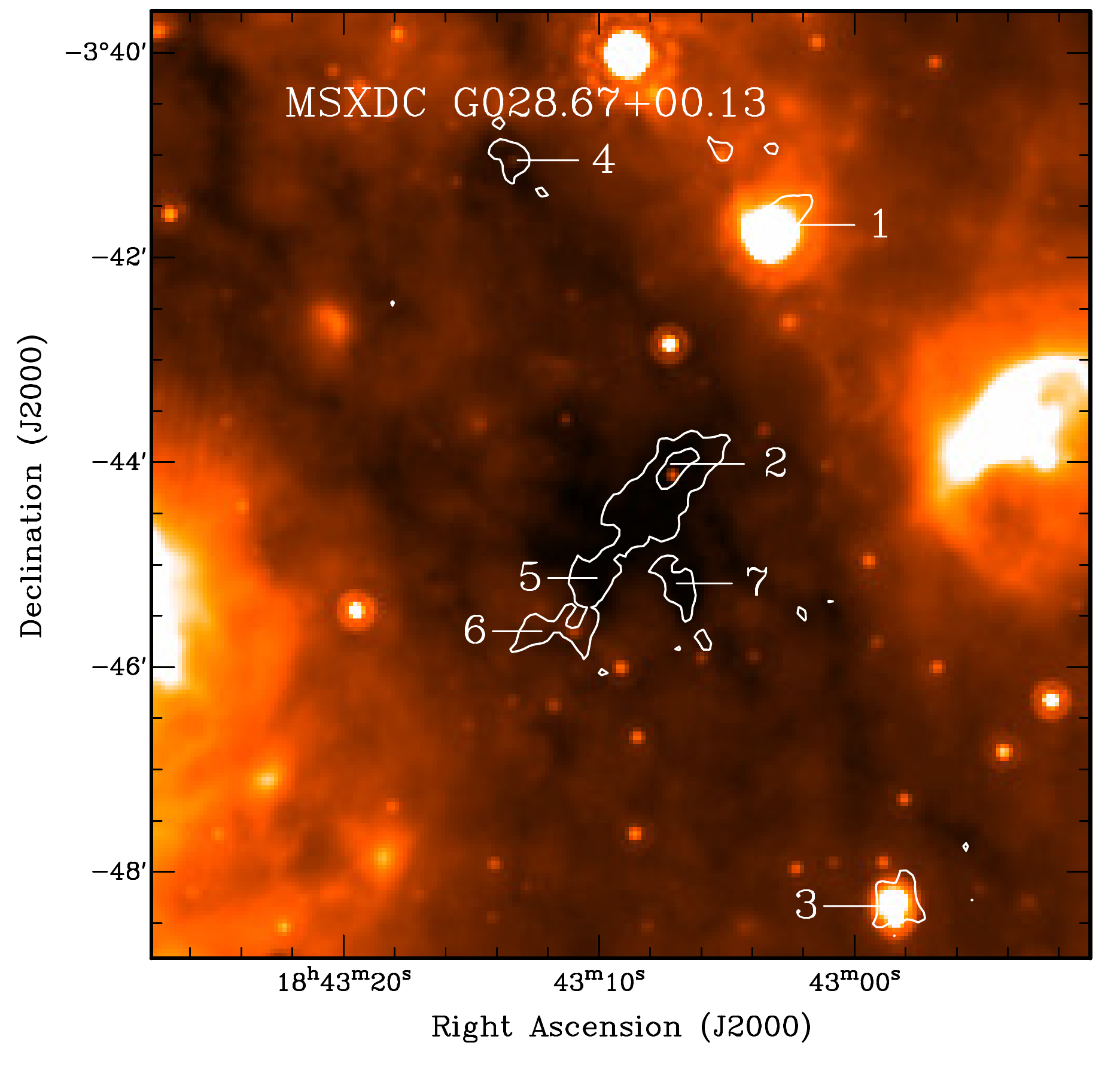}\\
\end{center}
\includegraphics[angle=90,width=0.5\textwidth]{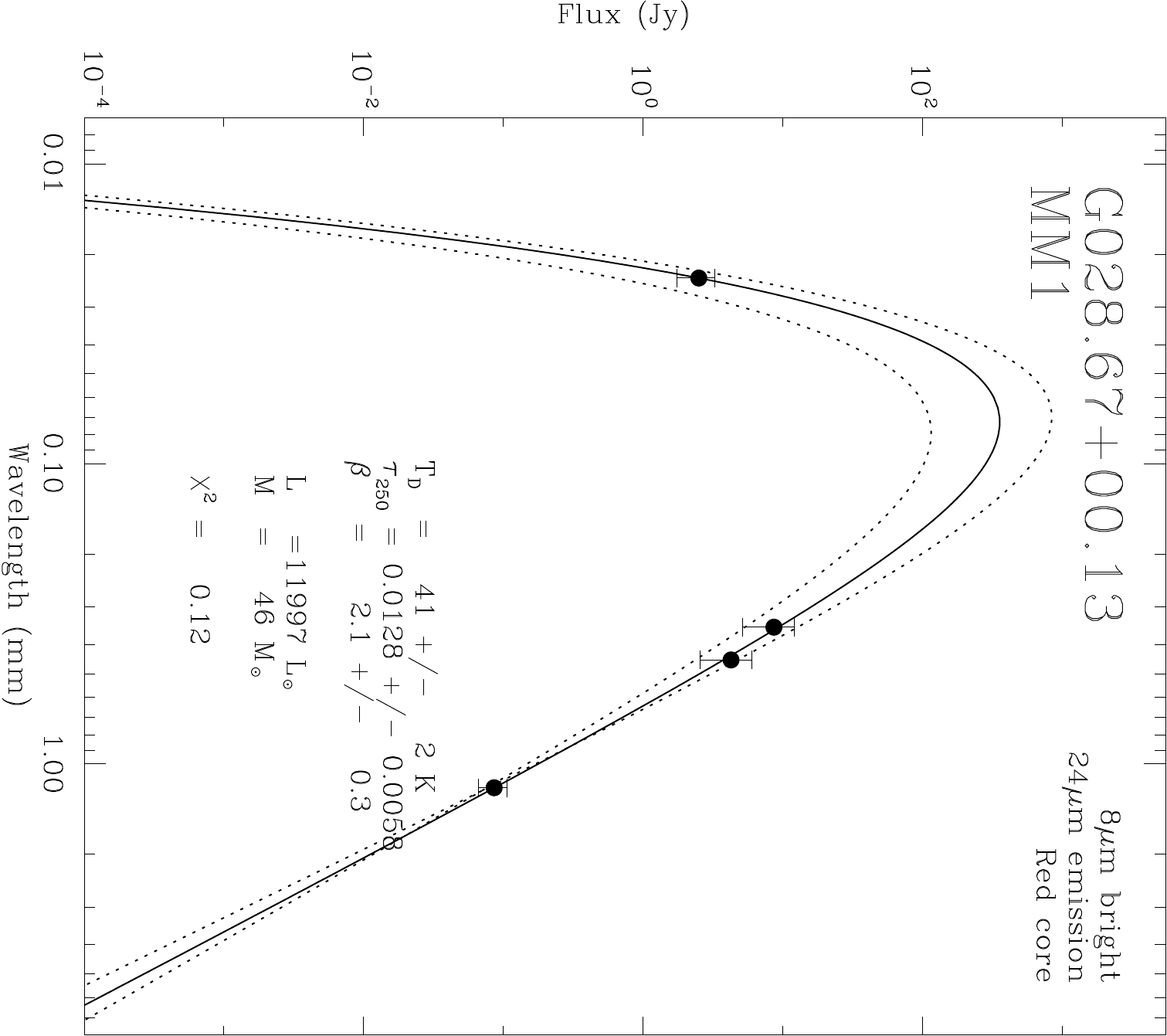}
\includegraphics[angle=90,width=0.5\textwidth]{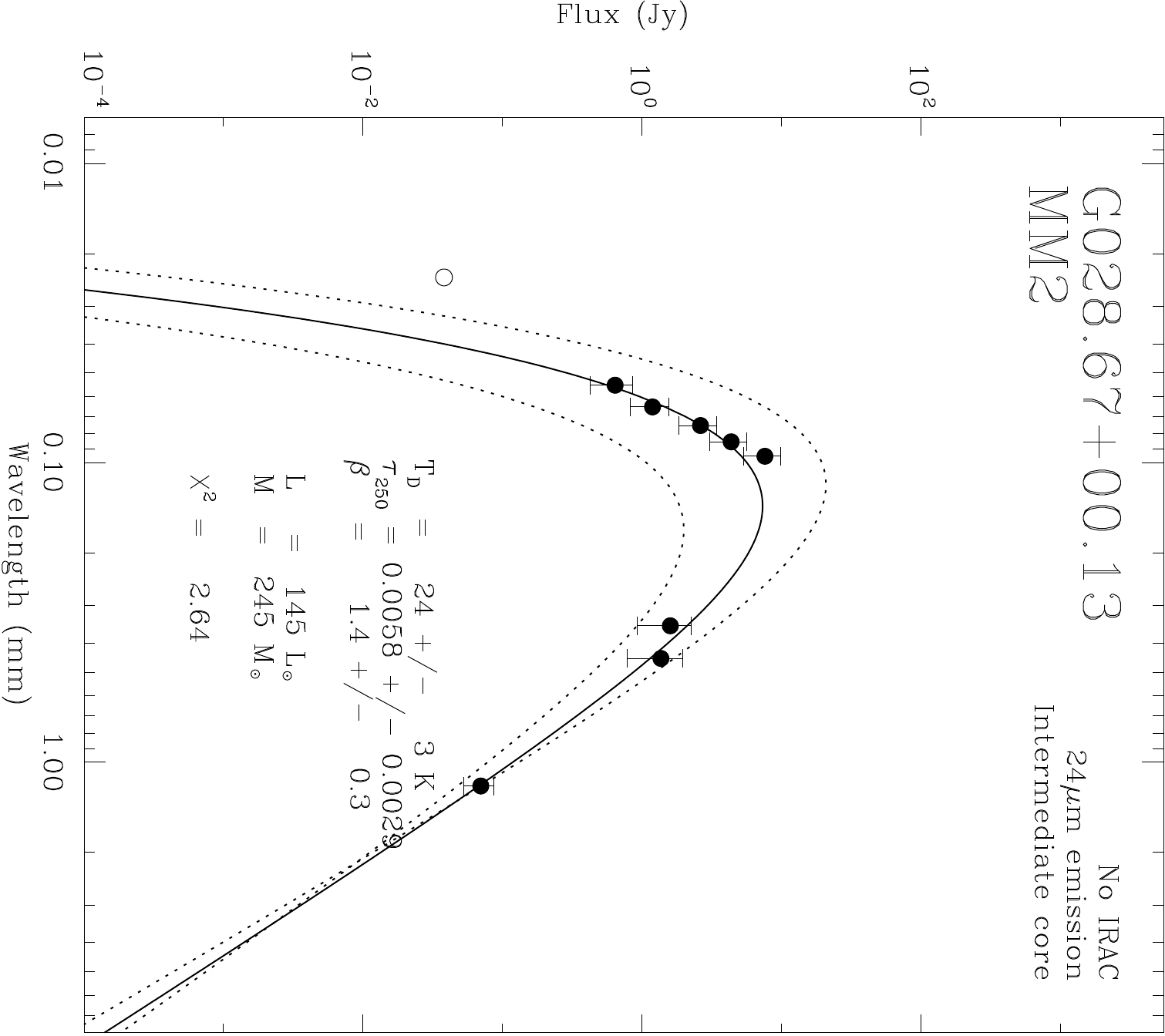}\\
\end{figure}
\clearpage 
\begin{figure}
\includegraphics[angle=90,width=0.5\textwidth]{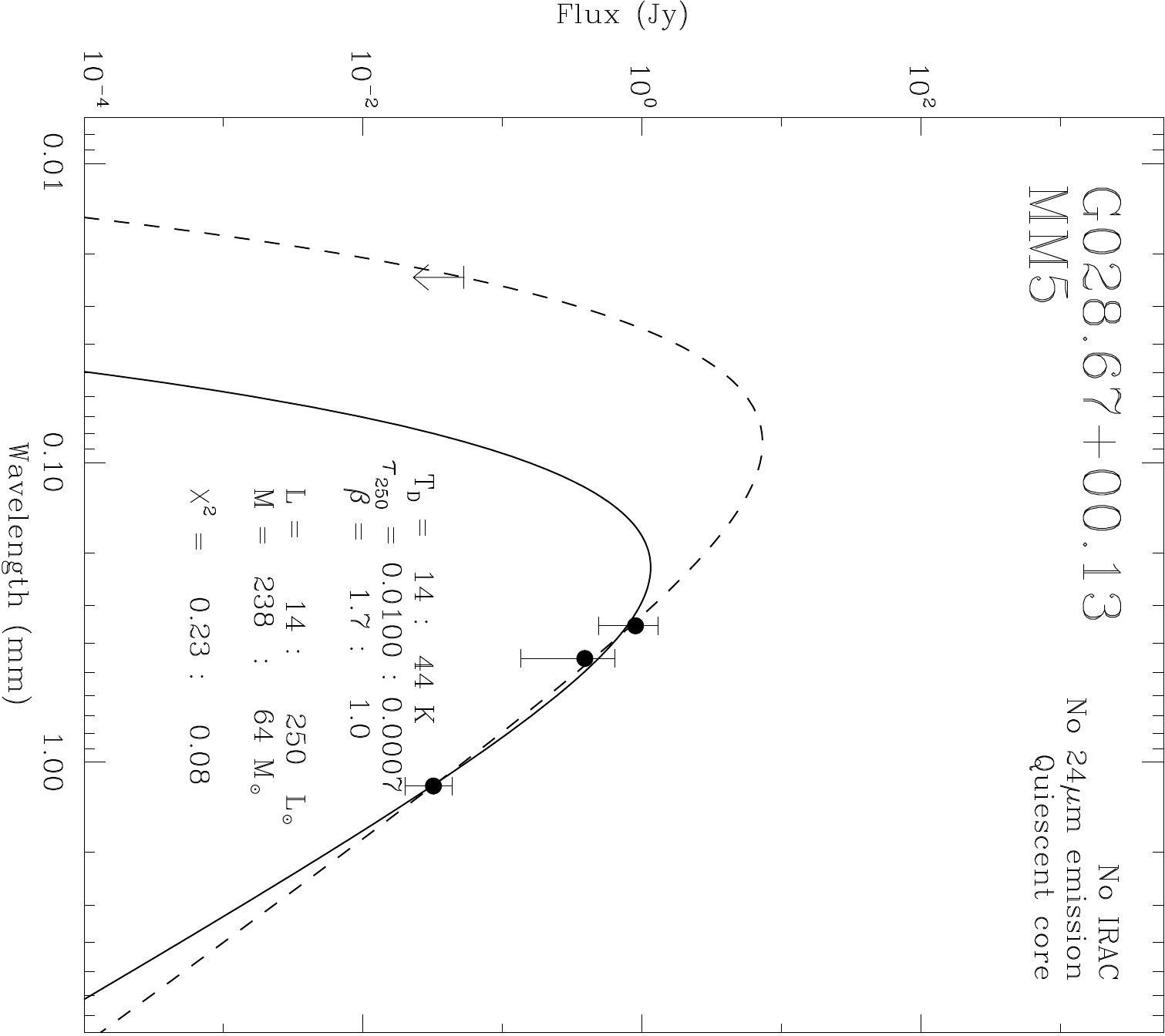}
\includegraphics[angle=90,width=0.5\textwidth]{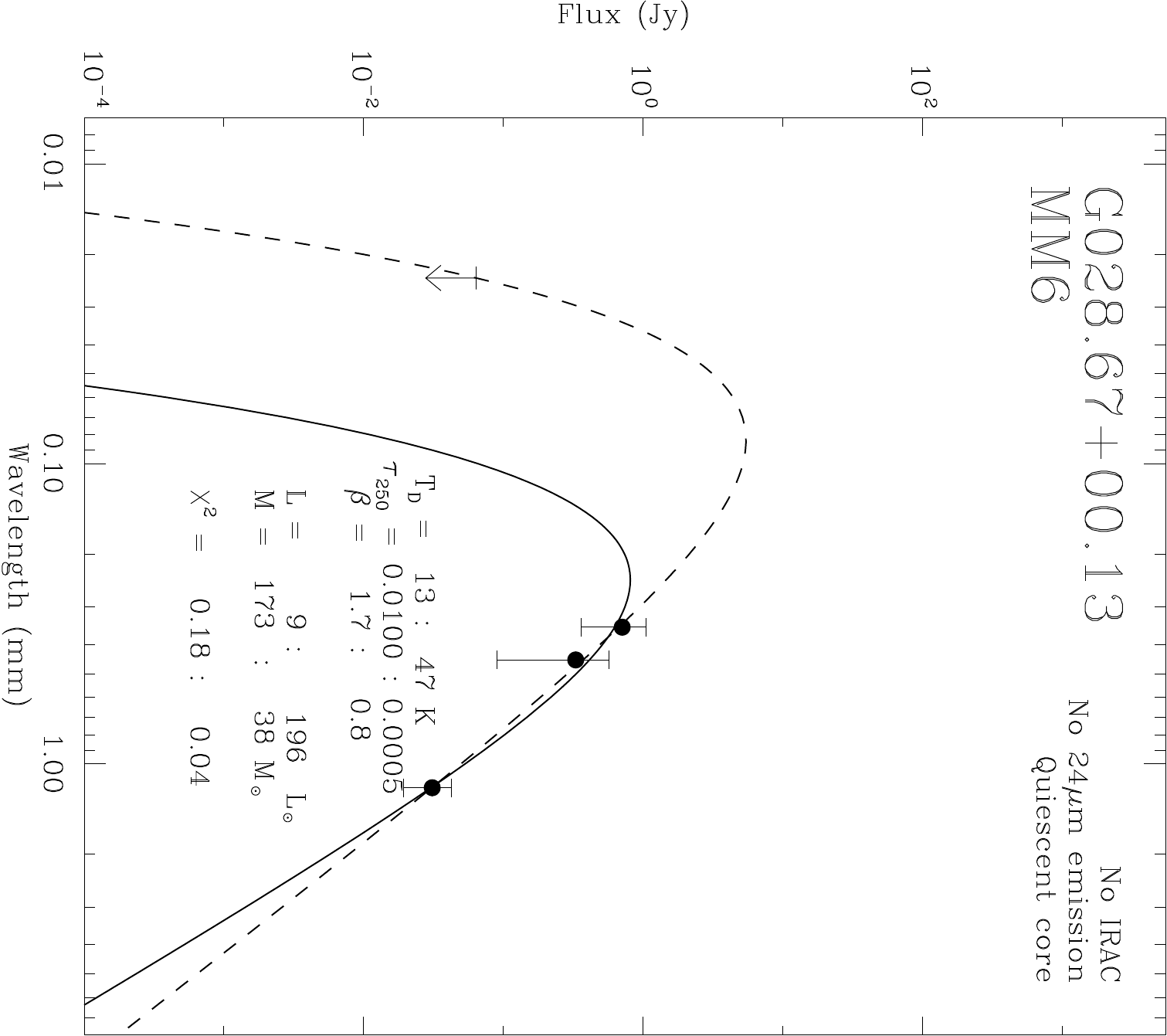}\\
\caption{\label{seds-19}\Spitzer\, 24\,\um\, image overlaid  
   with 1.2\,mm continuum emission for \irdcnineteen\, (contour levels
   are 30, 60, 90\,mJy beam$^{-1}$). The lower panels show the broadband
   SEDs for cores within this IRDC.  The fluxes derived from the
   millimeter, sub-millimeter, and far-IR  continuum data are shown as filled
   circles (with the corresponding error bars), while the 24\,\um\, fluxes are shown as  either a filled circle (when included within the fit), an open circle (when excluded from the fit),  or as an upper limit arrow. For cores that have measured fluxes only in the millimeter/sub-millimeter regime (i.e.\, a limit at 24\,\um), we show the results from two fits: one using only the measured fluxes (solid line; lower limit), while the other includes the 24\,\um\, limit as a real data (dashed line; upper limit). In all other cases, the solid line is the best fit gray-body, while the dotted lines correspond to the functions determined using the errors for the T$_{D}$, $\tau$, and $\beta$ output from the fitting.  Labeled on each plot is the IRDC and core name,  classification, and the derived parameters.}
\end{figure}
\clearpage 
\begin{figure}
\begin{center}
\includegraphics[angle=0,width=0.6\textwidth]{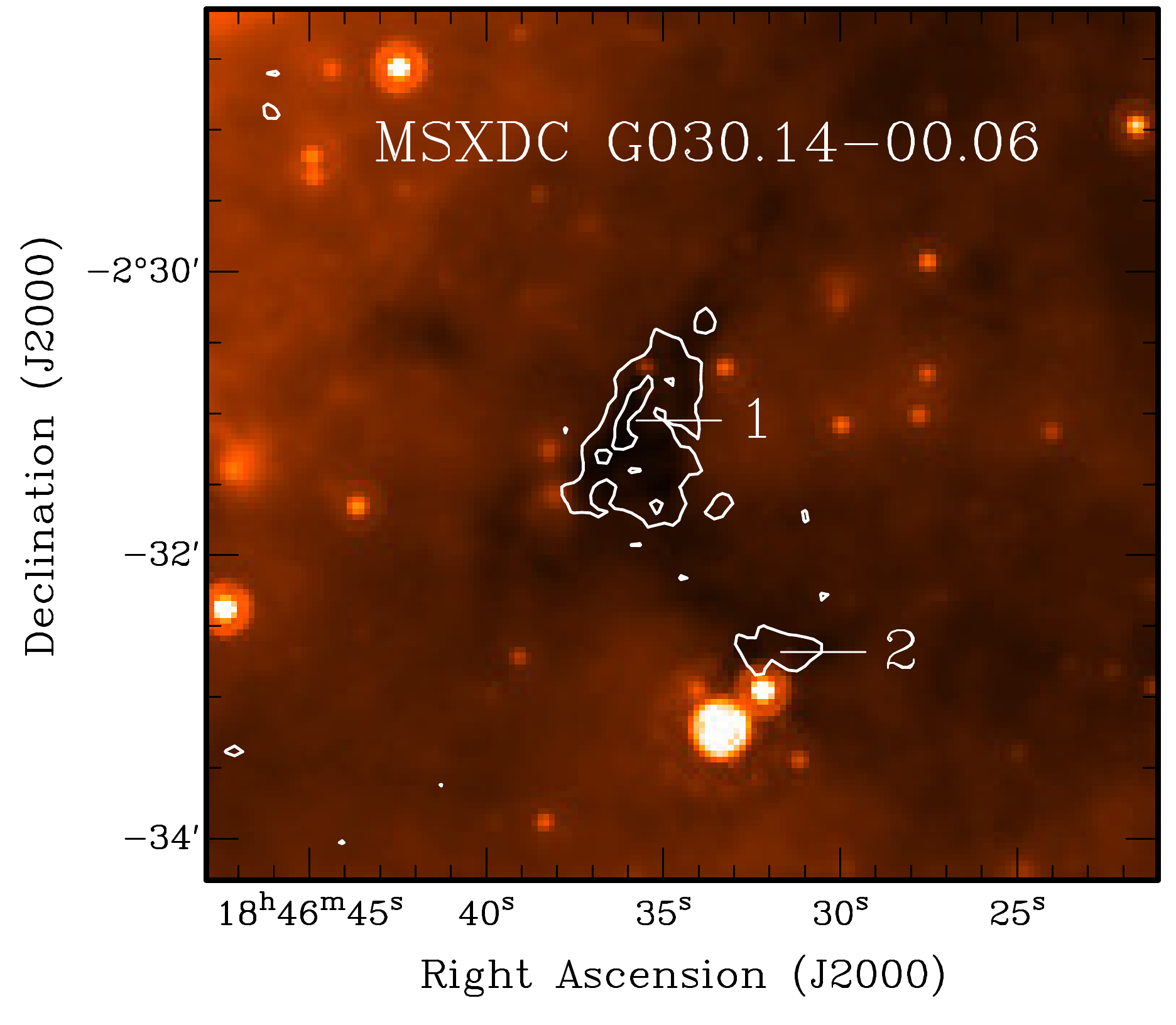}\\
\end{center}
\includegraphics[angle=90,width=0.5\textwidth]{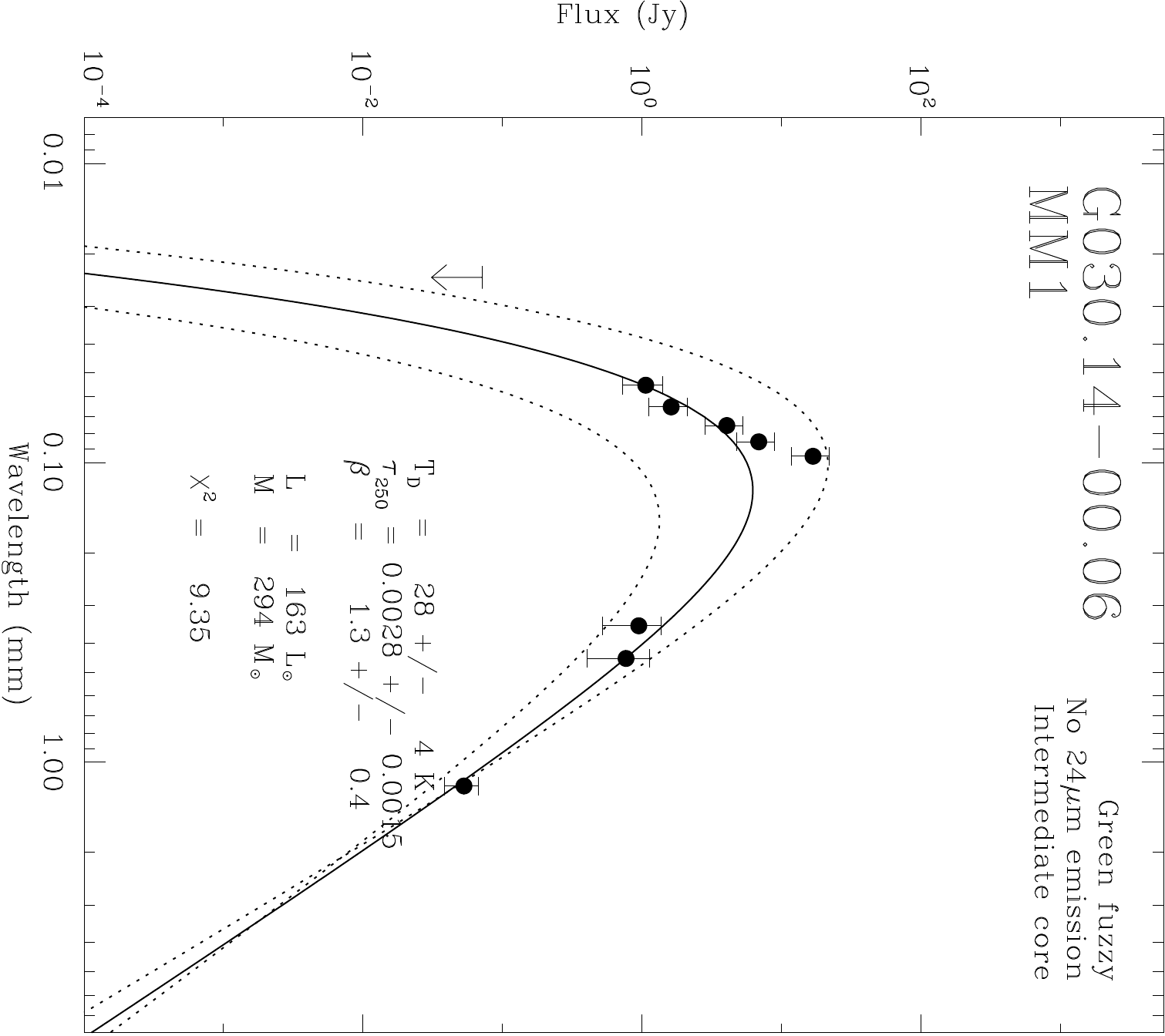}
\includegraphics[angle=90,width=0.5\textwidth]{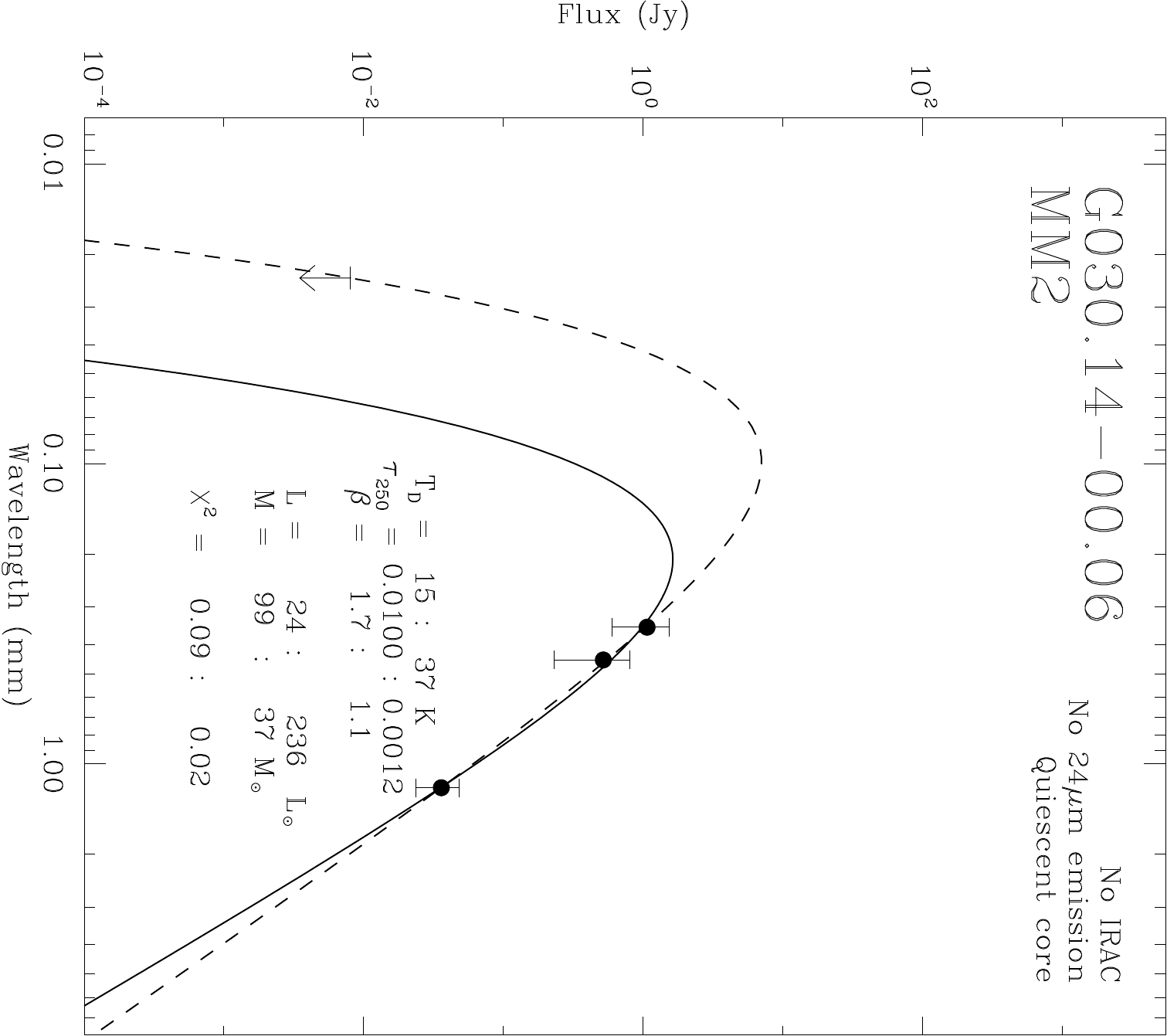}\\
\caption{\label{seds-57}\Spitzer\, 24\,\um\, image overlaid  
   with 1.2\,mm continuum emission for \irdcfiftyseven\, (contour
   levels are 30, 60\,mJy beam$^{-1}$). The lower panels show the broadband
   SEDs for cores within this IRDC.  The fluxes derived from the
   millimeter, sub-millimeter, and far-IR  continuum data are shown as filled
   circles (with the corresponding error bars), while the 24\,\um\, fluxes are shown as  either a filled circle (when included within the fit), an open circle (when excluded from the fit),  or as an upper limit arrow. For cores that have measured fluxes only in the millimeter/sub-millimeter regime (i.e.\, a limit at 24\,\um), we show the results from two fits: one using only the measured fluxes (solid line; lower limit), while the other includes the 24\,\um\, limit as a real data (dashed line; upper limit). In all other cases, the solid line is the best fit gray-body, while the dotted lines correspond to the functions determined using the errors for the T$_{D}$, $\tau$, and $\beta$ output from the fitting.  Labeled on each plot is the IRDC and core name,  classification, and the derived parameters.}
\end{figure}
\clearpage 
\begin{figure}
\begin{center}
\includegraphics[angle=0,width=0.6\textwidth]{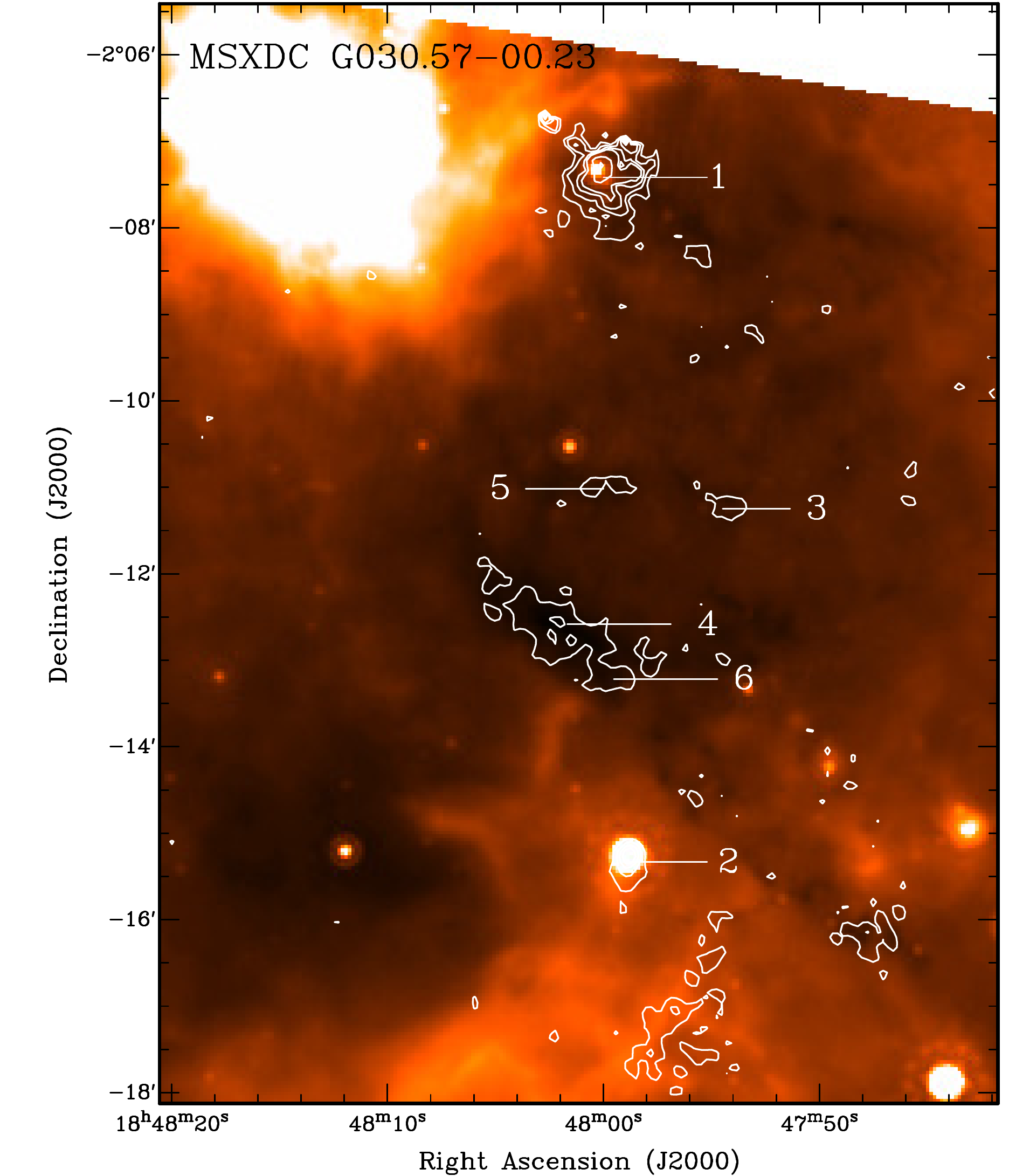}\\
\end{center}
\includegraphics[angle=90,width=0.5\textwidth]{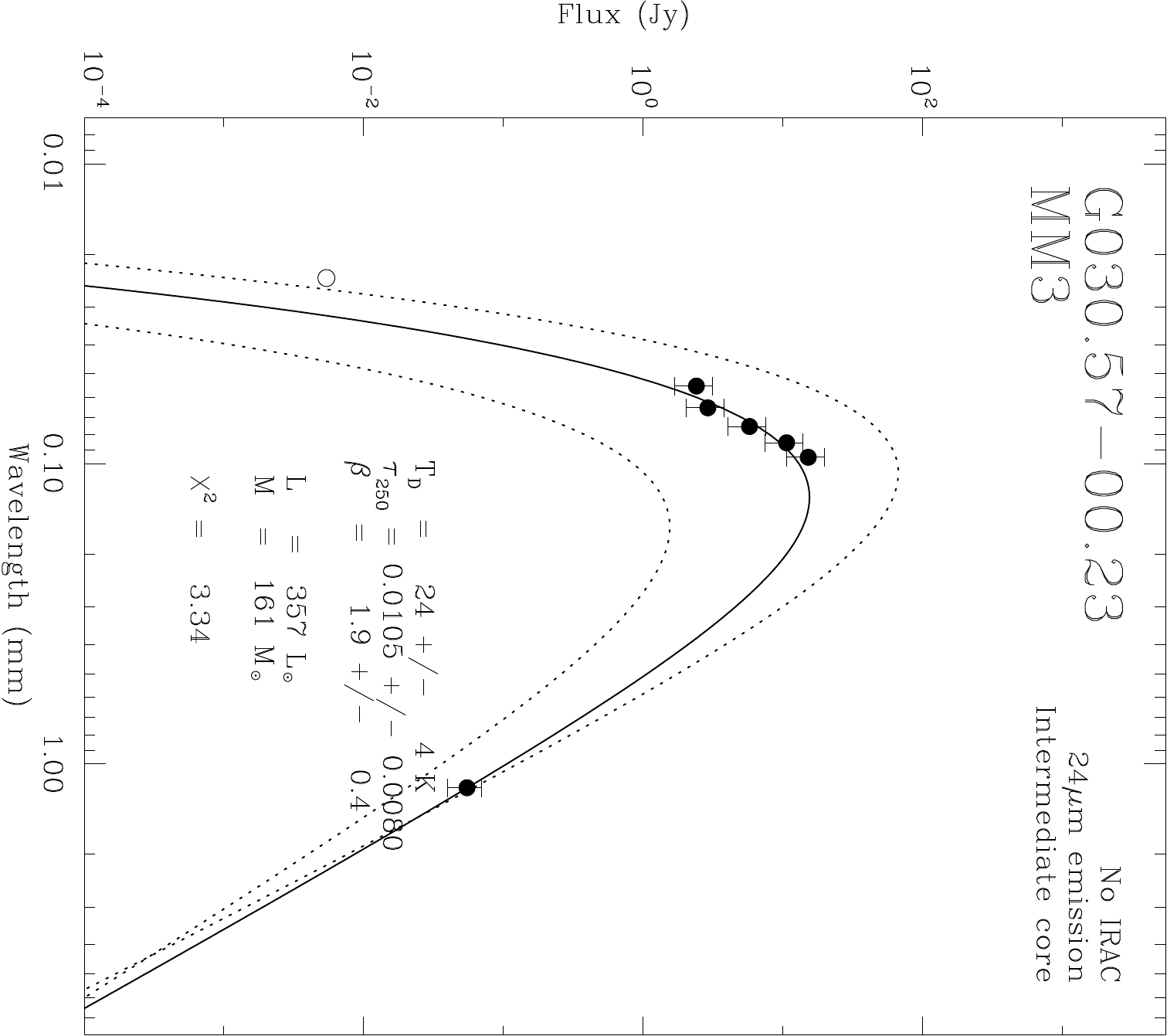}
\includegraphics[angle=90,width=0.5\textwidth]{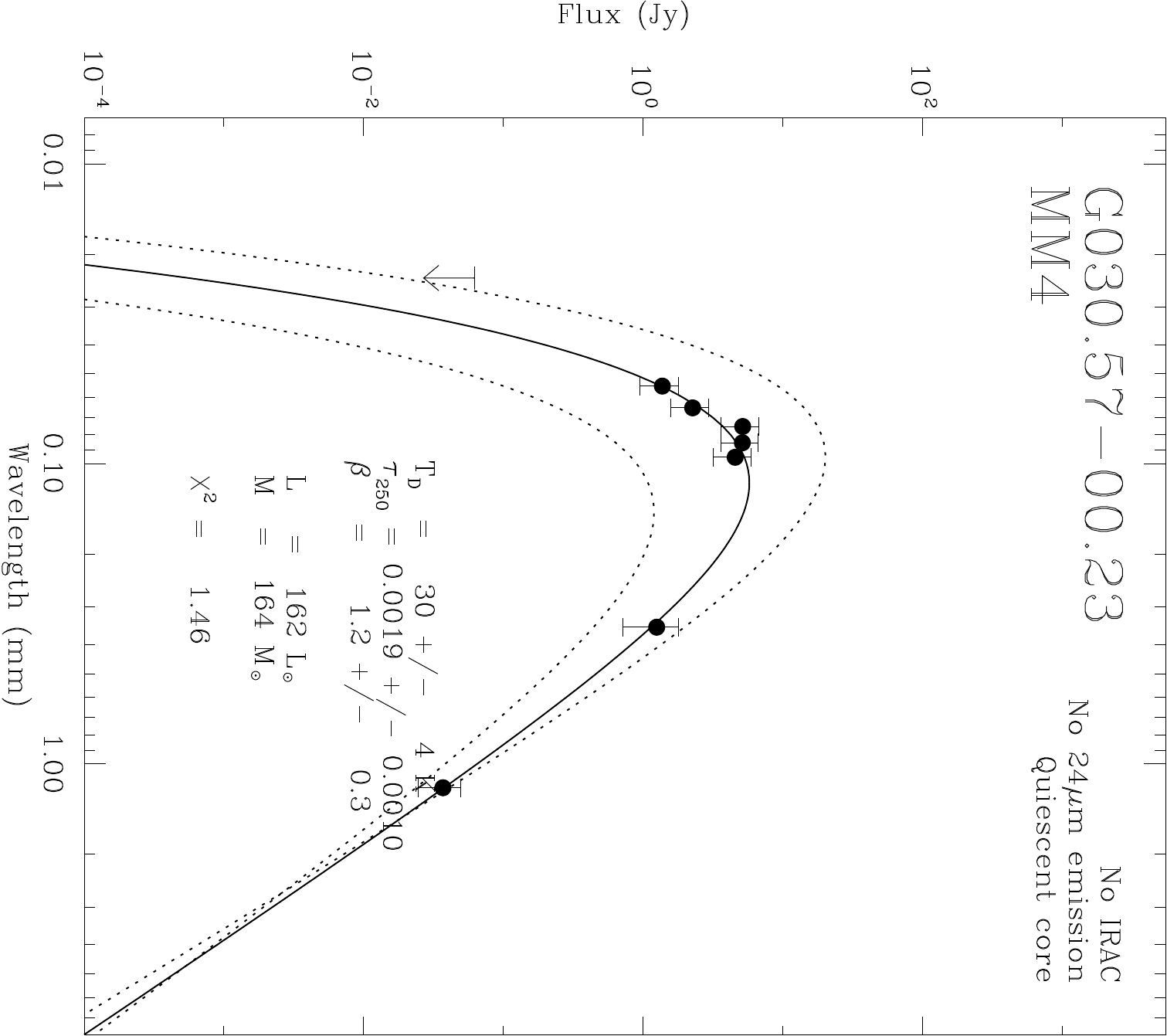}\\
\caption{\label{seds-54}\Spitzer\, 24\,\um\, image overlaid  
   with 1.2\,mm continuum emission for \irdcfiftyfour\, (contour
   levels are 30, 60, 90, 120, 240, 360, 480\,mJy beam$^{-1}$). The lower
   panels show the broadband SEDs for the cores within this IRDC. The lower panels show the broadband
   SEDs for cores within this IRDC.  The fluxes derived from the
   millimeter, sub-millimeter, and far-IR  continuum data are shown as filled
   circles (with the corresponding error bars), while the 24\,\um\, fluxes are shown as  either a filled circle (when included within the fit), an open circle (when excluded from the fit),  or as an upper limit arrow. For cores that have measured fluxes only in the millimeter/sub-millimeter regime (i.e.\, a limit at 24\,\um), we show the results from two fits: one using only the measured fluxes (solid line; lower limit), while the other includes the 24\,\um\, limit as a real data (dashed line; upper limit). In all other cases, the solid line is the best fit gray-body, while the dotted lines correspond to the functions determined using the errors for the T$_{D}$, $\tau$, and $\beta$ output from the fitting.  Labeled on each plot is the IRDC and core name,  classification, and the derived parameters.}
\end{figure}
\clearpage 
\begin{figure}
\begin{center}
\includegraphics[angle=0,width=0.6\textwidth]{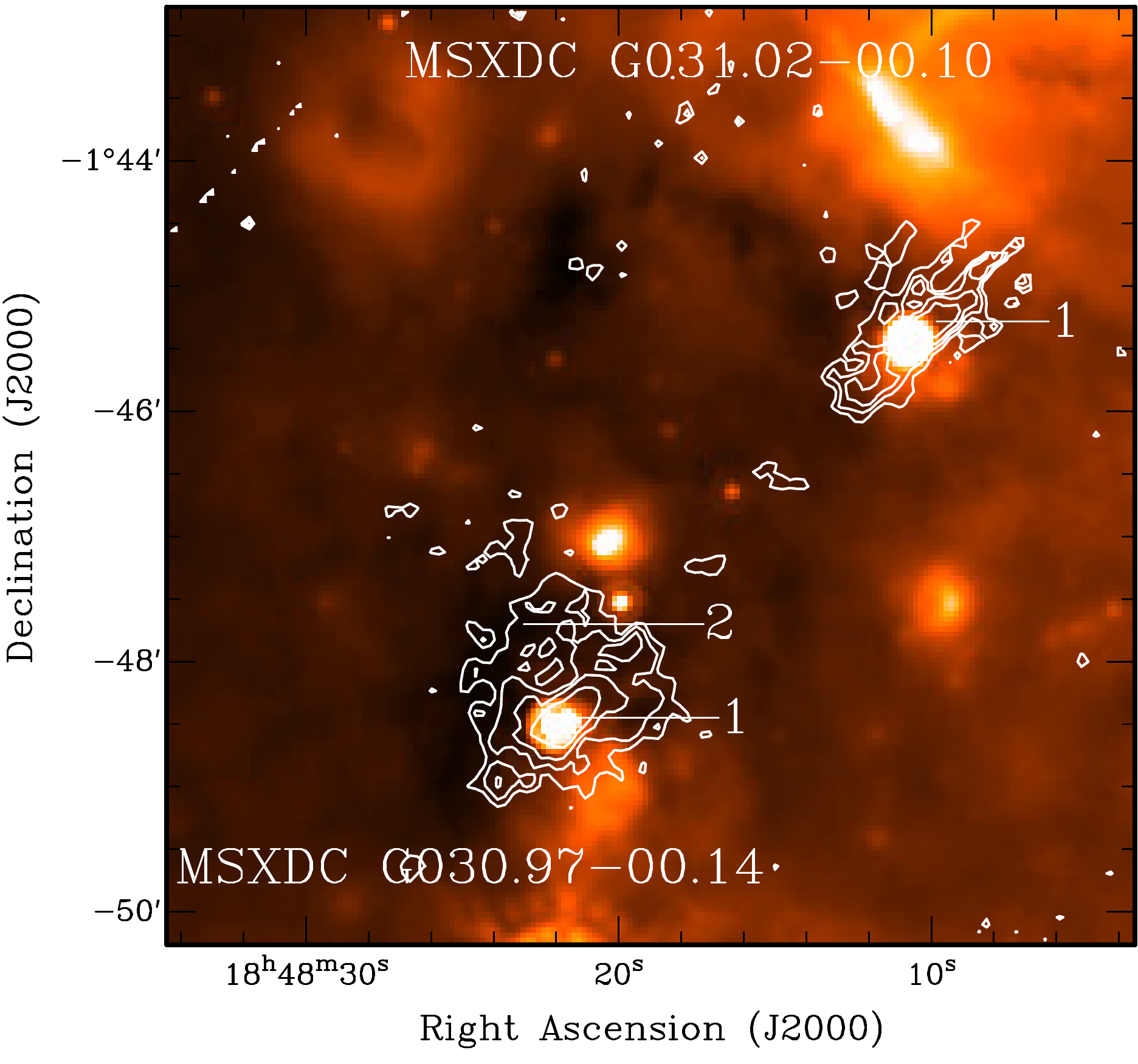}\\
\end{center}
\includegraphics[angle=90,width=0.5\textwidth]{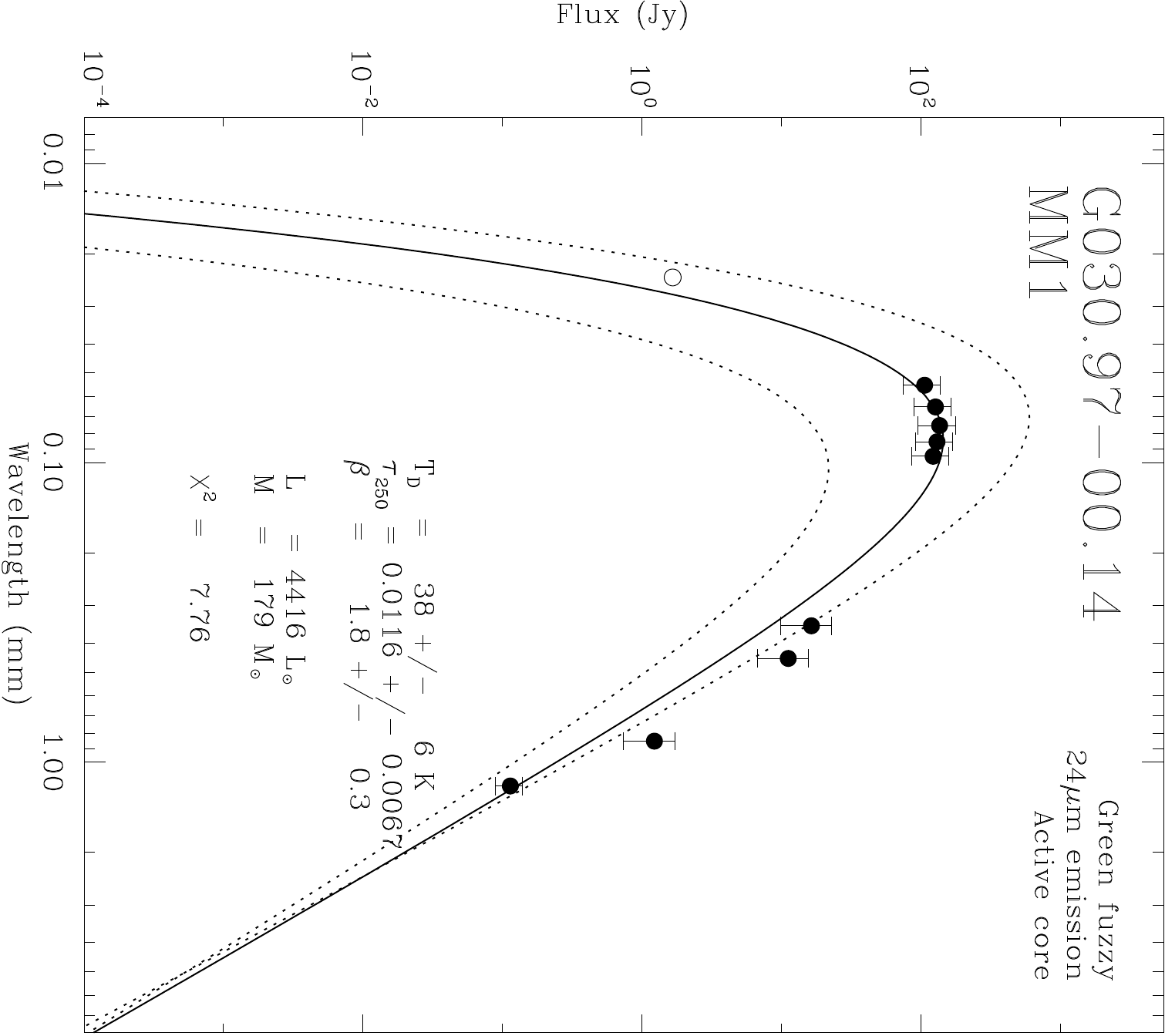}
\caption{\label{seds-291} \Spitzer\, 24\,\um\, image overlaid  
   with 1.2\,mm continuum emission for \irdctwentynineb\, and
   \irdctwentynine\, (contour levels are 40, 60, 80\,mJy
   beam$^{-1}$). The lower panels show the broadband
   SEDs for cores within this IRDC.  The fluxes derived from the
   millimeter, sub-millimeter, and far-IR  continuum data are shown as filled
   circles (with the corresponding error bars), while the 24\,\um\, fluxes are shown as  either a filled circle (when included within the fit), an open circle (when excluded from the fit),  or as an upper limit arrow. For cores that have measured fluxes only in the millimeter/sub-millimeter regime (i.e.\, a limit at 24\,\um), we show the results from two fits: one using only the measured fluxes (solid line; lower limit), while the other includes the 24\,\um\, limit as a real data (dashed line; upper limit). In all other cases, the solid line is the best fit gray-body, while the dotted lines correspond to the functions determined using the errors for the T$_{D}$, $\tau$, and $\beta$ output from the fitting.  Labeled on each plot is the IRDC and core name,  classification, and the derived parameters.}
\end{figure}
\clearpage 
\begin{figure}
\begin{center}
\includegraphics[angle=0,width=0.6\textwidth]{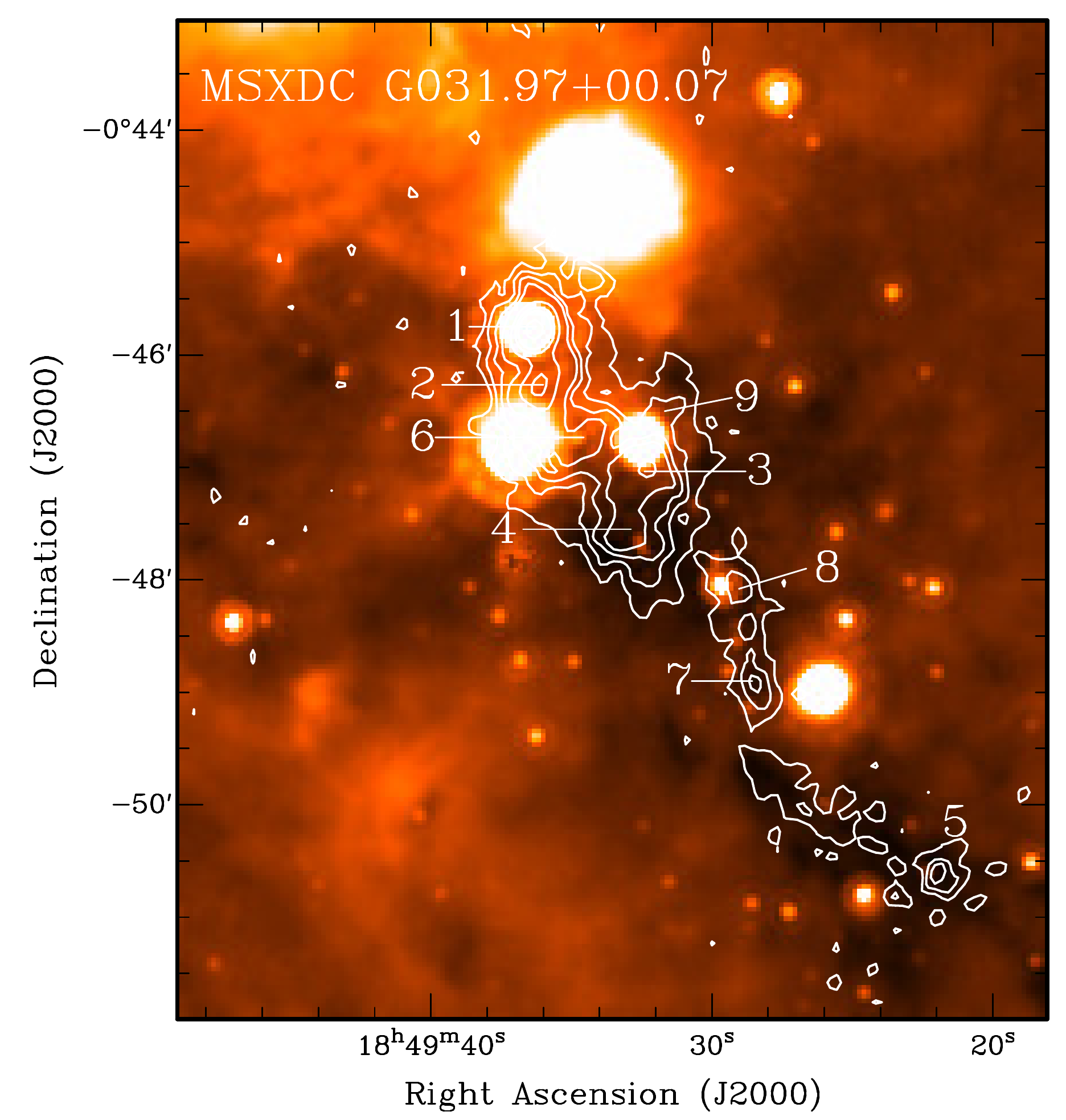}\\
\end{center}
\includegraphics[angle=90,width=0.5\textwidth]{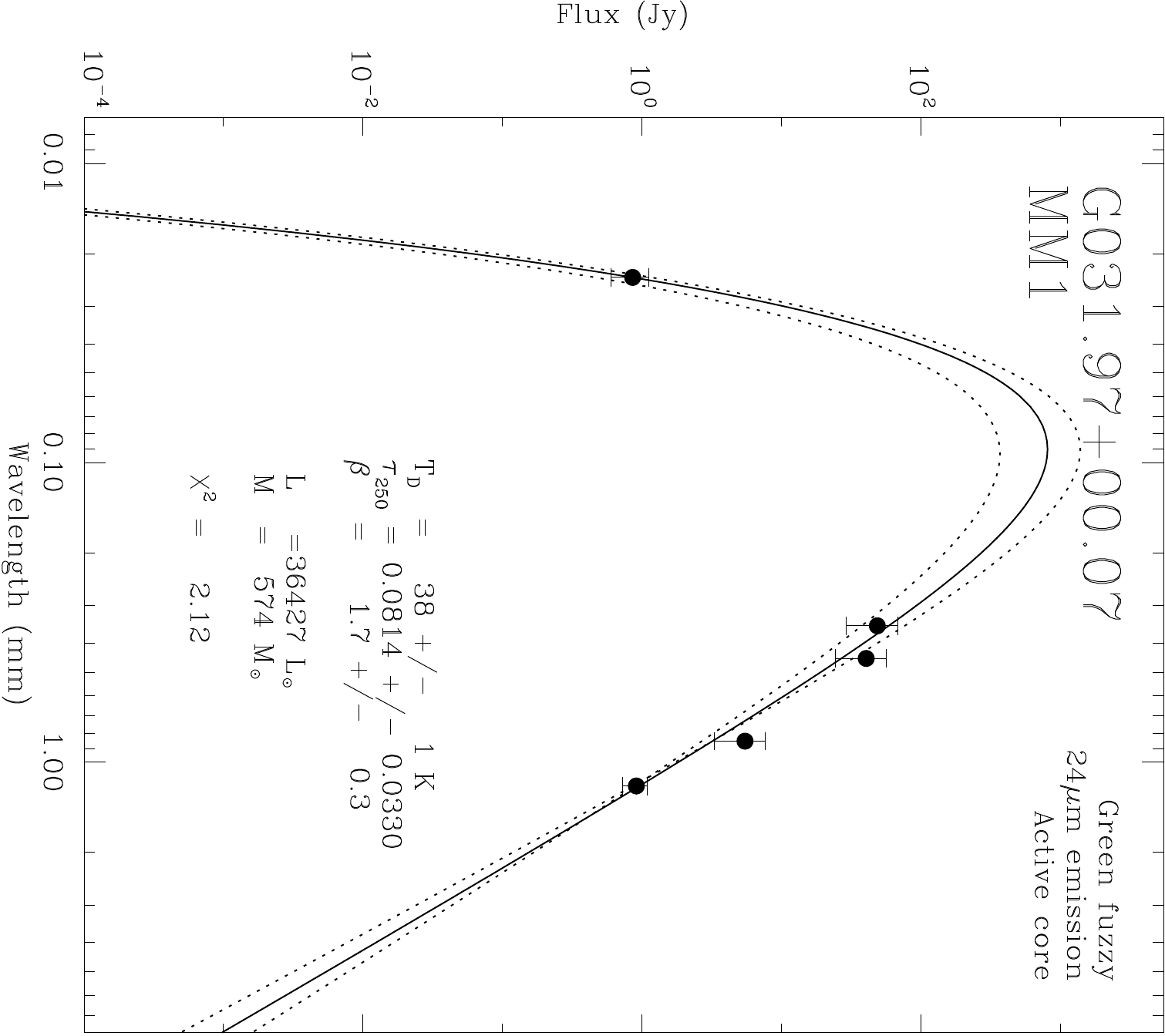}
\includegraphics[angle=90,width=0.5\textwidth]{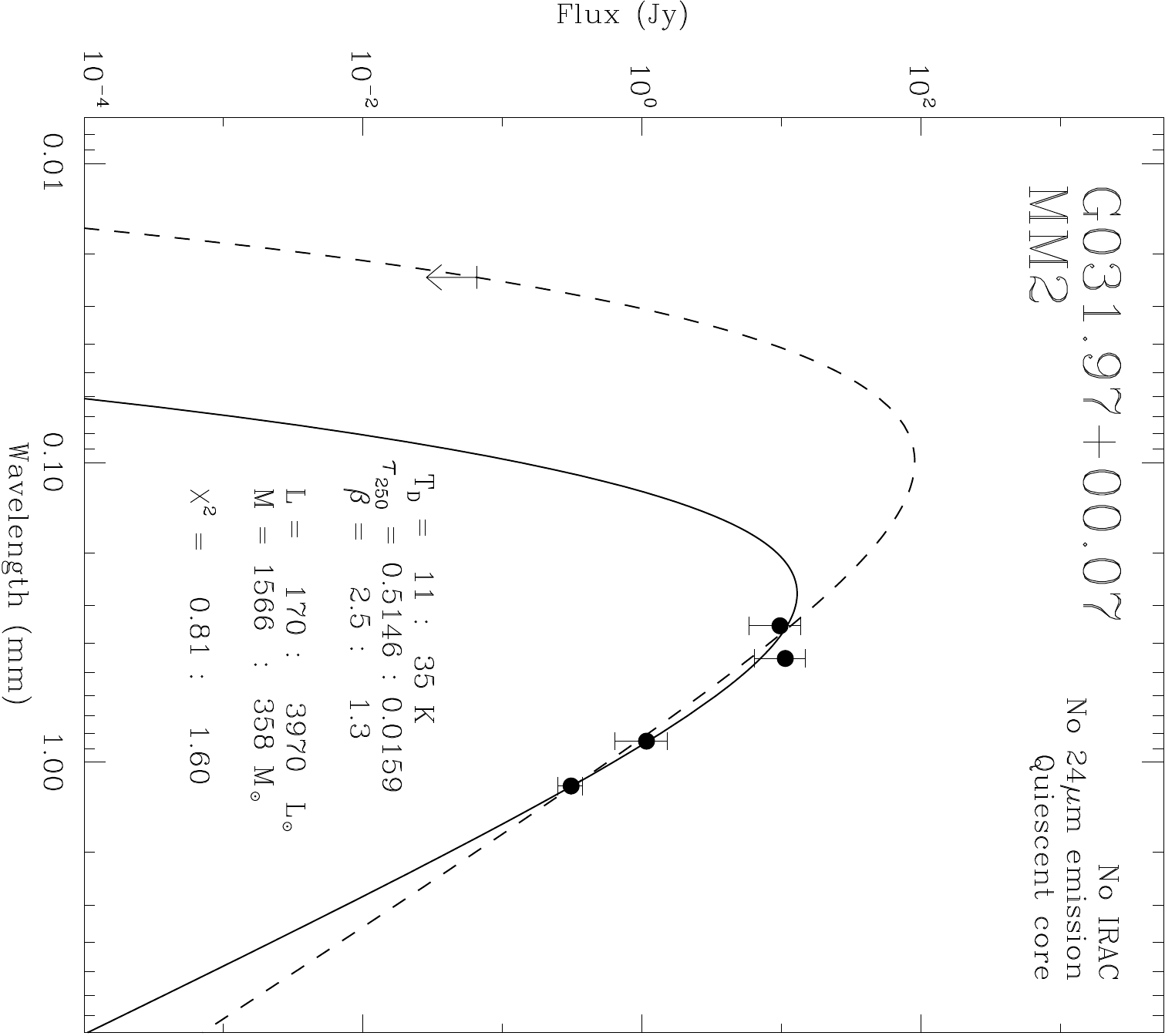}\\
\end{figure}
\clearpage 
\begin{figure}
\includegraphics[angle=90,width=0.5\textwidth]{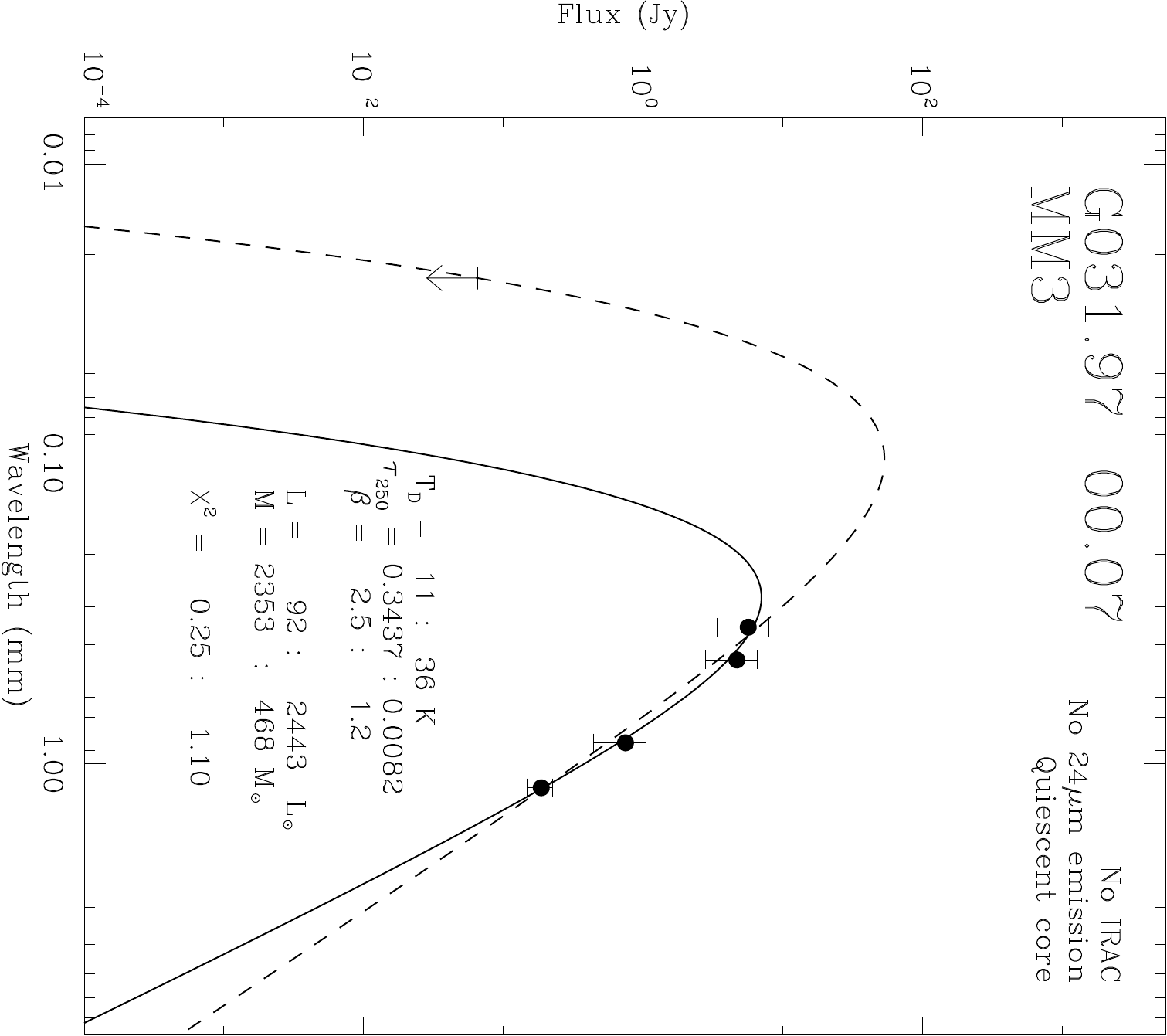}
\includegraphics[angle=90,width=0.5\textwidth]{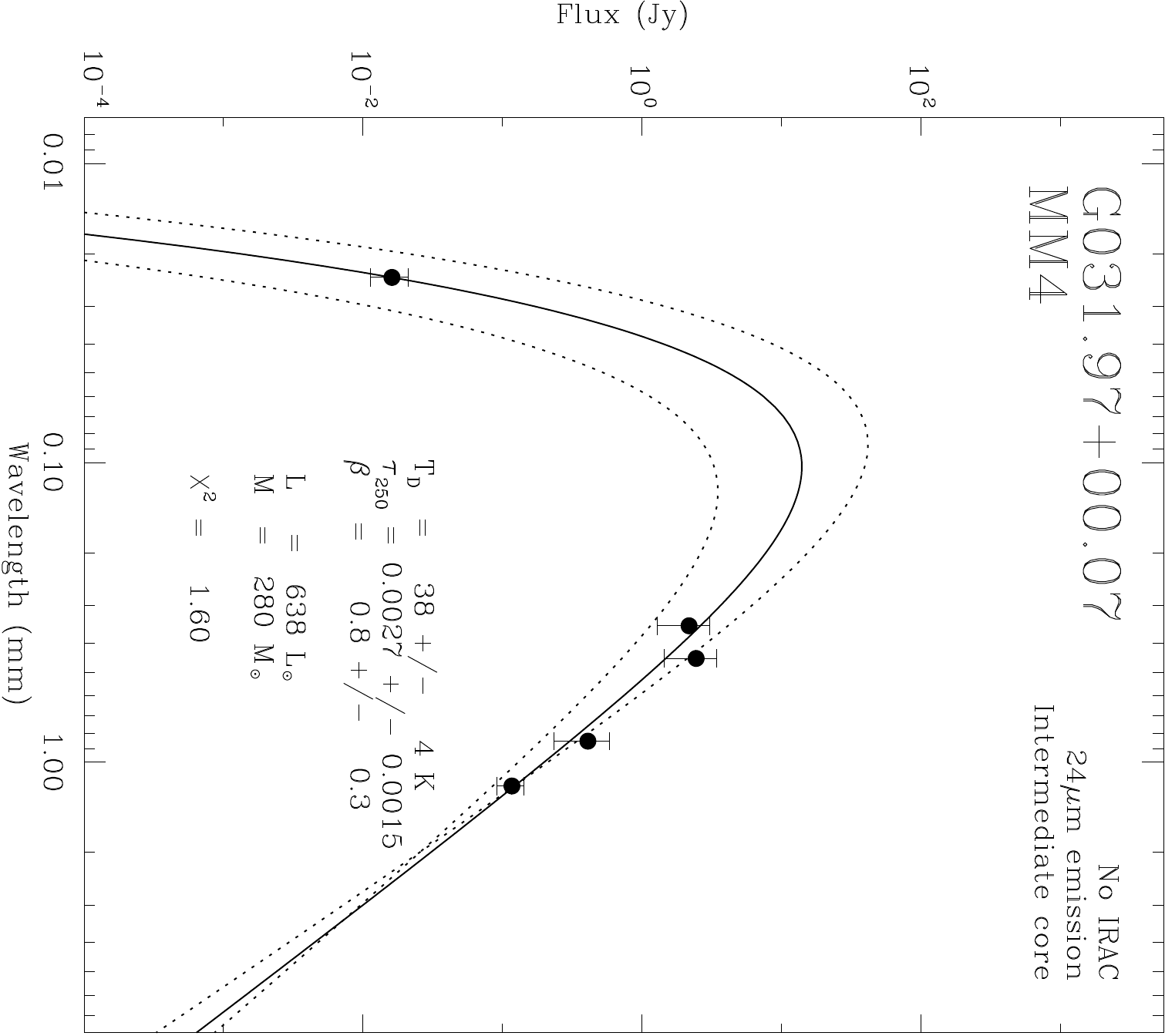}\\
\includegraphics[angle=90,width=0.5\textwidth]{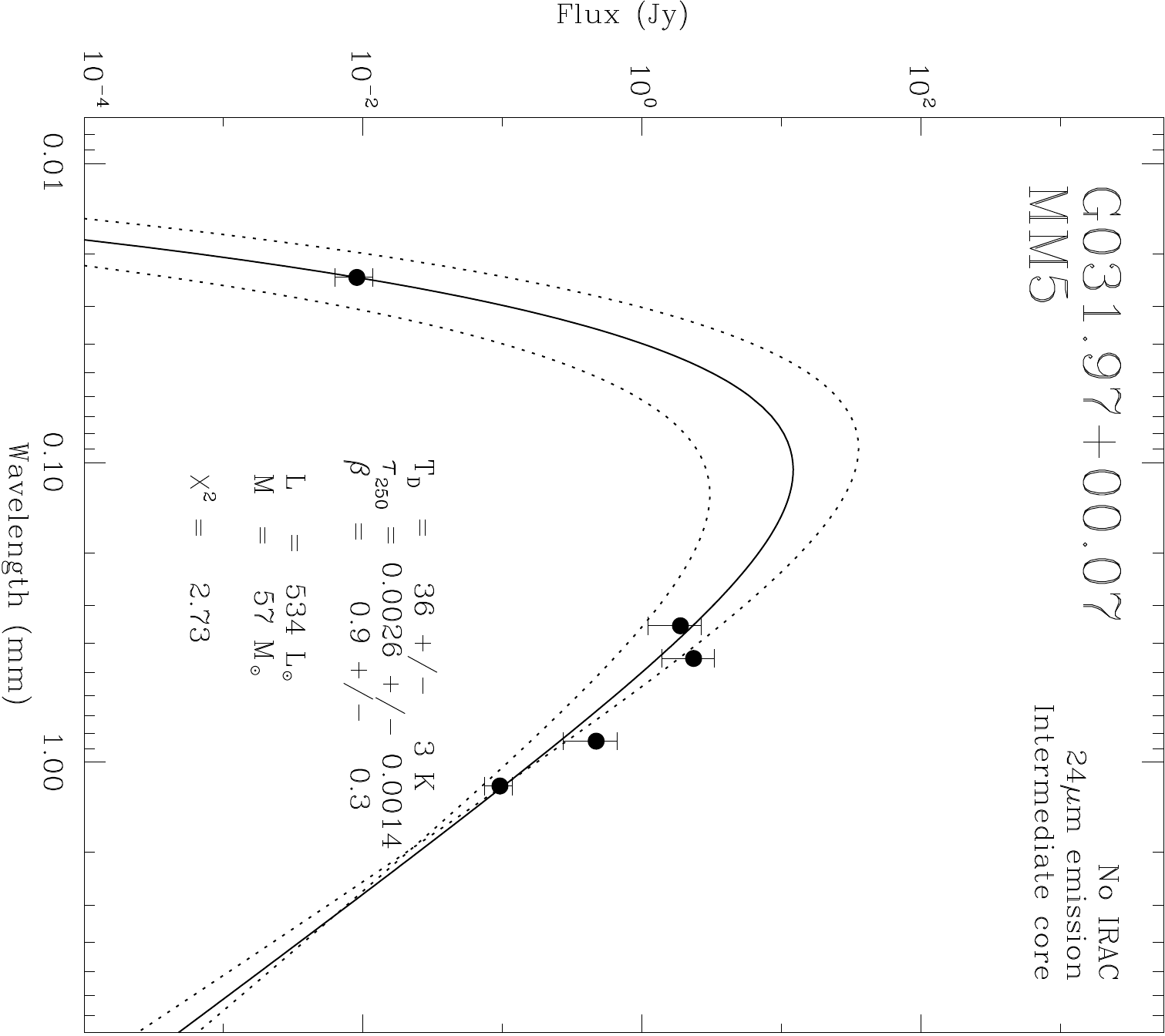}
\includegraphics[angle=90,width=0.5\textwidth]{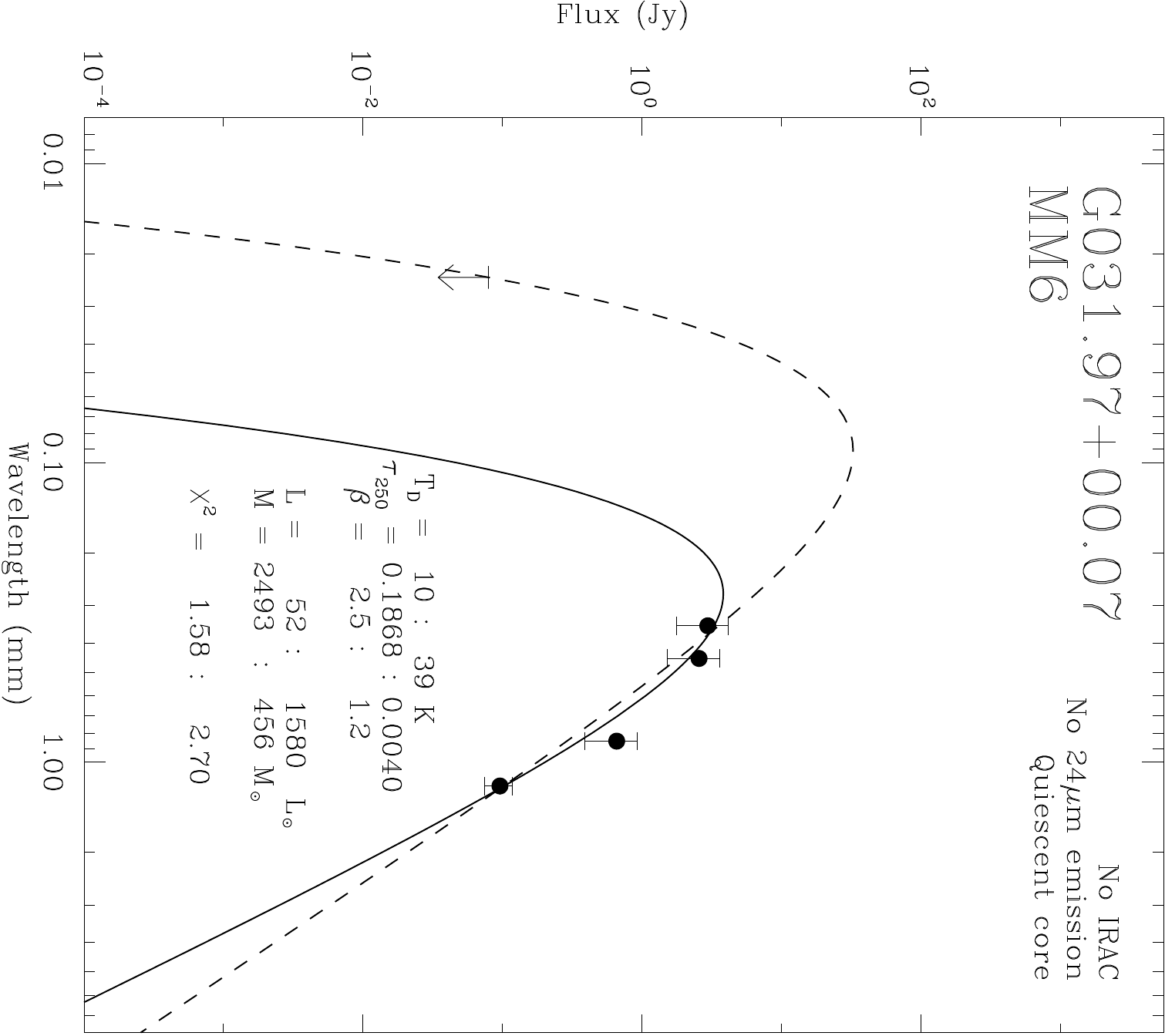}\\
\end{figure}
\clearpage 
\begin{figure}
\includegraphics[angle=90,width=0.5\textwidth]{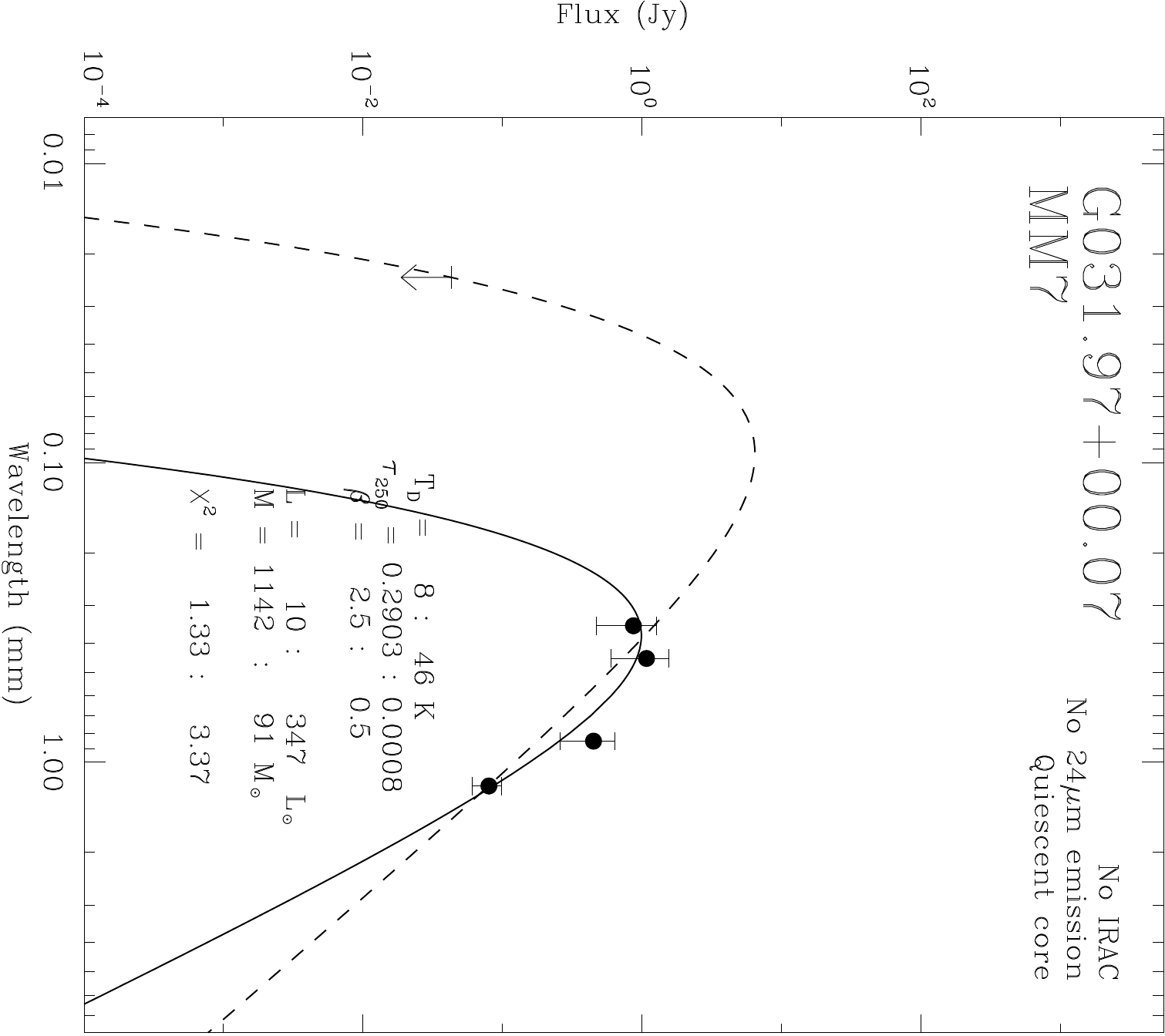}
\includegraphics[angle=90,width=0.5\textwidth]{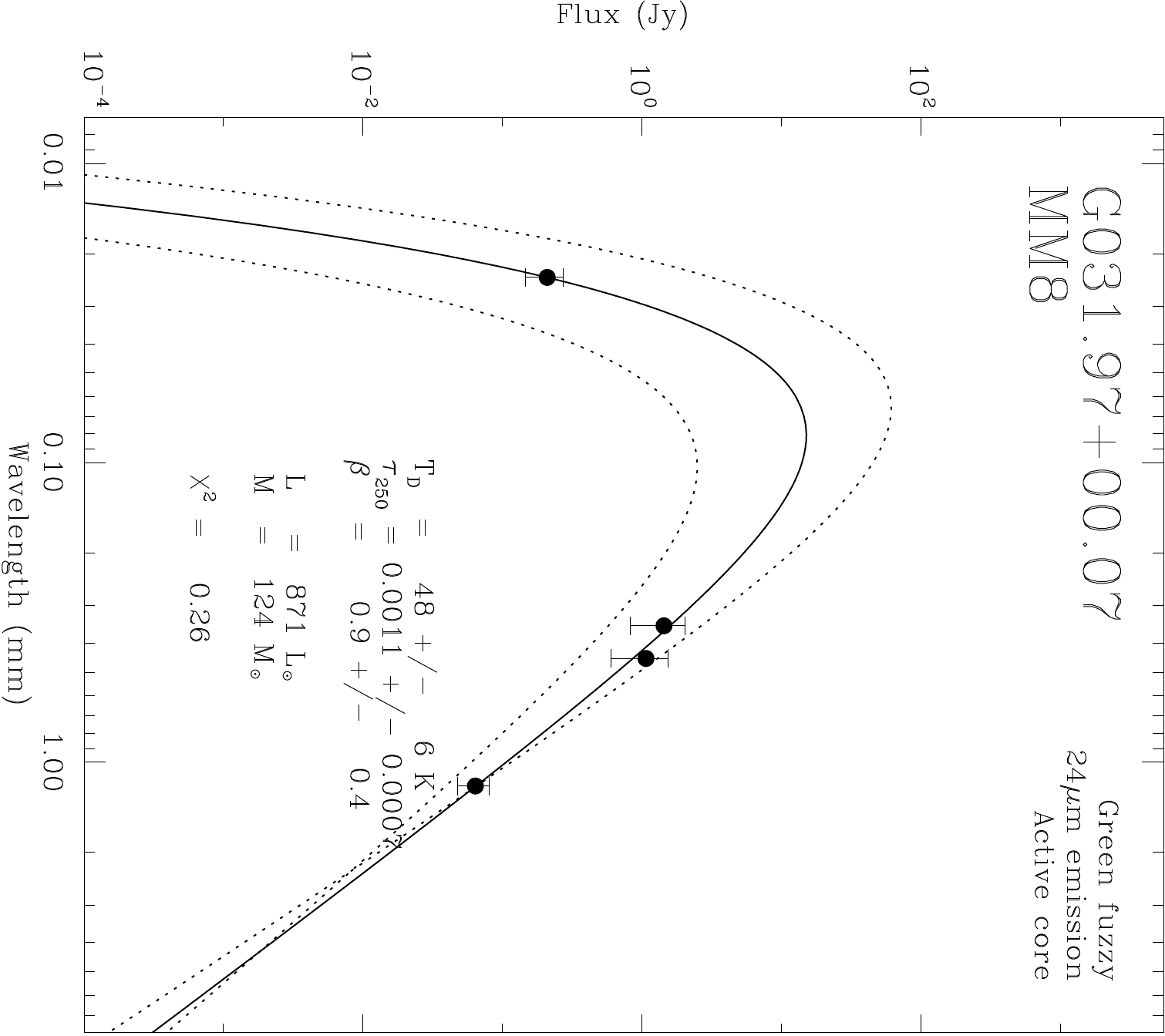}\\
\caption{\label{seds-4}\Spitzer\, 24\,\um\, image overlaid  
   with 1.2\,mm continuum emission for \irdcfour\, (contour levels are
   30, 60, 90,120, 240, 360, 480\,mJy beam$^{-1}$). The lower panels show the broadband
   SEDs for cores within this IRDC.  The fluxes derived from the
   millimeter, sub-millimeter, and far-IR  continuum data are shown as filled
   circles (with the corresponding error bars), while the 24\,\um\, fluxes are shown as  either a filled circle (when included within the fit), an open circle (when excluded from the fit),  or as an upper limit arrow. For cores that have measured fluxes only in the millimeter/sub-millimeter regime (i.e.\, a limit at 24\,\um), we show the results from two fits: one using only the measured fluxes (solid line; lower limit), while the other includes the 24\,\um\, limit as a real data (dashed line; upper limit). In all other cases, the solid line is the best fit gray-body, while the dotted lines correspond to the functions determined using the errors for the T$_{D}$, $\tau$, and $\beta$ output from the fitting.  Labeled on each plot is the IRDC and core name,  classification, and the derived parameters.}
\end{figure}
\clearpage 
\begin{figure}
\begin{center}
\includegraphics[angle=0,width=0.6\textwidth]{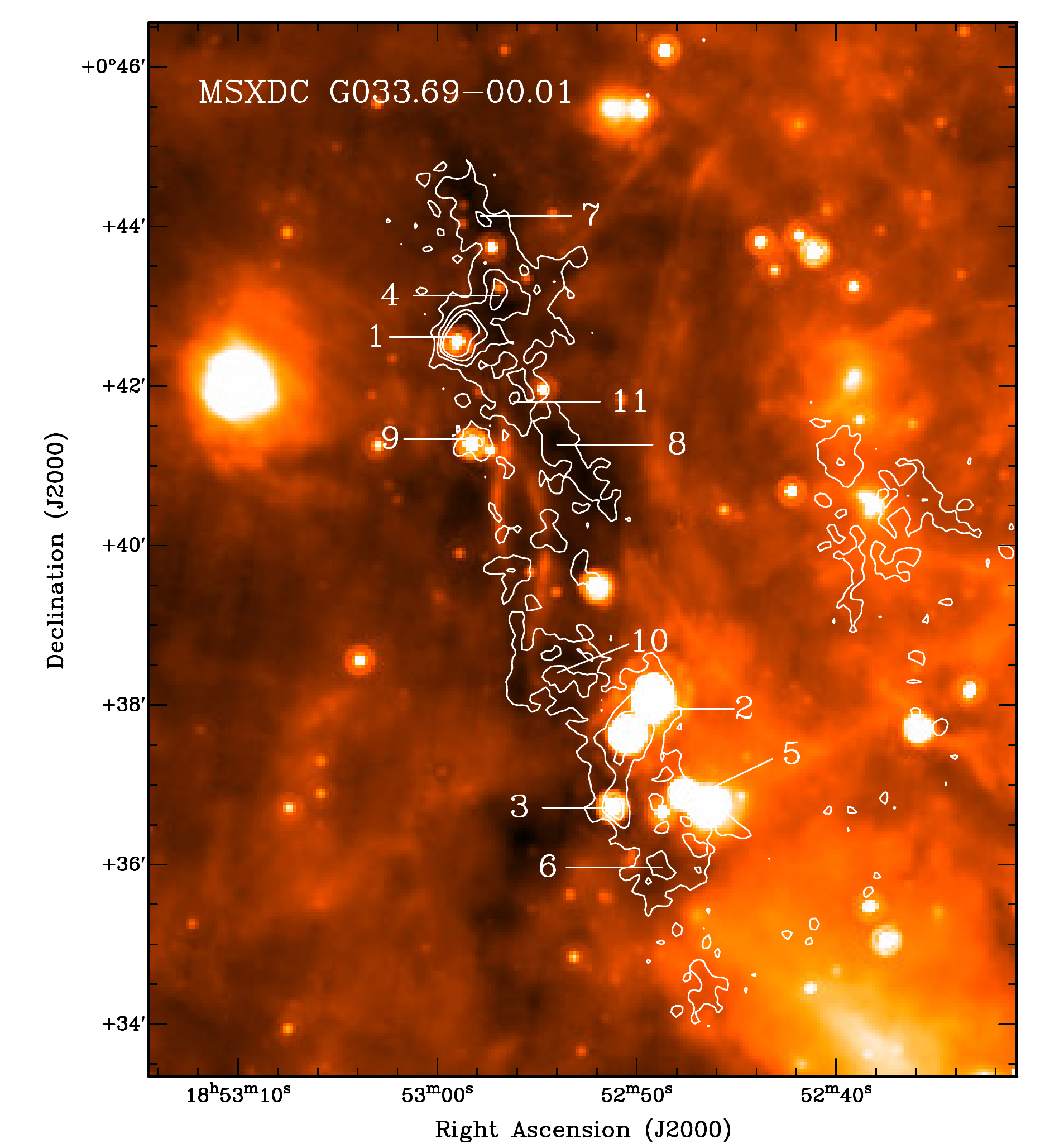}\\
\end{center}
\includegraphics[angle=90,width=0.5\textwidth]{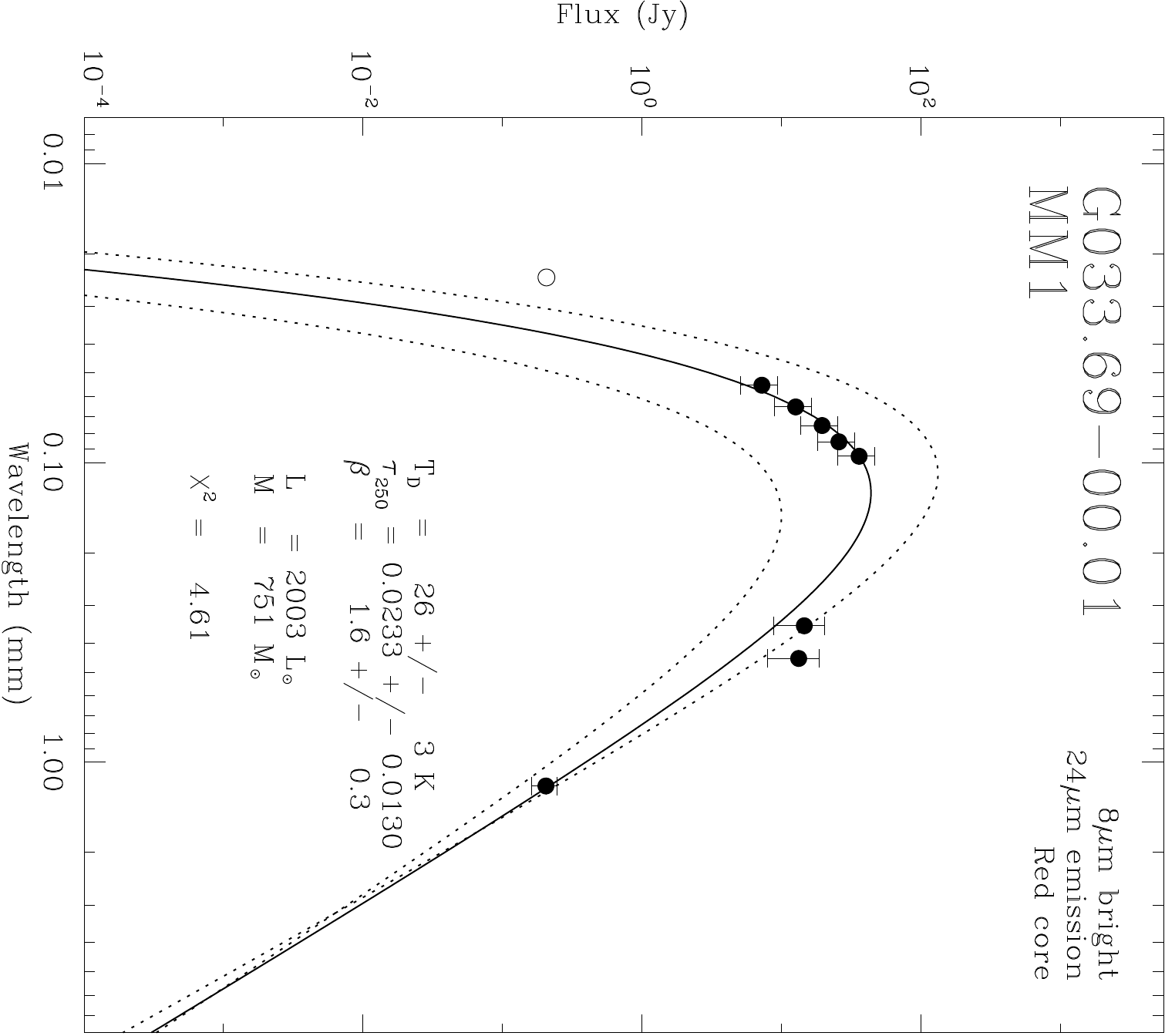}
\includegraphics[angle=90,width=0.5\textwidth]{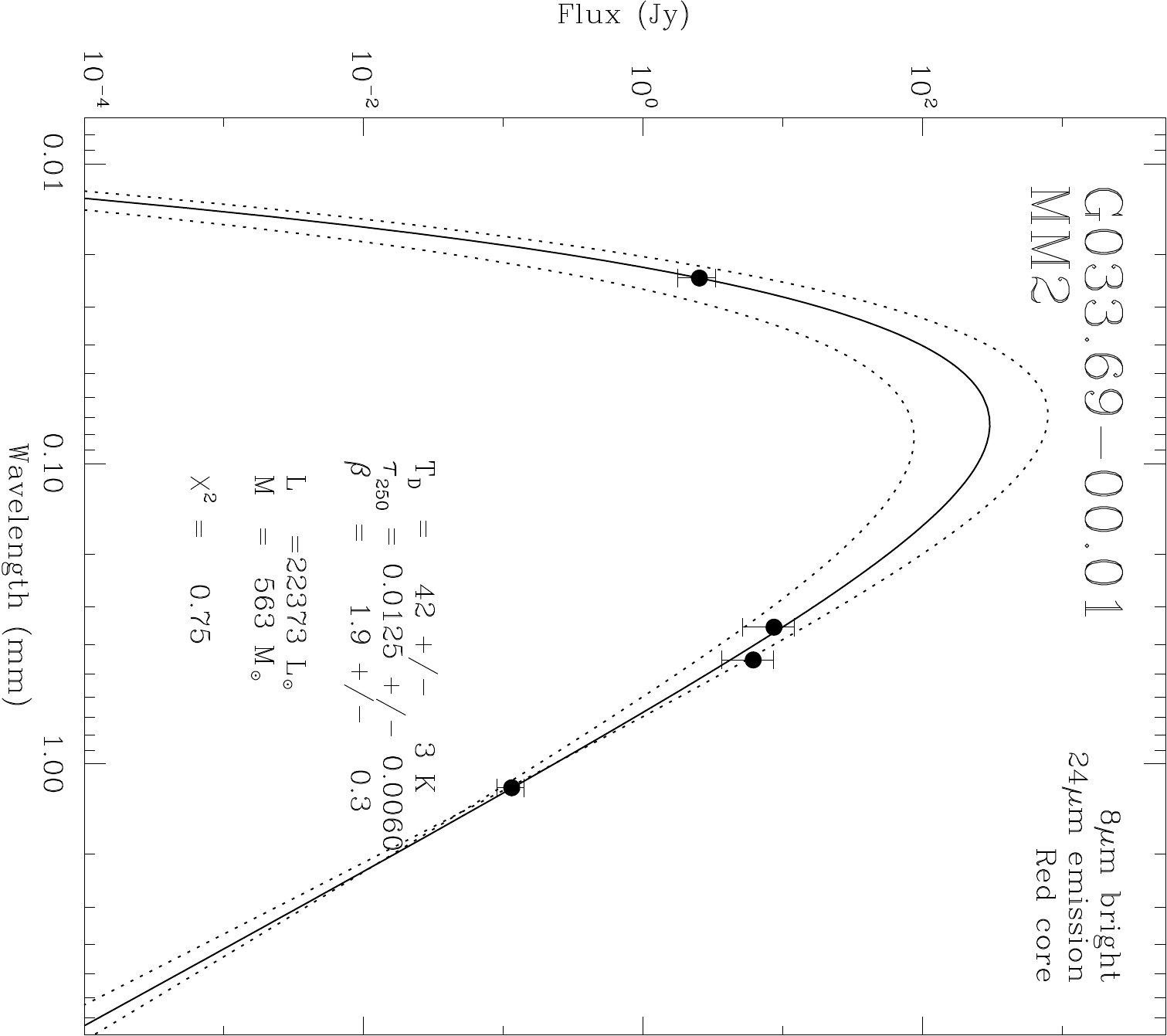}\\
\end{figure}
\clearpage 
\begin{figure}
\includegraphics[angle=90,width=0.5\textwidth]{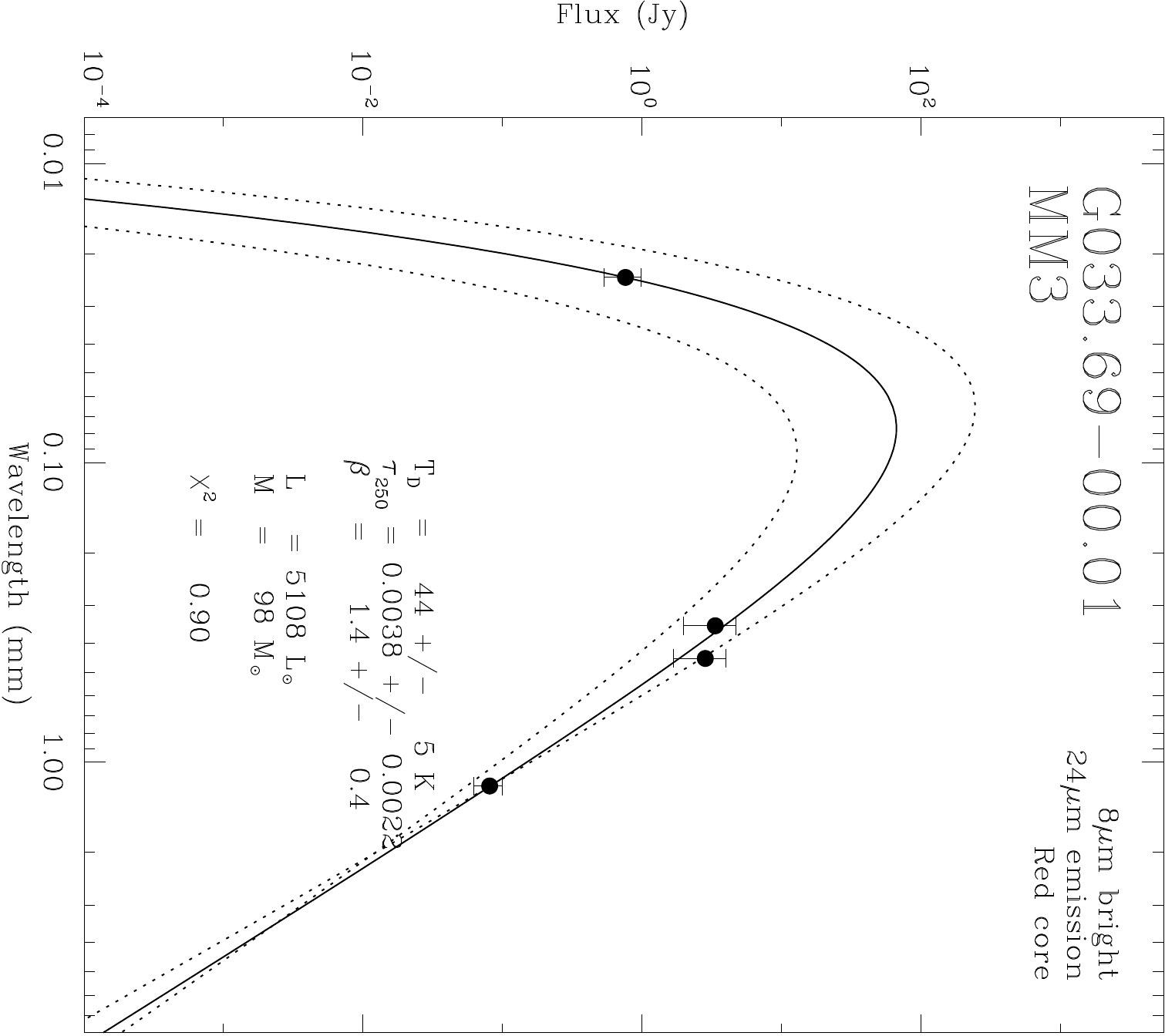}
\includegraphics[angle=90,width=0.5\textwidth]{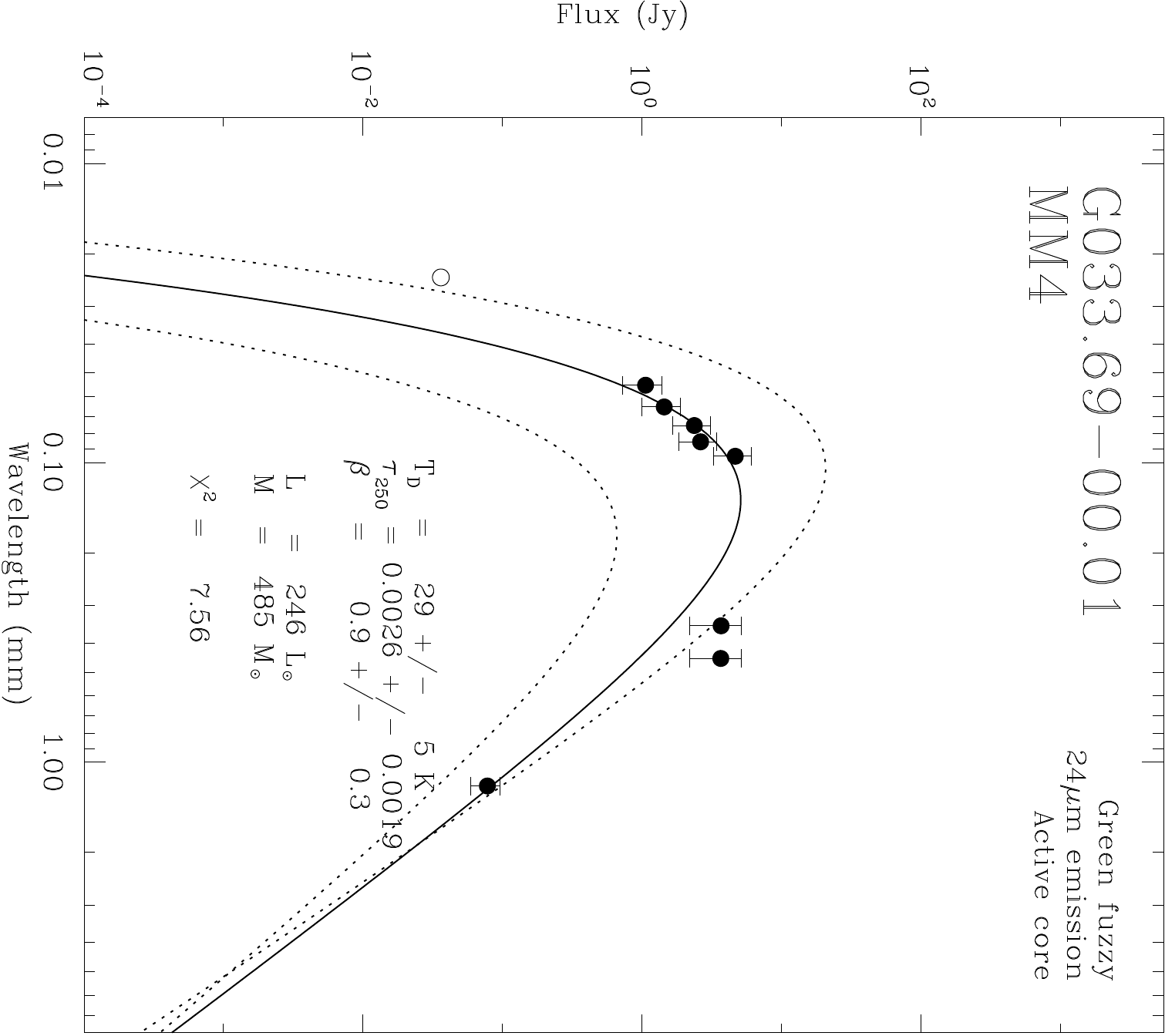}\\
\includegraphics[angle=90,width=0.5\textwidth]{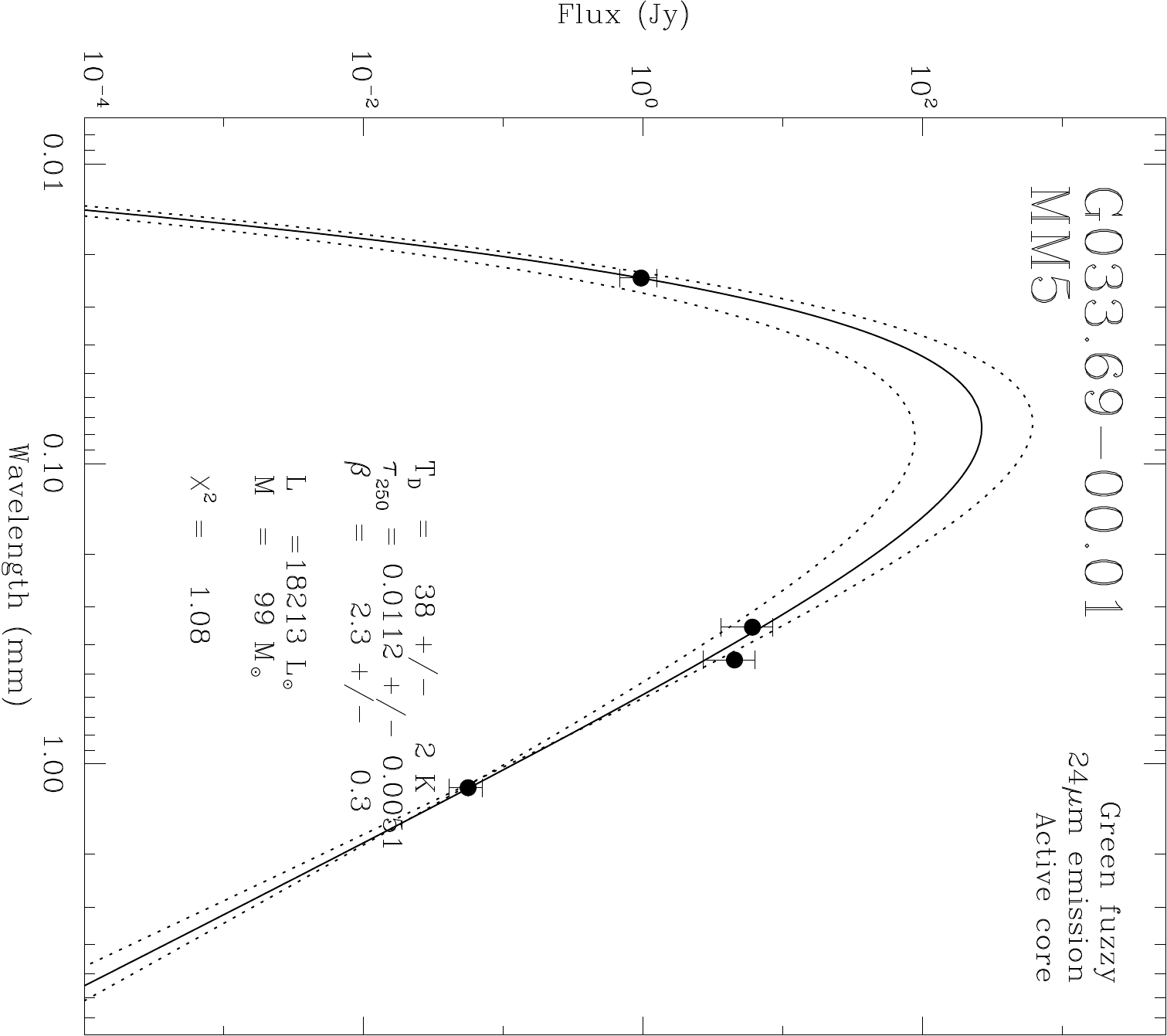}
\includegraphics[angle=90,width=0.5\textwidth]{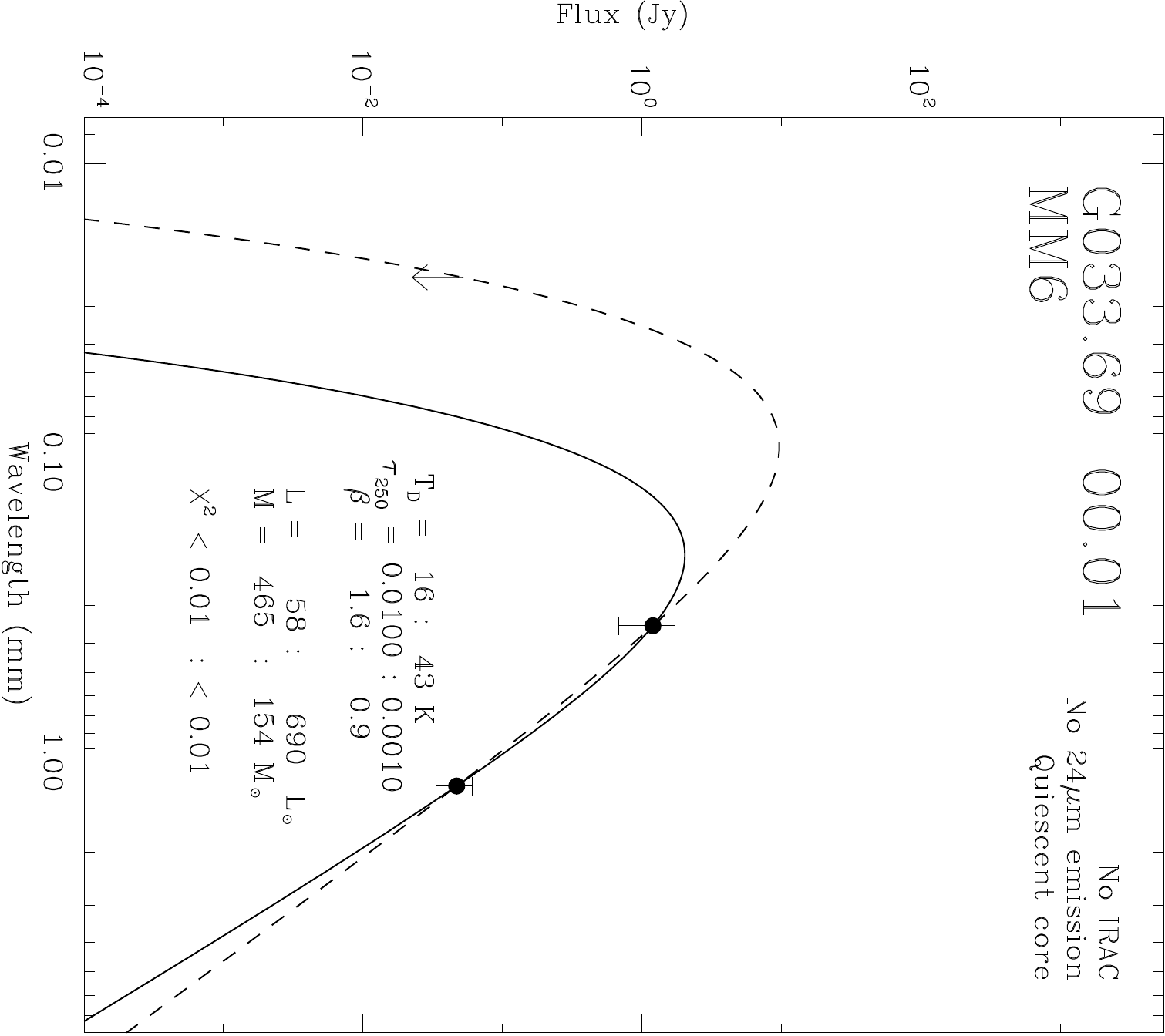}\\
\end{figure}
\clearpage 
\begin{figure}
\includegraphics[angle=90,width=0.5\textwidth]{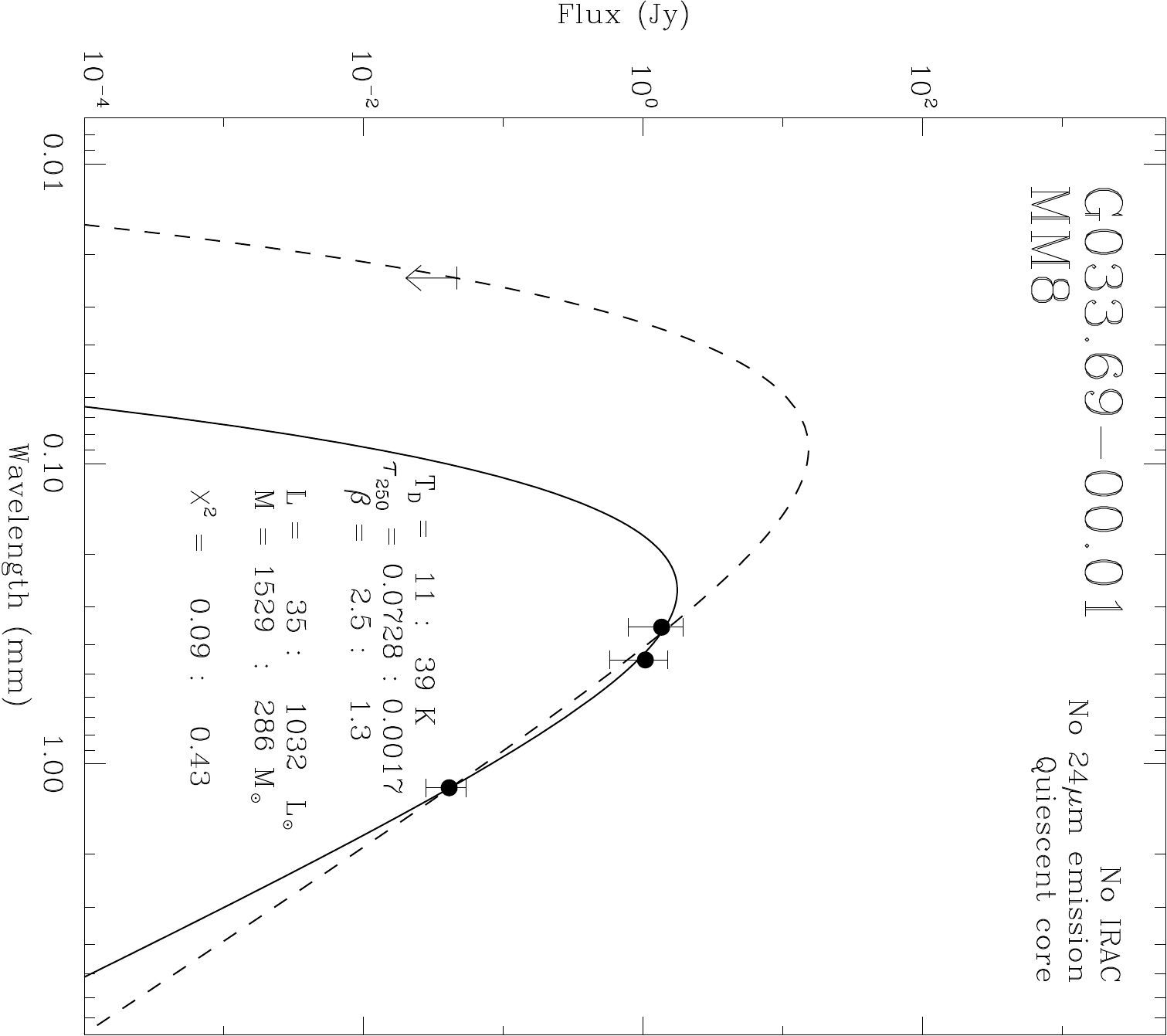}
\includegraphics[angle=90,width=0.5\textwidth]{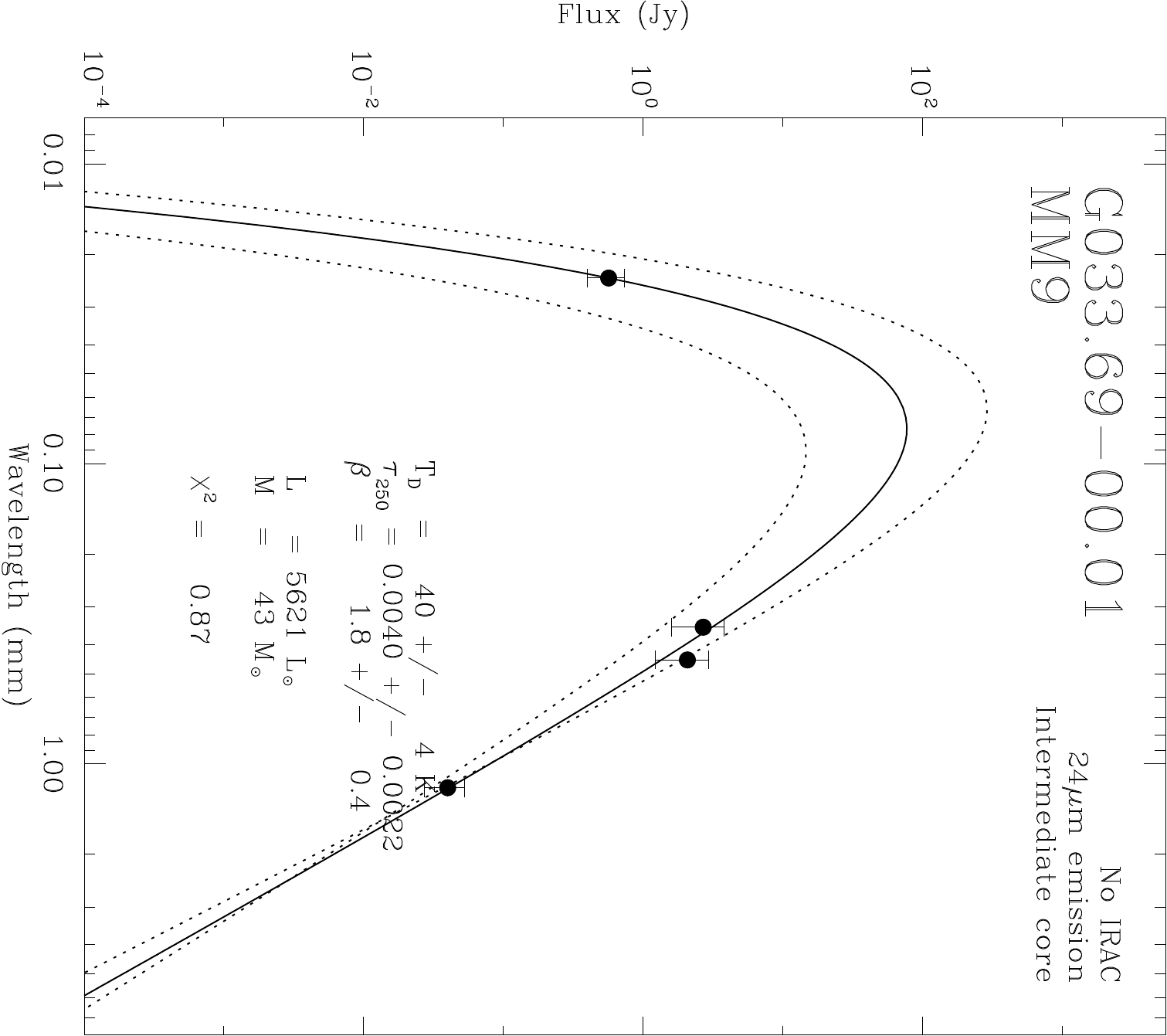}\\
\includegraphics[angle=90,width=0.5\textwidth]{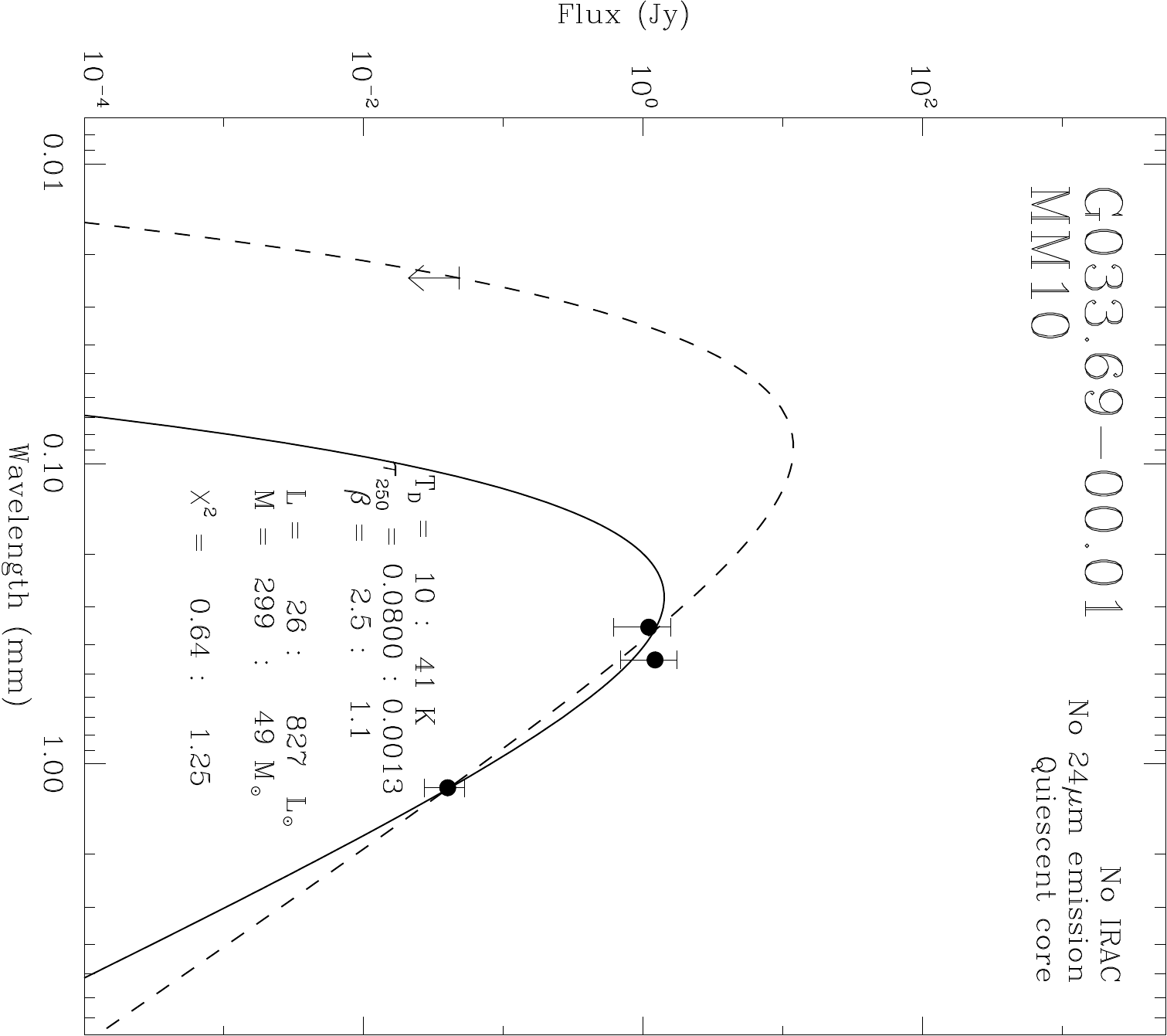}
\includegraphics[angle=90,width=0.5\textwidth]{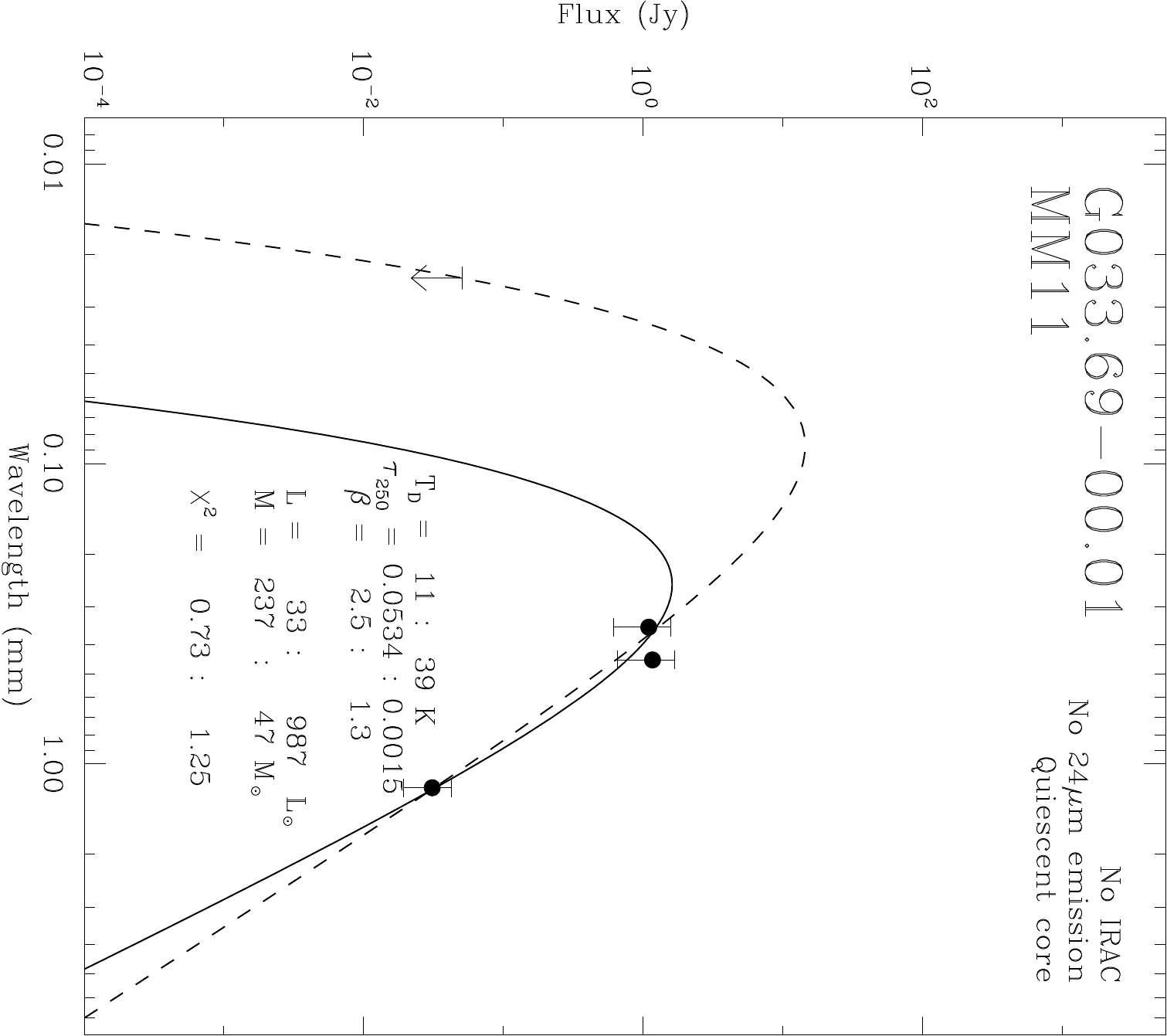}\\
\caption{\label{seds-39}\Spitzer\, 24\,\um\, image overlaid  
   with 1.2\,mm continuum emission for \irdcthirtynine\, (contour
   levels are 30, 60, 90, 120, 240\,mJy beam$^{-1}$). The lower panels show the broadband
   SEDs for cores within this IRDC.  The fluxes derived from the
   millimeter, sub-millimeter, and far-IR  continuum data are shown as filled
   circles (with the corresponding error bars), while the 24\,\um\, fluxes are shown as  either a filled circle (when included within the fit), an open circle (when excluded from the fit),  or as an upper limit arrow. For cores that have measured fluxes only in the millimeter/sub-millimeter regime (i.e.\, a limit at 24\,\um), we show the results from two fits: one using only the measured fluxes (solid line; lower limit), while the other includes the 24\,\um\, limit as a real data (dashed line; upper limit). In all other cases, the solid line is the best fit gray-body, while the dotted lines correspond to the functions determined using the errors for the T$_{D}$, $\tau$, and $\beta$ output from the fitting.  Labeled on each plot is the IRDC and core name,  classification, and the derived parameters.}
\end{figure}
\clearpage 
\begin{figure}
\begin{center}
\includegraphics[angle=0,width=0.6\textwidth]{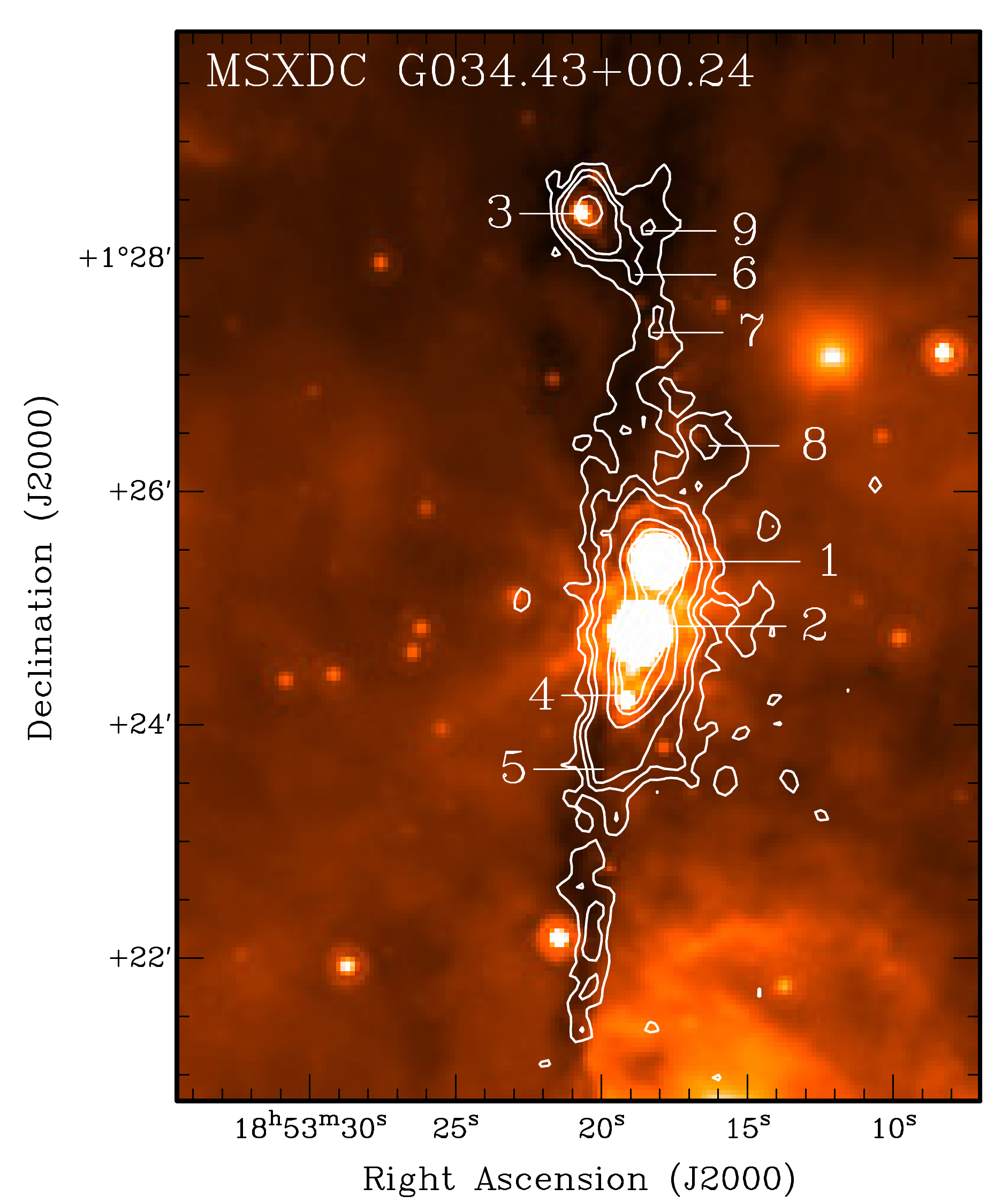}\\
\end{center}
\includegraphics[angle=90,width=0.5\textwidth]{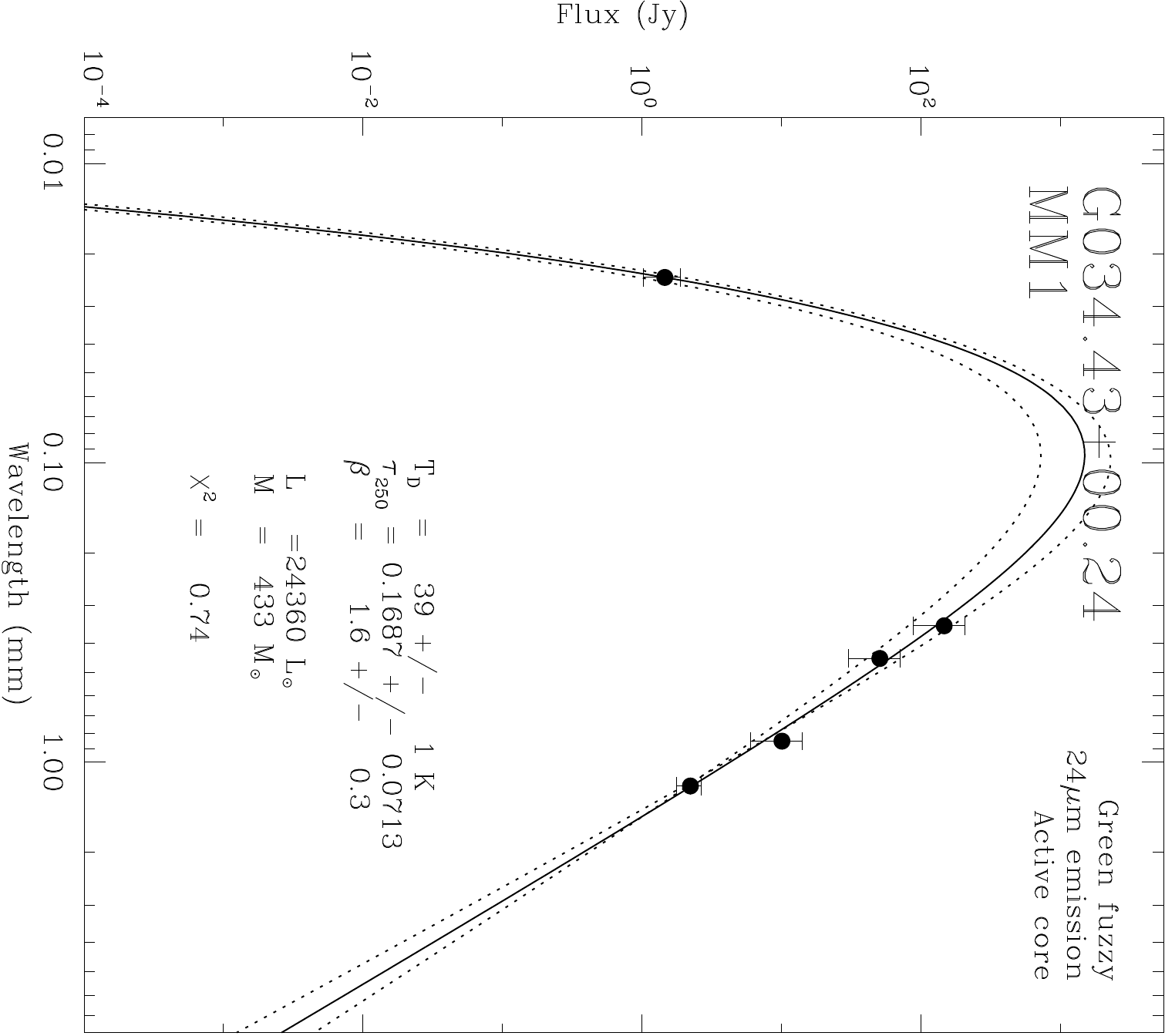}
\includegraphics[angle=90,width=0.5\textwidth]{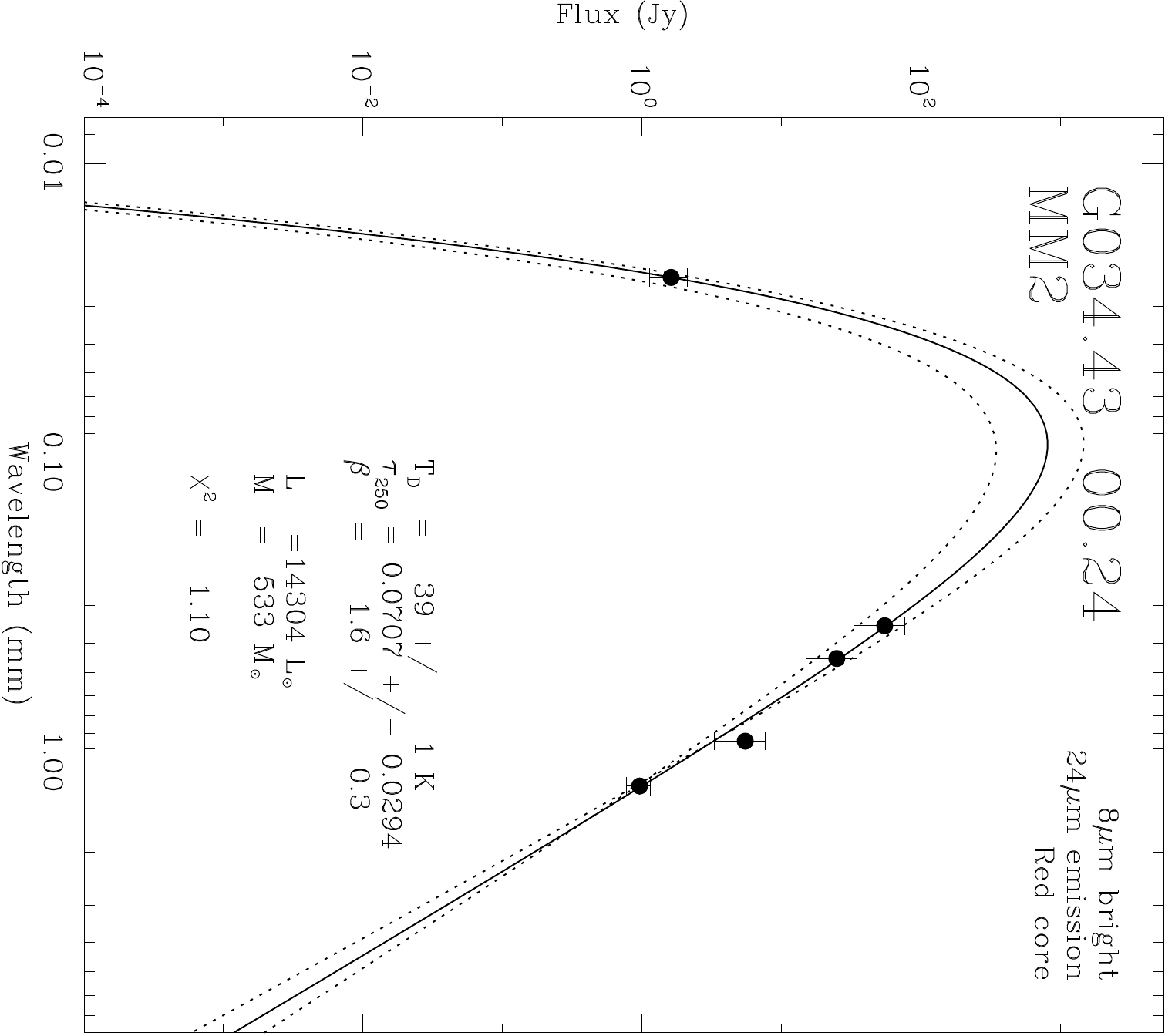}\\
\end{figure}
\clearpage 
\begin{figure}
\includegraphics[angle=90,width=0.5\textwidth]{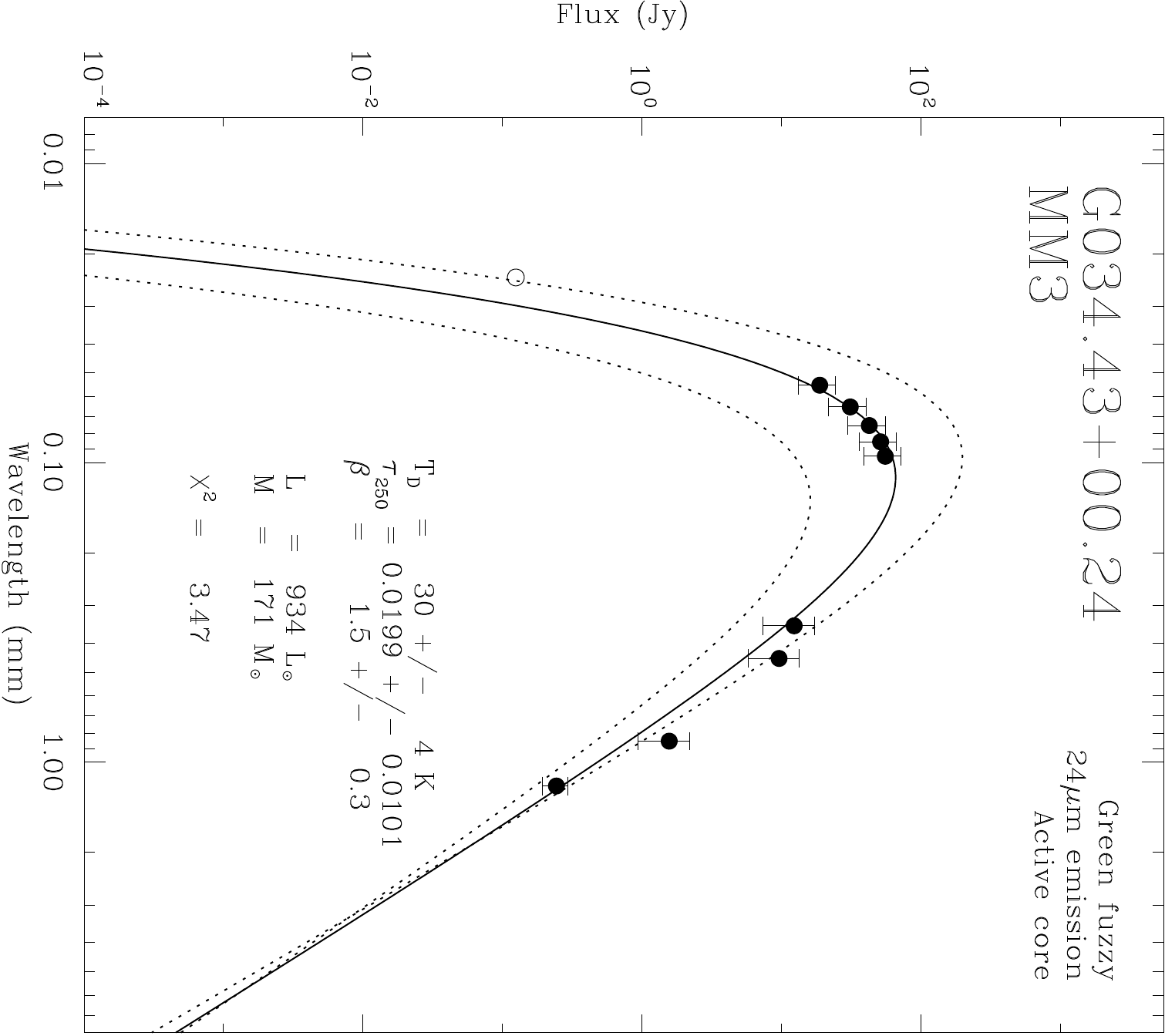}
\includegraphics[angle=90,width=0.5\textwidth]{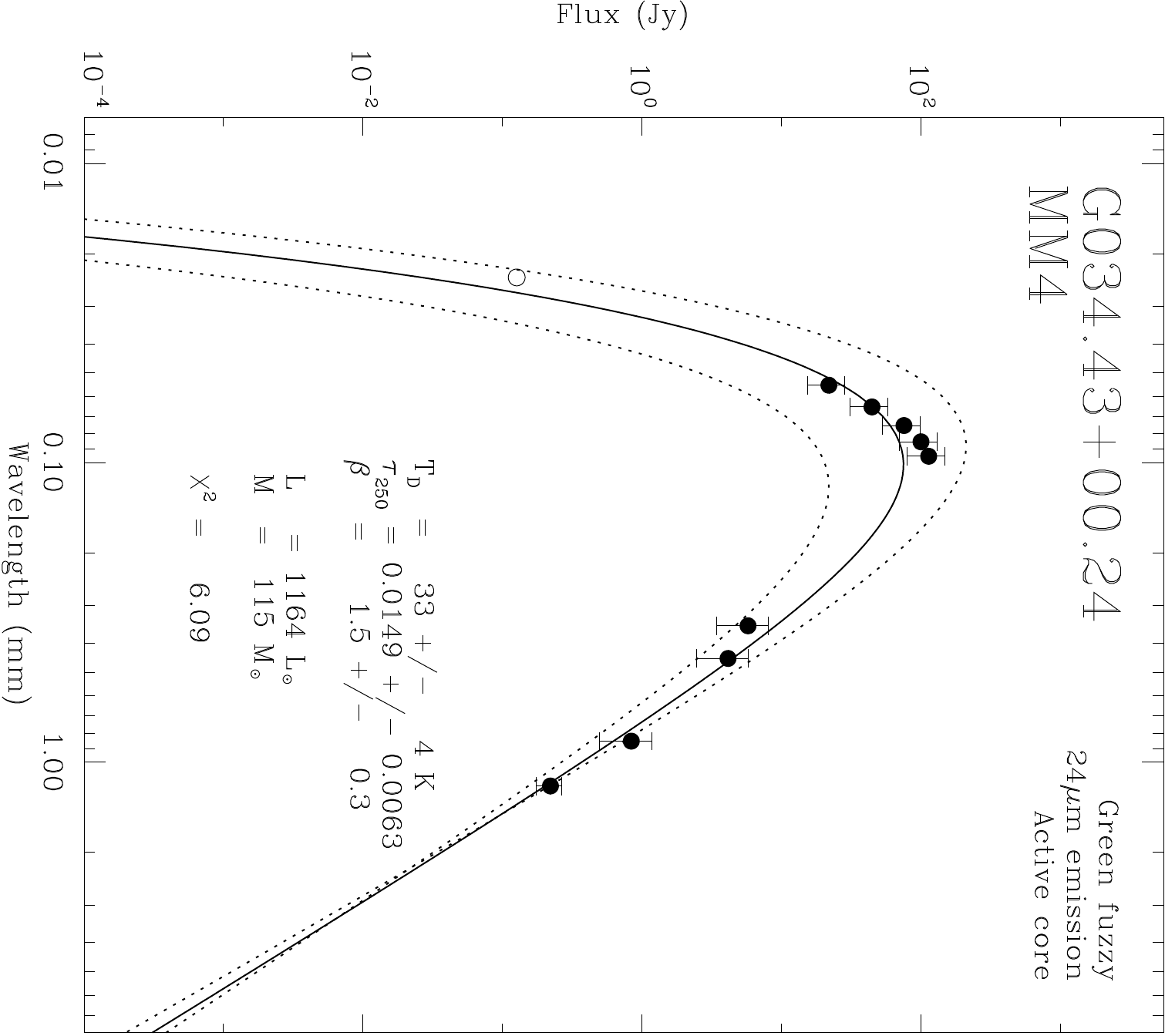}\\
\includegraphics[angle=90,width=0.5\textwidth]{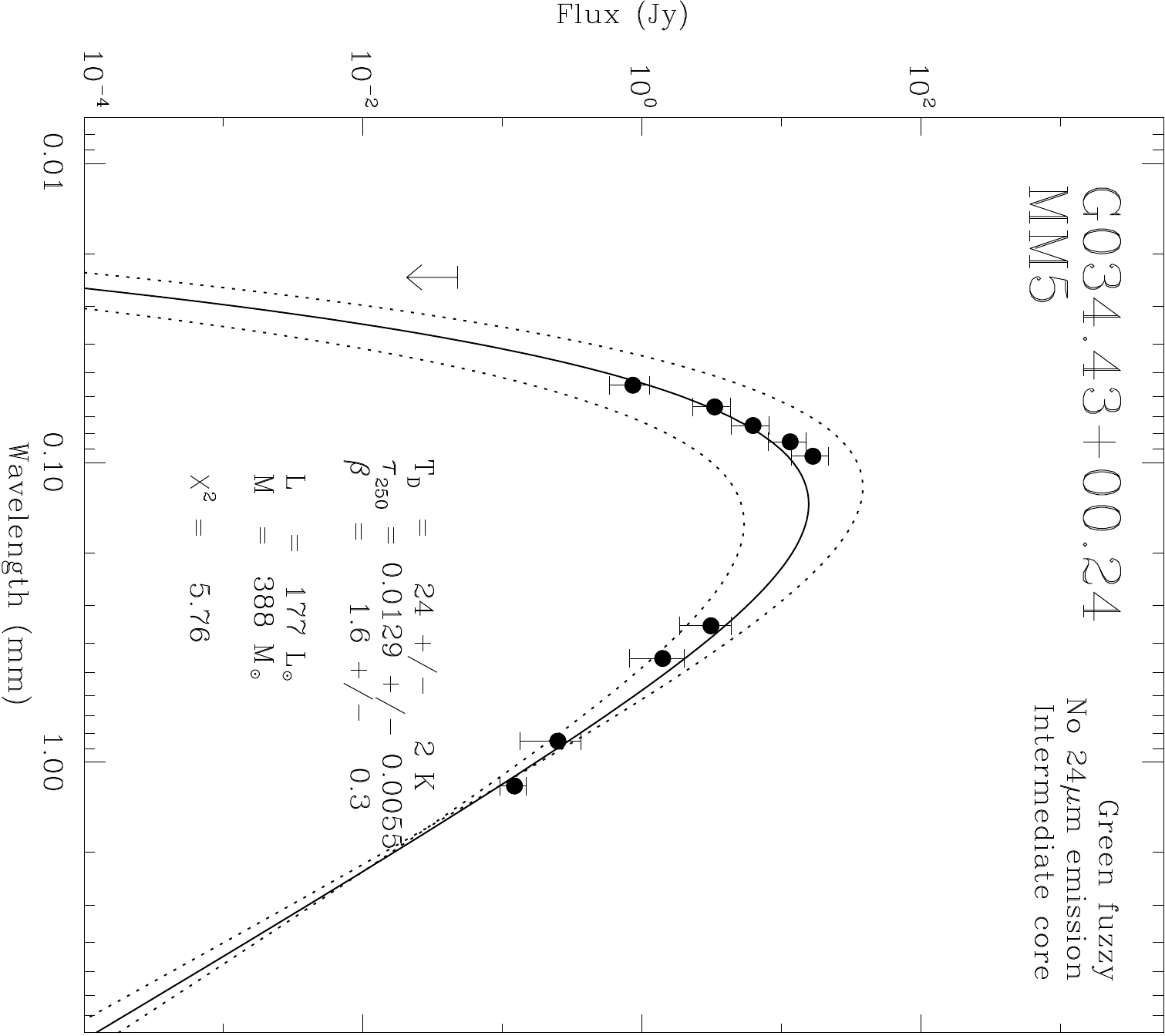}
\includegraphics[angle=90,width=0.5\textwidth]{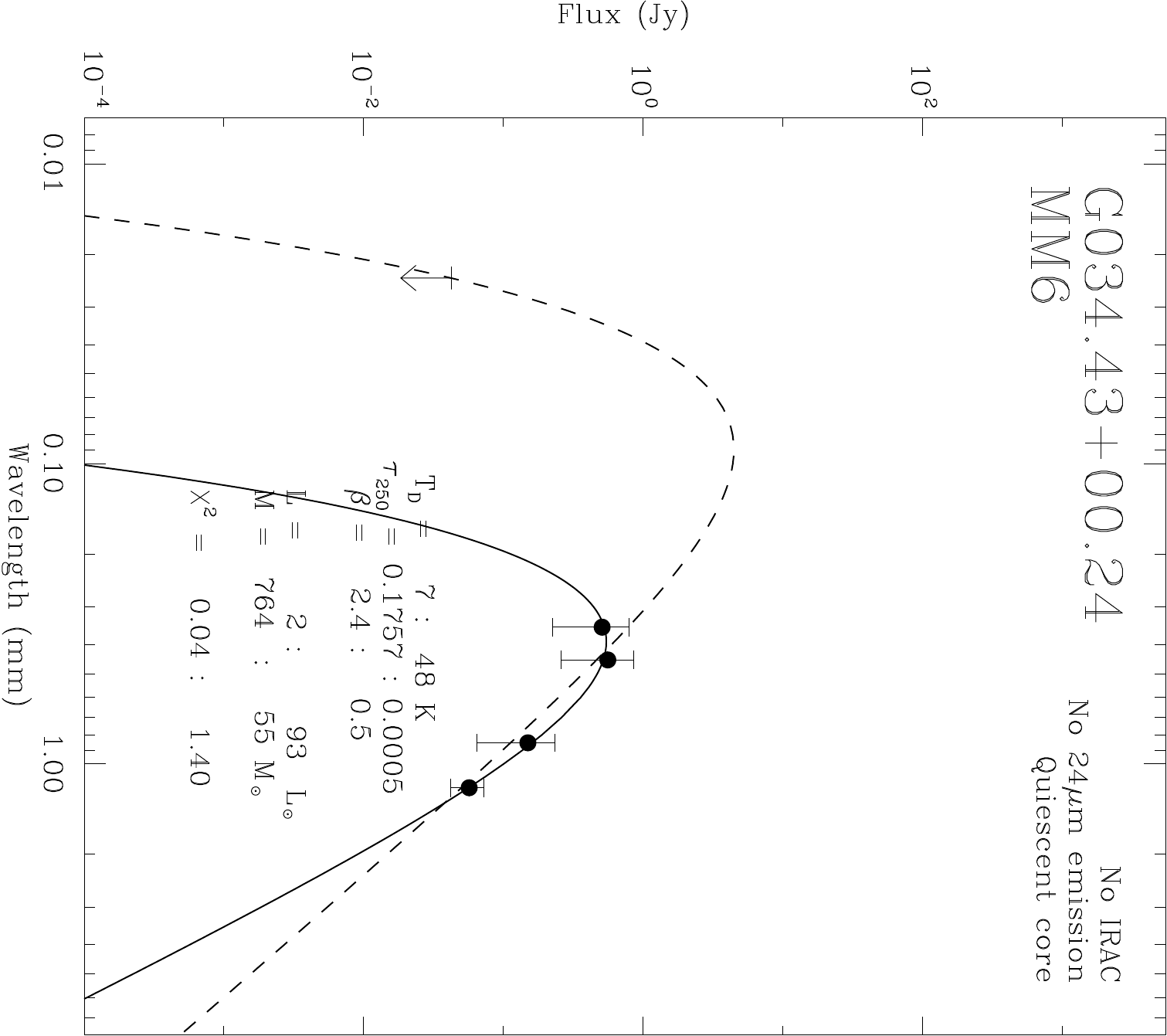}\\
\end{figure}
\clearpage 
\begin{figure}
\includegraphics[angle=90,width=0.5\textwidth]{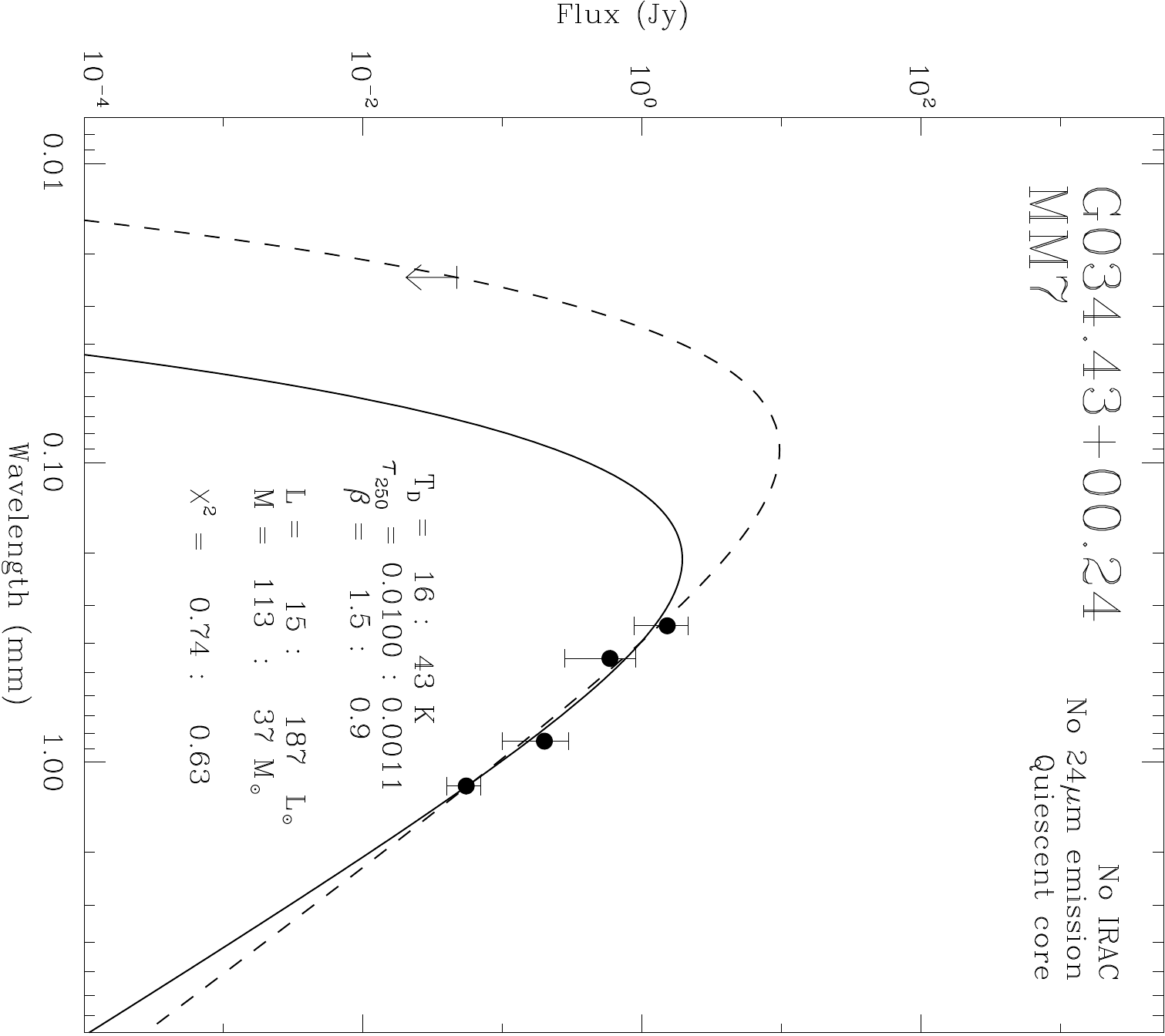}
\includegraphics[angle=90,width=0.5\textwidth]{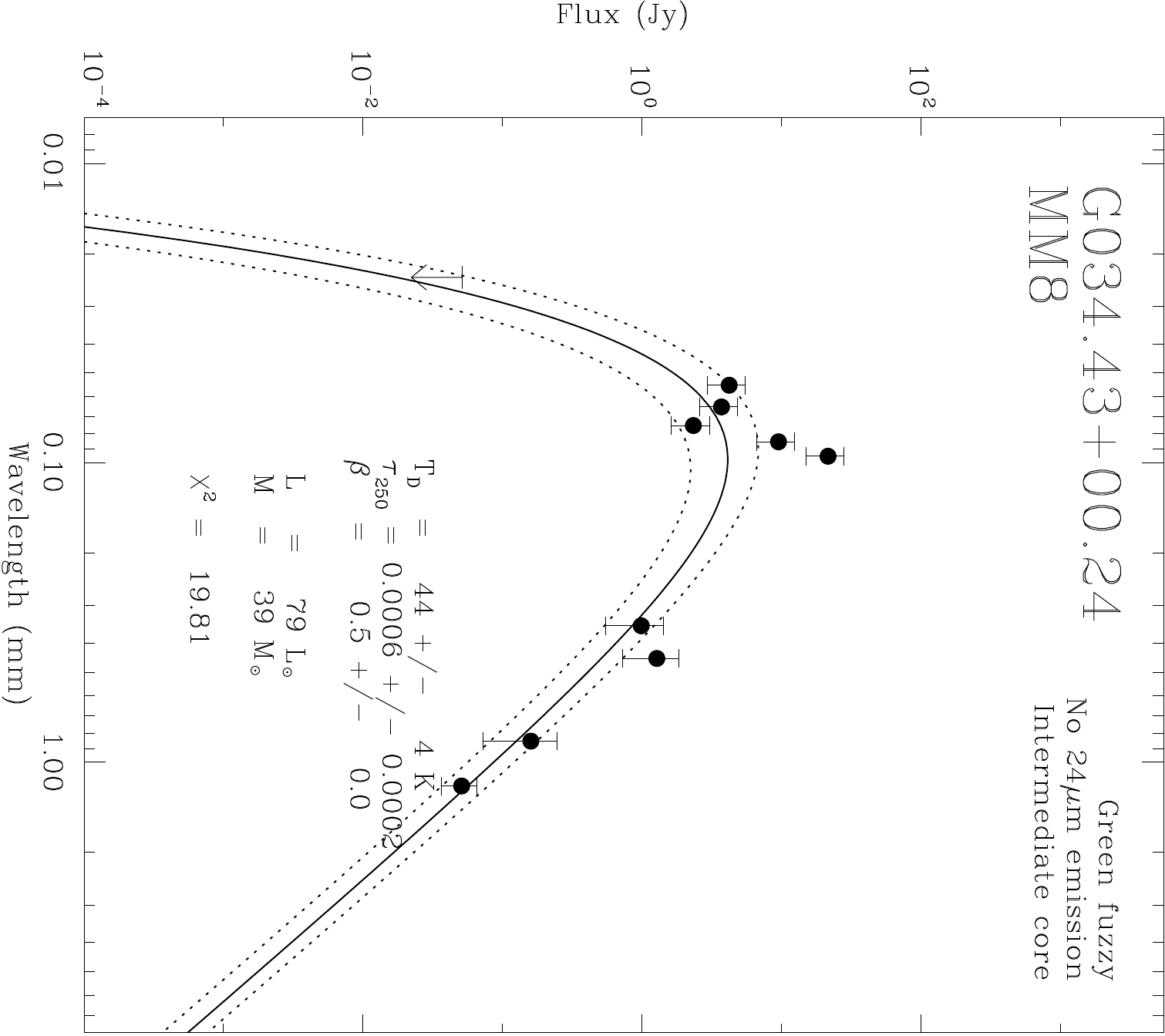}\\
\caption{\label{seds-43}\Spitzer\, 24\,\um\, image overlaid  
   with 1.2\,mm continuum emission for \irdcfortythree\, (contour
   levels are 60, 90, 120, 240, 360, 480, 840, 1200\,mJy
   beam$^{-1}$). The lower panels show the broadband
   SEDs for cores within this IRDC.  The fluxes derived from the
   millimeter, sub-millimeter, and far-IR  continuum data are shown as filled
   circles (with the corresponding error bars), while the 24\,\um\, fluxes are shown as  either a filled circle (when included within the fit), an open circle (when excluded from the fit),  or as an upper limit arrow. For cores that have measured fluxes only in the millimeter/sub-millimeter regime (i.e.\, a limit at 24\,\um), we show the results from two fits: one using only the measured fluxes (solid line; lower limit), while the other includes the 24\,\um\, limit as a real data (dashed line; upper limit). In all other cases, the solid line is the best fit gray-body, while the dotted lines correspond to the functions determined using the errors for the T$_{D}$, $\tau$, and $\beta$ output from the fitting.  Labeled on each plot is the IRDC and core name,  classification, and the derived parameters.}
\end{figure}
\clearpage 
\begin{figure}
\begin{center}
\includegraphics[angle=0,width=0.6\textwidth]{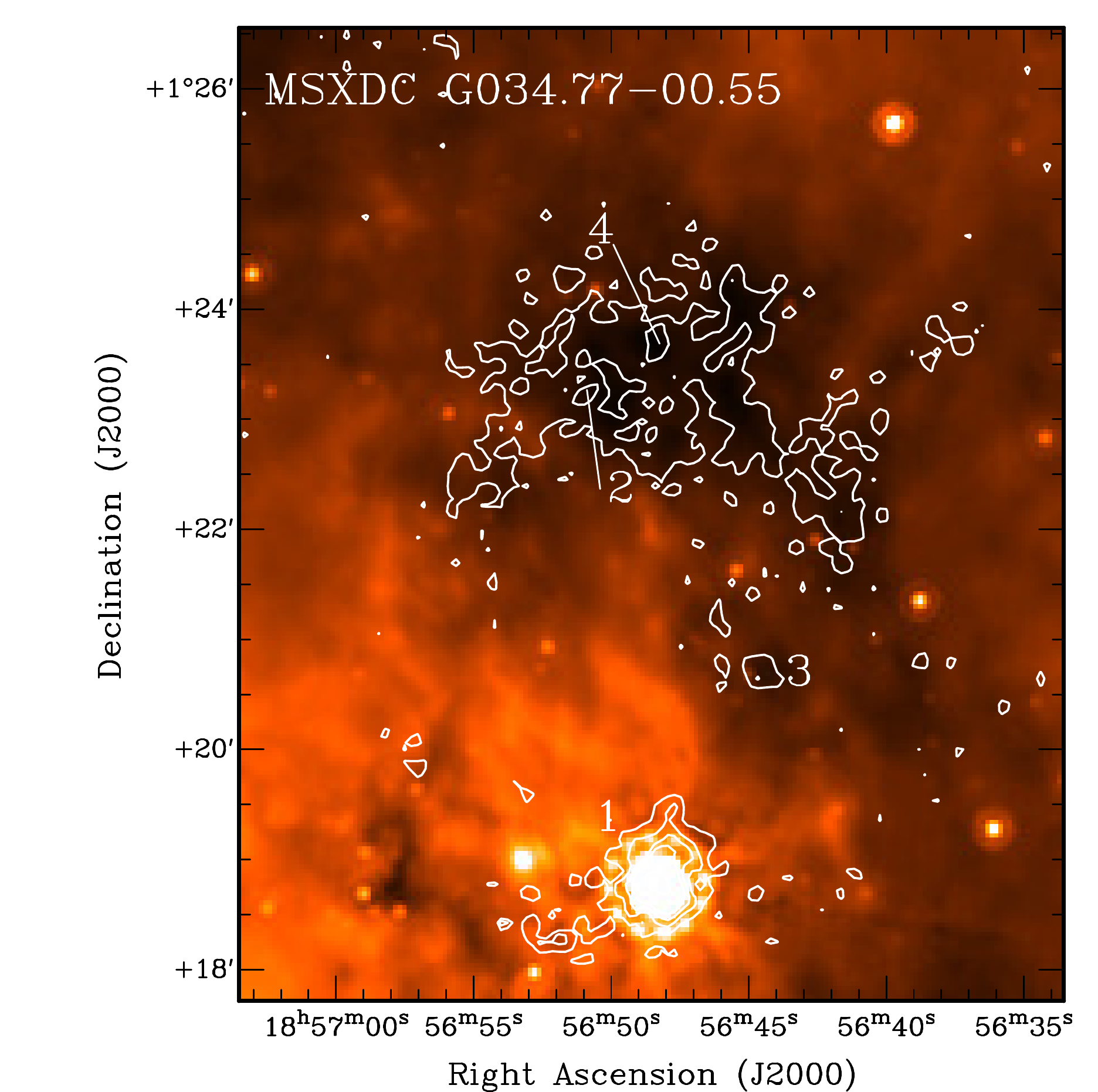}\\
\end{center}
\includegraphics[angle=90,width=0.5\textwidth]{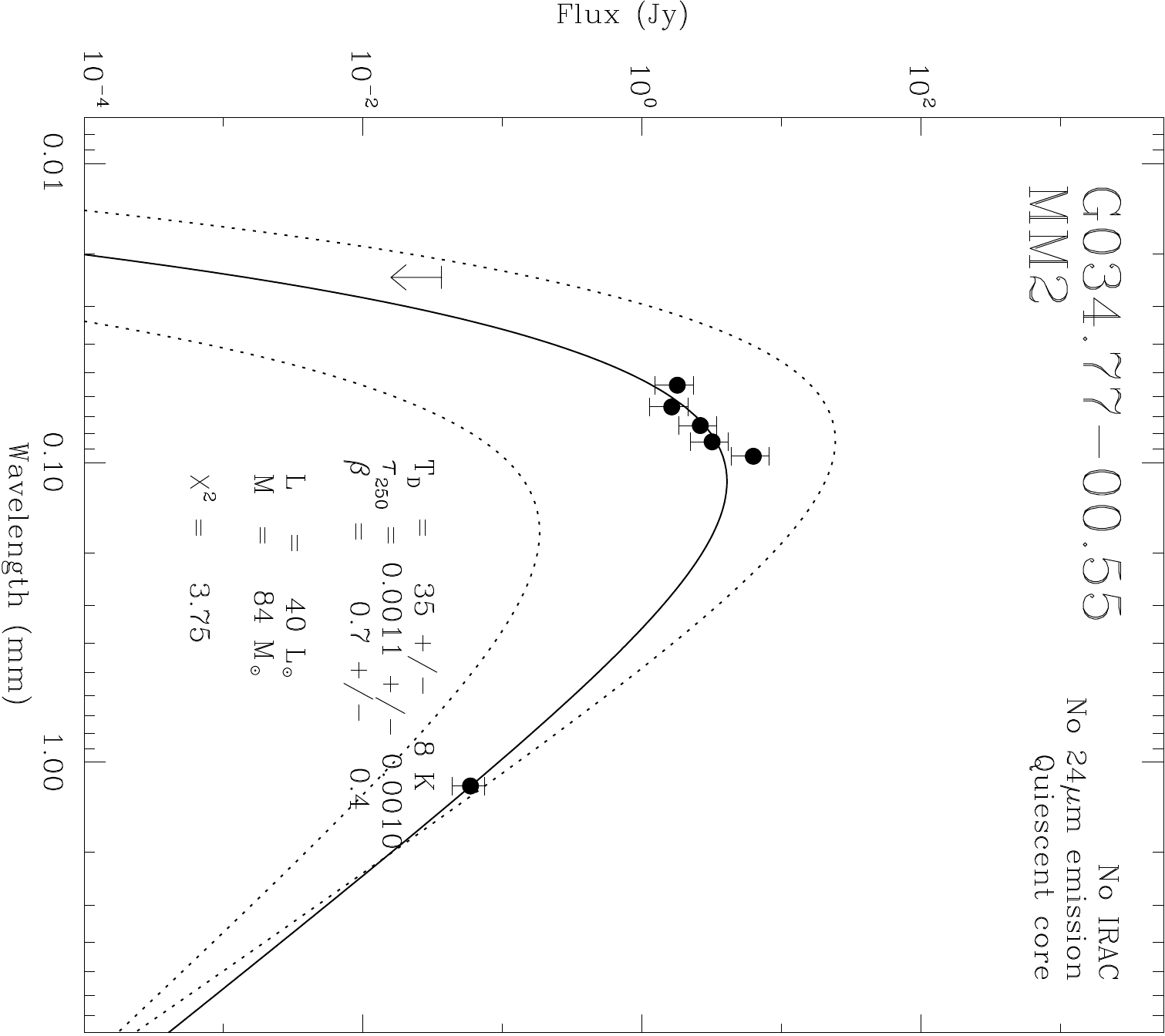}
\caption{\label{seds-21} \Spitzer\, 24\,\um\, image overlaid  
   with 1.2\,mm continuum emission for \irdctwentyone\, (contour
   levels are 30, 60, 90, 120, 240\,mJy beam$^{-1}$). The lower panels show the broadband
   SEDs for cores within this IRDC.  The fluxes derived from the
   millimeter, sub-millimeter, and far-IR  continuum data are shown as filled
   circles (with the corresponding error bars), while the 24\,\um\, fluxes are shown as  either a filled circle (when included within the fit), an open circle (when excluded from the fit),  or as an upper limit arrow. For cores that have measured fluxes only in the millimeter/sub-millimeter regime (i.e.\, a limit at 24\,\um), we show the results from two fits: one using only the measured fluxes (solid line; lower limit), while the other includes the 24\,\um\, limit as a real data (dashed line; upper limit). In all other cases, the solid line is the best fit gray-body, while the dotted lines correspond to the functions determined using the errors for the T$_{D}$, $\tau$, and $\beta$ output from the fitting.  Labeled on each plot is the IRDC and core name,  classification, and the derived parameters.}
\end{figure}
\clearpage 
\begin{figure}
\begin{center}
\includegraphics[angle=0,width=0.5\textwidth]{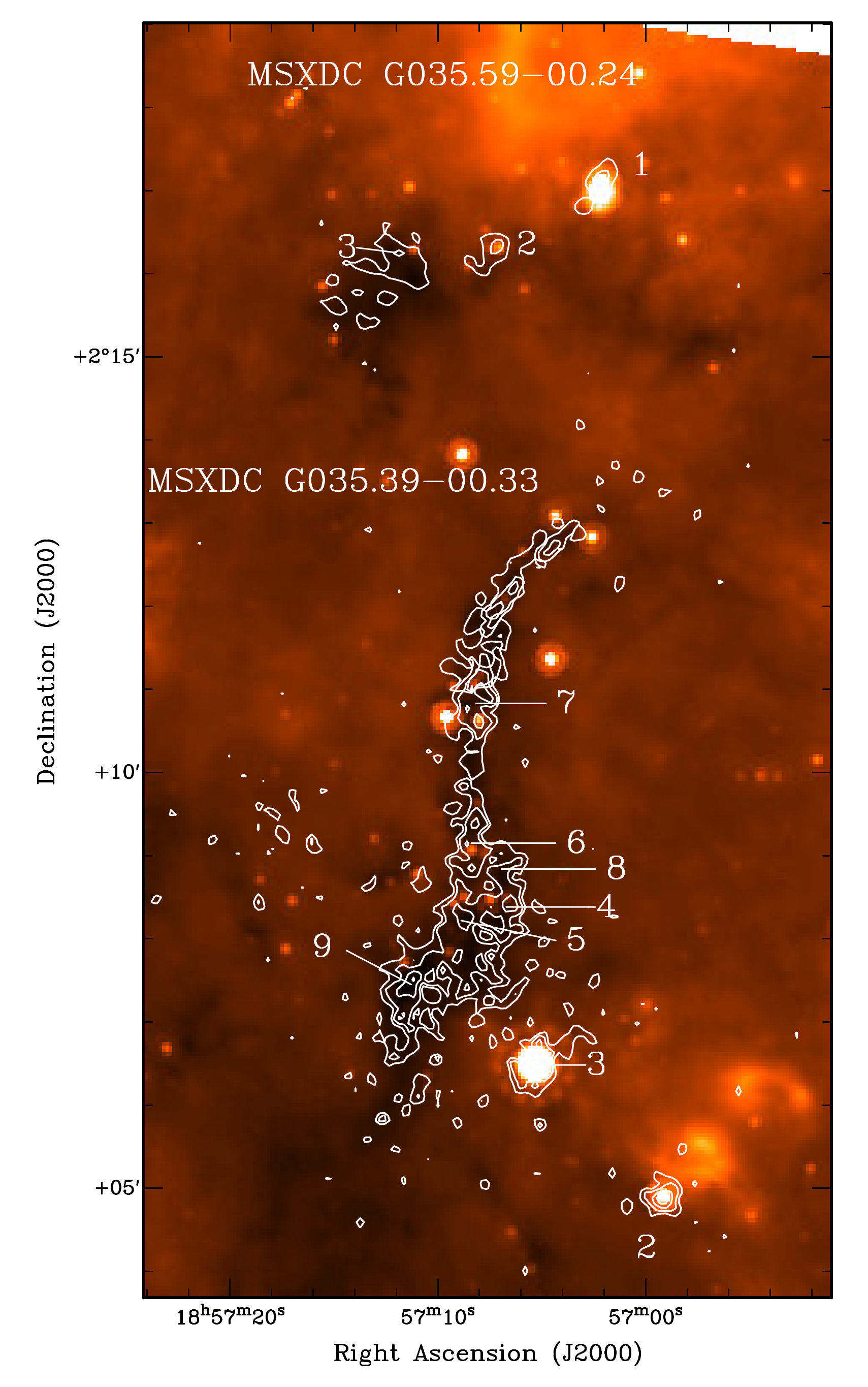}\\
\end{center}
\includegraphics[angle=90,width=0.5\textwidth]{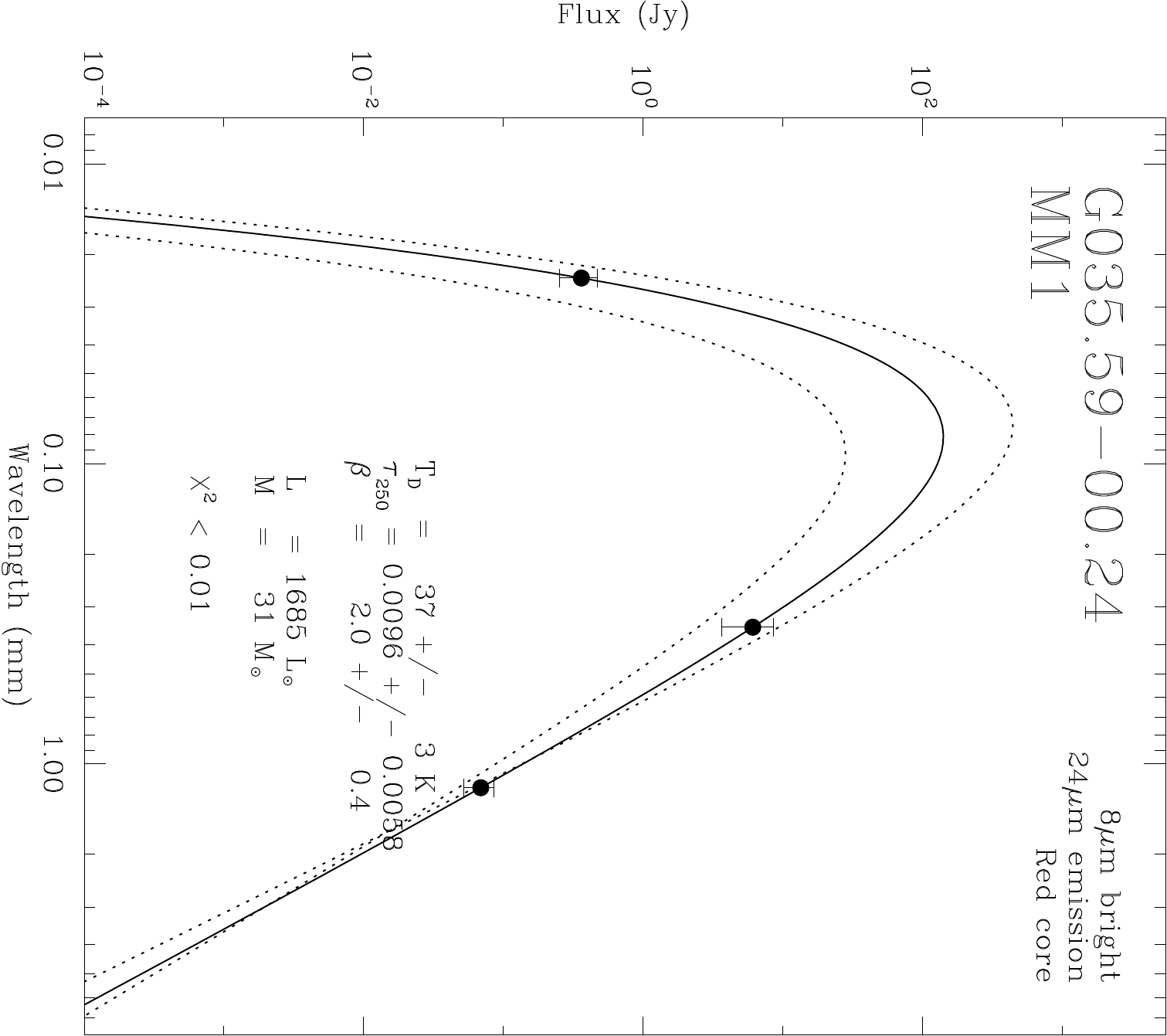}
\includegraphics[angle=90,width=0.5\textwidth]{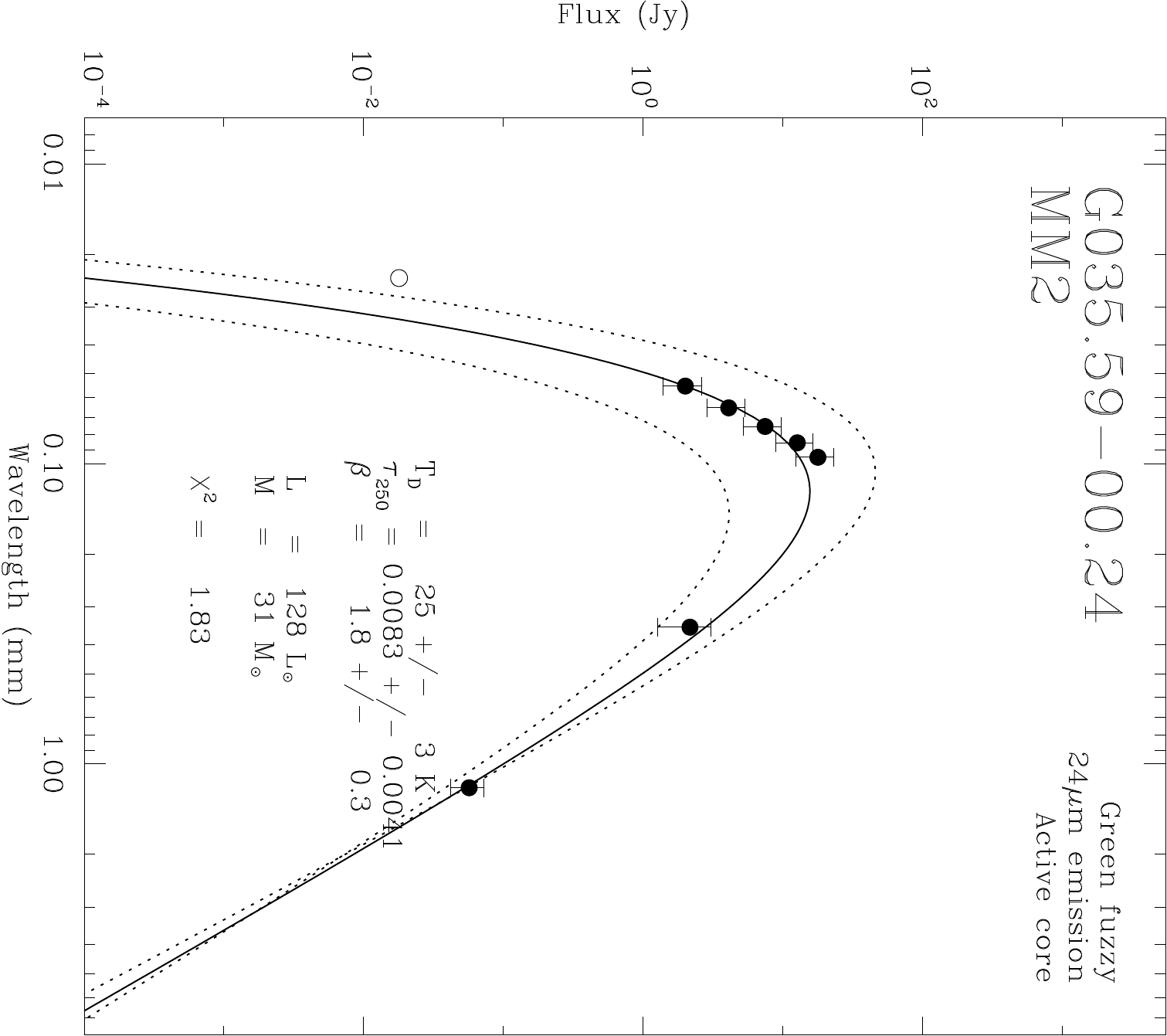}\\
\end{figure}
\clearpage 
\begin{figure}
\includegraphics[angle=90,width=0.5\textwidth]{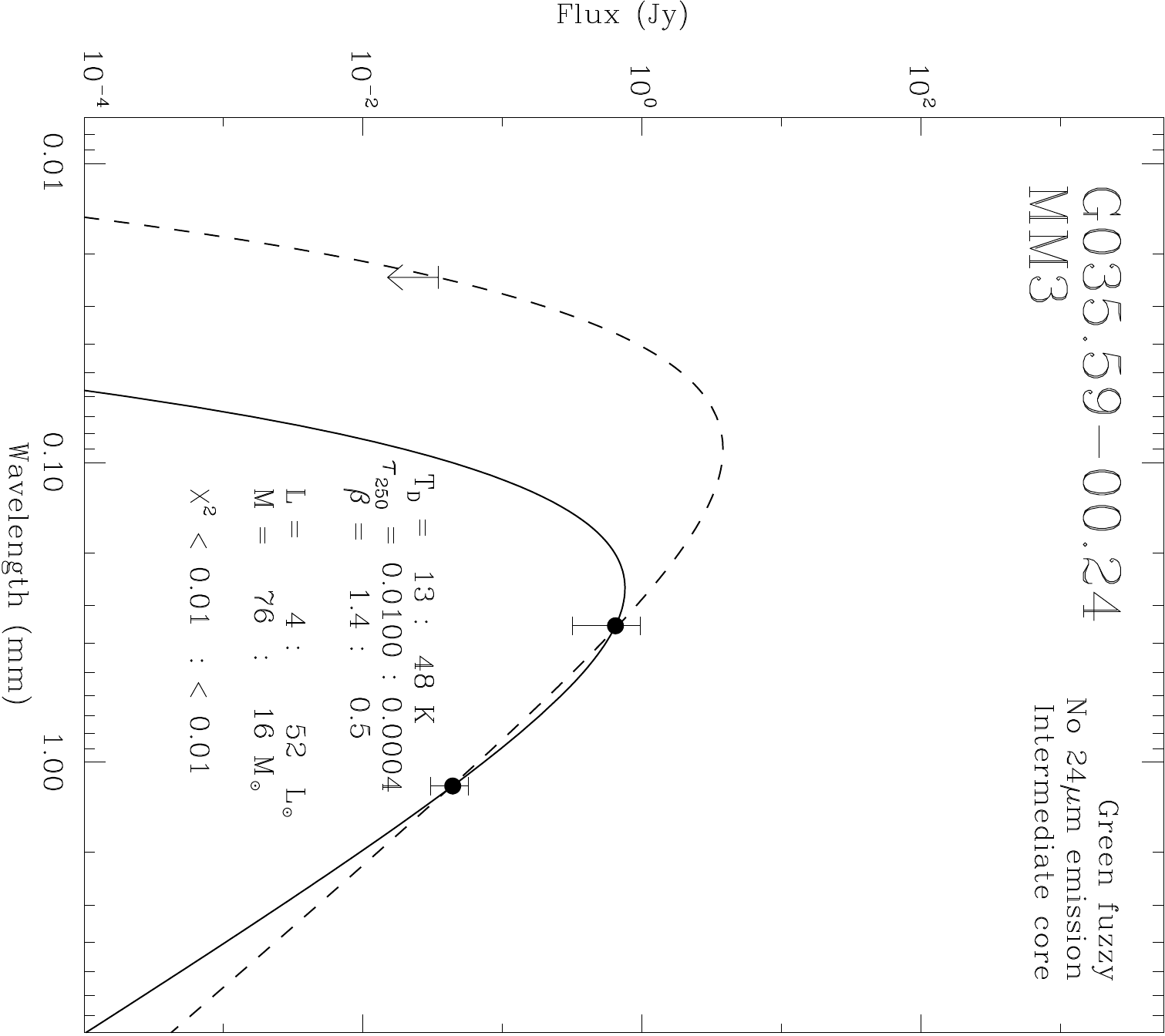}
\includegraphics[angle=90,width=0.5\textwidth]{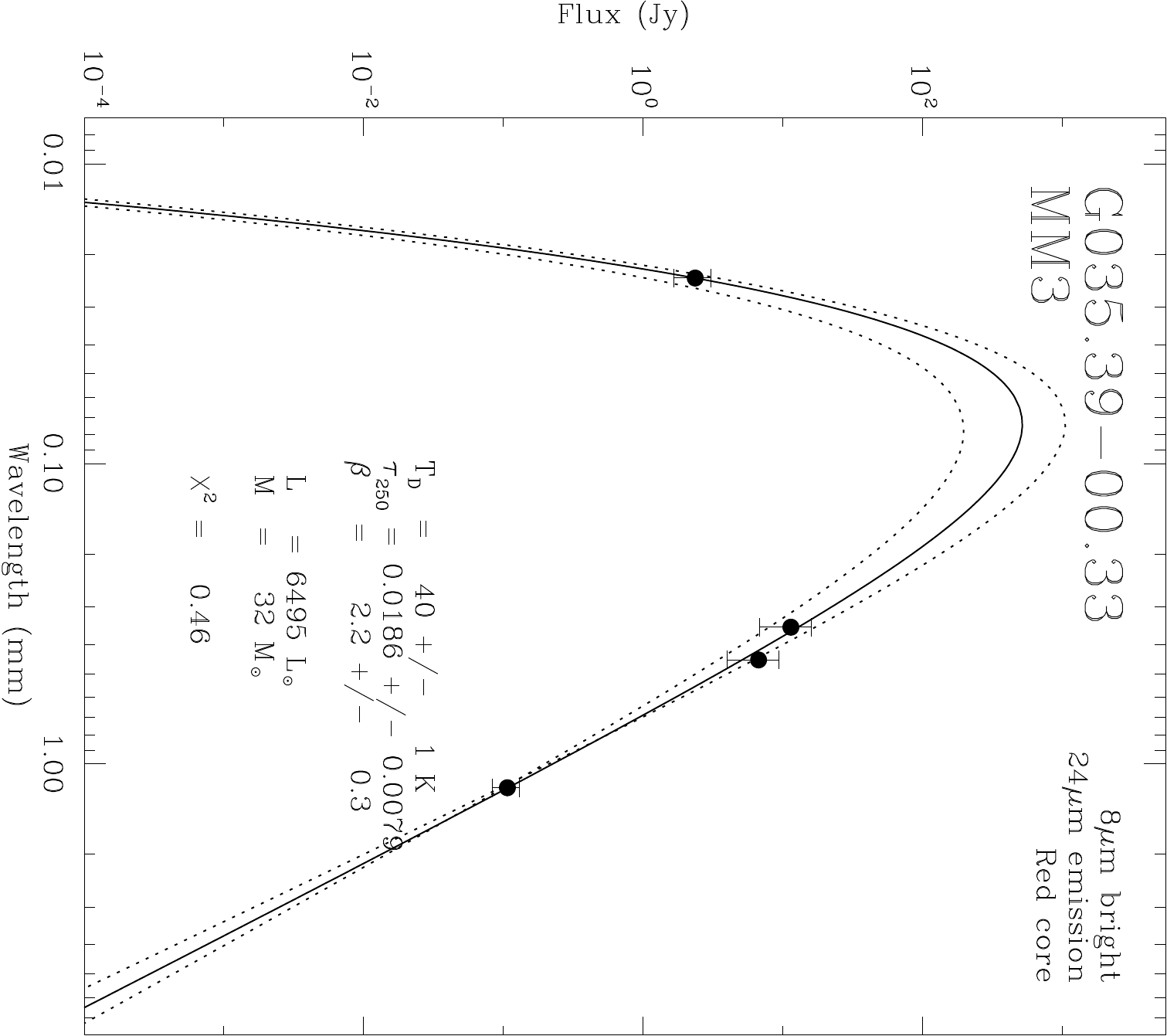}\\
\includegraphics[angle=90,width=0.5\textwidth]{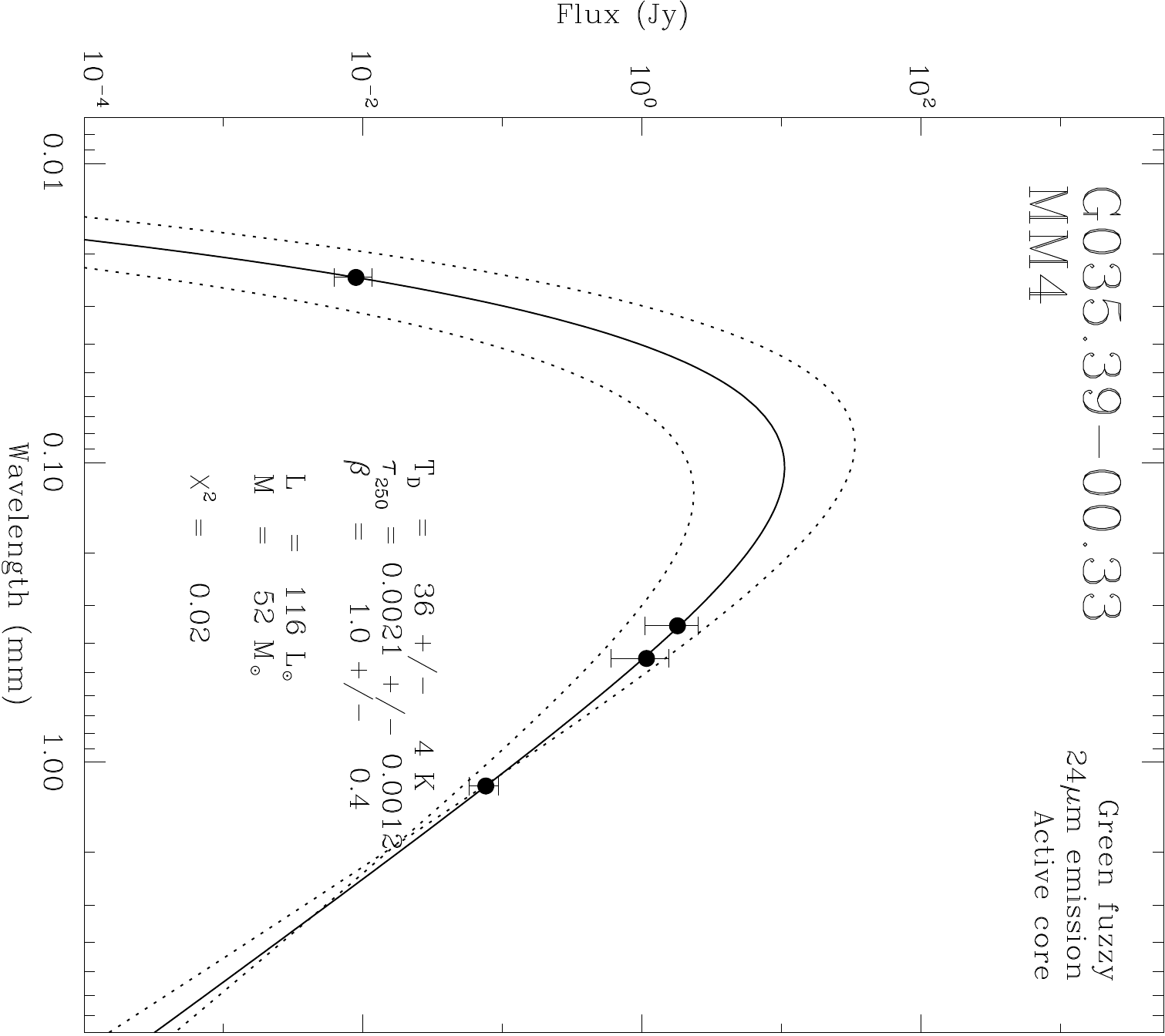}
\includegraphics[angle=90,width=0.5\textwidth]{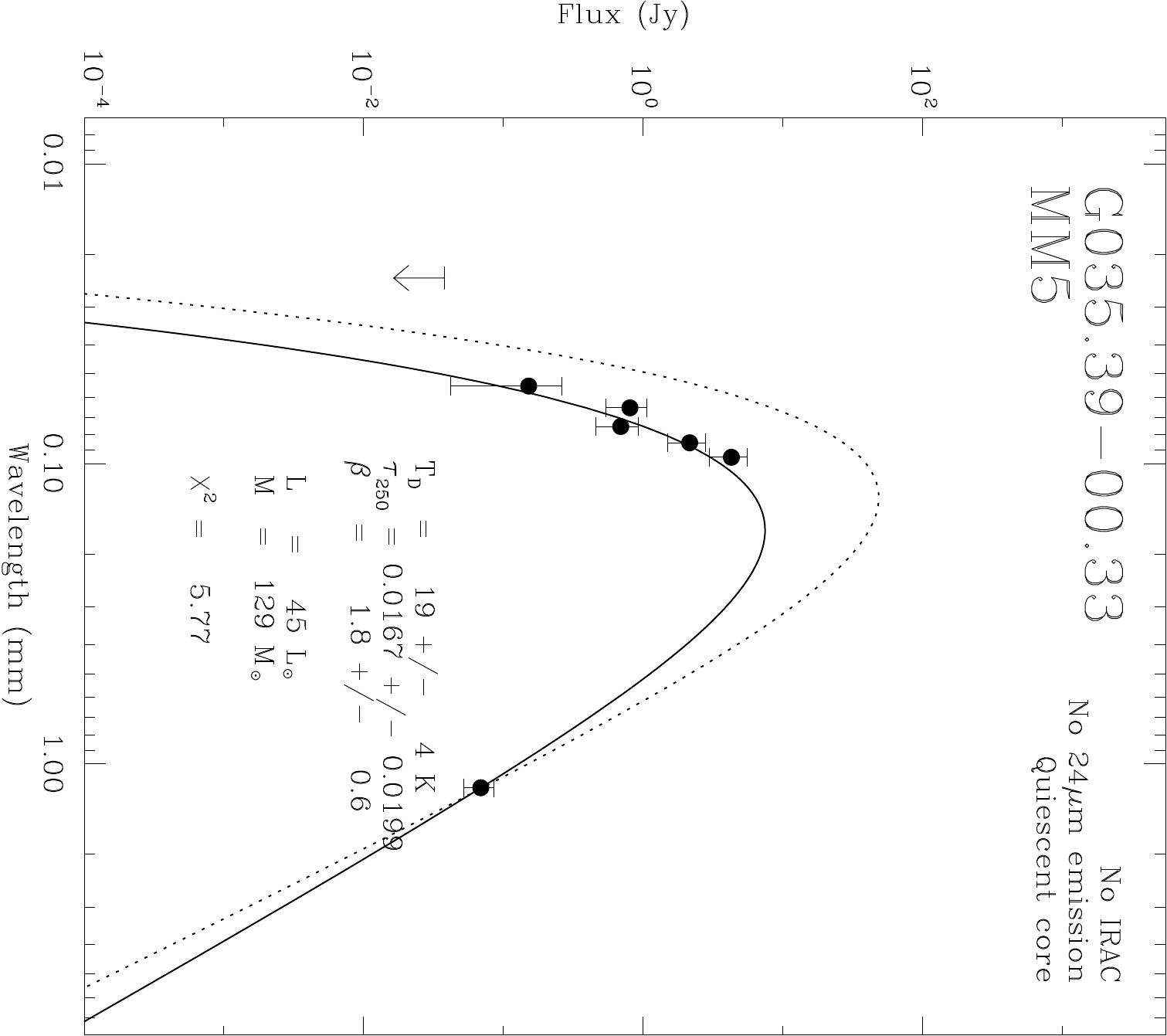}\\
\end{figure}
\clearpage 
\begin{figure}
\includegraphics[angle=90,width=0.5\textwidth]{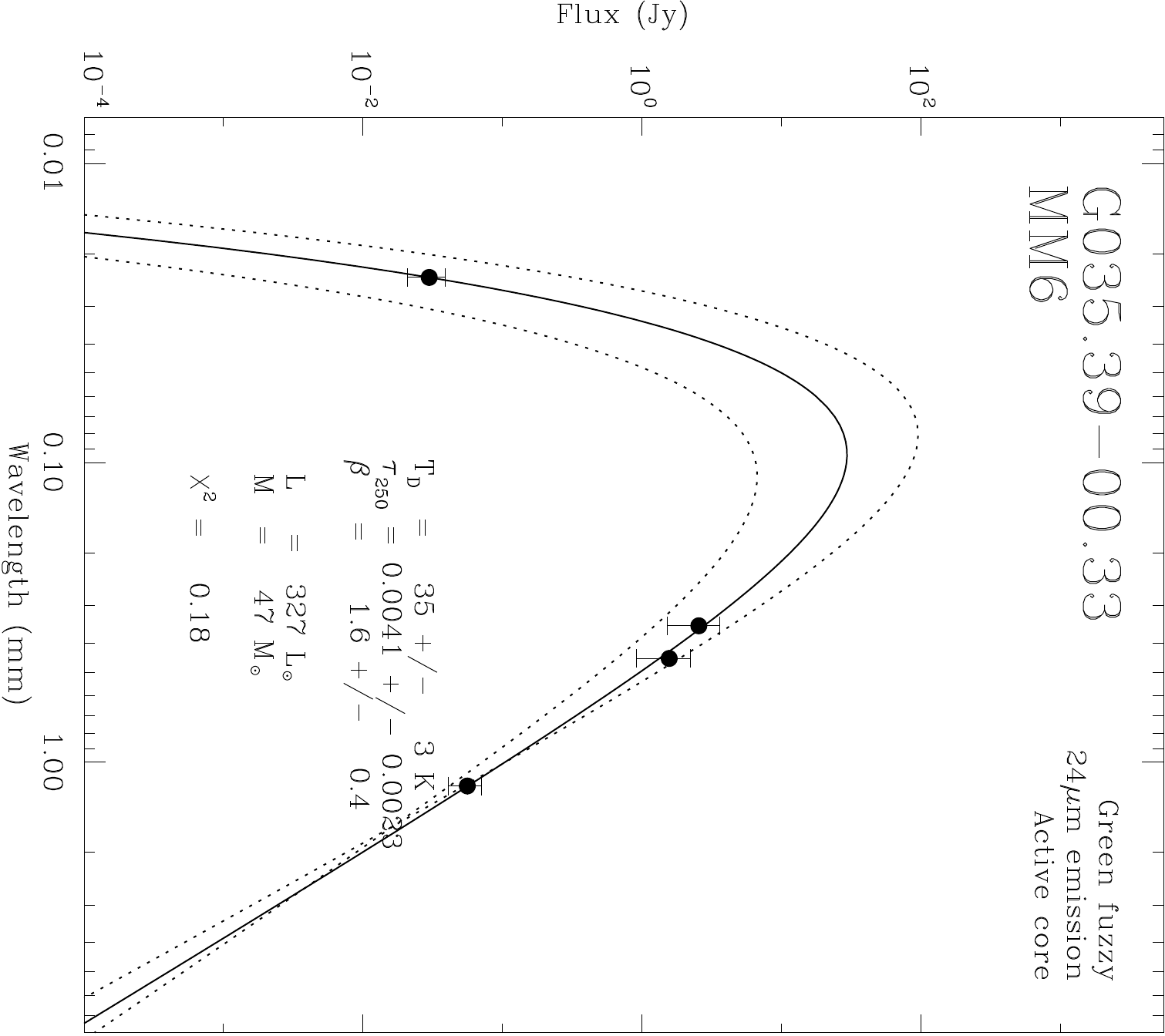}
\includegraphics[angle=90,width=0.5\textwidth]{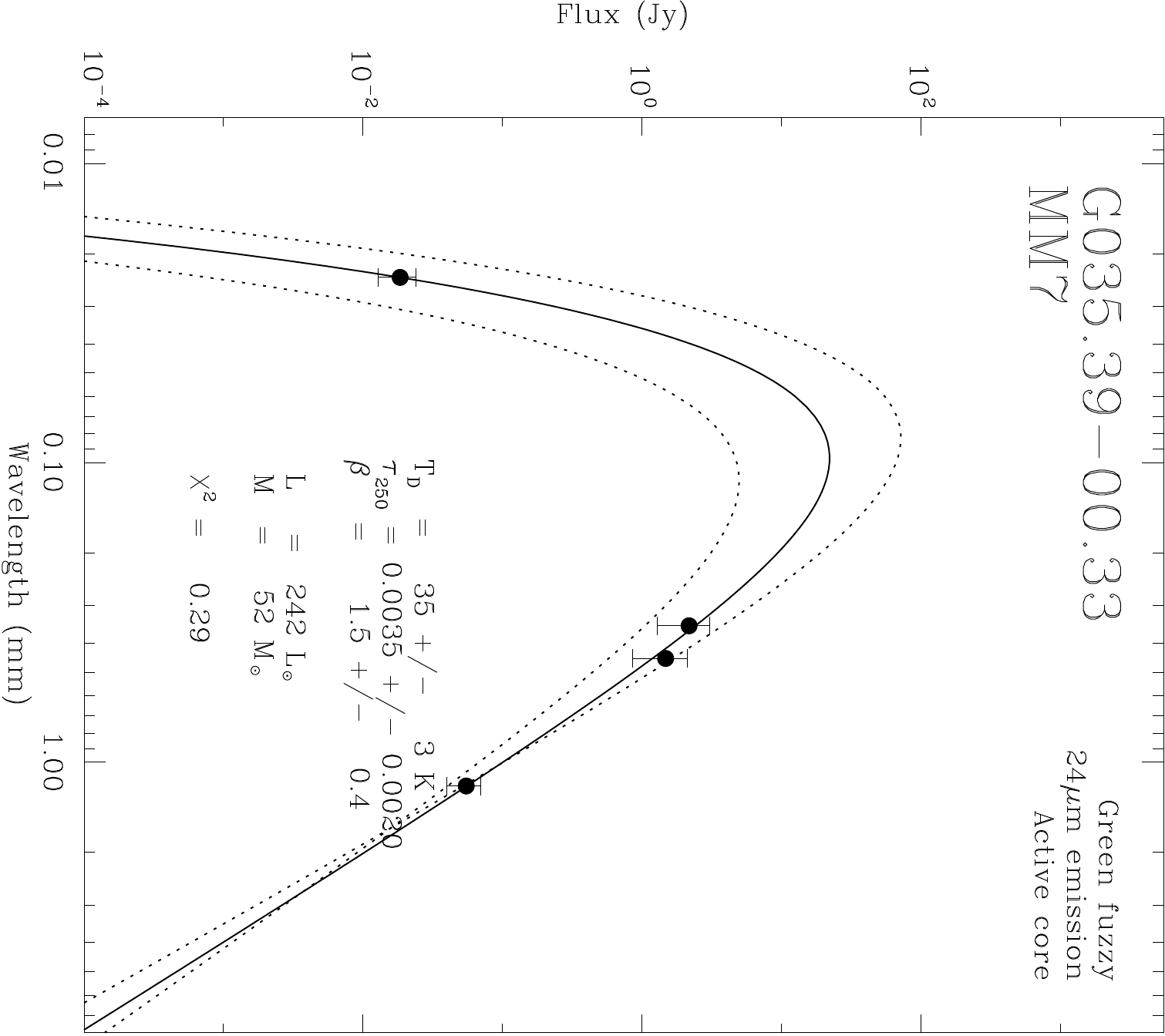}\\
\includegraphics[angle=90,width=0.5\textwidth]{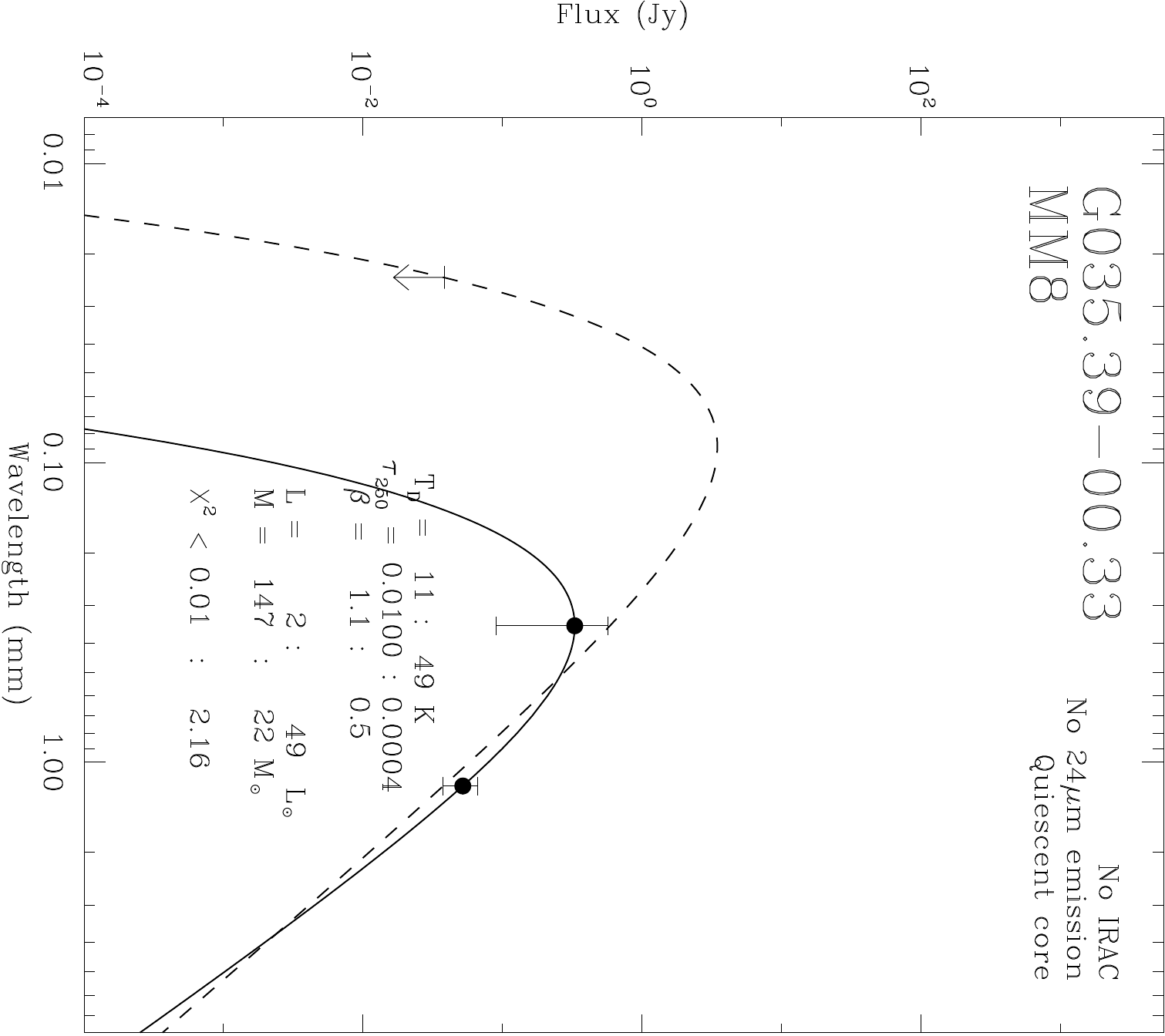}
\includegraphics[angle=90,width=0.5\textwidth]{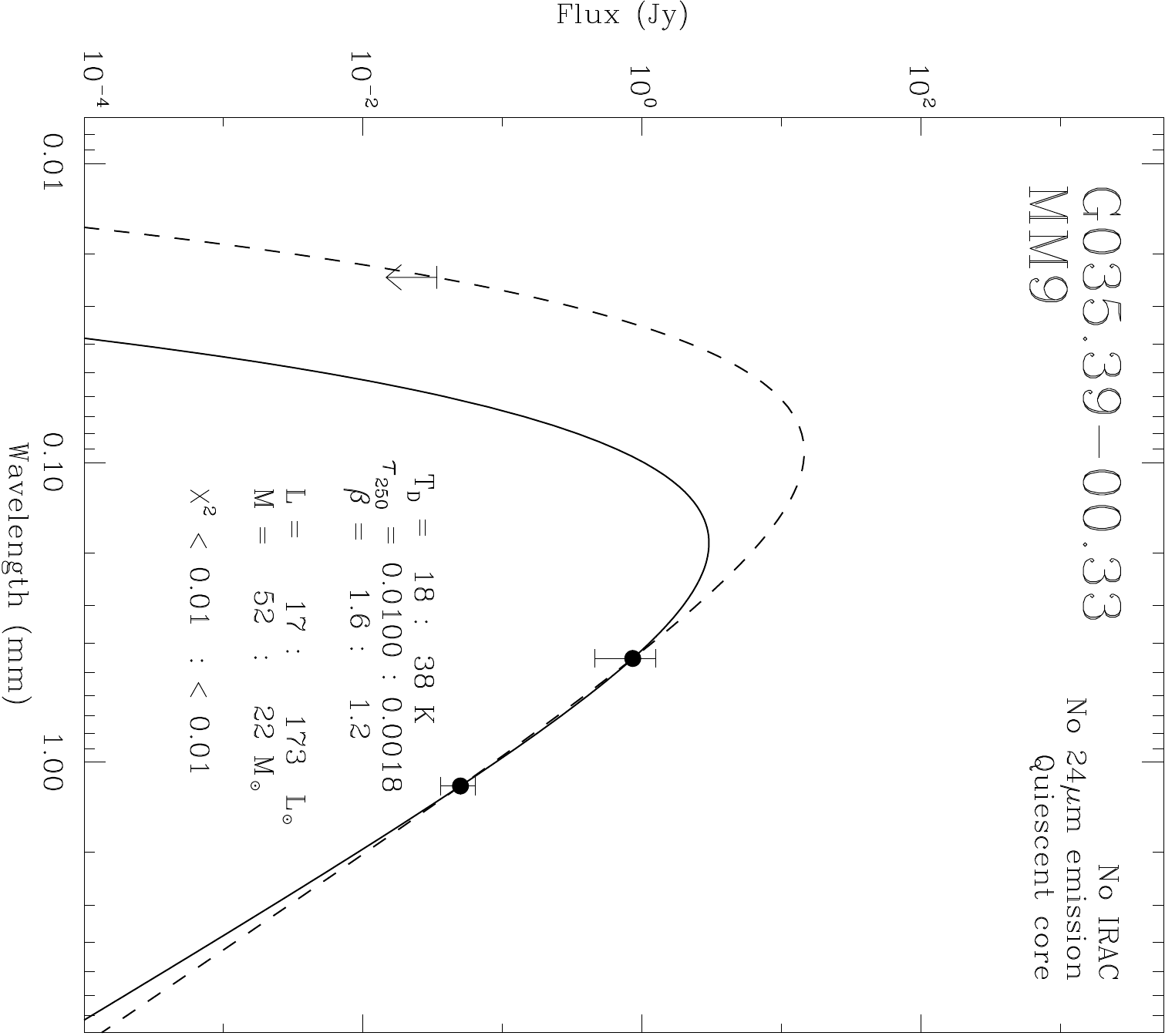}\\
\caption{\label{seds-3}\Spitzer\, 24\,\um\, image overlaid  
   with 1.2\,mm continuum emission for \irdceight\, and \irdcthree\,
   (contour levels are 30, 60, 90, 120\,mJy beam$^{-1}$). The lower panels show the broadband
   SEDs for cores within this IRDC.  The fluxes derived from the
   millimeter, sub-millimeter, and far-IR  continuum data are shown as filled
   circles (with the corresponding error bars), while the 24\,\um\, fluxes are shown as  either a filled circle (when included within the fit), an open circle (when excluded from the fit),  or as an upper limit arrow. For cores that have measured fluxes only in the millimeter/sub-millimeter regime (i.e.\, a limit at 24\,\um), we show the results from two fits: one using only the measured fluxes (solid line; lower limit), while the other includes the 24\,\um\, limit as a real data (dashed line; upper limit). In all other cases, the solid line is the best fit gray-body, while the dotted lines correspond to the functions determined using the errors for the T$_{D}$, $\tau$, and $\beta$ output from the fitting.  Labeled on each plot is the IRDC and core name,  classification, and the derived parameters.}
\end{figure}
\clearpage 
\begin{figure}
\begin{center}
\includegraphics[angle=0,width=0.6\textwidth]{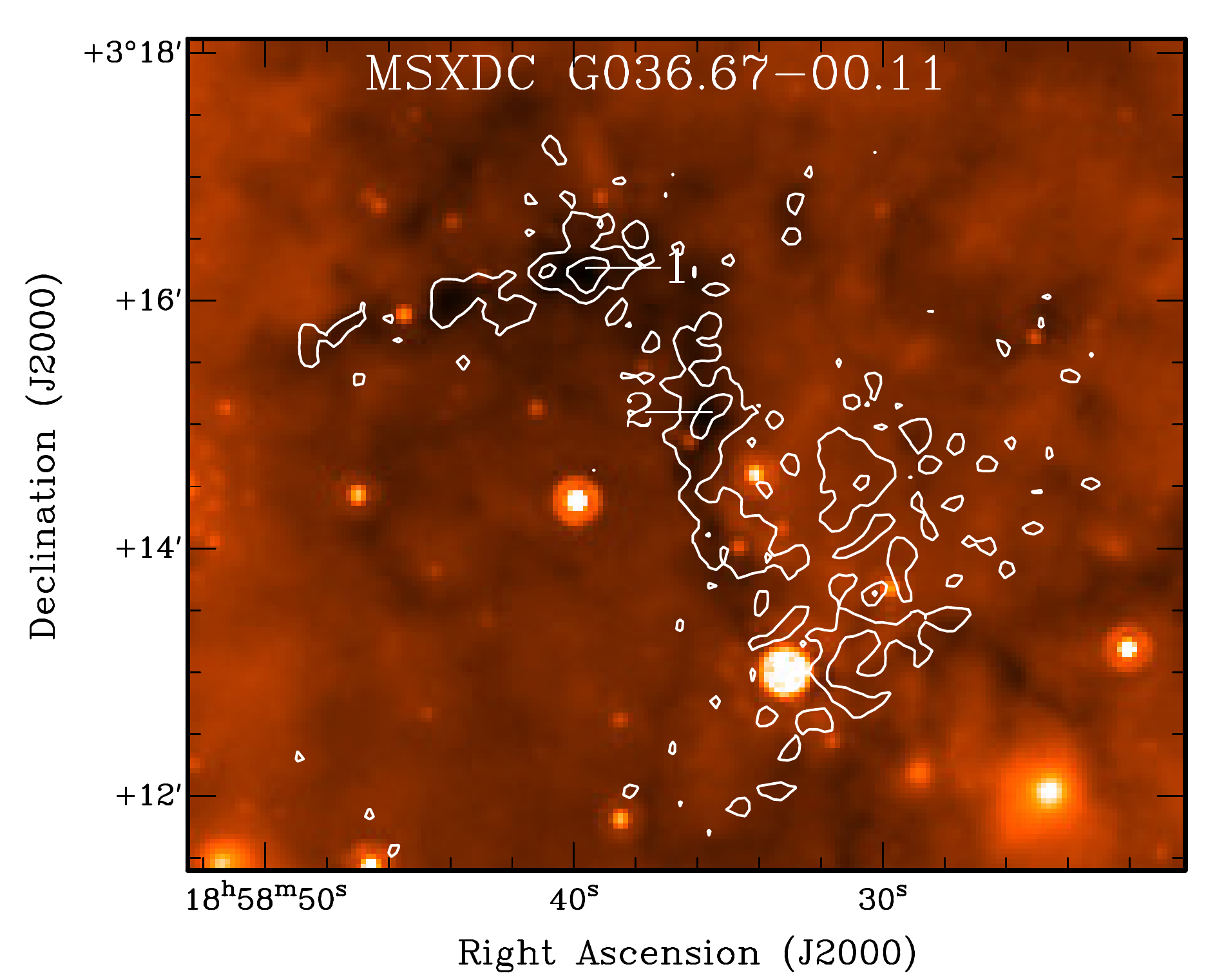}\\
\end{center}
\includegraphics[angle=90,width=0.5\textwidth]{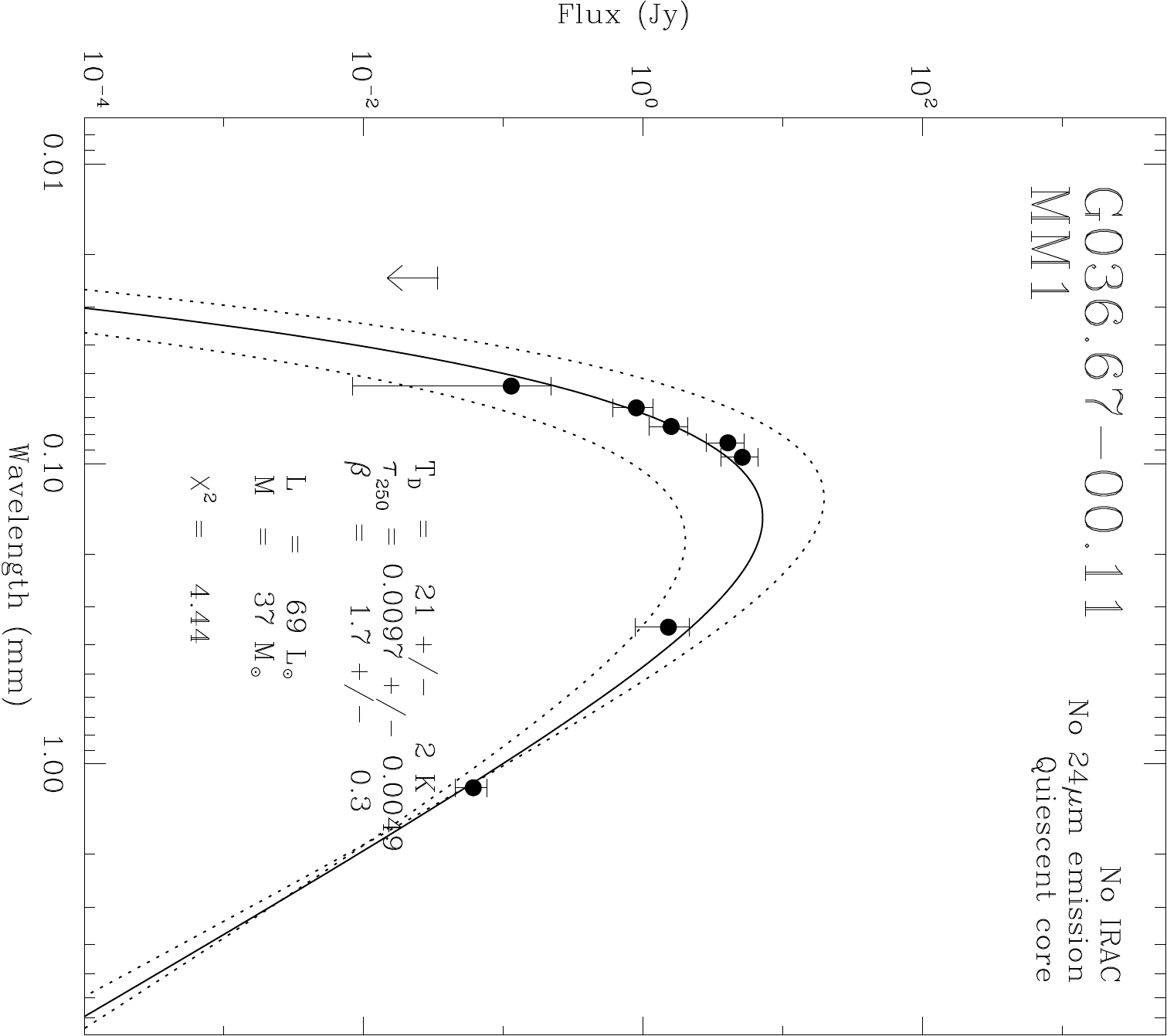}
\includegraphics[angle=90,width=0.5\textwidth]{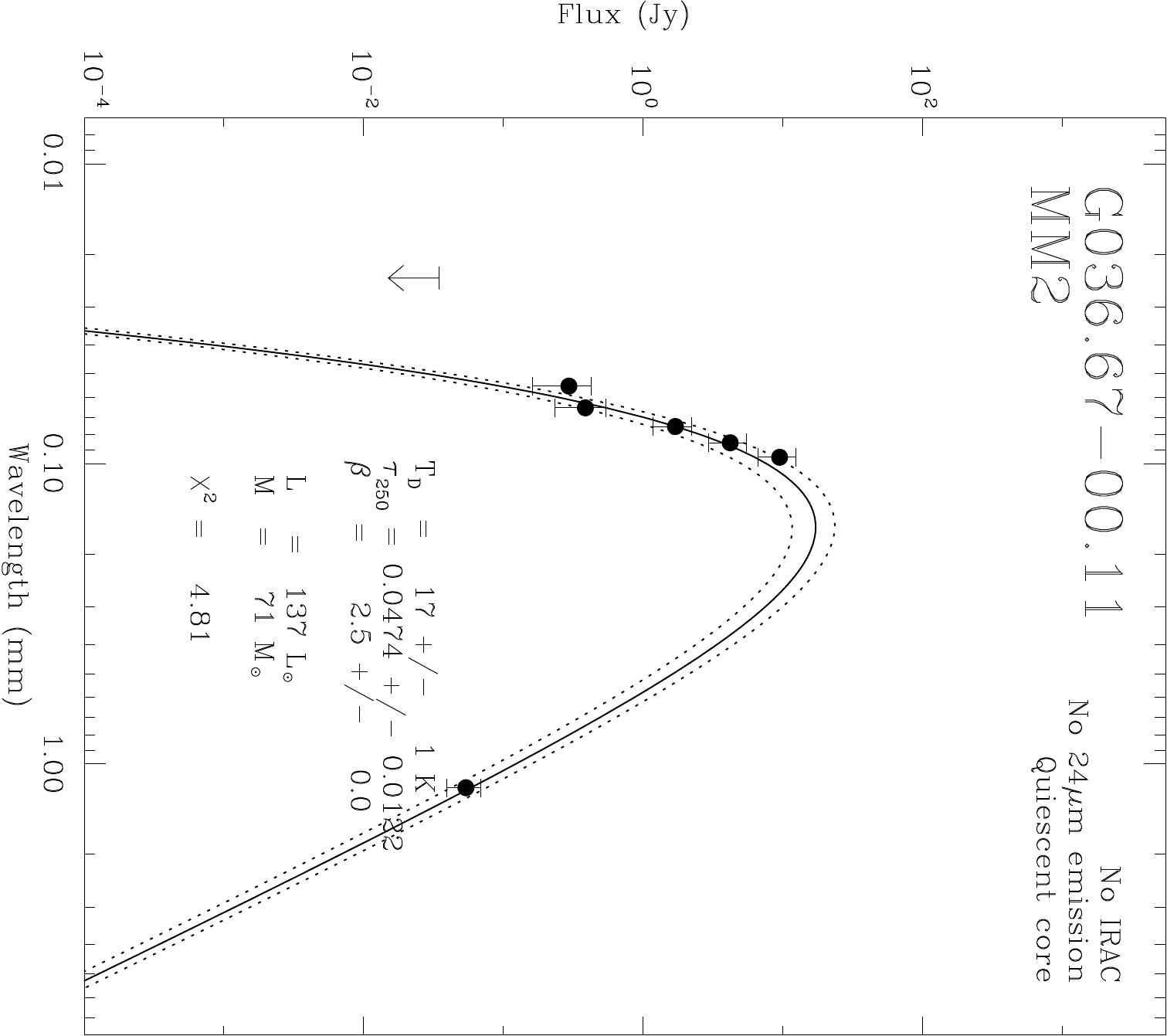}\\
\caption{\label{seds-31}\Spitzer\, 24\,\um\, image overlaid  
   with 1.2\,mm continuum emission for \irdcthirtyone\, (contour
   levels are 30, 60, 90, 120\,mJy beam$^{-1}$). The lower panels show the broadband
   SEDs for cores within this IRDC.  The fluxes derived from the
   millimeter, sub-millimeter, and far-IR  continuum data are shown as filled
   circles (with the corresponding error bars), while the 24\,\um\, fluxes are shown as  either a filled circle (when included within the fit), an open circle (when excluded from the fit),  or as an upper limit arrow. For cores that have measured fluxes only in the millimeter/sub-millimeter regime (i.e.\, a limit at 24\,\um), we show the results from two fits: one using only the measured fluxes (solid line; lower limit), while the other includes the 24\,\um\, limit as a real data (dashed line; upper limit). In all other cases, the solid line is the best fit gray-body, while the dotted lines correspond to the functions determined using the errors for the T$_{D}$, $\tau$, and $\beta$ output from the fitting.  Labeled on each plot is the IRDC and core name,  classification, and the derived parameters.}
\end{figure}
\clearpage 
\begin{figure}
\begin{center}
\includegraphics[angle=0,width=0.4\textwidth]{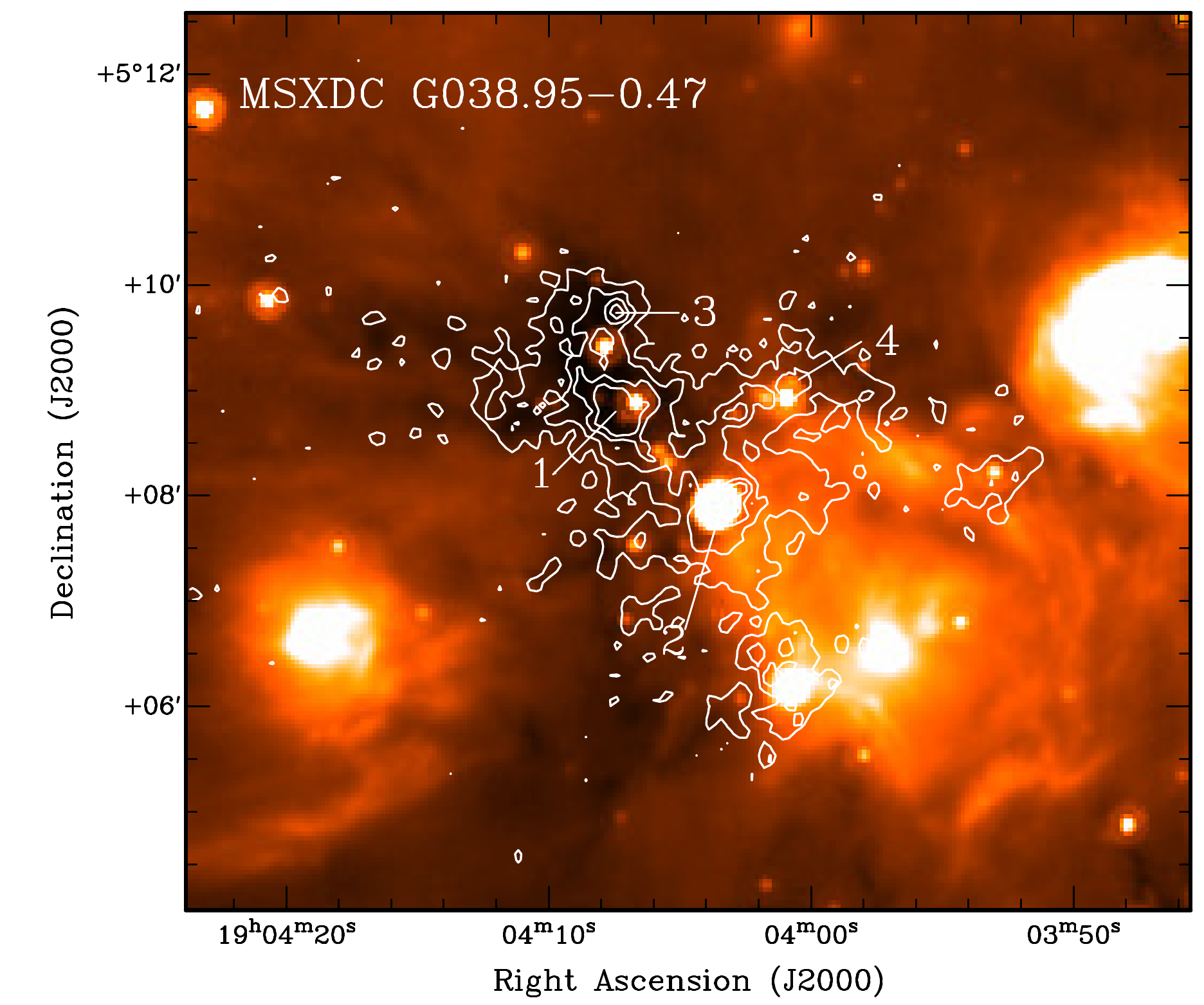}\\
\end{center}
\includegraphics[angle=90,width=0.5\textwidth]{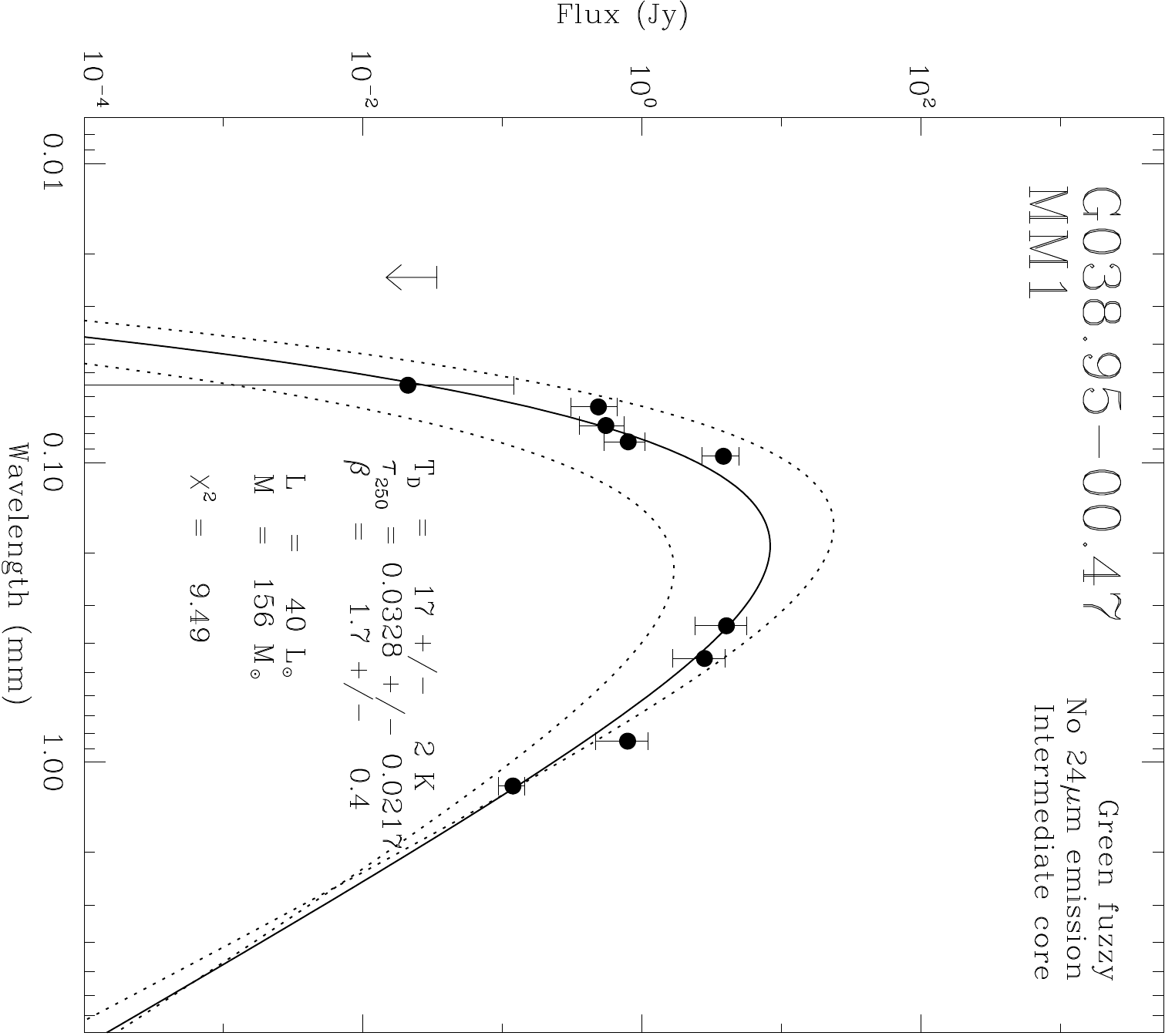}
\includegraphics[angle=90,width=0.5\textwidth]{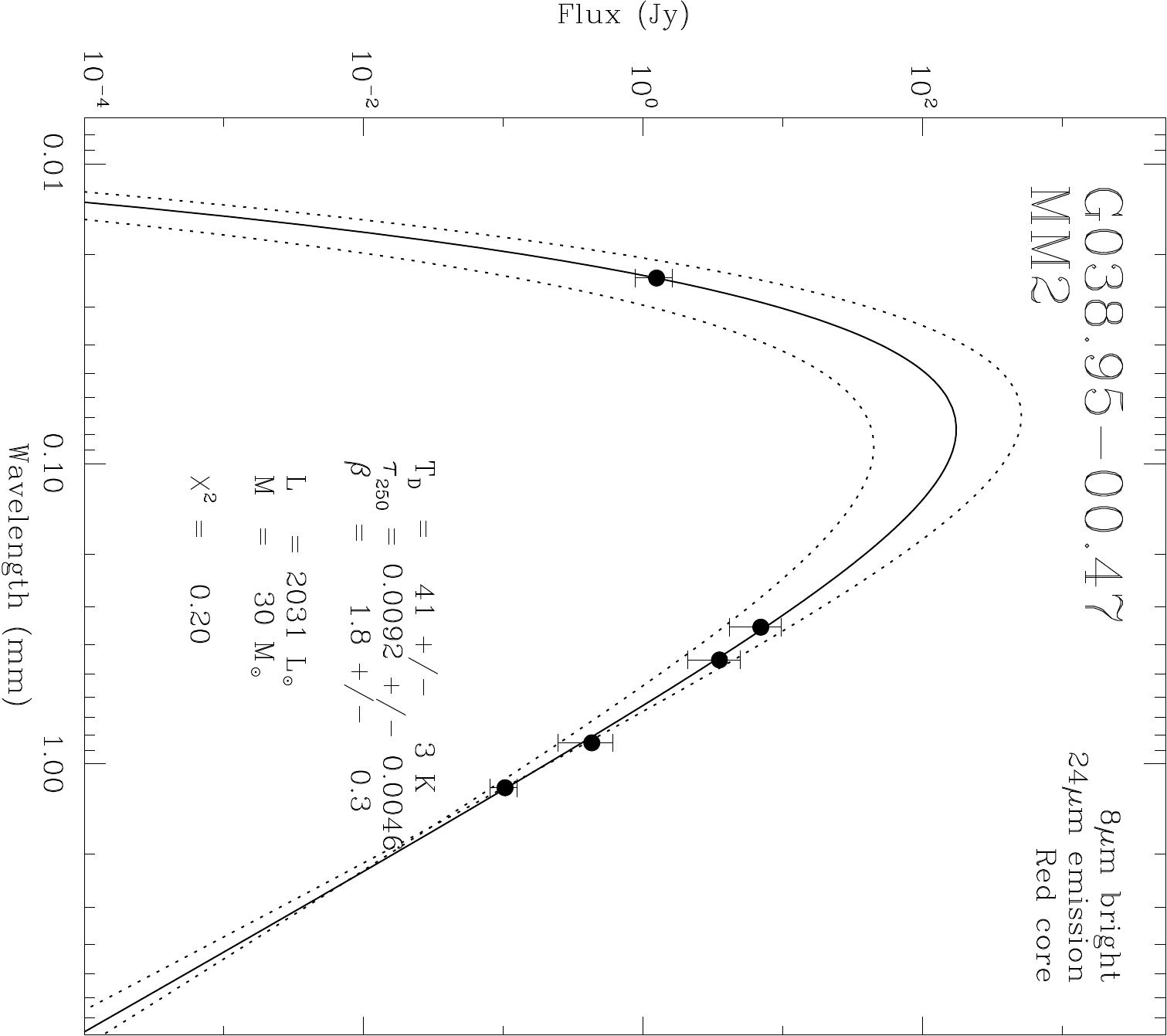}\\
\end{figure}
\clearpage 
\begin{figure}
\includegraphics[angle=90,width=0.5\textwidth]{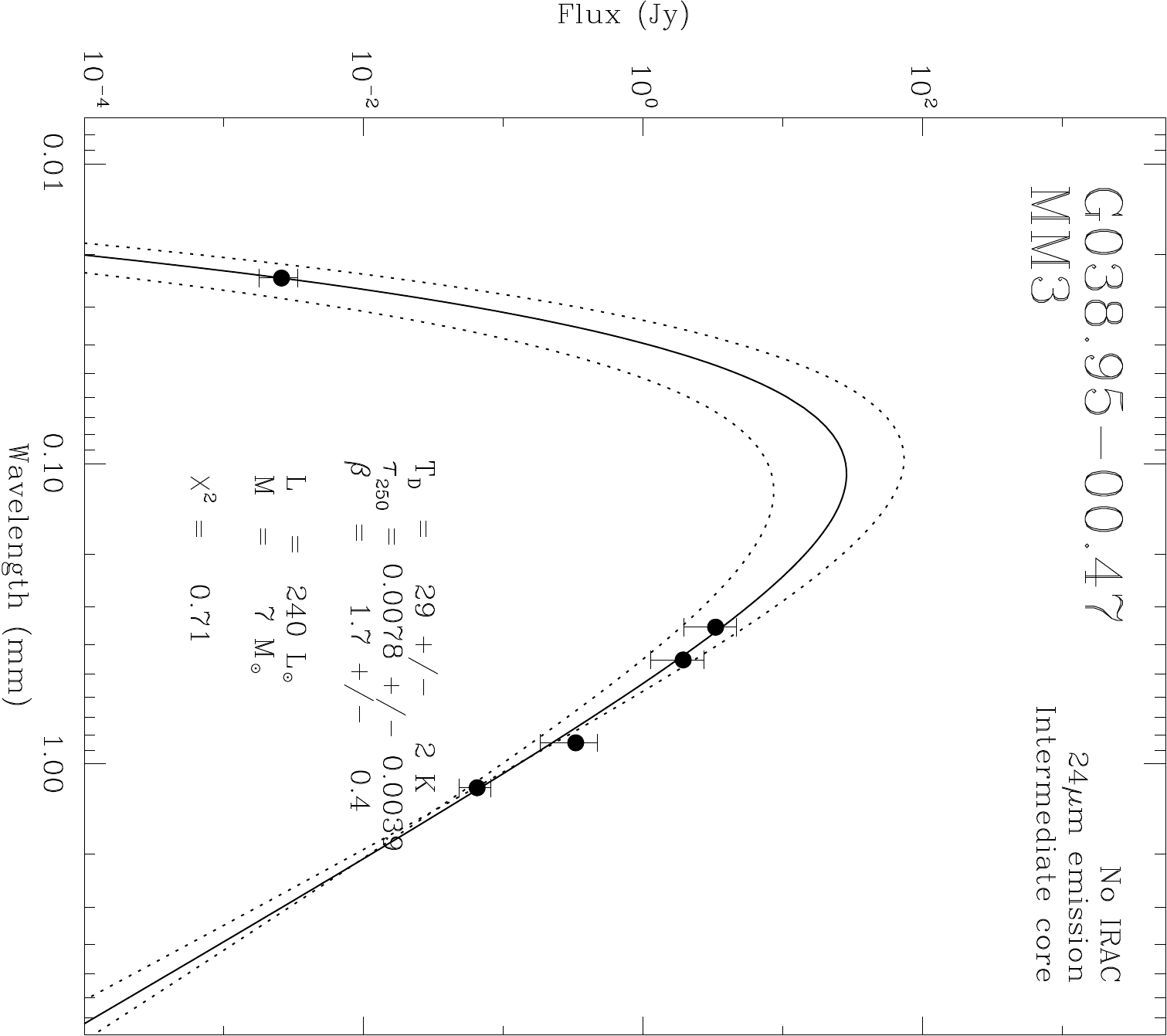}
\includegraphics[angle=90,width=0.5\textwidth]{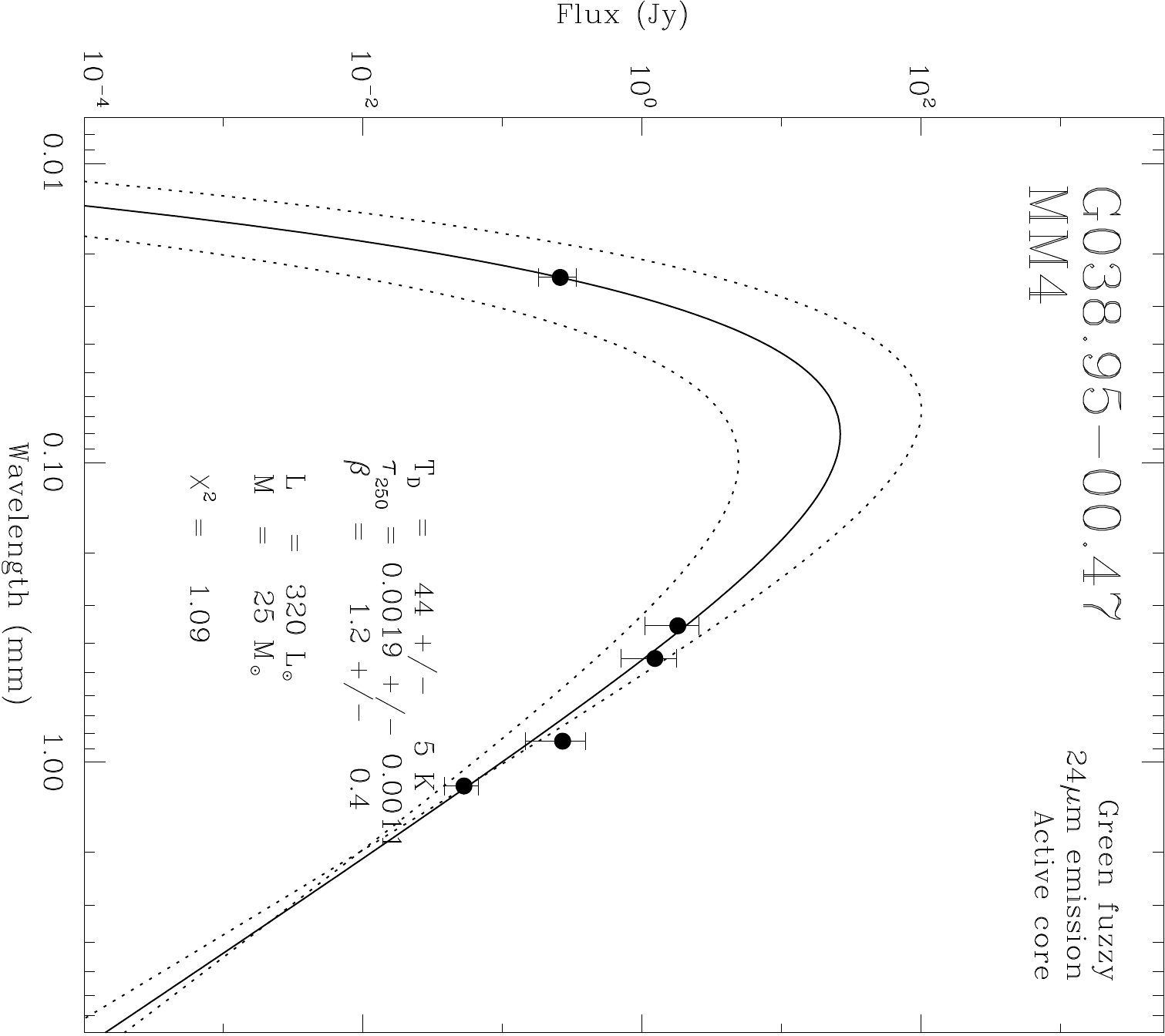}\\
\caption{\label{seds-5}\Spitzer\, 24\,\um\, image overlaid  
   with 1.2\,mm continuum emission for \irdcfive\, (contour levels are
   30, 60, 90, 120\,mJy beam$^{-1}$). The lower panels show the broadband
   SEDs for cores within this IRDC.  The fluxes derived from the
   millimeter, sub-millimeter, and far-IR  continuum data are shown as filled
   circles (with the corresponding error bars), while the 24\,\um\, fluxes are shown as  either a filled circle (when included within the fit), an open circle (when excluded from the fit),  or as an upper limit arrow. For cores that have measured fluxes only in the millimeter/sub-millimeter regime (i.e.\, a limit at 24\,\um), we show the results from two fits: one using only the measured fluxes (solid line; lower limit), while the other includes the 24\,\um\, limit as a real data (dashed line; upper limit). In all other cases, the solid line is the best fit gray-body, while the dotted lines correspond to the functions determined using the errors for the T$_{D}$, $\tau$, and $\beta$ output from the fitting.  Labeled on each plot is the IRDC and core name,  classification, and the derived parameters.}
\end{figure}
\clearpage 
\begin{figure}
\begin{center}
\includegraphics[angle=0,width=0.6\textwidth]{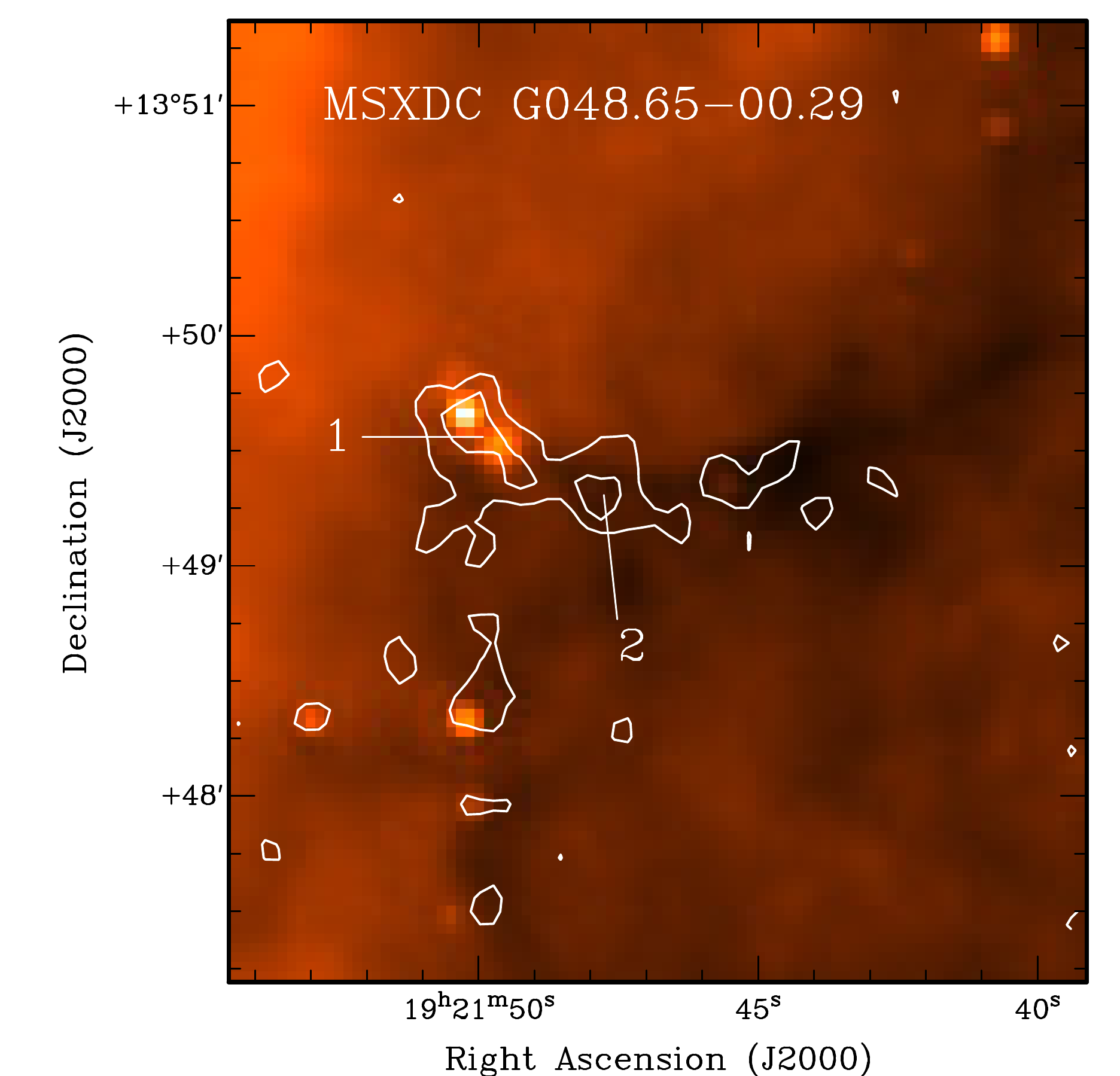}\\
\end{center}
\includegraphics[angle=90,width=0.5\textwidth]{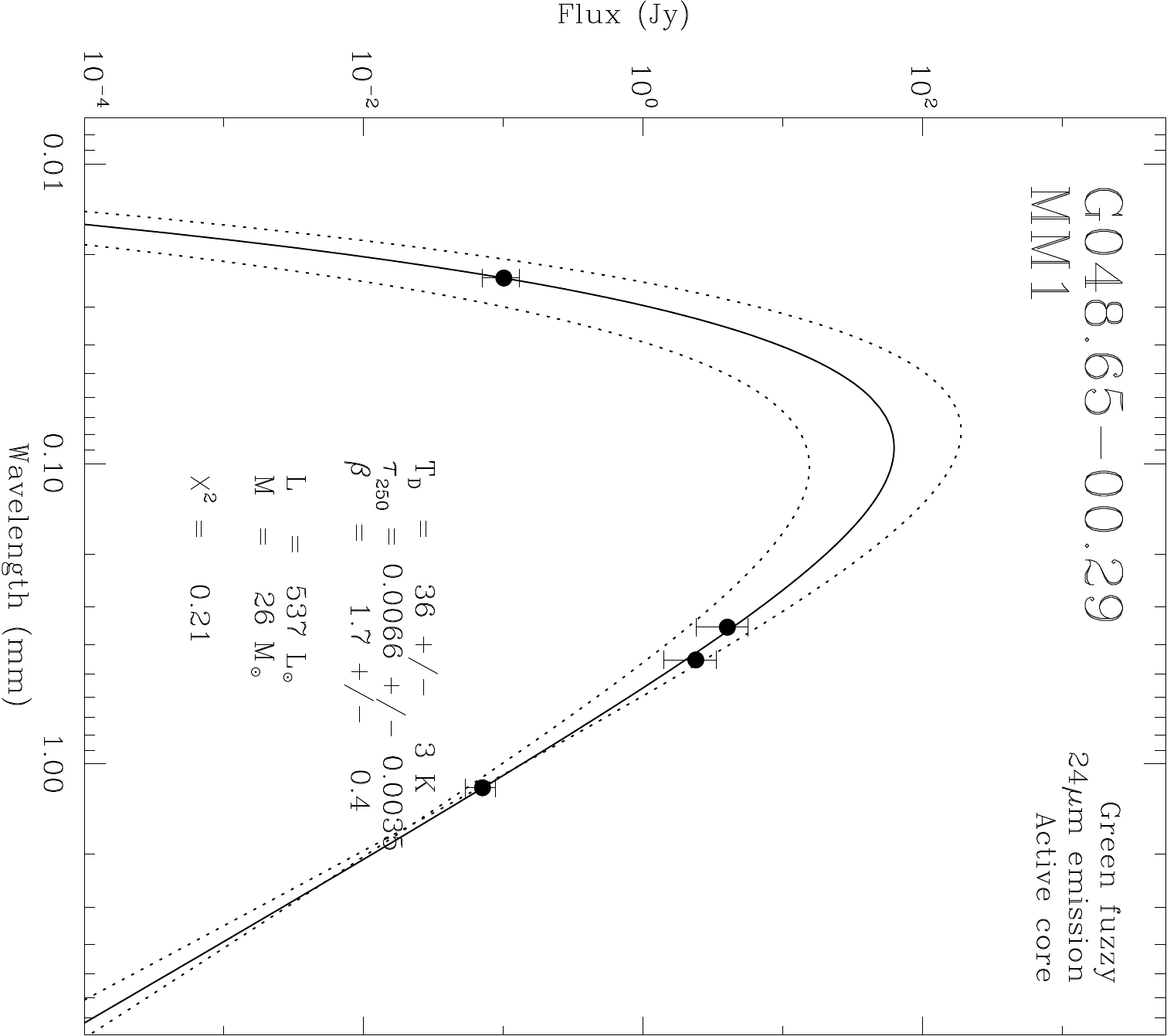}
\includegraphics[angle=90,width=0.5\textwidth]{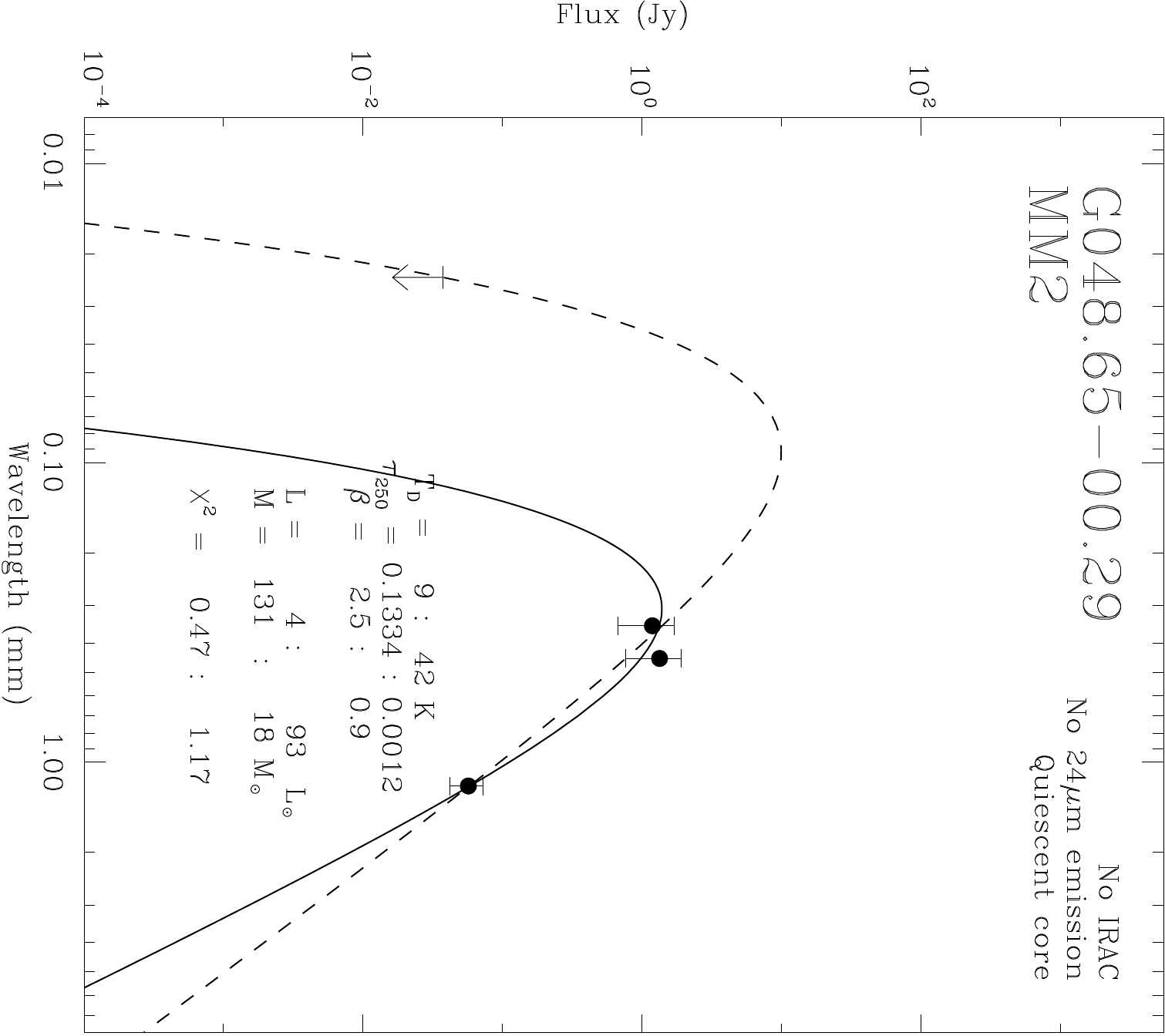}\\
\caption{\label{seds-44}\Spitzer\, 24\,\um\, image overlaid  
   with 1.2\,mm continuum emission for \irdcfortyfour\, (contour
   levels are 30, 60, 90\,mJy beam$^{-1}$). The lower panels show the broadband
   SEDs for cores within this IRDC.  The fluxes derived from the
   millimeter, sub-millimeter, and far-IR  continuum data are shown as filled
   circles (with the corresponding error bars), while the 24\,\um\, fluxes are shown as  either a filled circle (when included within the fit), an open circle (when excluded from the fit),  or as an upper limit arrow. For cores that have measured fluxes only in the millimeter/sub-millimeter regime (i.e.\, a limit at 24\,\um), we show the results from two fits: one using only the measured fluxes (solid line; lower limit), while the other includes the 24\,\um\, limit as a real data (dashed line; upper limit). In all other cases, the solid line is the best fit gray-body, while the dotted lines correspond to the functions determined using the errors for the T$_{D}$, $\tau$, and $\beta$ output from the fitting.  Labeled on each plot is the IRDC and core name,  classification, and the derived parameters.}
\end{figure}
\clearpage 
\begin{figure}
\begin{center}
\includegraphics[angle=0,width=0.4\textwidth]{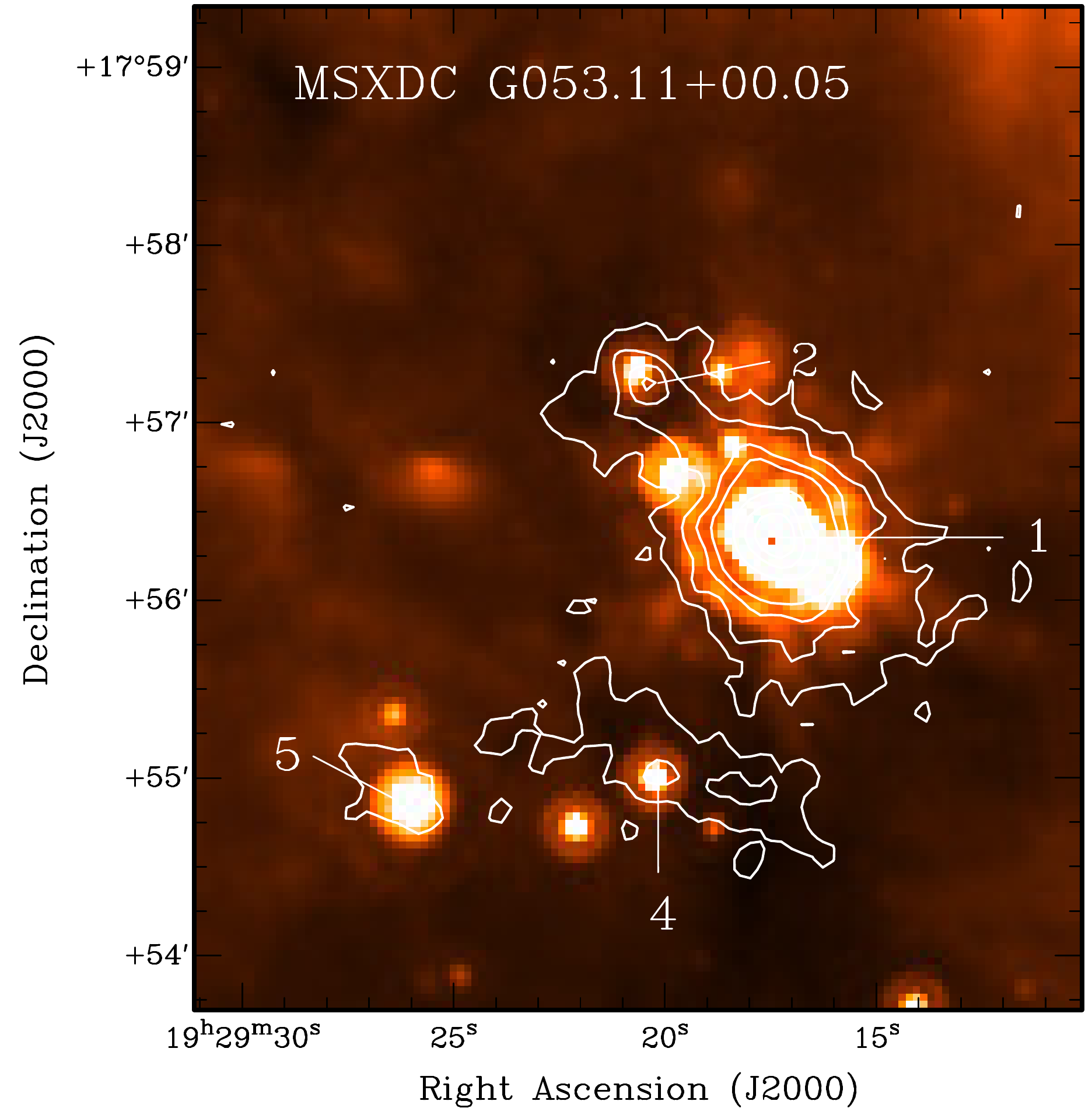}\\
\end{center}
\includegraphics[angle=90,width=0.5\textwidth]{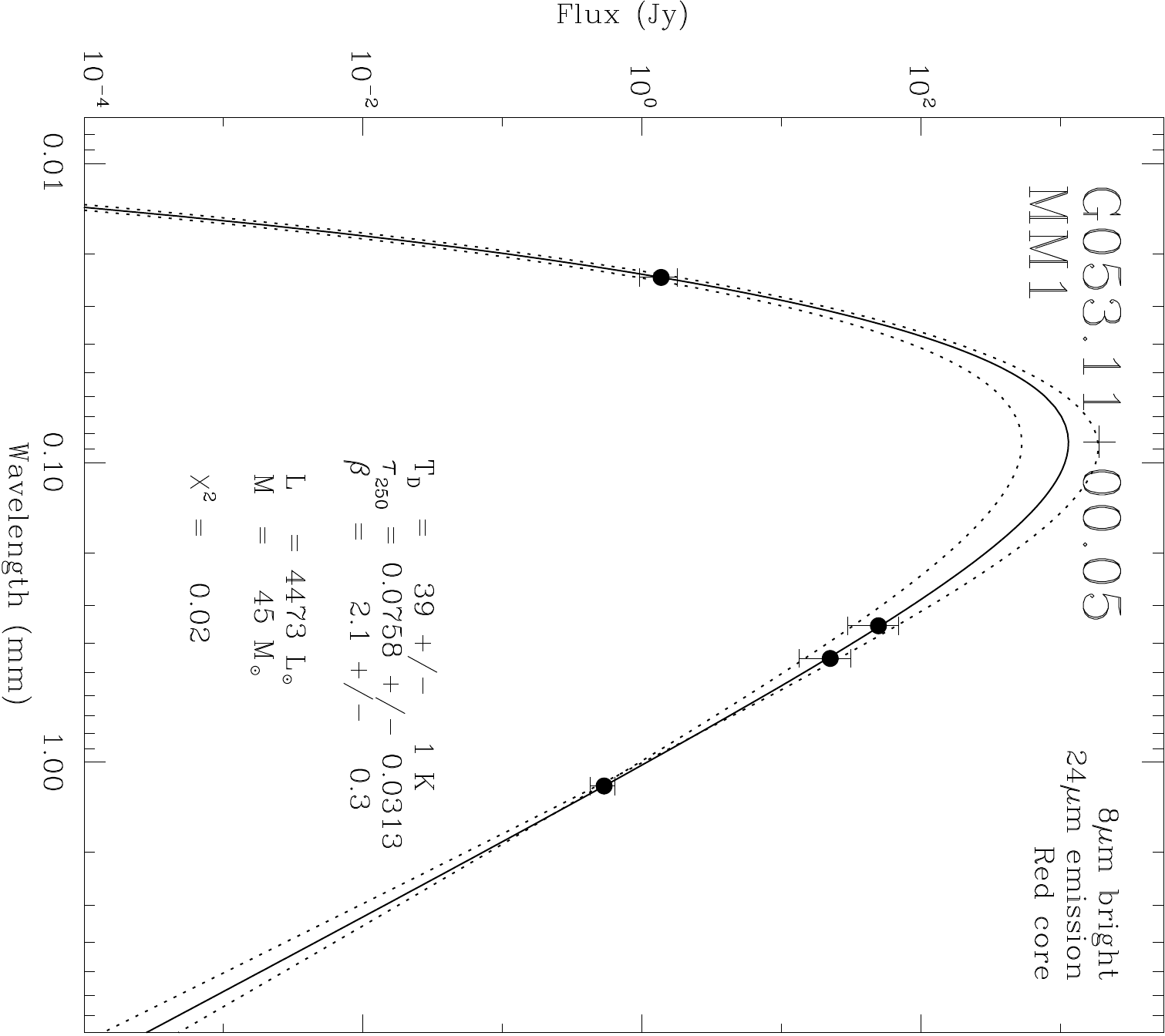}
\includegraphics[angle=90,width=0.5\textwidth]{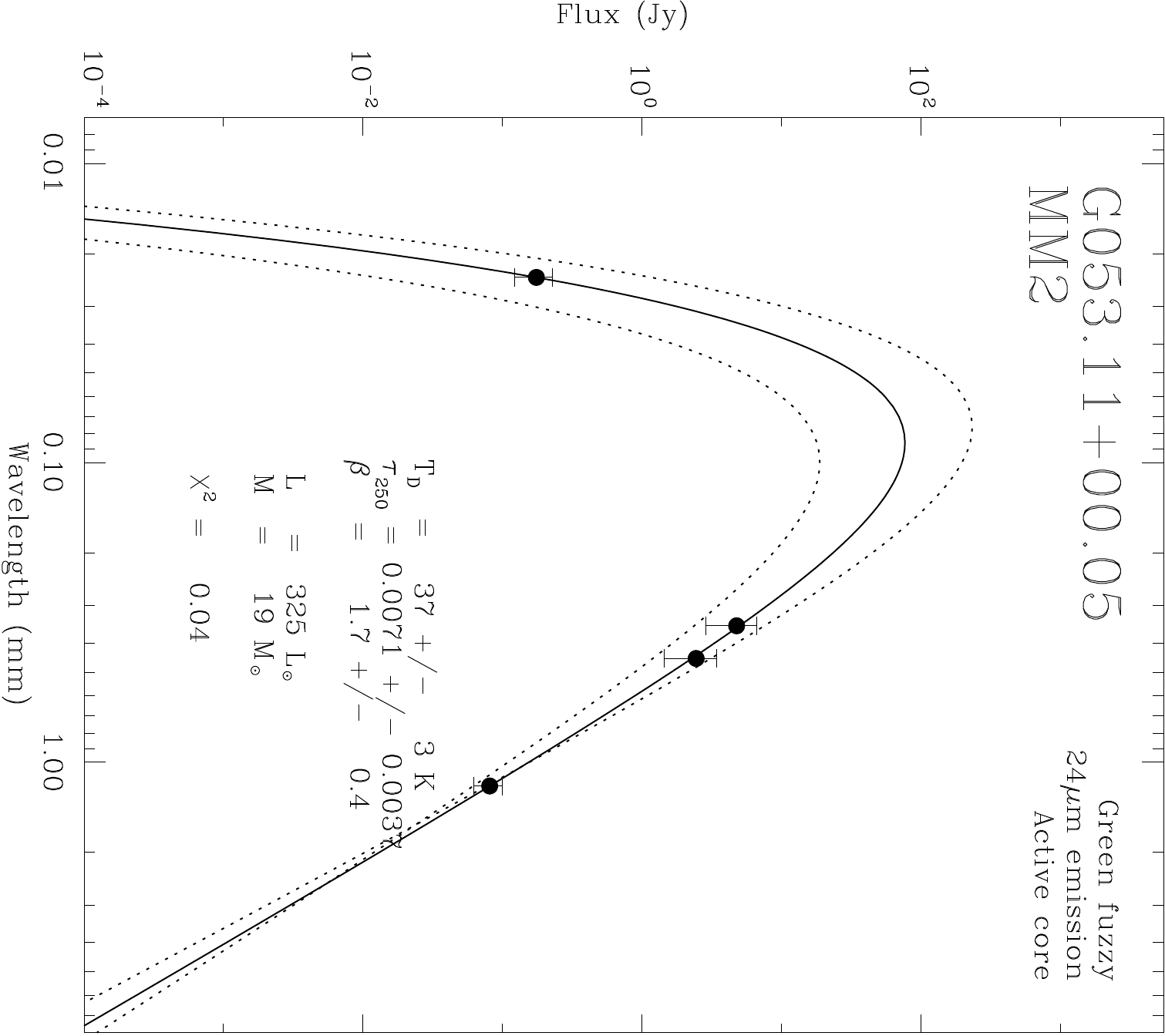}\\
\end{figure}
\clearpage 
\begin{figure}
\includegraphics[angle=90,width=0.5\textwidth]{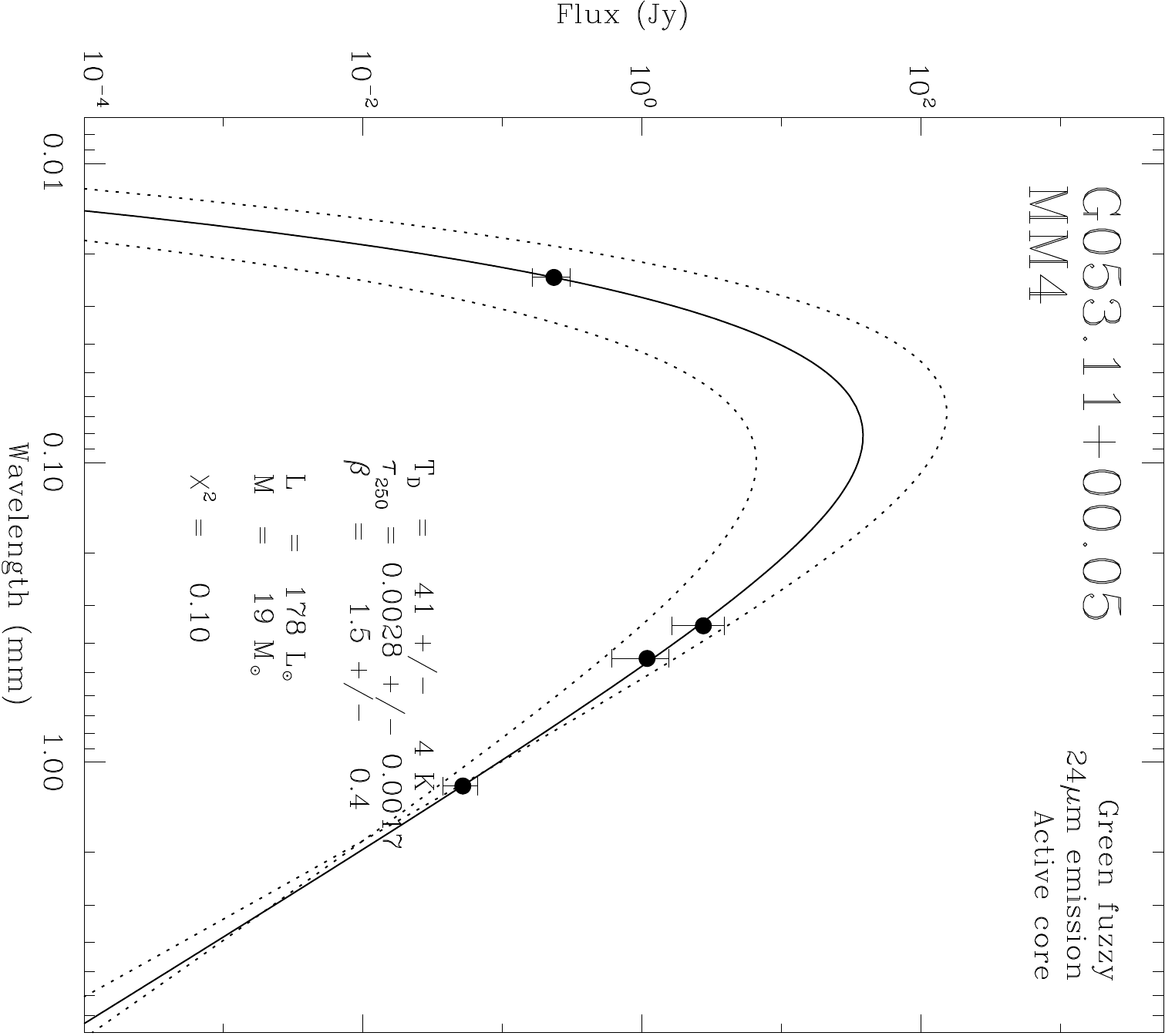}
\includegraphics[angle=90,width=0.5\textwidth]{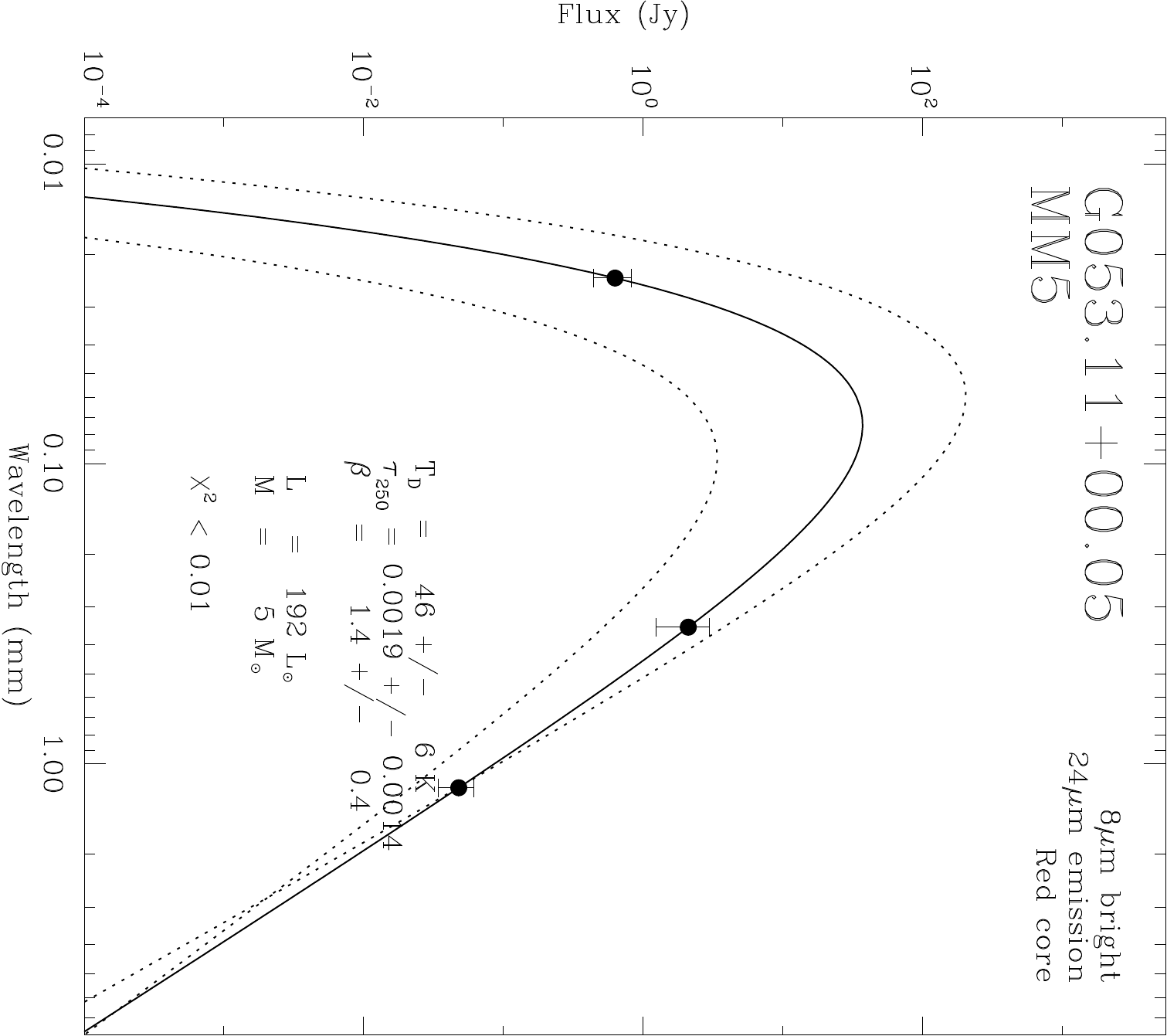}\\
\caption{\label{seds-18}\Spitzer\, 24\,\um\, image overlaid  
   with 1.2\,mm continuum emission for \irdceighteen\, (contour levels
   are 30, 60, 90, 120, 240, 360, 480\,mJy beam$^{-1}$). The lower panels show the broadband
   SEDs for cores within this IRDC.  The fluxes derived from the
   millimeter, sub-millimeter, and far-IR  continuum data are shown as filled
   circles (with the corresponding error bars), while the 24\,\um\, fluxes are shown as  either a filled circle (when included within the fit), an open circle (when excluded from the fit),  or as an upper limit arrow. For cores that have measured fluxes only in the millimeter/sub-millimeter regime (i.e.\, a limit at 24\,\um), we show the results from two fits: one using only the measured fluxes (solid line; lower limit), while the other includes the 24\,\um\, limit as a real data (dashed line; upper limit). In all other cases, the solid line is the best fit gray-body, while the dotted lines correspond to the functions determined using the errors for the T$_{D}$, $\tau$, and $\beta$ output from the fitting.  Labeled on each plot is the IRDC and core name,  classification, and the derived parameters.}
\end{figure}
\clearpage 
\begin{figure}
\begin{center}
\includegraphics[angle=0,width=0.6\textwidth]{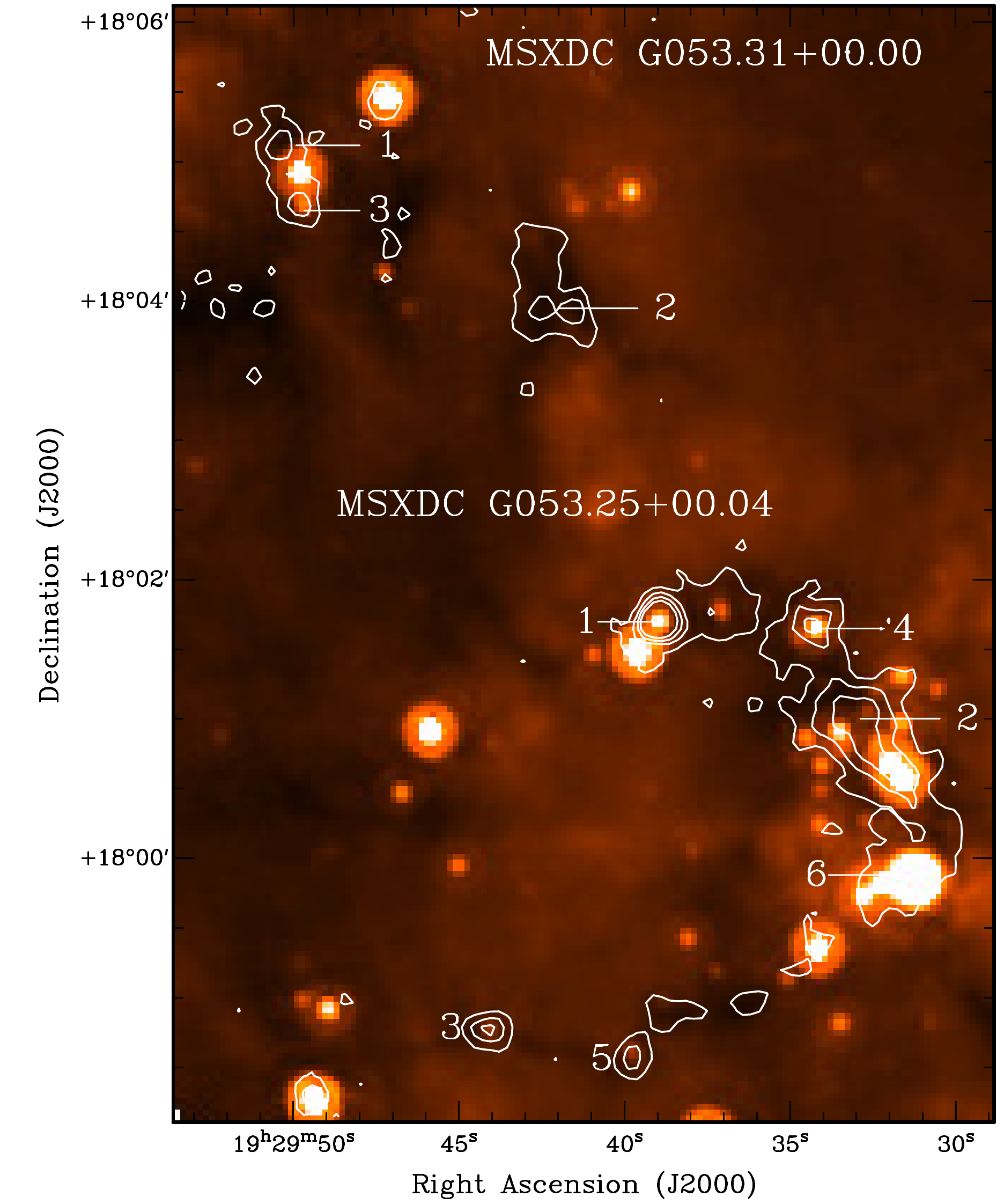}\\
\end{center}
\includegraphics[angle=90,width=0.5\textwidth]{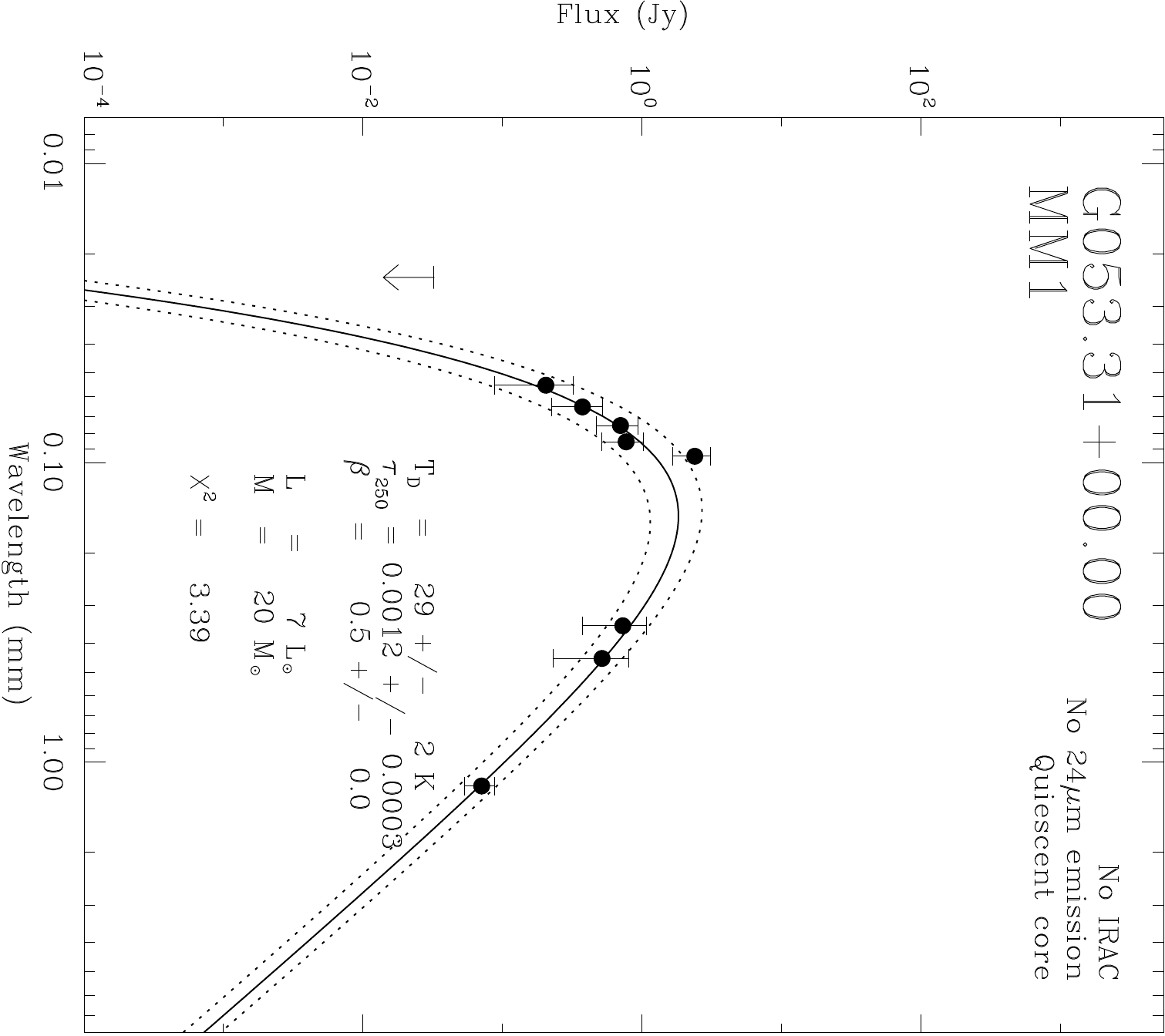}
\includegraphics[angle=90,width=0.5\textwidth]{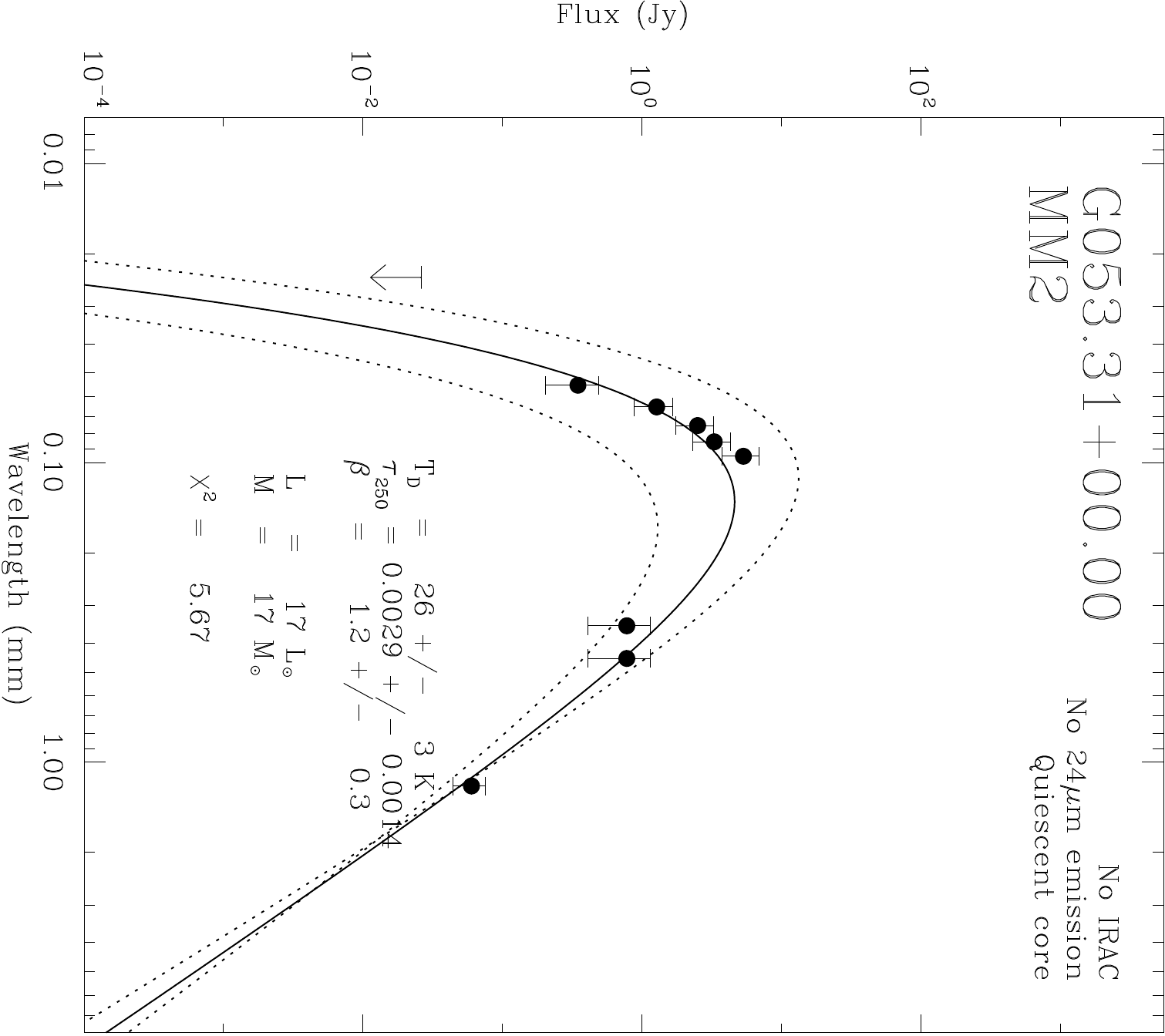}\\
\end{figure}
\clearpage 
\begin{figure}
\includegraphics[angle=90,width=0.5\textwidth]{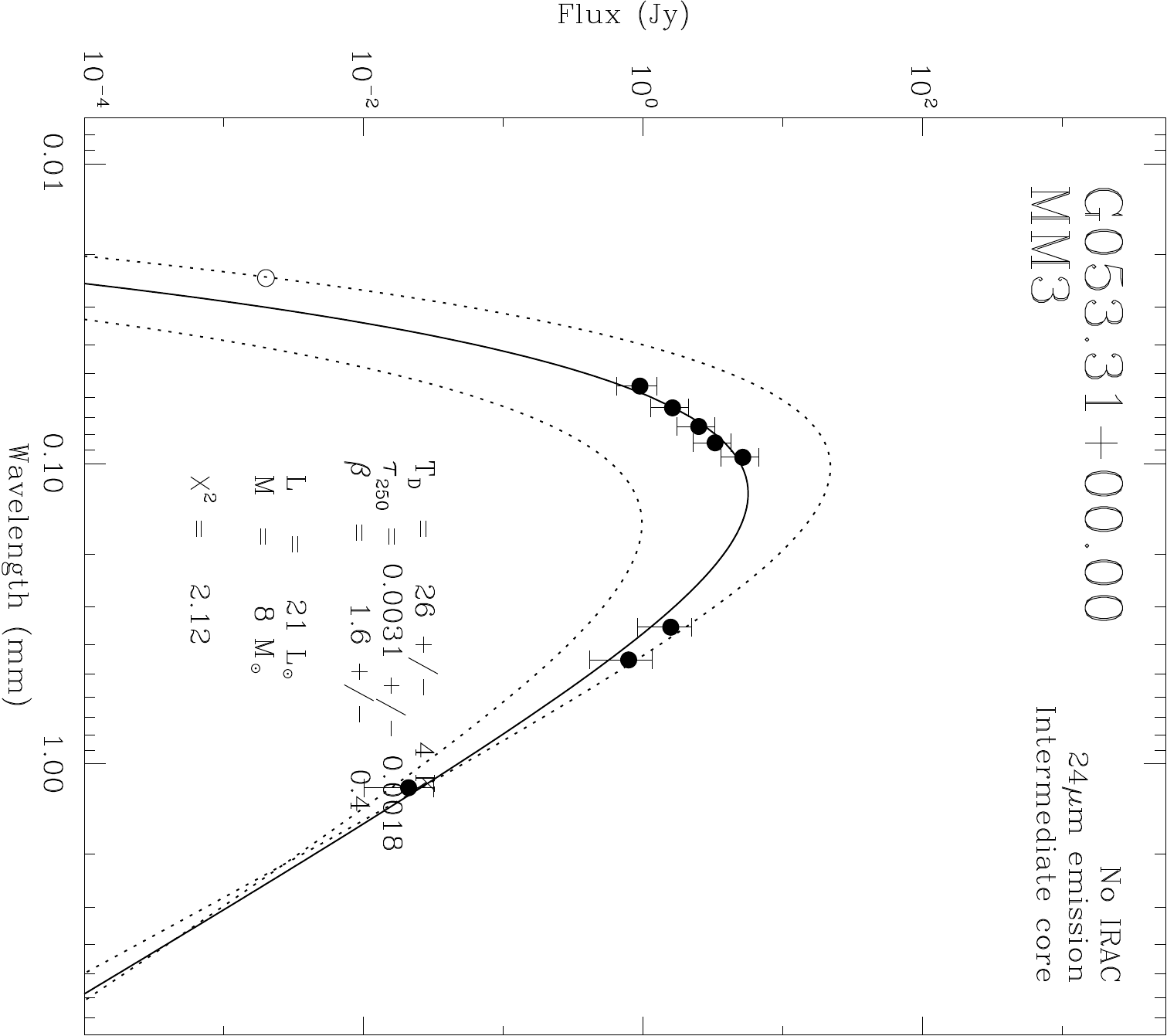}
\includegraphics[angle=90,width=0.5\textwidth]{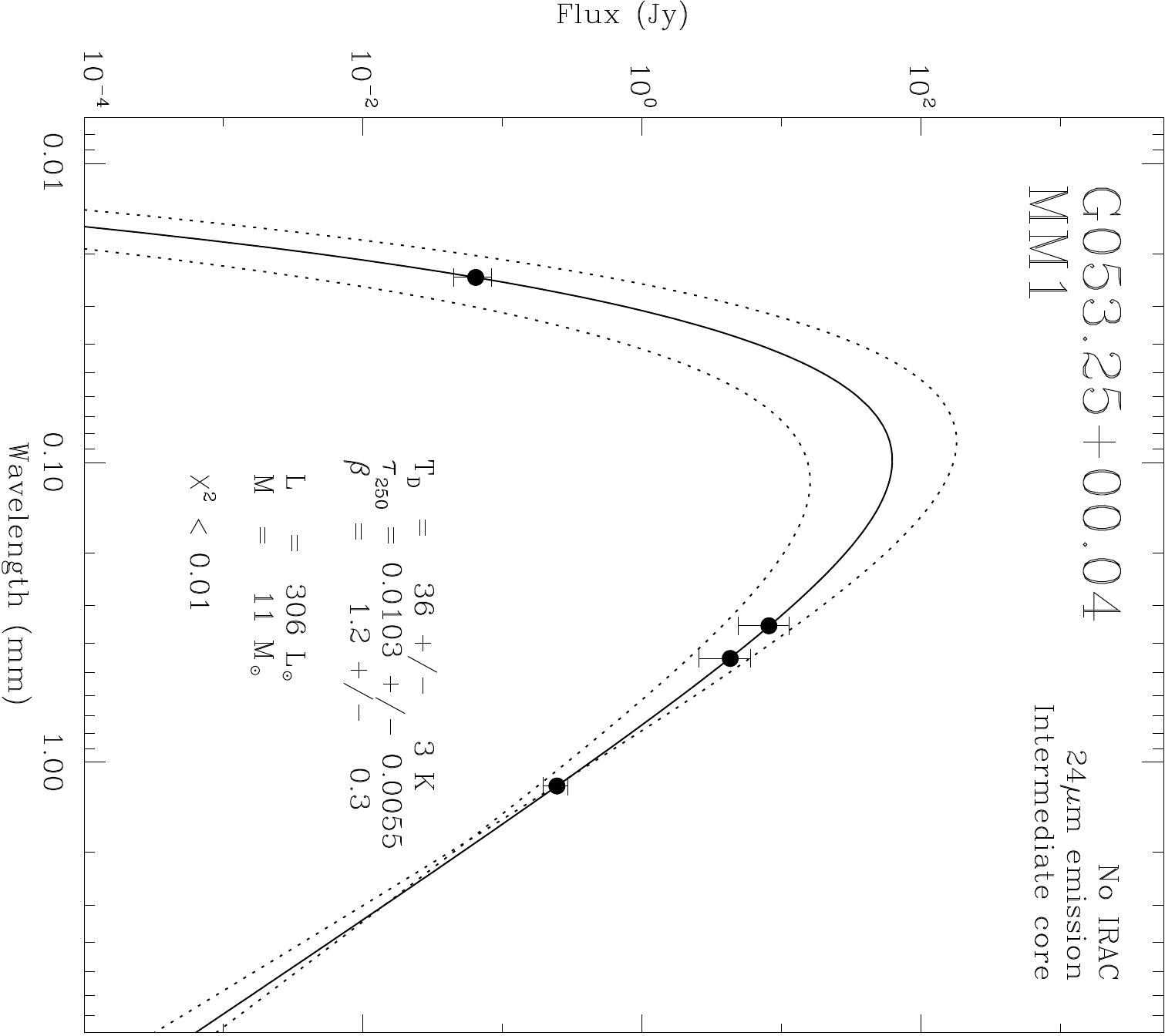}\\
\includegraphics[angle=90,width=0.5\textwidth]{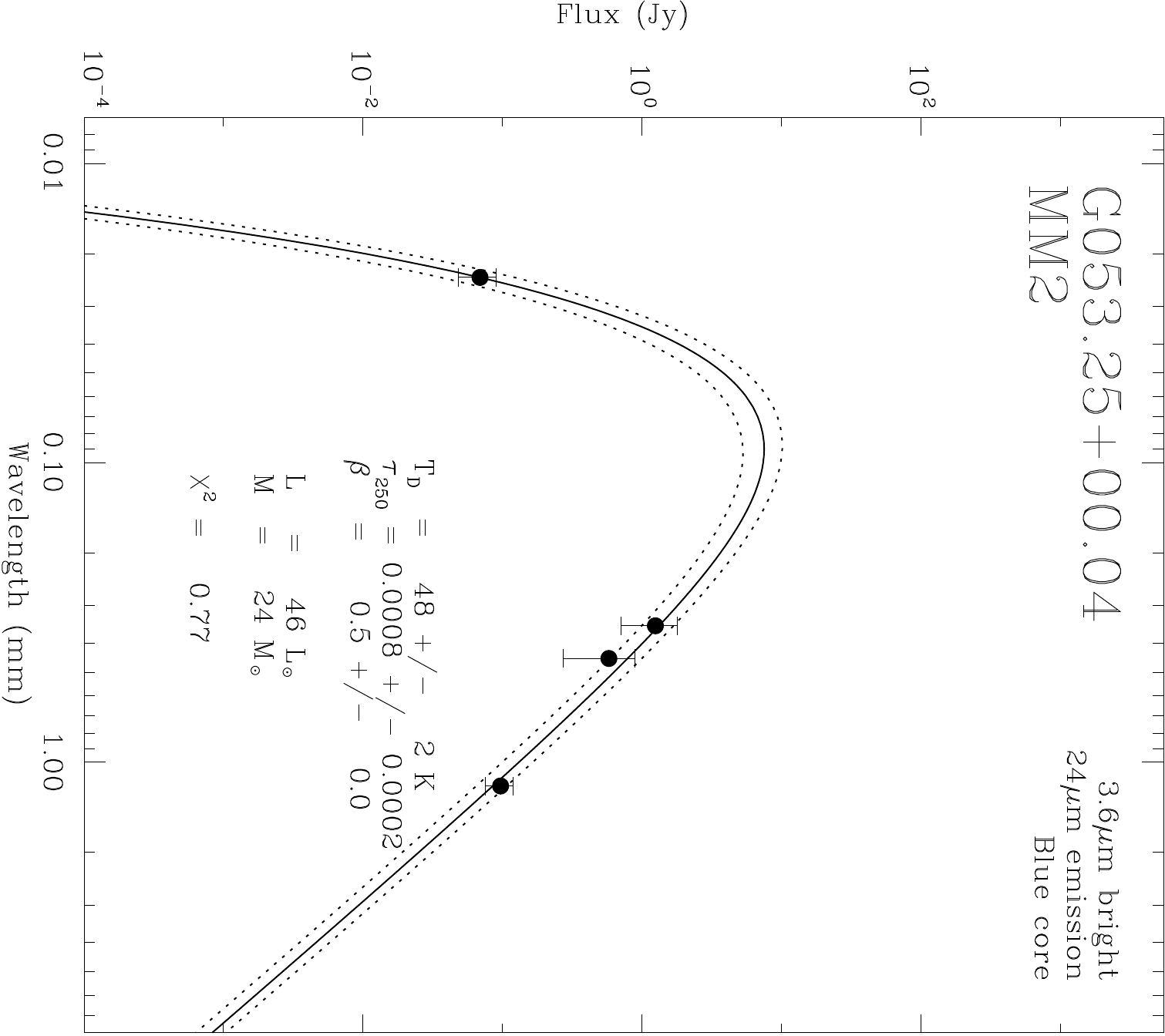}
\includegraphics[angle=90,width=0.5\textwidth]{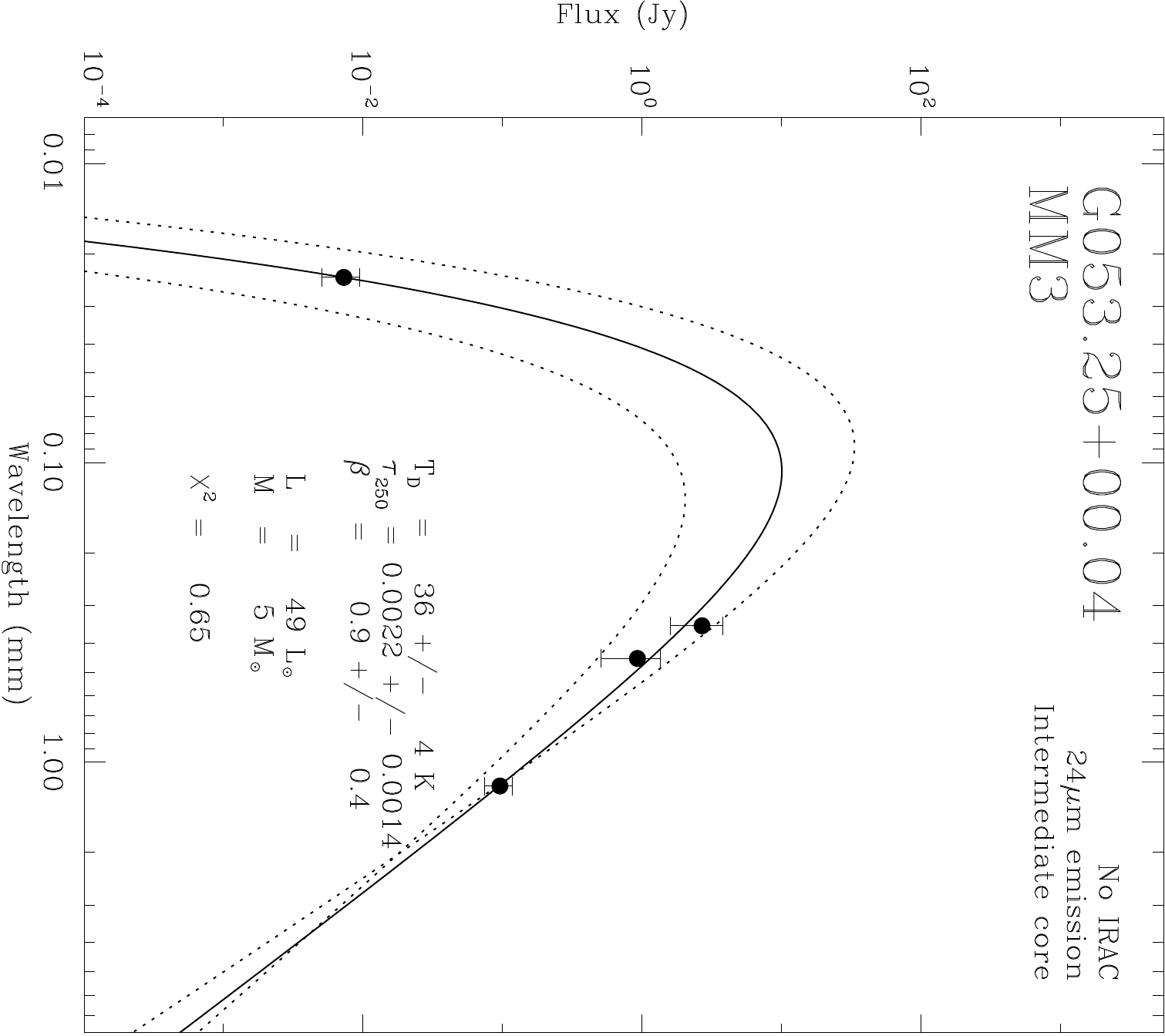}\\
\end{figure}
\clearpage 
\begin{figure}
\includegraphics[angle=90,width=0.5\textwidth]{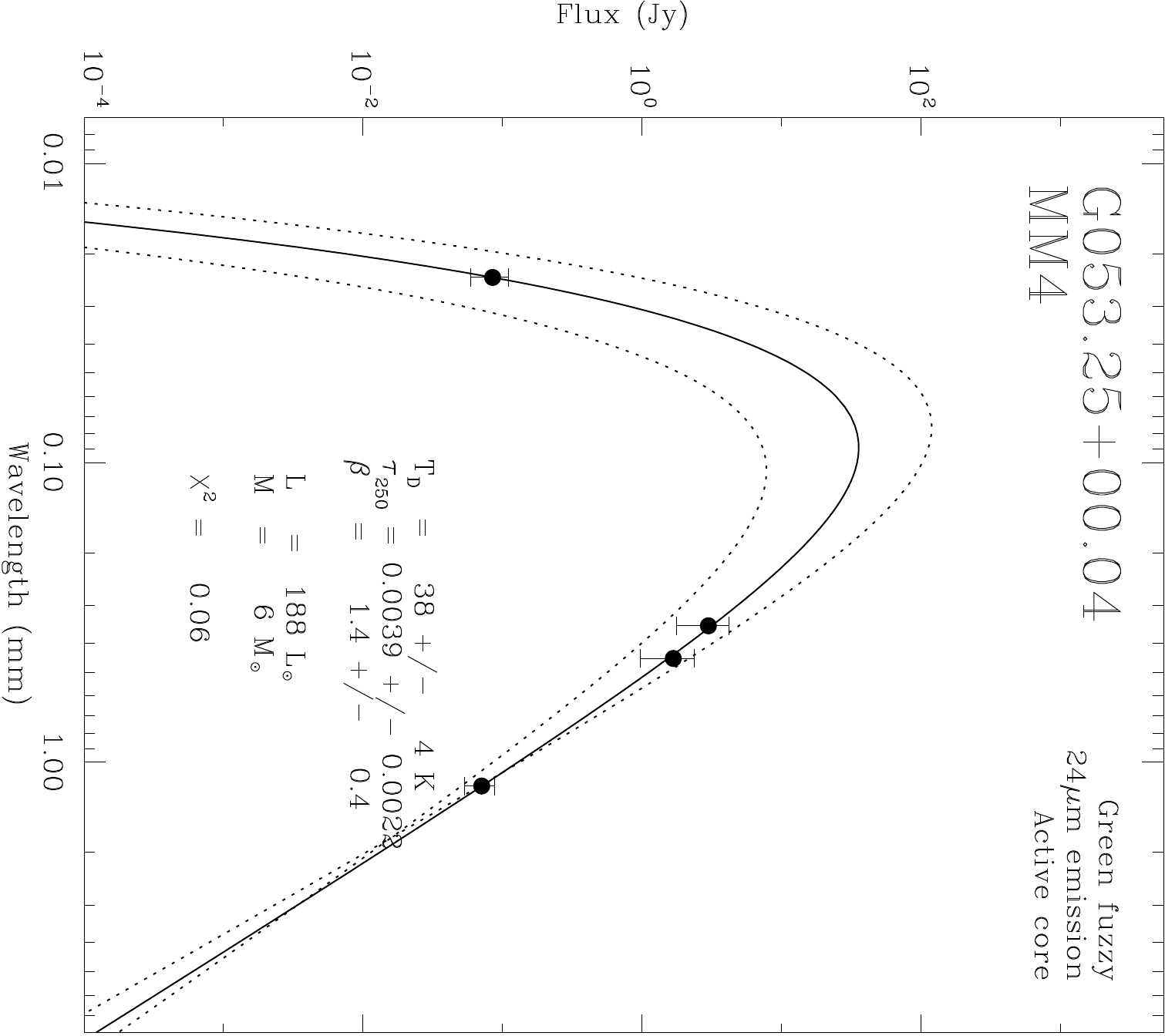}
\includegraphics[angle=90,width=0.5\textwidth]{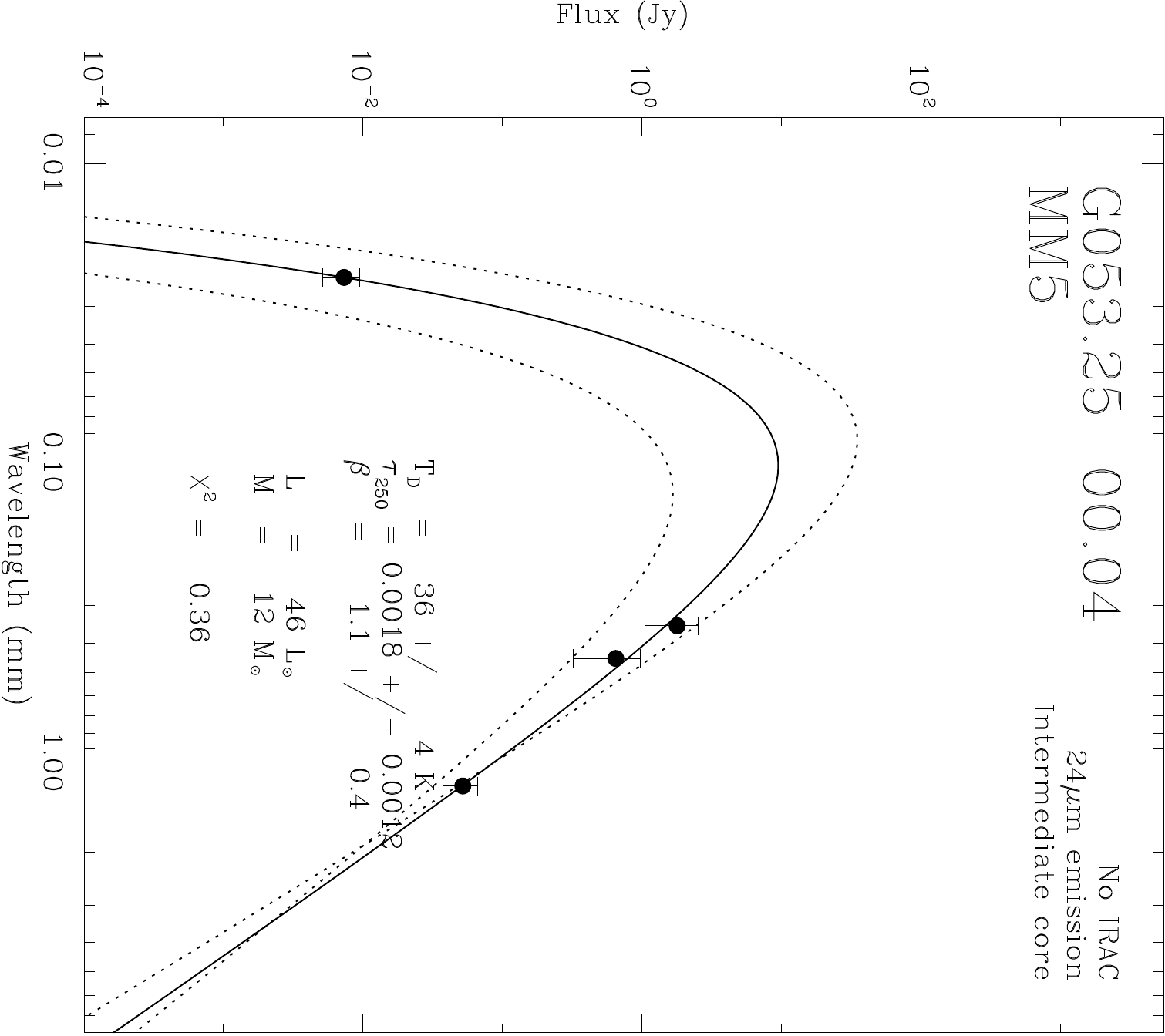}\\
\includegraphics[angle=90,width=0.5\textwidth]{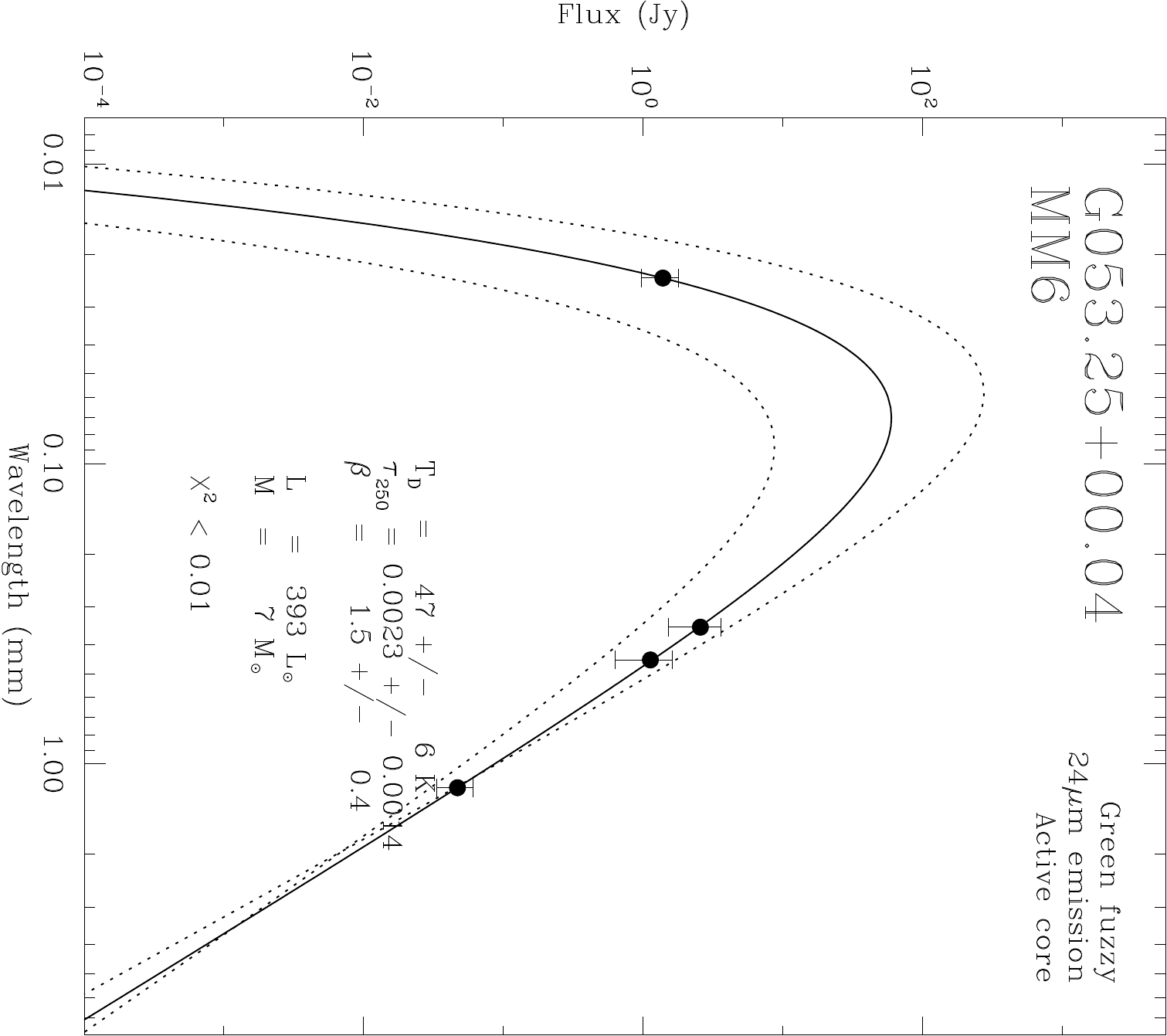}
\caption{\label{seds-34}\Spitzer\, 24\,\um\, image overlaid  
   with 1.2\,mm continuum emission for \irdcfiftyfive\, and
   \irdcthirtyfour\, (contour levels are 30, 60, 90, 120, 240\,mJy
   beam$^{-1}$). The lower panels show the broadband
   SEDs for cores within this IRDC.  The fluxes derived from the
   millimeter, sub-millimeter, and far-IR  continuum data are shown as filled
   circles (with the corresponding error bars), while the 24\,\um\, fluxes are shown as  either a filled circle (when included within the fit), an open circle (when excluded from the fit),  or as an upper limit arrow. For cores that have measured fluxes only in the millimeter/sub-millimeter regime (i.e.\, a limit at 24\,\um), we show the results from two fits: one using only the measured fluxes (solid line; lower limit), while the other includes the 24\,\um\, limit as a real data (dashed line; upper limit). In all other cases, the solid line is the best fit gray-body, while the dotted lines correspond to the functions determined using the errors for the T$_{D}$, $\tau$, and $\beta$ output from the fitting.  Labeled on each plot is the IRDC and core name,  classification, and the derived parameters.}
\end{figure}

  \renewcommand{\thetable}{A-\arabic{table}}
  \setcounter{table}{0}  

\clearpage
\begin{deluxetable}{llrcrrrrrrrrrr}
\tabletypesize{\tiny}
\tablecaption{Core coordinates, distances, and measured fluxes \label{cores}}
\tablewidth{0pt}
\tablehead{
\colhead{IRDC core} & \multicolumn{2}{c}{Coordinates}       & \colhead{D}     & \colhead{F$_{24}$} & \colhead{F$_{55}$}& \colhead{F$_{65}$}& \colhead{F$_{75}$}& \colhead{F$_{85}$}& \colhead{F$_{95}$} & \colhead{F$_{350}$} & \colhead{F$_{450}$} & \colhead{F$_{850}$} & \colhead{F$_{1.2}$}  \\
\colhead{}      & \colhead{(J2000)} & \colhead{(J2000)} &\colhead{(kpc)}  & \colhead{(Jy)}  & \colhead{(Jy)}  & \colhead{(Jy)}  & \colhead{(Jy)}  & \colhead{(Jy)}& \colhead{(Jy)}  & \colhead{(Jy)}  & \colhead{(Jy)}  & \colhead{(Jy)}  & \colhead{(Jy)}}
\startdata
\input{table1.tex}
\enddata
\end{deluxetable}


\clearpage
\begin{deluxetable}{lcccrrccrc}
\tabletypesize{\scriptsize}
\tablecaption{Properties of the cores derived from gray-body fits to their broadband SEDs.\label{core-properties-good}}
\tablewidth{0pt}
\tablehead{
\colhead{IRDC core} & \colhead{T$_{D}$}  & \colhead{$\tau_{250}$}   & \colhead{$\beta$} & \multicolumn{1}{c}{L}     & \colhead{M}     & \colhead{Log[N(\hh)]}            & \colhead{Log[n(\hh)]} &  \colhead{$\chi^{2}$} & \colhead{Classification\tablenotemark{a}} \\
\colhead{}  & \colhead{(K)}    & \colhead{}         & \colhead{}        & \colhead{(\Lsun)} & \colhead{(\Msun)} & \colhead{(\cms)} & \colhead{(\cmc)} & \colhead{}& \colhead{}}
\startdata
\input{table2.tex}
\enddata
\tablenotetext{a}{denotes the classification based on the IRAC colors and the presence or absence of 24\,\um\, emission.}
\end{deluxetable}


\clearpage
\begin{deluxetable}{lrccrcccc}
\tabletypesize{\scriptsize}
\tablecaption{Properties of the cores for which limits where estimated from the gray-body fits to their broadband SEDs.\label{core-properties-limits}}
\tablewidth{0pt}
\tablehead{
\colhead{IRDC core} &\colhead{T$_{D}$}  & \colhead{$\tau_{250}$}   & \colhead{$\beta$} & \multicolumn{1}{c}{L}     & \colhead{M}     &  \colhead{$\chi^{2}$} & \colhead{Classification\tablenotemark{a}} \\
\colhead{}  & \colhead{(K)}    & \colhead{}         & \colhead{}        & \colhead{(\Lsun)} & \colhead{(\Msun)}  & \colhead{}& \colhead{}}
\startdata
\input{table3.tex}
\enddata
\tablenotetext{a}{denotes the classification based on the IRAC colors and the presence or absence of 24\,\um\, emission.}
\end{deluxetable}


\end{document}

%% file: table1.tex
        G015.05+00.07 MM1  &   18:17:50.4  &  $-$15:53:38  &  3.1    & $<$   0.05  &        ...  &        ...  &        ...  &        ...  &        ...   &      3.92  &       3.96  &        ...  &       0.12 \\ 
        G015.05+00.07 MM2  &   18:17:40.0  &  $-$15:48:55  &  3.1    & $<$   0.05  &        ...  &        ...  &        ...  &        ...  &        ...   &      0.29  &        ...  &        ...  &       0.05 \\ 
        G015.05+00.07 MM3  &   18:17:42.4  &  $-$15:47:03  &  3.1    & $<$   0.06  &        ...  &        ...  &        ...  &        ...  &        ...   &       ...  &        ...  &        ...  &       0.04 \\ 
        G015.05+00.07 MM4  &   18:17:32.0  &  $-$15:46:35  &  3.1    & $<$   0.07  &        ...  &        ...  &        ...  &        ...  &        ...   &      0.78  &        ...  &        ...  &       0.04 \\ 
        G015.05+00.07 MM5  &   18:17:40.2  &  $-$15:49:47  &  3.1    & $<$   0.06  &        ...  &        ...  &        ...  &        ...  &        ...   &      0.56  &        ...  &        ...  &       0.03 \\ 
        G015.31$-$00.16 MM1  &   18:18:56.4  &  $-$15:45:00  &  1.9    &       0.19  &       4.83  &       5.72  &       9.62  &      14.11  &      17.00   &      1.64  &        ...  &        ...  &       0.04 \\ 
        G015.31$-$00.16 MM2  &   18:18:50.4  &  $-$15:43:19  &  1.9    &       0.02  &       0.52  &       0.75  &       3.09  &       3.86  &       8.91   &      0.80  &        ...  &        ...  &       0.04 \\ 
        G015.31$-$00.16 MM3  &   18:18:45.3  &  $-$15:41:58  &  1.9    & $<$   0.06  &        ...  &        ...  &        ...  &        ...  &        ...   &       ...  &        ...  &        ...  &       0.04 \\

%% file: table2.tex
        G015.31$-$00.16 MM1  &  29  &  0.004  &   1.6    &       56  &     7  & 22.41  &  5.69  &       1.91  &                 R \\
        G015.31$-$00.16 MM2  &  23  &  0.003  &   1.5    &       16  &    21  & 22.88  &  6.16  &       9.72  &                 I \\
        G018.82$-$00.28 MM1  &  26  &  0.091  &   2.0    &     4526  &   495  & 23.51  &  6.42  &       1.44  &                 I \\
        G018.82$-$00.28 MM2  &  38  &  0.020  &   1.9    &     8990  &    87  & 22.76  &  5.66  &       1.65  &                 R \\
        G018.82$-$00.28 MM3  &  34  &  0.034  &   2.4    &    11356  &   103  & 22.83  &  5.73  &       0.61  &                 A \\
        G018.82$-$00.28 MM4  &  17  &  0.031  &   2.1    &      198  &   228  & 23.18  &  6.08  &       4.94  &                 I \\
        G019.27+00.07 MM1  &  33  &  0.027  &   2.1    &     1998  &    52  & 23.09  &  6.27  &       1.50  &                 B \\
        G019.27+00.07 MM2  &  30  &  0.017  &   2.1    &      760  &    60  & 23.15  &  6.33  &       1.27  &                 A \\

%% file: table3.tex
        G015.05+00.07 MM1  &             11 -  36  &       0.190 -  0.006  &          2.5 -   1.4    &        22 -      515  &          225 -    50  &        0.68 -   1.27  &                 Q \\
        G015.05+00.07 MM2  &             10 -  51  &       0.012 -  0.000  &          1.2 -   0.5    &         1 -       57  &          209 -    27  &    $<$ 0.01 -   2.45  &                 R \\
        G015.05+00.07 MM4  &             14 -  47  &       0.010 -  0.001  &          1.6 -   0.8    &         6 -       96  &           61 -    15  &  $<$ 0.01 - $<$ 0.01  &                 Q \\
        G015.05+00.07 MM5  &             13 -  49  &       0.010 -  0.000  &          1.6 -   0.6    &         4 -       64  &           47 -    10  &  $<$ 0.01 - $<$ 0.01  &                 Q \\
        G015.31$-$00.16 MM4  &             13 -  48  &       0.010 -  0.000  &          1.8 -   0.8    &         2 -       27  &           13 -     3  &  $<$ 0.01 - $<$ 0.01  &                 Q \\
        G015.31$-$00.16 MM5  &             15 -  41  &       0.010 -  0.001  &          2.1 -   1.4    &         4 -       60  &           24 -     8  &  $<$ 0.01 - $<$ 0.01  &                 Q \\
        G018.82$-$00.28 MM5  &             18 -  41  &       0.010 -  0.002  &          1.6 -   1.1    &        42 -      432  &          169 -    67  &        0.21 -   0.09  &                 Q \\
        G022.35+00.41 MM3  &             19 -  38  &       0.010 -  0.003  &          1.6 -   1.4    &        38 -      457  &          101 -    48  &  $<$ 0.01 - $<$ 0.01  &                 B \\